\documentclass[twocolumn]{aastex62}

\turnoffediting

\usepackage{miller}
\usepackage{epstopdf}
\usepackage{natbib}
\usepackage{amsmath}

\DeclareMathOperator{\sinc}{sinc}

\received{August 22, 2018}
\accepted{January 07, 2019}
\submitjournal{PASP}

\shorttitle{APOGEE Spectrographs}
\shortauthors{Wilson et al.}

\renewcommand{\figurename}{Figure}

\begin{document}


\title{The Apache Point Observatory Galactic Evolution Experiment (APOGEE) Spectrographs}


\correspondingauthor{John C. Wilson}
\email{jcw6z@virginia.edu}

\author{J.~C.~Wilson}
\affiliation{Astronomy Department, University of Virginia,
Charlottesville, VA 22901, USA}

\author{F.~R.~Hearty}
\affiliation{Department of Astronomy and Astrophysics, The Pennsylvania State University, University Park, PA 16802, USA}

\author{M.~F.~Skrutskie}
\affiliation{Astronomy Department, University of Virginia,
Charlottesville, VA 22901, USA}

\author{S.~R.~Majewski}
\affiliation{Astronomy Department, University of Virginia,
Charlottesville, VA 22901, USA}

\author{J.~A.~Holtzman}
\affiliation{Dept of Astronomy, New Mexico State Univ, P. O. Box 30001, MSC 4500, Las Cruces, NM 88003, USA}

\author{D.~Eisenstein}
\affiliation{Harvard-Smithsonian Center for Astrophysics, 60 Garden Street, MS 20, Cambridge, MA 02138, USA}

\author{J.~Gunn}
\affiliation{Department of Astrophysical Sciences, Princeton University, Princeton, NJ 08544, USA}

\author{B.~Blank}
\affiliation{PulseRay, 4583 State Route 414, Beaver Dams, NY 14812, USA}

\author{C.~Henderson}
\affiliation{PulseRay, 4583 State Route 414, Beaver Dams, NY 14812, USA}

\author{S.~Smee}
\affiliation{Department of Physics and Astronomy, Johns Hopkins University, Baltimore, MD 21218, USA}

\author{M.~Nelson}
\affiliation{Astronomy Department, University of Virginia,
Charlottesville, VA 22901, USA}

\author{D.~Nidever}
\affiliation{National Optical Astronomy Observatory, 950 North Cherry Avenue, Tucson, AZ 85719, USA}

\author{J.~Arns}
\affiliation{Kaiser Optical Systems, Inc., 371 Parkland Plaza, Ann Arbor, MI 48103, USA}

\author{R.~Barkhouser}
\affiliation{Department of Physics and Astronomy, Johns Hopkins University, Baltimore, MD 21218, USA}

\author{J.~Barr}
\affiliation{Astronomy Department, University of Virginia,
Charlottesville, VA 22901, USA}

\author{S.~Beland}
\affiliation{Laboratory for Atmospheric and Space Physics, University of Colorado, 3665 Discovery Dr., Boulder, CO 80303, USA}

\author{M.~A.~Bershady}
\affiliation{Department of Astronomy, University of Wisconsin-Madison, 475 N. Charter St., Madison, WI 53726, USA}
\affiliation{South African Astronomical Observatory, P.O. Box 9, Observatory 7935, Cape Town, South Africa}

\author{M.~R.~Blanton}
\affiliation{Center for Cosmology and Particle Physics, Department of Physics, New York University, 726 Broadway Rm. 1005, New York, NY 10003, USA}

\author{S.~Brunner}
\affiliation{Astronomy Department, University of Virginia,
Charlottesville, VA 22901, USA}

\author{A.~Burton}
\affiliation{Astronomy Department, University of Virginia,
Charlottesville, VA 22901, USA}

\author{L.~Carey}
\affiliation{Department of Astronomy, University of Washington, Box 351580, Seattle, WA 98195, USA}

\author{M.~Carr}
\affiliation{Department of Astrophysical Sciences, Princeton University, Princeton, NJ 08544, USA}

\author{J.~P.~Colque}
\affiliation{Centro de Astronom\'{i}a (CITEVA), Universidad de Antofagasta, Avenida Angamos 601, Antofagasta, Chile}

\author{J.~Crane}
\affiliation{Observatories of the Carnegie Institution for Science, 813 Santa Barbara Street, Pasadena, CA 91101, USA}

\author{G.~J.~Damke}
\affiliation{AURA Observatory in Chile, Cisternas 1500, La Serena, Chile}
\affiliation{Centro Multidisciplinario de Ciencia y Tecnolog\'ia, Universidad de La Serena, Cisternas 1200, La Serena, Chile}

\author{J.~W.~Davidson~Jr.}
\affiliation{Astronomy Department, University of Virginia,
Charlottesville, VA 22901, USA}

\author{J.~Dean}
\affiliation{Astronomy Department, University of Virginia,
Charlottesville, VA 22901, USA}

\author{F.~Di~Mille}
\affiliation{Las Campanas Observatory, Colina El Pino Casilla 601 La Serena, Chile}

\author{K.~W.~Don}
\affiliation{Steward Observatory, University of Arizona, 933 N. Cherry Ave, Tucson, AZ 85721, USA}

\author{G.~Ebelke}
\affiliation{Astronomy Department, University of Virginia,
Charlottesville, VA 22901, USA}

\author{M.~Evans}
\affiliation{Department of Astronomy, University of Washington, Box 351580, Seattle, WA 98195, USA}

\author{G.~Fitzgerald}
\affiliation{New England Optical Systems, Inc., 237 Cedar Hill St., Marlborough, MA 01752, USA}

\author{B.~Gillespie}
\affiliation{Apache Point Observatory, P.O. Box 59, Sunspot, NM 88349, USA}

\author{M.~Hall}
\affiliation{Astronomy Department, University of Virginia,
Charlottesville, VA 22901, USA}

\author{A.~Harding}
\affiliation{Department of Physics and Astronomy, Johns Hopkins University, Baltimore, MD 21218, USA}

\author{P.~Harding}
\affiliation{Department of Astronomy, Case Western Reserve University, Cleveland, OH 44106, USA}

\author{R.~Hammond}
\affiliation{Department of Physics and Astronomy, Johns Hopkins University, Baltimore, MD 21218, USA}

\author{D.~Hancock}
\affiliation{Astronomy Department, University of Virginia,
Charlottesville, VA 22901, USA}

\author{C.~Harrison}
\affiliation{C Technologies, 757 Route 202/206, Bridgewater, NJ 08807, USA}

\author{S.~Hope}
\affiliation{Department of Physics and Astronomy, Johns Hopkins University, Baltimore, MD 21218, USA}

\author{T.~Horne}
\affiliation{Meinel 733, College of Optical Sciences, Univ of Arizona, 1630 East Univ Blvd, Tucson, AZ 85721, USA}

\author{J.~Karakla}
\affiliation{Department of Physics and Astronomy, Johns Hopkins University, Baltimore, MD 21218, USA}

\author{C.~Lam}
\affiliation{Astronomy Department, University of Virginia,
Charlottesville, VA 22901, USA}

\author{F.~Leger}
\affiliation{Department of Astronomy, University of Washington, Box 351580, Seattle, WA 98195, USA}

\author{N.~MacDonald}
\affiliation{University of California Observatories, UC Santa Cruz, 1156 High St., Santa Cruz, CA 95064, USA}

\author{P.~Maseman}
\affiliation{Steward Observatory, University of Arizona, 933 N. Cherry Ave, Tucson, AZ 85721, USA}

\author{J.~Matsunari}
\affiliation{THK America, Inc., 200 East Commerce Dr., Schaumburg, IL 60173, USA}

\author{S.~Melton}
\affiliation{US Conec, Ltd., PO Box 2306, 1555 4th Ave SE, Hickory, NC 28602, USA}

\author{T.~Mitcheltree}
\affiliation{US Conec, Ltd., PO Box 2306, 1555 4th Ave SE, Hickory, NC 28602, USA}

\author{T.~O'Brien}
\affiliation{Department of Astronomy, Ohio State University, Columbus, OH 43210, USA}

\author{R.~W.~O'Connell}
\affiliation{Astronomy Department, University of Virginia,
Charlottesville, VA 22901, USA}

\author{A.~Patten}
\affiliation{Department of Astronomy, University of Washington, Box 351580, Seattle, WA 98195, USA}

\author{W.~Richardson}
\affiliation{Astronomy Department, University of Virginia,
Charlottesville, VA 22901, USA}

\author{G.~Rieke}
\affiliation{Steward Observatory, University of Arizona, 933 N. Cherry Ave, Tucson, AZ 85721, USA}

\author{M.~Rieke}
\affiliation{Steward Observatory, University of Arizona, 933 N. Cherry Ave, Tucson, AZ 85721, USA}

\author{A.~Roman-Lopes}
\affiliation{Departamento de F\'{i}sica, Facultad de Ciencias, Universidad de La Serena, Cisternas 1200, La Serena, Chile}

\author{R.~P.~Schiavon}
\affiliation{Astrophysics Research Institute, Liverpool John Moores University, 146 Brownlow Hill, Liverpool, L3 5RF, UK}

\author{J.~S.~Sobeck}
\affiliation{Department of Astronomy, University of Washington, Box 351580, Seattle, WA 98195, USA}

\author{T.~Stolberg}
\affiliation{New England Optical Systems, Inc., 237 Cedar Hill St., Marlborough, MA 01752, USA}

\author{R.~Stoll}
\affiliation{C Technologies, 757 Route 202/206, Bridgewater, NJ 08807, USA}

\author{M.~Tembe}
\affiliation{Astronomy Department, University of Virginia,
Charlottesville, VA 22901, USA}

\author{J.~D.~Trujillo}
\affiliation{Department of Astronomy, University of Washington, Box 351580, Seattle, WA 98195, USA}

\author{A.~Uomoto}
\affiliation{Observatories of the Carnegie Institution for Science, 813 Santa Barbara Street, Pasadena, CA 91101, USA}

\author{M.~Vernieri}
\affiliation{C Technologies, 757 Route 202/206, Bridgewater, NJ 08807, USA}

\author{E.~Walker}
\affiliation{Astronomy Department, University of Virginia,
Charlottesville, VA 22901, USA}

\author{D.~H.~Weinberg}
\affiliation{Department of Astronomy, Ohio State University, Columbus, OH 43210, USA}

\author{E.~Young}
\affiliation{USRA, NASA Ames Research Center, Moffett Field, CA 94035, USA}

\author{B.~Anthony-Brumfield}
\affiliation{Astronomy Department, University of Virginia,
Charlottesville, VA 22901, USA}

\author{D.~Bizyaev}
\affiliation{Apache Point Observatory, P.O. Box 59, Sunspot, NM 88349, USA}
\affiliation{Sternberg Astronomical Institute, Moscow State University, Moscow, Russia}

\author{B.~Breslauer}
\affiliation{Astronomy Department, University of Virginia,
Charlottesville, VA 22901, USA}

\author{N.~De~Lee}
\affiliation{Department of Physics, Geology, and Engineering Tech, Northern Kentucky University, Highland Heights, KY 41099, USA}
\affiliation{Vanderbilt University, Department of Physics \& Astronomy, 6301 Stevenson Center Ln., Nashville, TN 37235, USA}

\author{J.~Downey}
\affiliation{Apache Point Observatory, P.O. Box 59, Sunspot, NM 88349, USA}

\author{S.~Halverson}
\altaffiliation{NASA Sagan Postdoctoral Fellow}
\affiliation{MIT Kavli Institute for Astrophysics and Space Research, 77 Massachusetts Ave., 37-241, Cambridge, MA 02139, USA}

\author{J.~Huehnerhoff}
\affiliation{Hindsight Imaging, Inc., 1 Harvard St., Suite 302, Brookline, MA 02445, USA}

\author{M.~Klaene}
\affiliation{Apache Point Observatory, P.O. Box 59, Sunspot, NM 88349, USA}

\author{E.~Leon}
\affiliation{Apache Point Observatory, P.O. Box 59, Sunspot, NM 88349, USA}

\author{D.~Long}
\affiliation{Apache Point Observatory, P.O. Box 59, Sunspot, NM 88349, USA}

\author{S.~Mahadevan}
\affiliation{Department of Astronomy and Astrophysics, The Pennsylvania State University, University Park, PA 16802, USA}

\author{E.~Malanushenko}
\affiliation{Apache Point Observatory, P.O. Box 59, Sunspot, NM 88349, USA}

\author{D.~C.~Nguyen}
\affiliation{Department of Computer Science, University of Illinois at Urbana-Champaign, Thomas M. Siebel Center for Computer Science, 201 North Goodwin Ave., Urbana, IL 61801, USA}

\author{R.~Owen}
\affiliation{Department of Astronomy, University of Washington, Box 351580, Seattle, WA 98195, USA}

\author{J.~R.~S{\'a}nchez-Gallego}
\affiliation{Department of Astronomy, University of Washington, Box 351580, Seattle, WA 98195, USA}

\author{C.~Sayres}
\affiliation{Department of Astronomy, University of Washington, Box 351580, Seattle, WA 98195, USA}

\author{N.~Shane}
\affiliation{Lamont-Doherty Earth Observatory of Columbia University, 61 Route 9W, Palisades, NY 10960, USA}

\author{S.~A.~Shectman}
\affiliation{Observatories of the Carnegie Institution for Science, 813 Santa Barbara Street, Pasadena, CA 91101, USA}

\author{M.~Shetrone}
\affiliation{McDonald Observatory, University of Texas at Austin, Fort Davis, TX 79734, USA}

\author{D.~Skinner}
\affiliation{Center for Relativistic Astrophysics, School of Physics, Georgia Institute of Technology, 837 State St., Atlanta, GA 30332, USA}

\author{F.~Stauffer}
\affiliation{Apache Point Observatory, P.O. Box 59, Sunspot, NM 88349, USA}

\author{B.~Zhao}
\affiliation{Dept of Astronomy, Univ of Florida, 211 Bryant Space Science Center, Gainesville, FL 32611, USA}



\begin{abstract}
We describe the design and performance of the near-infrared (1.51--$1.70 \micron$), fiber-fed, multi-object (300 fibers), high resolution ($R = \lambda/{\Delta \lambda} \sim 22{,}500$) spectrograph built for the Apache Point Observatory Galactic Evolution Experiment (APOGEE). APOGEE is a survey of $\sim10^5$ red giant stars that systematically sampled all Milky Way populations (bulge, disk, and halo) to study the Galaxy's chemical and kinematical history.  It was part of the Sloan Digital Sky Survey III (SDSS-III) from 2011 -- 2014 using the $2.5\,\rm{m}$ Sloan Foundation Telescope at Apache Point Observatory, New Mexico.  The APOGEE-2 survey is now using the spectrograph as part of SDSS-IV, as well as a second spectrograph, a close copy of the first, operating at the $2.5\,\rm{m}$ du Pont Telescope at Las Campanas Observatory in Chile.  Although several fiber-fed, multi-object, high resolution spectrographs have been built for visual wavelength spectroscopy, the APOGEE spectrograph is one of the first such instruments built for observations in the near-infrared.  The instrument's successful development was enabled by several key innovations, including a ``gang connector'' to allow simultaneous connections of 300 fibers; hermetically sealed feedthroughs to allow fibers to pass through the cryostat wall continuously; the first cryogenically deployed mosaic volume phase holographic grating; and a large refractive camera that includes mono-crystalline silicon and fused silica elements with diameters as large as $\sim400\, \rm{mm}$.  This paper contains a comprehensive description of all aspects of the instrument including the fiber system, optics and opto-mechanics, detector arrays, mechanics and cryogenics, instrument control, calibration system, optical performance and stability, lessons learned, and design changes for the second instrument.
\end{abstract}





\section{Introduction}

The Apache Point Observatory Galactic Evolution Experiment \citep[APOGEE;][]{maj17} is a large-scale survey of $\sim10^5$ red giant stars in the Milky Way with a fiber-fed multi-object near-infrared spectrograph that provides $R = \lambda/{\Delta \lambda} \sim22{,}500$ spectra throughout the 1.51--$1.70 \micron$ wavelength range.  \edit1{The survey is} the first comprehensive, uniform, high precision study of the kinematical and chemical abundance history in all Milky Way populations (bulge, disk, and halo). One of four surveys of the Sloan Digital Sky Survey III \citep[SDSS-III;][]{eis11} on the $2.5\,\rm{m}$ Sloan Foundation Telescope \citep{gun06} at Apache Point Observatory (APO), New Mexico, APOGEE conducted survey operations from 2011 to 2014.  Data from commissioning and the first year of operation were publicly released as part of the SDSS Data Release 10 \citep[DR10;][]{ahn14} in 2013 July and the full survey results were publicly released as part of the SDSS Data Release 12 \citep[DR12;][]{ala15, hol15} in 2015 January.  The instrument is currently in operation for the APOGEE-2 survey as part of SDSS-IV \citep{bla17} and the first data from that survey were publicly released as part of the SDSS Data Release 14 \citep[DR14;][]{abo17}.  A second spectrograph, a close copy of the first, is now operating at the $2.5\,\rm{m}$ du Pont Telescope \citep{bow73} at Las Campanas Observatory in Chile.  The second spectrograph was delivered and commissioned in early 2017 and future SDSS-IV data releases will include data taken with that instrument.

Comprehensive studies of all Galactic populations, including the highly-extinguished bulge and low latitude disk, are enabled by observations in the near-infrared where dust extinction is significantly reduced compared to visual wavelengths.  The $H$-band was specifically chosen as the spectral range to be sampled by APOGEE given that region's plentiful number of atomic and molecular lines useful for chemical abundance analysis, the brightness of giant stars (the primary APOGEE target) at these wavelengths, and low thermal background.


\begin{deluxetable*}{lc}
\tabletypesize{\scriptsize}
\tablewidth{0pt}
\tablecaption{APOGEE-North Spectrograph Performance\tablenotemark{a} \label{tbl-parameters}}
\tablehead{\colhead{Parameter} & \colhead{Performance}}

\startdata
Median\tablenotemark{b} native ($\lambda/\rm{FWHM}$) resolution (Blue; Green; Red) & $22{,}800$; $23{,}700$; $22{,}000$ \\
Wavelength Coverage (Blue, Green, Red) & 1.514 -- $1.581\,\micron$; 1.585 -- $1.644\,\micron$; 1.647 -- $1.696\,\micron$ \\
Total Spectrograph Fibers & 300 \\
Fiber Core Diameter & $120\,\micron$ \\
Fiber Size on Sky & $2.0\,\arcsec$ \\
Dispersion (at 1.54; 1.61; $1.66\,\micron$) & 0.326; 0.282; 0.235 $\AA/\rm{pixel}$ \\
Point Spread Function (spatial FWHM) (at 1.54; 1.61; $1.66\,\micron$) & 2.16; 2.14; $2.24\,\rm{pixels}$ \\
Predicted throughput\tablenotemark{c} (at 1.54; 1.61; $1.66\,\micron$) & 11; 12; $8\,\%$ \\
Measured throughput\tablenotemark{d} ($1.61\,\micron$) & $18\,\%$ \\
Detector Arrays & Teledyne HgCdTe H2RG, $2.5\,\micron$ cutoff \\
Detector Pixel Size & $18\,\micron$ \\
$\rm{LN_2}$ Hold Time (Days) & $\sim 6$ \\
\enddata

\tablenotetext{a}{Table format from \citet{maj17}.}
\tablenotetext{b}{The average of the medians for fiber sub-samples at the top, middle and bottom of the detector arrays (see \S~\ref{apogee_north_resolution}).}
\tablenotetext{c}{Including the atmosphere through the detector array QE.}
\tablenotetext{d}{Based on measured flux for stars of known $H$ magnitude.}

\end{deluxetable*}

Several trades between competing requirements were considered during development of the science requirements for the survey and instrument.  Given finite detector area, higher spectral resolution improves abundance determination accuracy at the expense of the wavelength coverage necessary to measure large numbers of atomic or molecular species.  Given a practical maximum of $3 \times 2048$ pixels of spectral coverage per target, APOGEE adopted a requirement of $R = \lambda/{\Delta \lambda} \ge 22{,}500$ and signal-to-noise $\rm{S/N} \ge 100$ per pixel in $3\,\rm{hours}$ integration time (nominally divided into three separate one-hour ``visits'') for a target magnitude of $H = 12.2$ for final co-added spectra.  Doing so enabled determination of stellar parameters (e.g., effective temperature, metallicity, and surface gravity) and abundances of at least 15 known chemical elements represented by lines in the $H$-band (e.g., C, N, alpha, odd-Z, and iron peak elements) to 0.1 dex precision \citep{hol15}, the goal for providing strong constraints on Galactic chemical evolution models.  Detection of the \ion{K}{1} line at $1.5160\,\micron$, the only suitable \ion{K}{1} line in the $H$-band for abundance determination, along with the \ion{Mn}{1} lines at 1.5157--$1.5263\,\micron$, established the short wavelength limit.  The three \ion{Al}{1} lines at 1.6720--$1.6770\,\micron$ defined the long wavelength goals.  In fact the use of three $2048 \times 2048$ pixel detector arrays enabled detection to $1.696\,\micron$.  The survey required a radial velocity precision $\le 0.5\,\rm{km\,s^{-1}}$ to constrain dynamical models of the Galaxy and enable determination of whether targets are spectroscopic binaries (when comparing spectra from multiple visits).

\citet{maj17} gives an overview of the survey and science case, the technical requirements, and the instrument design drivers.  And it provides introductions to all aspects of the survey including the instrument, nightly operations, target and field selection, the automated data reduction pipeline, the automated stellar parameters and abundance pipelines, achieved performance, and examples of early science results.  \citet{nid15} discusses the data reduction software and instrument performance from a scientific perspective.

This paper presents details of the spectrograph design, fabrication and performance primarily from an opto-mechanical standpoint.  In general the instrument performance met the technical requirements as summarized in Table~\ref{tbl-parameters}.


The APOGEE spectrograph (hereafter we will refer to both the instrument and survey as APOGEE) is one of the first of a class of high resolution near-infrared multi-object spectrographs.  It is best seen as an evolutionary extension of the SDSS spectrographs \citep{sme13} but implemented at higher resolutions comparable to those of visual instrument counterparts such as \edit1{Hectochelle \citep{sze11} for the MMT, the WIYN fiber-fed Bench Spectrograph \citep{ber08}, the Michigan/MIKE Fiber System \citep{wal07} for the Magellan Telescopes, and FLAMES \citep{pas00} for the VLT}.


\begin{figure*}
\epsscale{1.0}
\plotone{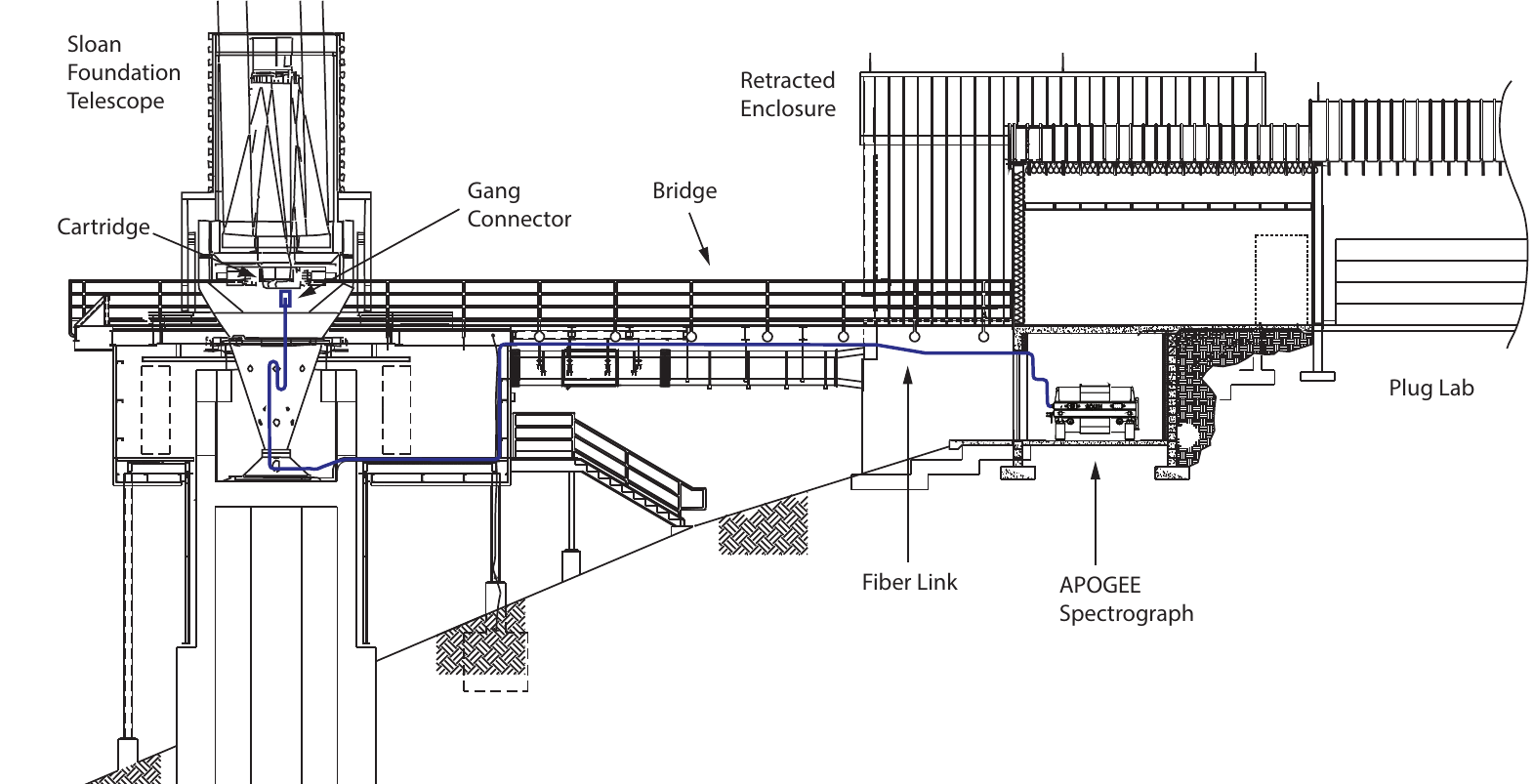}
\caption{The fiber routing from the telescope to the instrument in the adjacent warm support building.  Three hundred fibers, each about $46\,\rm{m}$ long and made up of two segments, transfer the light imaged by the telescope onto the fibers at the plug plate to the instrument pseudo-slit.  Figure includes a portion of the telescope schematic originally shown in \citet{gun06}.}\label{fig_fiber_routing}
\end{figure*}


APOGEE is housed in a building adjacent to the Sloan Foundation Telescope.  A permanent installation of 300 fibers, each $\approx 45\,\rm{m}$ in length, starts at a ``gang connector'' near the telescope focal plane, descends along the telescope cone bearing, follows cable trays connecting the telescope and adjacent warm building, and enters the APOGEE instrument room (Figure~\ref{fig_fiber_routing}).  The fibers then enter the cryogenically cooled instrument through hermetic feedthroughs at the cryostat wall and terminate at a curved ``pseudo-slit'' constructed by aligning the fiber ends along the appropriately shaped curve as a ``long-slit.''  Multiple fiber cartridges, each containing standard pre-drilled SDSS plates corresponding to different fields on the sky and pre-plugged with $2\,\rm{m}$ long fibers that terminate at a gang connector socket, are observed throughout the night.  The gang connector, a key technology required to make APOGEE possible, allows simultaneous disconnection and reconnection of the 300 fibers from the instrument as cartridges are changed over the course of a night's observing.\footnote{This multi-fiber connection system between cartridge and instrument in a separate building is a major change from the method used for the BOSS spectrographs \citep{sme13}, which mount on the back of the telescope and have cartridges with direct fiber connection to slitheads mounted to the interchangeable cartridges.  These changes are discussed fully in \S~\ref{cartridge_and_fibers}.}


\begin{figure}
\epsscale{0.9}
\plotone{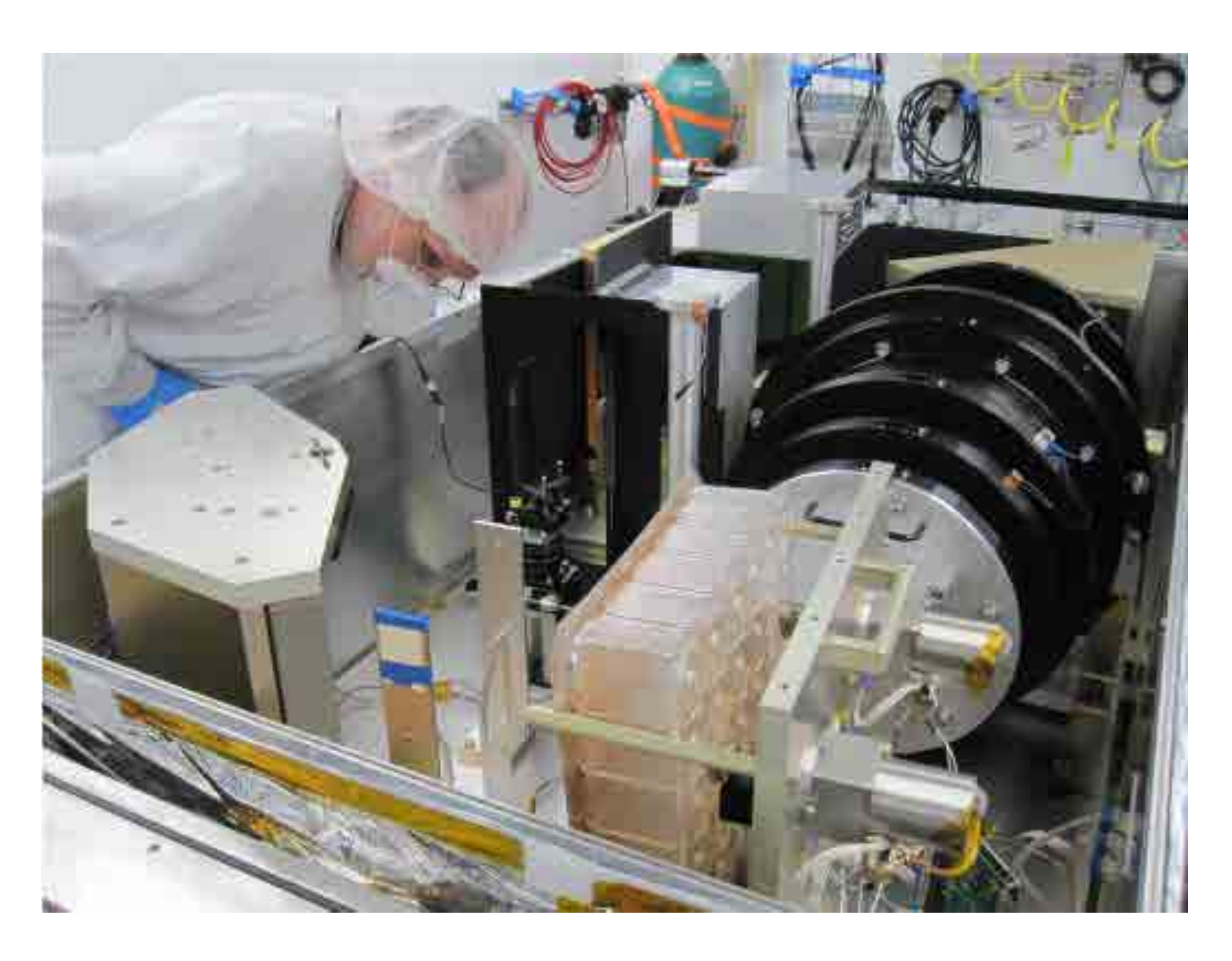}
\caption{The APOGEE Spectrograph during laser alignment of the fore-optics inside the instrument room at the Apache Point Observatory.}\label{fig_apogee_inside}
\end{figure}


Inside the cryostat (Figure~\ref{fig_apogee_inside}), a spherical mirror intercepts and collimates each $\sim f/3.5$ beam emanating from the fiber tips that form the pseudo-slit.  The collimator is an uncorrected Schmidt design where spherical aberration correction is accomplished by the first element of the camera.  Placed near the pupil, a mosaic volume phase holographic (VPH) grating used in Littrow mode disperses the light.  This grating is the first deployed cryogenic mosaic VPH grating in an astronomy instrument and was another key technology enabling the instrument's success.  Past the grating, a six-element \edit1{$f/1.3$} refractive camera focuses the 300 parallel spectra onto the detector arrays.  Featuring elements as large as $\sim 400\,\rm{mm}$, the camera employs four mono-crystalline silicon elements and two fused silica elements.  Three Teledyne 2048 x 2048 HAWAII-2RG \edit1{(H2RG)} sensor chip assemblies mounted side-by-side in the dispersion direction form the detector array mosaic.  A precision linear actuator, which translates the three arrays together in the spectral direction in 0.5 pixel dither (i.e., back and forth) steps, enables critical sampling of the native spectra, which are purposefully undersampled at the blue end of the spectrum to maximize instrument wavelength coverage.  All of the optics and a $100\, \rm{liter}$ $\rm{LN_2}$ tank are contained within a large ($1.4 \times 2.3 \times 1.3\,\rm{m}$), $\approx 1{,}800\,\rm{kg}$, custom stainless steel cryostat supported on four vibration isolation stands.

An Astronomical Research Cameras (San Diego, CA; also known as ``Leach'') Generation III Controller reads out the arrays in sample-up-the-ramp mode \citep[see, e.g.,][]{rau07}.  In addition to the spectral dithering mechanism, three actuators enable tip-tilt-piston positioning of the collimating mirror and a cold shutter arm can be positioned to cover the pseudo-slit to prevent inadvertent illumination of the detectors when not observing.  This is the extent of moving parts in the system, deliberately minimized to maximize reliability and instrument stability over multiple thermal cycles.

The gang connector system also enables illumination of the spectrograph with several calibration light sources fed from an integrating sphere in a calibration box.  \edit1{Instead of} plugging the gang connector into a fiber cartridge on the telescope, the gang connector can be plugged into sockets on a podium adjacent to the telescope that connect the instrument to fiber bundles that originate at the calibration box. \edit1{There is also a socket for a fiber feed from the New Mexico State University (NMSU) $1\,\rm{m}$ telescope \citep{hol10} located adjacent to the Sloan Foundation Telescope.  The connection to the NMSU $1\,\rm{m}$ telescope, consisting of ten fibers, has enabled single-object observations of bright stars, along with illumination of adjacent sky, for abundance calibration and science \citep{hol15} when APOGEE is not fed by light from the Sloan Foundation Telescope.}

APOGEE had a fast-paced instrument development schedule:  Following approval as an SDSS-III project in 2006 November, a Conceptual Design Review was held in April 2008 in which the goal of collecting data on the telescope by the second quarter of 2011 was set.  A Preliminary Design Review took place in 2009 May and a Critical Design Review (CDR) in 2009 August.  With the exception of the camera, which was given approval to start fabrication in 2009 June to meet schedule, approval to start instrument fabrication was given after successful completion of the CDR.  Spectrograph first light at the Sloan Foundation Telescope occurred 21 months later, with commissioning in 2011 May-June.

After discussing use of the Sloan Foundation Telescope in the near-infrared (\S~\ref{near_ir_telescope}), we review the cartridge and fiber system (\S~\ref{cartridge_and_fibers}); spectrograph optics and opto-mechanics (\S~\ref{optics_section}); detector arrays and their electronic control (\S~\ref{detector_section}); mechanics and cryogenics (\S~\ref{mechanics section}); moving parts (\S~\ref{mechanism_design}); instrument control (\S~\ref{instrument_control}); calibration system (\S~\ref{calibration section}); shipping, installation, and instrument room setup (\S~\ref{logistics section}); instrument optical performance and stability (\S~\ref{performance section}); lessons learned (\S~\ref{lessons learned}); and design changes for the second instrument in use at the $2.5\,\rm{m}$ du Pont Telescope as part of SDSS-IV (\S~\ref{second_spectrograph}).  The Appendices address fiber alignment errors at the telescope and the configuration for the first instrument's commissioning period.

In most cases the instrument performance during SDSS-III is reported.  When noteworthy, performance at the beginning of SDSS-IV is also discussed.


\section{The Sloan Foundation Telescope at Near Infrared Wavelengths\label{near_ir_telescope}}

\subsection{Imaging Performance}

The \edit1{$2.5\,\rm{m}$} Sloan Foundation Telescope \citep{gun06} is a modified two-corrector Ritchey-Chr\'{e}tien design with a net focal ratio of $f/5.0$. A Gascoigne corrector corrects astigmatism and a highly aspheric corrector (``spectrograph corrector'') corrects lateral color.  Designed for optimal performance in the wavelength range $3{,}000$ -- $10{,}600\,$\AA\  accessible to CCDs (hereafter called the visual band), the telescope has good image quality in the visual band across a $3\degr$ diameter field of view.

When optimally focused for $1.6\,\micron$ light with the same correctors, the intrinsic image quality is essentially the same as for visual wavelengths for field angles up to $30\,\arcmin$ radius, i.e., with $0.4\,\arcsec$ RMS diameter spot sizes.  For larger field angles image performance degrades linearly to a $1.25\,\arcsec$ RMS diameter at a $90\,\arcmin$ radius.  This degradation with field also occurs in the visual for wavelengths other than $5300\,$\AA\ (the central design wavelength), but it is less severe.

The optimal focal plane location for $1.6\,\micron$, relative to the visual focal plane, varies by over $800\,\micron$ as field radius \edit1{changes} from 0 -- $90\,\arcmin$, and is only confocal with $5300\,$\AA\ light for the $\sim 50\,\arcmin$ radial position.  Radial positions smaller than this are out of focus by up to \edit1{$\approx +240\,\micron$} (positive defocus means farther from the secondary mirror) which corresponds to a defocus blur disk of $\approx 0.65\,\arcsec$ RMS image size.  Radial positions $> 50\,\arcmin$ are out of focus by up to \edit1{$\approx -620\,\micron$} which corresponds to a defocus blur disk of $\approx 1.65\,\arcsec$ RMS image size.


\begin{figure}
\epsscale{1.15}
\plotone{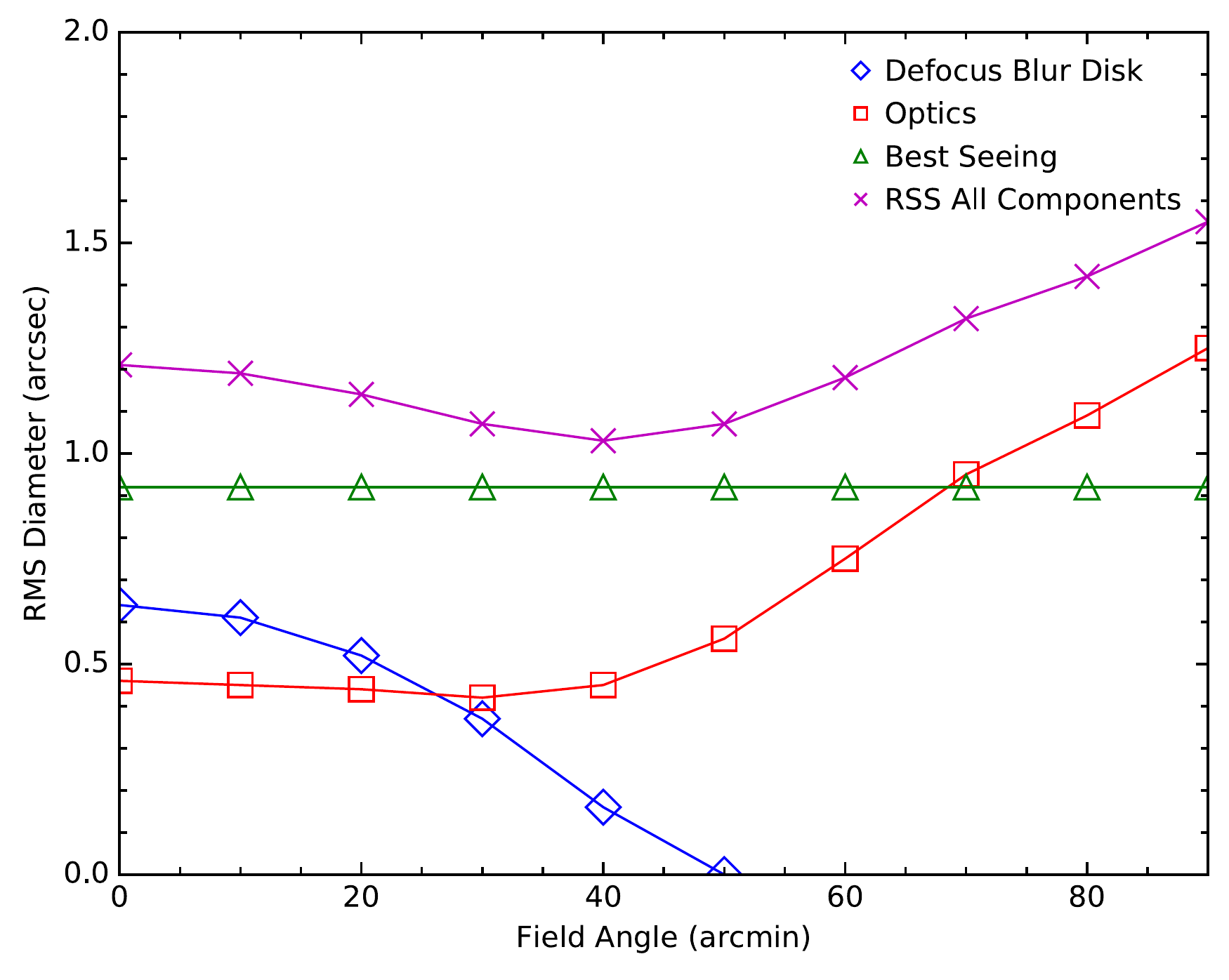}
\caption{Predicted telescope image size (root sum square of intrinsic image quality at $1.6\,\micron$, defocus due to the compromise plug plate position, and seeing blur) as a function of field angle.  The defocus blur is zero beyond $50\,\arcmin$ because the back of the plug plates are counter-bored to bring the ferrules into focus for $1.6\,\micron$ light.}\label{fig_image_quality}
\end{figure}


\edit1{Two constraints governed potential focal plane modifications to maximize image quality and light input into the APOGEE fibers at near-infrared wavelengths.  First, it was desired that both visual and near-infrared surveys co-observe during SDSS-III to maximize telescope efficiency.  During the initial year of the APOGEE survey numerous fields were co-observed with the Multi-Object APO Radial Velocity Exoplanet Large-area Survey (MARVELS) spectrograph \citep{ge08} that operated at $5400\,$\AA\ with a $900\,$\AA\ wavelength coverage.  And throughout SDSS-IV there has been co-observing with the Mapping Nearby Galaxies at Apache Point Observatory \citep[MaNGA;][]{bun15} survey, which uses the Baryon Oscillation Spectroscopic Survey (BOSS) spectrographs \citep{sme13} at visual wavelengths.  This co-observing constrained the plug-plate to be positioned at a focus that accommodated the science goals of both APOGEE and visual surveys simultaneously.}

\edit1{Secondly, one of the functions of the fiber cartridges (described below) is to mechanically bend plug-plates so they assume curvatures that conform to the ideal focal surface for $5300\,$\AA, the design wavelength of the visual spectrographs.  The SDSS plug plates, $0.125\,\rm{in}$ ($\approx 3.2\,\rm{mm}$) thick and fabricated from 6061-T6 aluminum alloy, are flat when not in use.  Bending rings within the cartridges mechanically clamp the perimeters of the plates to hold them in place.  And these bending rings, together with a central rod that pushes up the center of the plate, induce the desired plate curvature when mounted in the cartridge.  Given the desire to co-observe between surveys to maximize survey efficiency, the development of cartridges with bending ring assemblies tailored to bend the plug-plates to curvatures optimized for APOGEE was not a viable solution.  Fortunately, as described below, a solution using counterbored fiber holes in plug plates with the existing cartridge system was developed so that the APOGEE fiber ferrules could be more accurately positioned in focus to regain acceptable imaging performance.}

Given the good telescope image performance at $1.6\,\micron$ for small field radii the plug plate was positioned at the visual-use position and the central focus errors for radial positions $< 50\,\arcmin$ were accepted.  For radial positions greater than this, focus is corrected by counterboring the backside of the plug-plates so the fiber ferrule tips are located closer to the telescope secondary.  Drilled counterbore depths increase linearly with radial position for optimal correction.

Given these accommodations, the predicted telescope RMS image diameter for $1.6\,\micron$ light at the fiber tips as a function of field angle is shown in Figure~\ref{fig_image_quality}, where the RMS diameter is the root sum square of the intrinsic telescope image quality at best focus, defocus blur due to the compromise focal plane position but mitigated for outer field angles by counterboring, and good seeing ($0.92\,\arcsec$ RMS diameter at $1.6\,\micron$).  The median seeing\footnote{This is apt to be a conservative estimate due to extra broadening from cross-talk in the coherent fibers used for guiding the Sloan Foundation Telescope.} is $1.37\,\arcsec$ RMS diameter at $1.6\,\micron$.  The median seeing value derives from the $1.42\,\arcsec$ median FWHM of visual guider images recorded during APOGEE survey observations between 2011 June -- 2013 December and is adjusted for near-infrared wavelengths assuming Gaussian seeing that scales as $\lambda^{-0.2}$.

\subsection{Fiber Encircled Energy\label{ee}}

\begin{figure}
\epsscale{1.15}
\plotone{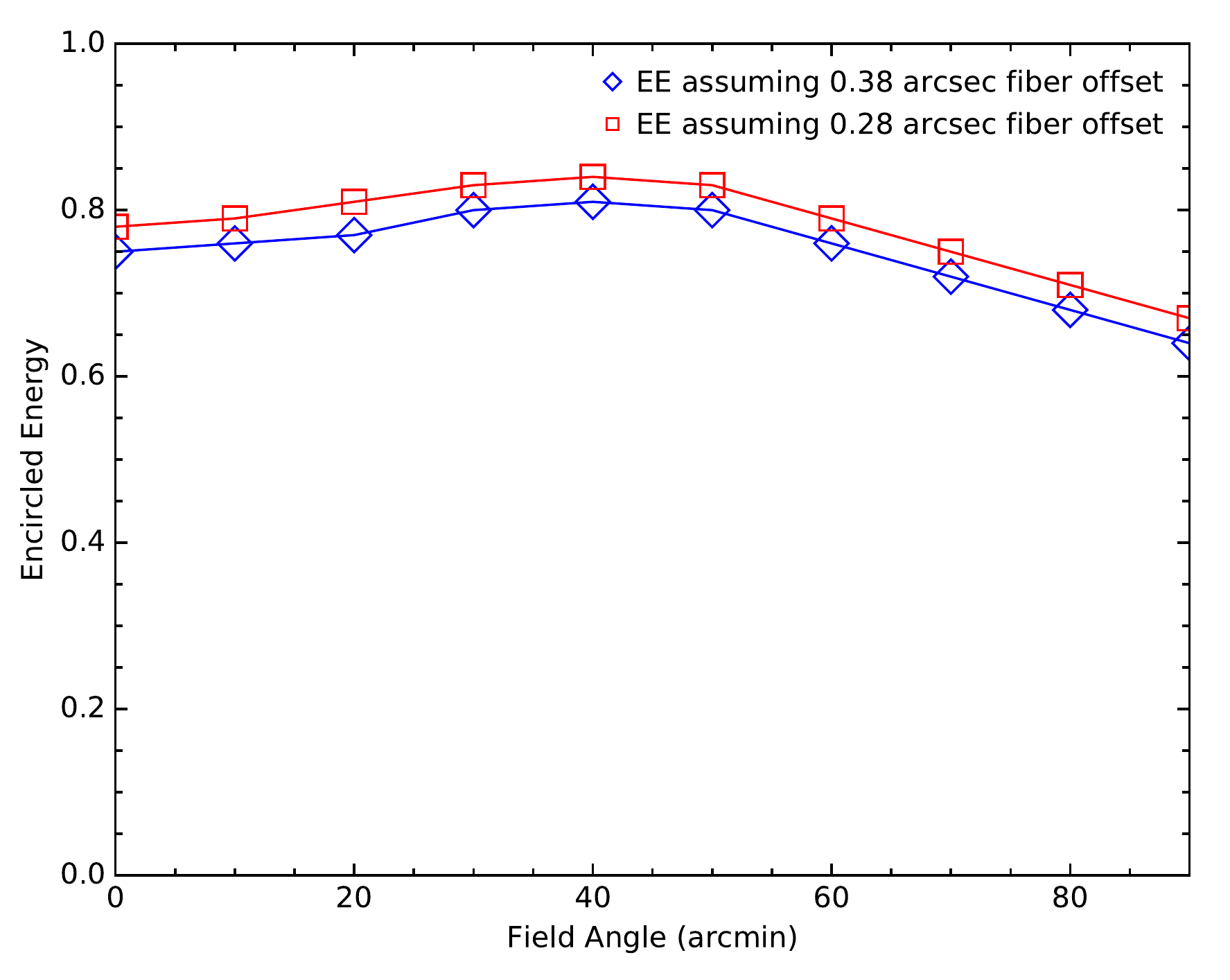}
\caption{Predicted encircled energy accepted by the $2.0\,\arcsec$ fibers as a function of field angle given the predicted image diameter from Figure~\ref{fig_image_quality} after taking into account errors (see Appendix \ref{appendix_errors}) that offset the fiber from the image location in the focal plane.}\label{fig_ee}
\end{figure}

The APOGEE fibers subtend $2\,\arcsec$ on the sky.  Figure~\ref{fig_ee} shows the predicted encircled energy intercepted by the APOGEE fibers given the predicted image diameter from Figure~\ref{fig_image_quality} after taking into account errors that offset the fiber from the image location in the focal plane (see Appendix \ref{appendix_errors} for details).  The model telescope PSF uses a bi-Gaussian function, specifically Equation 14 of \citet{ani11}.


\subsection{Telescope Corrector Throughput}\label{corrector_throughput}

\edit1{The telescope correctors' anti-reflection (AR) coatings are optimized for visual transmission and are not favorable for transmission of near-infrared wavelengths.  While the correctors are made from fused silica and thus have excellent near-infrared intrinsic transmission, taken together the pair of correctors only transmit $\sim 60\, \%$ of the light at APOGEE wavelengths because of the AR coatings.}  This transmission was determined by measuring the on-axis illumination of a light source through each corrector individually when removed from the telescope using a Goodrich room temperature InGaAs camera.  While a new corrector could have been fabricated and AR coated to optimize throughput simultaneously for both $1.6\,\micron$ (for APOGEE) and $5400\,$\AA\ (for MARVELS) and installed in the telescope when those instruments were in use, the costs were difficult to justify since the APOGEE system already had sufficient sensitivity to meet its science requirements.

\subsection{Guiding}

Guiding is accomplished in the visual (effective wavelength $\sim 5400\,$\AA) using the SDSS architecture described in \citet{sme13} with the only exception that the science fiber holes lie at the predicted positions of the targets' $1.66\,\micron$ light,\footnote{Use of $1.66\,\micron$ was inadvertent -- $1.60\,\micron$, close to the center of the science wavelength span of the instrument, would have been a better choice.  But the effect from atmospheric differential refraction, between $1.50\,\micron$, the shortest science wavelength, and $1.66\,\micron$, is $< 0.05\,\arcsec$, even for $\rm{airmass} = 2$ based on the index of refraction measurements of \citet{pec72}.  This error is much smaller than the typical fiber RMS radial offset error of $0.28\,\arcsec$ (see \S~\ref{appendix_fiber_offset_errors}).} accounting for differential refraction at the design hour angle (HA).  To efficiently capture the near-infrared light within the APOGEE fibers while guiding in the visible a model of the differential color refraction, \edit1{calculated using formulae from \citet{pec72},} is used to adjust the reference positions of the guide stars with changing hour angle.\footnote{The guiding code also permits changing the intended science wavelength during observations.  This was used, for instance, to provide compromise guiding during APOGEE and MARVELS co-observing.}

\edit1{Guiding corrections are implemented by changing telescope pointing, rotating the cartridge, and changing the focal plane scale.  There is an unavoidable quadrupole term in the residuals when correcting observations away from the design hour angle.  The hour angle range for observing is based on the condition (under the assumption of perfect guiding in overall rotation, shift and scale) that star centers may not drift further than $0.3\,\arcsec$ from any hole center.  For example, for plates on the equator, whose minimum airmass is 1.15 when observed from APO, the allowed observation range is $\pm 1\,\rm{hour}$ from the design hour angle.  In good conditions the guiding residuals (RMS of the amplitude of guiding errors calculated for all guide fibers) get down to 0.25 -- $0.30\,\arcsec$, but they can also reach $\ge 0.6\,\arcsec$.  Attributes of good conditions include observing near the designed hour angle, airmass $< 1.35$, good seeing, no windshake, and no hardware problems.}


\section{Cartridge and Fiber System}\label{cartridge_and_fibers}


\begin{figure*}
\epsscale{0.9}
\plotone{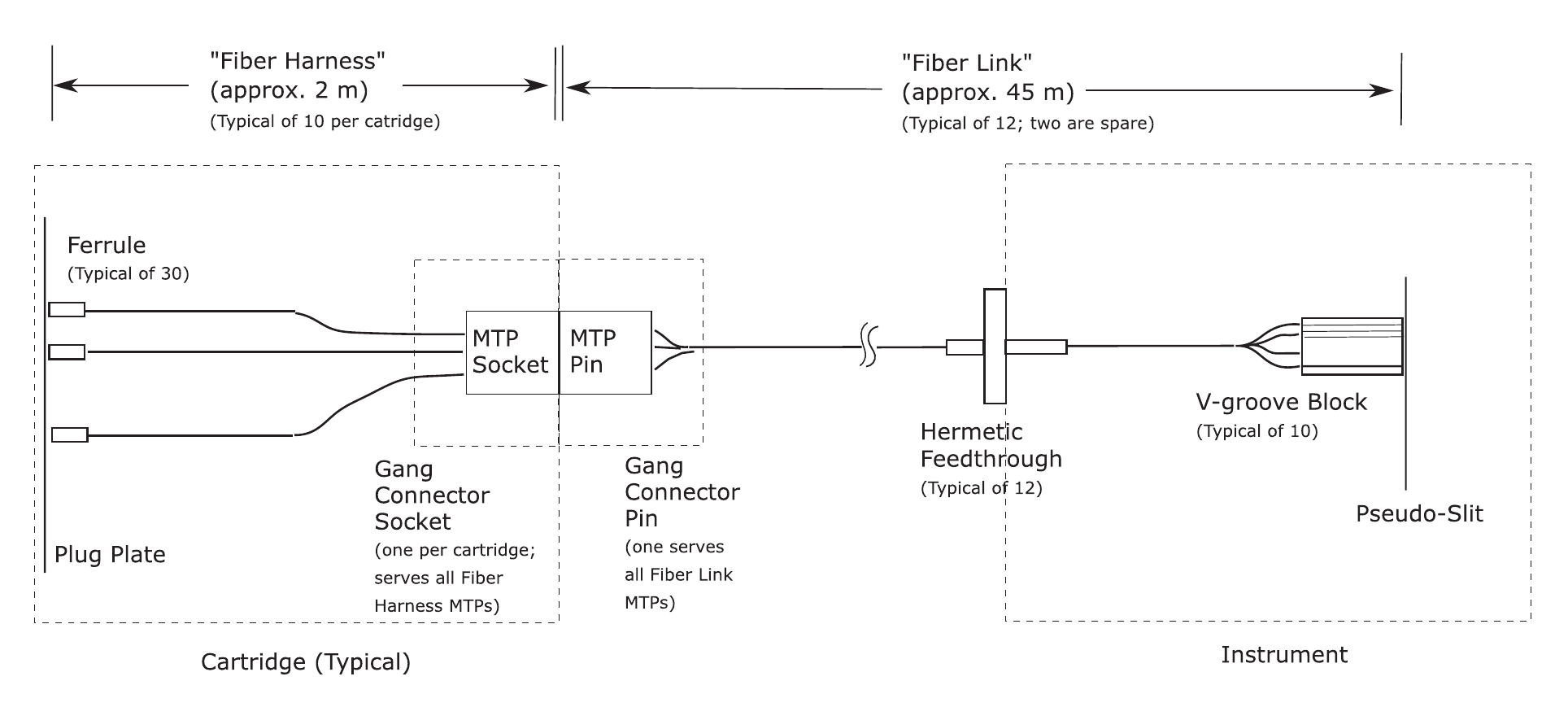}
\caption{\edit1{Fiber system schematic.  Cartridges, which are changed throughout the night and plugged for a specific field, contain ten fiber harnesses that consist of fiber assemblies that terminate in sets of individual ferrules for the plug plate on one end and are grouped within an MTP\textsuperscript{\textregistered} connector at the other.  The ten MTP\textsuperscript{\textregistered} connectors are together mounted within a gang connector socket that is accessible from below the cartridge.  A single gang connector pin plugs into the various cartridges to simultaneously connect the complementary MTP\textsuperscript{\textregistered} connectors, which are the terminations of the ten fiber links routed from the instrument. Each fiber link consists of 30 fibers, and each set of 30 enter the instrument without break through hermetic feedthroughs to minimize FRD.  Lastly, the fibers terminate at the instrument pseudo-slit in groups of 30 on v-groove blocks.}}\label{fig_fiber_block_diagram}
\end{figure*}


\edit1{The APOGEE fiber system, shown schematically in Figure~\ref{fig_fiber_block_diagram}, includes multiple components that are discussed in detail in subsequent sub-sections following this introductory overview.}

APOGEE uses the extensive and proven SDSS plug-plate and cartridge system \citep[see][]{gun06, sme13} that has enabled efficient fiber-fed spectroscopic observations of multiple fields nightly for nearly twenty years at the Sloan Foundation Telescope.  SDSS ``plug-plates,'' each approximately $800\,\rm{mm}$ in diameter covering $3\,\degr$ of sky, contain pre-drilled holes for fiber ferrules corresponding to the locations of astronomical objects of interest.   These plates mount on ``cartridges'' for rapid interchange of plates and associated fibers at the telescope focal plane.  Multiple cartridges, each containing a full complement of fibers that are manually plugged into plates during the day, are available at the telescope each night.


\begin{figure*}
\epsscale{0.9}
\plotone{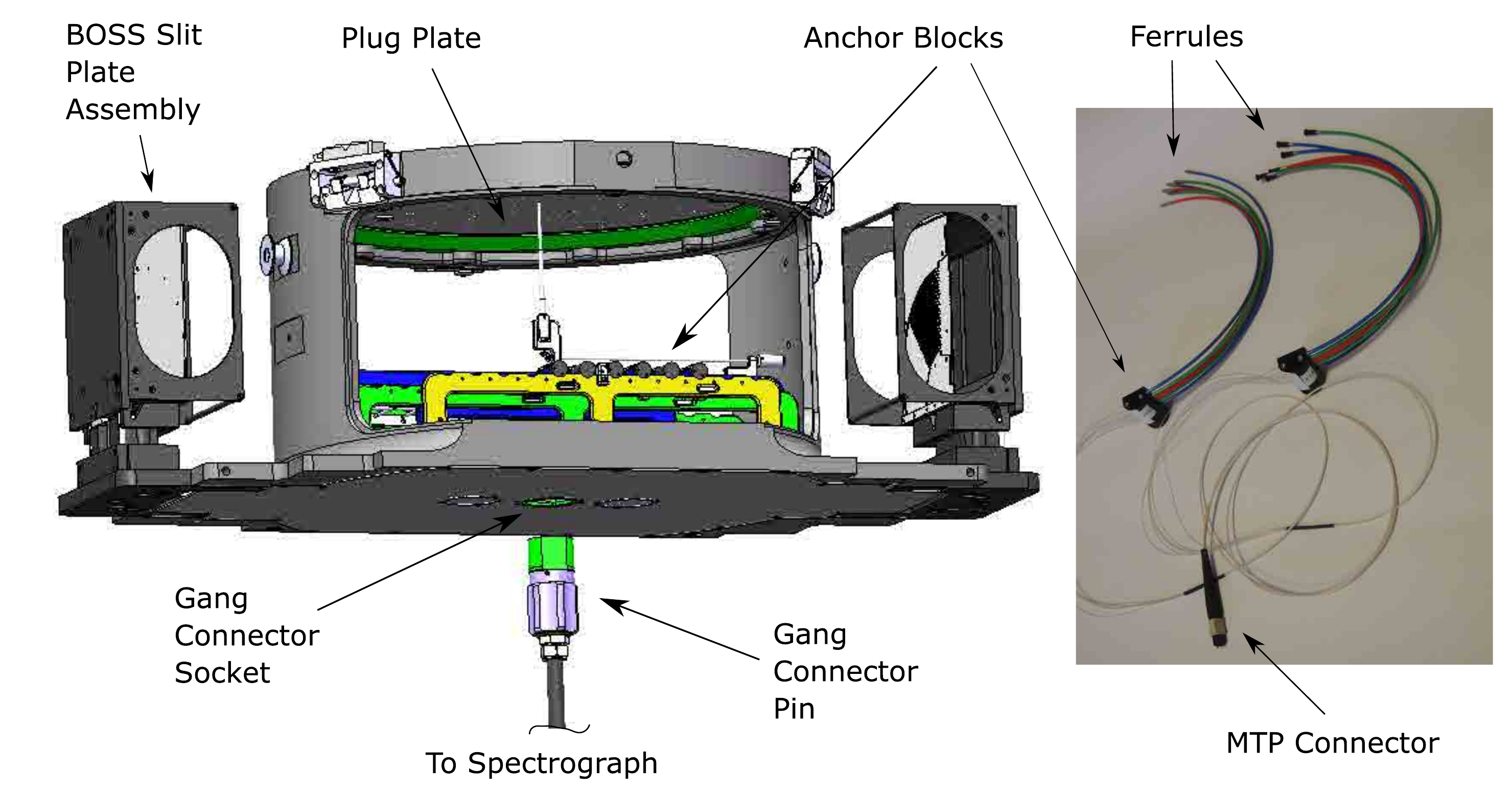}
\caption{\edit1{(Left) A mechanical rendering of the cartridges used for APOGEE.  The gang connector pin plugs into the gang connector socket, accessible from the bottom of the cartridge, after the cartridge is loaded on the telescope for observing.  (Right) A fiber harness.  The ferrules are plugged into the plug plate during the daytime.  The anchor blocks and MTP connectors are permanently secured within the cartridge and gang connector socket, respectively.}}\label{fig_cartridge}
\end{figure*}


\edit1{APOGEE cartridges and the fiber assemblies contained within them are shown in Figure~\ref{fig_cartridge}}.  The fibers, $\sim 2\,\rm{m}$ long and grouped in sets of \edit1{30 within} ``harnesses,'' terminate on one end in stainless steel ferrules, mentioned above, that are plugged into plug-plate holes.  The other end of the each group of 30 fibers terminate \edit1{at an MTP\textsuperscript{\textregistered} connector.  The cartridge contains ten harnesses and the ten MTP\textsuperscript{\textregistered} connectors of these harnesses are installed within a gang connector ``socket'' accessible from below the cartridge for easy interface to a corresponding gang connector ``pin'' that contains the set of fibers leading to the spectrograph.}  This gang connector \edit1{system} allows simultaneous disconnection and reconnection of the instrument fibers from any given cartridge.  A gang connector that accommodated 60-fibers at the cartridge had already been developed for the MARVELS survey (Ge et al., in prep).  So we chose to adopt that method and modify the gang connector design to accommodate 300-fibers.  Cartridges that had previously been used for SDSS visual spectroscopy were modified and fitted for both APOGEE and MARVELS operations, and new cartridges were fabricated for the BOSS survey.

\edit1{Note that the APOGEE cartridges still include two slit plate assemblies adjacent to the main cartridge body that insert into the BOSS spectrographs when the cartridge is installed on the telescope.  While not populated with fibers, these slit plate assemblies are necessary for optical guiding.}

\edit1{The fiber train from the gang connector pin to the instrument is composed of a single, $44\,\rm{m}$ pigtail of 300 fibers composed of 12 ``fiber links,'' two of which are spares.  The fiber links terminate at v-groove blocks that are positioned on the pseudo-slit inside the cryogenically cooled instrument.}

\edit1{Since the terminations of fibers can impact Focal Ratio Degradation (FRD) as a result of mechanical stress at connectors or vagaries in surface polishing methods \citep[see, e.g.,][]{cla89,lee01,oli05,eig12}, we purposefully minimized the number of couplings in the fiber train to minimize FRD.  We also developed a custom feedthrough at the cryostat wall to bring the fibers into the cryogenic environment without break.}

\subsection{Fibers and Fiber Routing}\label{fibers_and_fiber_routing}

To maximize throughput in the 1.5 -- $1.7\,\micron$ range, low-OH silica core, step-index, multi-mode fibers from Polymicro Technologies (Phoenix, AZ; part no. FIP120170190) were selected.  The nominal core, clad, and polyimide buffer outside diameters are $120\,\micron$, $170\,\micron$, and $190\,\micron$, respectively.  Given the Sloan Foundation Telescope's $f/5$ beam, the fiber core subtends $2\,\arcsec$ on the sky.  This fiber field of view was considered a good compromise to provide sufficient angular size on the sky to accommodate guiding, pointing, and plug-plate drilling errors while minimizing excessive sky noise.  Intrinsic fiber transmission through $\approx 46\,\rm{m}$, the length from the telescope plug-plates to the pseudo-slit, is $99\%$ (attenuation $\sim 1\,\rm{dB \, km^{-1}}$) throughout the APOGEE wavelength range based on vendor measurements of FIP120170190 fiber.


\begin{deluxetable*}{lc}
\tabletypesize{\scriptsize}
\tablewidth{0pt}
\tablecaption{Fiber Details\label{tbl-fiber1}}
\tablehead{\colhead{Item} & \colhead{Material}}

\startdata
Fiber & Polymicro FIP120170190 with Polyimide Buffer \\
Outside Cryostat Jacketing & Tyco Electronics Jacketing Material\tablenotemark{a} \\
Inside Cryostat Jacketing & Alpha Wire Uncoated Fiberglass Sleeving P/N PIF-240-18 \\
\enddata

\tablenotetext{a}{Inner Tube - Polypropylene impregnated
with carbon black, $1.8\,\rm{mm}$ Inner Diameter; Strength Member - Kevlar; Outer Jacket - PVC, $3.8\,\rm{mm}$ OD}

\end{deluxetable*}


A schematic of the fiber routing is shown in Figure~\ref{fig_fiber_routing} and the fiber and jacketing materials used to fabricate the system are listed in Table~\ref{tbl-fiber1}.  Starting from the gang connector, the fiber link runs through a hole in the telescope floor and into the cone bearing \edit1{below} where it is \edit1{draped over a semi-circular piece of Delrin\textsuperscript{\textregistered} with $3\,\rm{in}$ ($76\,\rm{mm}$) radius of curvature (a gentle radius from the standpoint of bending stress).  On one side of the Delrin\textsuperscript{\textregistered} assembly the fibers drop into the slack loop coming down from the telescope.  On the other side the fibers continue down to the bottom of the cone.  This system manages fiber payout with changes in telescope pointing with minimal fiber stress --- the only tension is from the weight of the fiber itself.  The fiber link then routes to the bottom of the cone and} into cable trays that connect the telescope building to the lower level of the adjacent plug lab where the instrument is located (Figure~\ref{fig_fiber_routing}).  Within the external cable trays the fiber link is inserted into flexible polyurethane sheathing to minimize environmental exposure and mechanical damage.

\subsection{Terminations and Feedthroughs}

Detailed information regarding materials and epoxies used for the fiber terminations in the plug-plate ferrules, within the gang connector, and at the v-groove blocks, are listed in Tables~\ref{tbl-fiber2a} and \ref{tbl-fiber2b}.  Our material and epoxy choices were based in large part on component FRD testing (\S~\ref{FRD_testing}; \citet{bru10}) \edit1{as well as vendor experience.  All fiber assemblies were fabricated and polished by C-Technologies (Bridgewater, NJ).}  None of the fiber terminations were AR coated.  This saved development time and complexity at the expense of Fresnel reflection losses.


\begin{deluxetable*}{lcccc}
\tabletypesize{\scriptsize}
\tablewidth{0pt}
\tablecaption{Fiber Harness\tablenotemark{a} Details\label{tbl-fiber2a}}
\tablehead{\colhead{Termination/Feature} & \colhead{Quantity} & \colhead{Material/Connector} & \colhead{Epoxy}}

\startdata
Plug-plate Ferrules & 300 total (30 per Harness) & 303 Stainless Steel & Master Bond EP21LV \\
Gang Connector Ferrules & 10 total (30 fibers per connector) & US Conec Custom MTP\textsuperscript{\textregistered} 32 (Socket Side) & Master Bond EP21LV \\
\enddata

\tablenotetext{a}{10 sets of $\sim 1.8\,\rm{m}$ long assemblies permanently mounted in each of 8 cartridges}

\end{deluxetable*}


\begin{deluxetable*}{lcccc}
\tabletypesize{\scriptsize}
\tablewidth{0pt}
\tablecaption{Fiber Link\tablenotemark{a} Details\label{tbl-fiber2b}}
\tablehead{\colhead{Termination/Feature} & \colhead{Quantity} & \colhead{Material/Connector} & \colhead{Epoxy}}

\startdata
Gang Connector Ferrules & 10\tablenotemark{b} & US Conec Custom MTP\textsuperscript{\textregistered} 32 (Pin Side) &  Master Bond EP21LV \\
Hermetic Fiber Feedthrough & 12\tablenotemark{c} & 304 stainless steel & Master Bond EP37-3FLFAO \\
V-groove Block & 10\tablenotemark{d} & Alloy 39 & Master Bond EP29LPSP \\
\enddata

\tablenotetext{a}{a $\sim 44\,\rm{m}$ long multi-fiber assembly that terminates at cryostat pseudo-slit}
\tablenotetext{b}{30 fibers per connector}
\tablenotetext{c}{30 fibers per feedthrough; 2 feedthroughs for installed spares}
\tablenotetext{d}{30 fibers per block}

\end{deluxetable*}


\subsubsection{Plug-plate Ferrules}

For commonality, the APOGEE plug-plate ferrules have identical designs and are manufactured and polished in the same manner as the ferrules used in the BOSS cartridges \citep{sme13}.  \edit1{Master Bond EP21LV epoxy was used to bond the fiber within the ferrule, as C-Technologies considered EP21LV to be an excellent epoxy for all-around fiber optic terminations due to its adhesion to both metal and plastic, polishing characteristics, a room temperature cure cycle that requires no added heat, and its service temperature range ($-65\,^{\circ}\,\rm{F}$ -- $+250\,^{\circ}\,\rm{F}$).}

\subsubsection{MTP\textsuperscript{\textregistered} Connectors and Gang Connectors}\label{MTP_section}


\begin{figure}
\epsscale{1.0}
\plotone{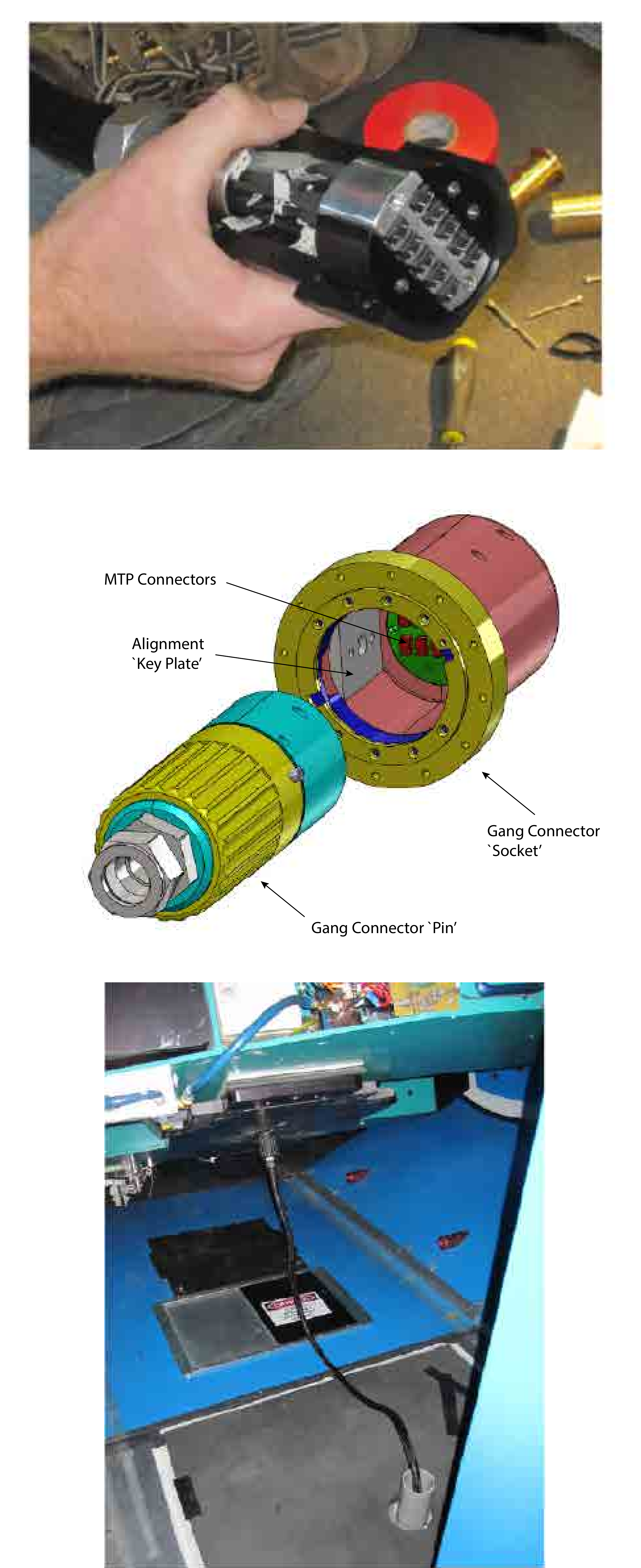}
\caption{(Top) Gang connector ``pin'' during final assembly at the telescope.  The ten US Conec MTP\textsuperscript{\textregistered} connectors are visible along with the interior cabling.  (Middle) Model views of the gang connector ``pin'' and ``socket.''  (Bottom) The gang connector plugged into the bottom of a cartridge mounted on the telescope.  The top and bottom images are from \url{http://www.sdss.org}.}\label{fig_gang_connector}
\end{figure}

\begin{figure}
\epsscale{0.8}
\plotone{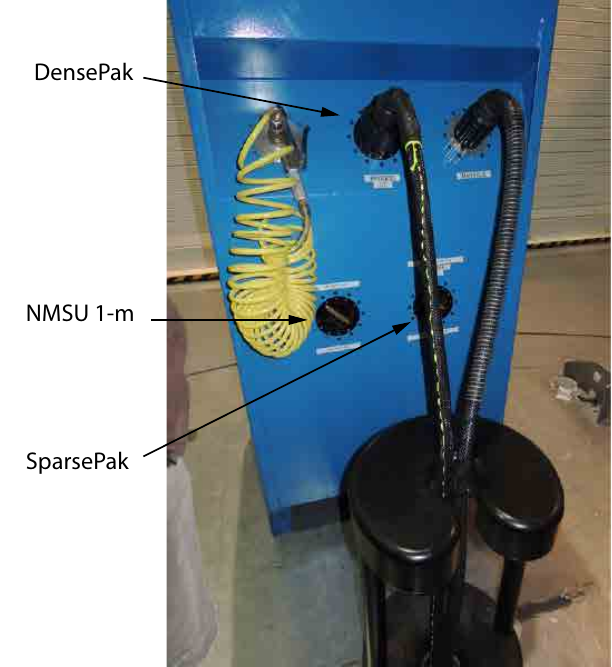}
\caption{When not plugged into a cartridge on the telescope, the gang connector can be plugged into several sockets in a podium adjacent to the telescope.}\label{fig_gang_connector_ports}
\end{figure}

In cartridges used for APOGEE, the fiber harnesses terminate in groups of 30 at custom US Conec (Hickory, NC) 32-fiber MTP\textsuperscript{\textregistered} connectors (described below).  \edit1{Master Bond EP21LV epoxy was also used to bond the fibers within the MTP\textsuperscript{\textregistered} ferrules.}  Ten MTP\textsuperscript{\textregistered} connectors are in turn grouped together in a gang connector socket at the bottom of each cartridge.  As cartridges are interchanged throughout the night, the single gang connector pin at the end of the fiber link from the instrument is inserted into each new cartridge gang connector socket.  Figure~\ref{fig_gang_connector} shows the gang connector system.  When not in use, the gang connector pin can be inserted in the following auxiliary sockets at a podium (Figure~\ref{fig_gang_connector_ports}) adjacent to the telescope:  (1) Densepak\footnote{The names ``DensePak'' \citep{bar98b} and ``SparsePak'' \citep{ber04} are adopted from the names of formatted fiber field units used with the WIYN Bench Spectrograph.}, a bundle of 300 calibration fibers illuminated by an integrating sphere within a calibration box located one floor below the telescope;  (2) Sparsepak, a bundle of 50 calibration fibers (every sixth fiber on the pseudo-slit) illuminated by the integrating sphere; and (3) \edit1{the} light feed from the NMSU 1-m telescope.  The calibration system is discussed in \S~\ref{calibration section}.


US Conec's MTP\textsuperscript{\textregistered} brand connector system utilizes tightly-toleranced stainless steel guide pins coupled with high precision multi-fiber rectangular ferrules to optimize fiber alignment.  For this application, US Conec fabricated a custom mold to manufacture specialized ferrules using a Polyphenylene Sulfide (PPS) based glass-filled thermoplastic material system; the true position of the fiber holes held to less than $3\,\micron$ of radial eccentricity.  Coupled with high force ($20\,\rm{N}$) compression springs and polishing processes that promote creation of slightly protruded fiber tips, this design is intended to yield fiber-fiber physical contact across each ferrule's $4 \times 8$ fiber array that should minimize optical insertion losses and Fresnel reflections.  In practice this performance is not achieved, as discussed in \S~\ref{MTP_coupling_performance}.

In this application, the normal MTP\textsuperscript{\textregistered} connector latching mechanisms were removed and we instead relied upon the gang connector system, which included a mechanical detent when fully seated to latch all connectors simultaneously.  MTP\textsuperscript{\textregistered} connection systems have been deployed for many years in high-reliability telecommunication systems exposed to a wide range of environmental conditions (typically $-40\,^{\circ}\,\rm{C}$ -- $+80\,^{\circ}\,\rm{C}$) and high levels of humidity without performance degradation.  Therefore we felt confident these connection systems would experience no environmental-related issues in our telescope application.  In fact, no problems have been seen from exposure to the mountain-top temperature swings.

Dust accumulation, on the other hand, is a problem given the observatory's proximity to White Sands National Monument and its very large sand dunes of gypsum crystal in the valley below.  The gang connector pin is blown with pressurized air every afternoon during afternoon checkout.  Also, the gang connector pin and cartridge gang connector sockets are blown at each cartridge change during the night.  Lastly, fiber tips in the gang connector pin, all eight cartridge gang connector sockets, and the fiber ferrules at the plug-plates are cleaned monthly with alcohol.

\subsubsection{Hermetic Fiber Feedthrough}


\begin{figure}
\epsscale{1.1}
\plotone{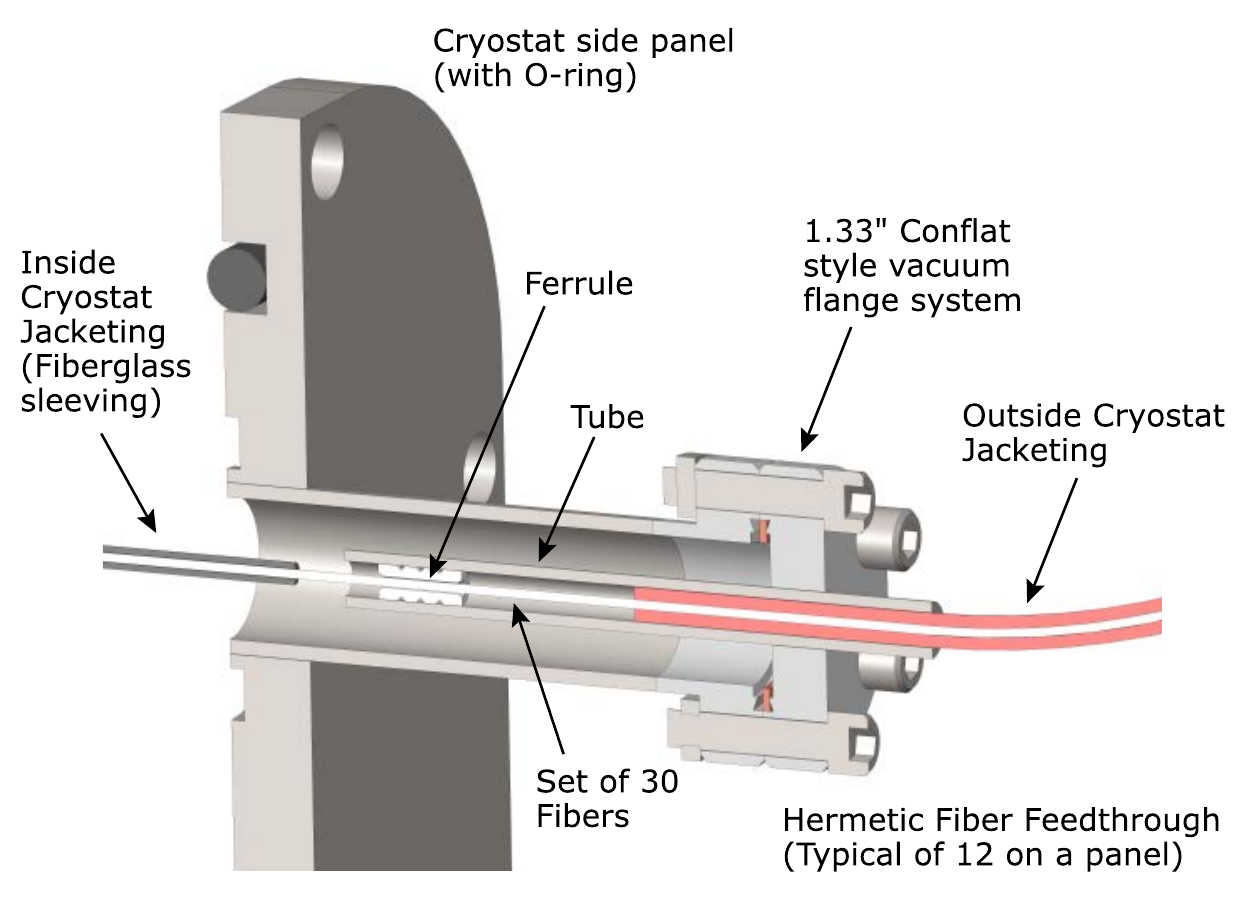}
\caption{A cut-away of the model of one of 12 custom hermetic feedthrough assemblies mounted on a panel at the cryostat wall.  Each feedthrough contains a set of 30 fibers.}\label{fig_fiber_feedthrough}
\end{figure}


As previously mentioned, rather than inserting another connector in the fiber chain, we decided to feed the fibers without break through the cryostat wall using custom feedthroughs (Figure~\ref{fig_fiber_feedthrough}) to minimize throughput losses and focal ratio degradation (FRD).  Each set of 30 fibers coming from the gang connector was fed together through a $1\,\rm{cm}$ long ferrule with an internal hole of $0.050\,\rm{in}$ ($1.3\,\rm{mm}$) diameter.  \edit1{The ferrule was bonded and sealed around the fiber bundle with Masterbond EP37-3FLFAO epoxy and allowed to cure.  This epoxy, which performed the best in FRD testing, has low-outgassing, a very slow room temperature cure (2-3 days), and maintains a flexible consistency after curing.  The latter two characteristics likely contribute to low stress on the fiber.}  Then the outside surface of the ferrule was coated with epoxy and positioned inside a $2.75\,\rm{in}$ ($69.9\,\rm{mm}$) long, $\frac{1}{4}\,\rm{in}$ ($6.4\,\rm{mm}$) diameter 304 stainless steel tube.  (The inside diameter (ID) of the tube is slightly larger than the outside diameter (OD) of the ferrule.)  With the ferrule positioned 3 -- $5\,\rm{mm}$ from the vacuum-end of the tube, a small amount of epoxy was added to cap and seal the ferrule within the tube while still leaving room for additional sealing if necessary.  (In some cases the initial vacuum seal did not meet specifications and additional epoxy was applied as discussed in \S~\ref{feedthrough_vacuum_integrity}.)

Conflat\textsuperscript{\textregistered} style vacuum flanges with $1.33\,\rm{inch}$ nominal OD, vacuum-welded to the tubes prior to the feedthrough assembly described above, were secured to complementary flanges on the cryostat wall during fiber installation.  The knife-edge seal is on the room temperature side of the cryostat wall so installation of fiber assemblies starts by feeding the v-groove terminated ends through an appropriate OFHC copper Conflat\textsuperscript{\textregistered} gasket and then through the hole of a receiving Conflat\textsuperscript{\textregistered} flange on the cryostat wall of the instrument.  Each set of 30 fibers was subsequently pulled through various guides inside the instrument and positioned on the pseudo-slit.  To accommodate differential shrinkage, a single loop of fiber was coiled prior to the pseudo-slit within a tray system (one tray for each fiber set) secured to the top of the Fold Mirror 2 mount.  After the v-groove was installed at the pseudo-slit, the Conflat\textsuperscript{\textregistered} flange was rotated until residual twist of the fiber within the cryostat was removed.  Only then was the flange tightened.  This process was repeated for all ten fiber assemblies and two spare assemblies.

\edit1{Inside the cryostat, uncoated fiberglass sleeving (Alpha Wire P/N PIF-240-18) surrounds the groups of 30 fibers to aid handling.  This material simply consists of braided glass fibers that have been heat treated to remove textile sizing and produce resistance to fraying when cut.\footnote{Alpha Wire, private communication.}  So we did not anticipate any outgassing issues and none have been encountered.}  The sleeving extends to within about 5 to $10\,\rm{cm}$ of the v-groove blocks and is left floating on that end to accommodate the differential coefficient of thermal expansion (CTE) between fiberglass and fibers.  The first batch of fiber links had the sleeving anchored to the vacuum-end of the feedthrough.  But for subsequent batches this end was left floating as well to avoid having to cut away the fiberglass in case it was necessary to add epoxy to improve the feedthrough vacuum seal.

\subsubsection{V-groove Blocks}

\begin{figure}
\epsscale{1.15}
\plotone{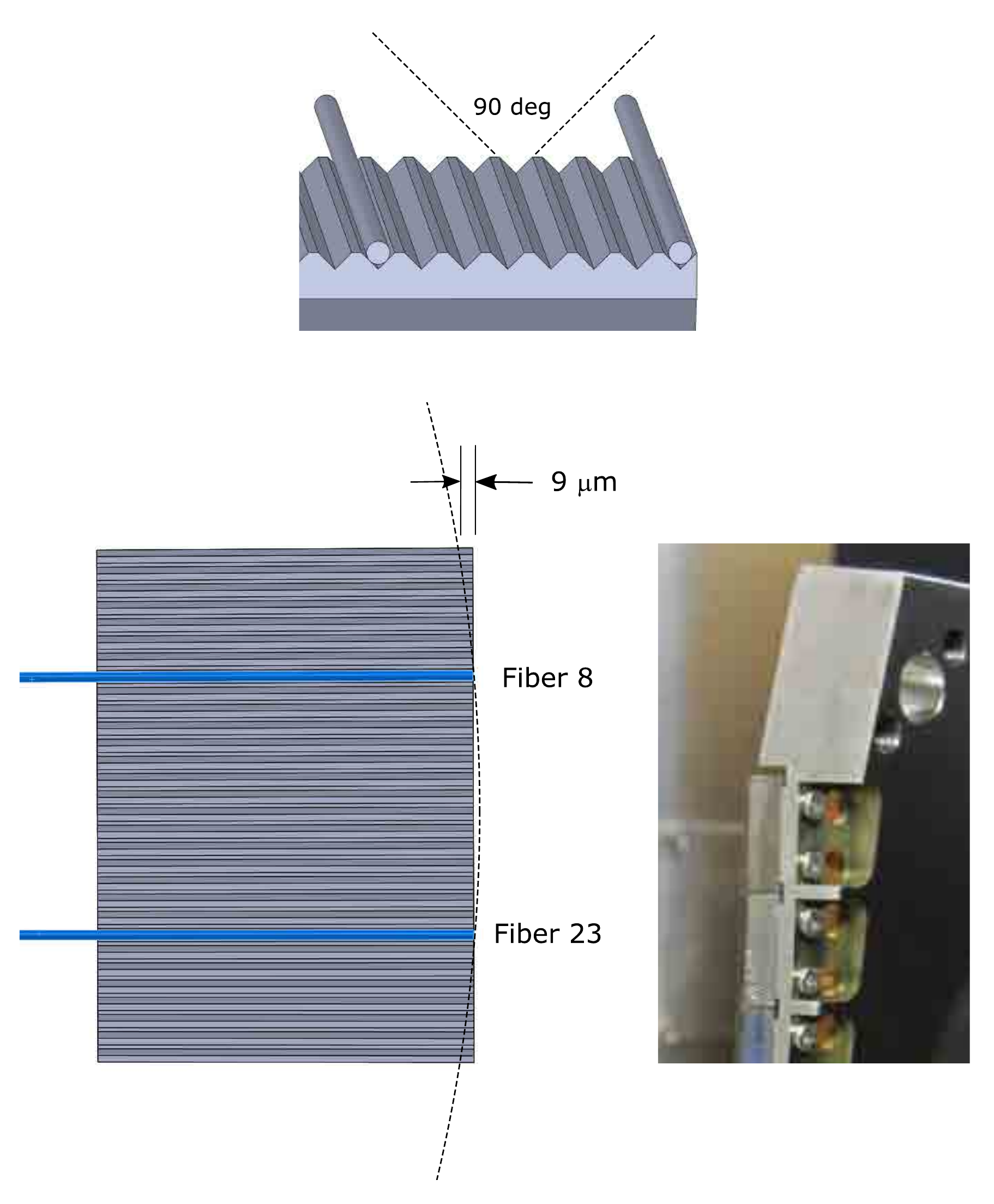}
\caption{\edit1{Details of the v-groove design.  (Top) Individual v-grooves have $90\,\degr$ opening angles.  (Bottom Left) V-groove blocks, polished with flat faces, are mounted on the slit bar (see \S~\ref{pseudo_slit}) so that the tips of Fibers \#8 and \#23 from each MTP\textsuperscript{\textregistered} are on the ideal focus curve (shown in the figure with highly exaggerated radius of curvature).  This results in at most $9\,\micron$ displacement for any fiber tip from the ideal focal position.  (Bottom Right) Back-side of a few v-groove blocks mounted to the slit bar.}}\label{fig_v-groove_design}
\end{figure}

Inside the instrument the fibers terminate in sets of 30 on \edit1{Alloy-39 (A39)} v-groove blocks (Figure~\ref{fig_v-groove_design}).  To minimize mechanical stress on the fibers, and thus FRD, it is important to fabricate the blocks using a material with a similarly low thermal contraction between room temperature and $80\,\rm{K}$ as the fused silica fibers \citep[see, e.g.,][]{lee01}.  While we did not empirically measure the aggregate contraction of our chosen fibers, fused silica is known to have negligible contraction (1 part per million (ppm); \citealt{hah72}) over this range.  In contrast, 6061-T6 aluminum, the major structural metal used elsewhere within the cryostat, contracts by $4{,}000\,\rm{ppm}$ over this range.\footnote{See \url{http://cryogenics.nist.gov} for thermal expansion information on a variety of metals, including 6061-T6 Aluminum and Invar discussed in this section.}  Instead of aluminum, we used Carpenter Technology Corp. Low Expansion ``39'' alloy (A-39) for the blocks.  This nickel alloy steel contracts by $\sim 800\,\rm{ppm}$ over this range.\footnote{Carpenter Technology Corp., private communication.}  While Invar-36 has even lower contraction ($380\,\rm{ppm}$), FRD testing showed that use of A-39 resulted in a marginally smaller FRD than Invar-36 \citep{bru10}.


While some glass ceramics such as MACOR\textsuperscript{\textregistered} and materials such as fused silica may have also been suitable for the v-groove blocks \citep{lee01} given their low thermal contraction, we did not fabricate and test any of these materials because expertise within SDSS-III (at the University of Washington) for fabricating accurate v-groove blocks was based on machining metals with Electric Discharge Machines (EDM).

Individual fibers are epoxied with Master Bond EP29LPSP into separate v-groove block grooves that have $90\,\degr$ opening angles.  \edit1{This epoxy, which has low out-gassing and is rated for service at cryogenic temperatures, did well in FRD testing when paired with A-39.}  The groove-to-groove spacing at the front of the v-groove block is 0.350 mm.  A cap of A-39 is also epoxied onto the assembly to cover the grooves embedded with fibers.  The front face of the assembly is then polished as a unit.  As discussed below, ideally the pseudo-slit and the face of each v-groove block would have a radius of curvature one-half that of the collimator and all fiber tips would be orthogonal to the continuously curving pseudo-slit surface.  In practice the individual v-groove blocks, with all fibers polished along a straight edge, are positioned on a slit bar to form a ``polygon approximation'' of the ideal pseudo-slit shape.  The bottom of each v-groove block has two 0-80 tapped holes so the block can be secured to the slit bar.  Also fabricated from A-39, the slit bar features ten different mounting pads for the v-groove blocks.  Each pad includes two keyhole-shaped features that allow individual v-groove blocks to be captured roughly in place before final positioning and bolt tightening --- the bolt heads of the 0-80 screws are inserted through the oversized holes and then the v-groove is slid forward so the bolt shanks are captured in the parallel keyways.  This v-groove capturing is important as the bolt heads and back side of the pseudo-slit are essentially in the middle of the large instrument so in practice it is difficult to get a bolt driver on them for tightening.

\edit1{Given its complexity, the pseudo-slit was manufactured using a combination of conventional Computer Numerical Control (CNC) mills and a 2D EDM.  The EDM was instrumental in fabricating the slit bar's complex shape needed to mount the v-groove blocks in their various orientations.  For assembly, after a Coordinate Measuring Machine (CMM) was used to check the accuracy of the assembly, the thicknesses of three cold plate mounting pads on the assembly base plate were milled as needed to compensate for manufacturing errors.  This method allowed the center of the slit plate to have $0.001\,\rm{in}$ ($25.4\,\micron$) positional accuracy.}

\subsection{FRD Testing}\label{FRD_testing}

\subsubsection{Component Testing}

Extensive component FRD testing was conducted at the University of Virginia (U.Va.) to choose the best materials, epoxies, and handling procedures for the hermetic feedthrough and v-groove blocks \citep{bru10}.  A far-field FRD testing station was developed in which a camera was placed a known distance from the fiber output.  This allowed determination of how the encircled energy as a function of radial image size (output $f/\#$) varied as test fiber assemblies were changed.  Tests were conducted at APOGEE wavelengths by filtering the input light of a halogen light source with a narrow band filter centered at $1.6\, \micron$ and using a Goodrich room-temperature InGaAs camera.  The station allowed injection of $f/5$ light into the fiber under test to mimic the beam of the Sloan Foundation Telescope, although it did not include a central obscuration to mimic the telescope secondary.

From this testing we learned, in accordance with intuition, that epoxy depth in the feedthroughs should be held to the minimum necessary to allow a sufficient hermetic seal.  FRD increased with epoxy depth, presumably due to increased stress on the fiber outer surface.  Regarding v-groove blocks, the choice of material for the v-groove and cap had the most influence on FRD; choice of epoxy had a secondary effect.

Lastly, lab testing gave us the practical experience that twisting and crossing of fibers, particularly within the potted feedthroughs, had to be minimized to reduce FRD.  Also, the geometric approach of fibers close to rigid, epoxied connections had to be gentle to minimize FRD.

\subsubsection{Spectrograph Input f/\#}\label{fno}


\begin{figure*}
\epsscale{0.85}
\plotone{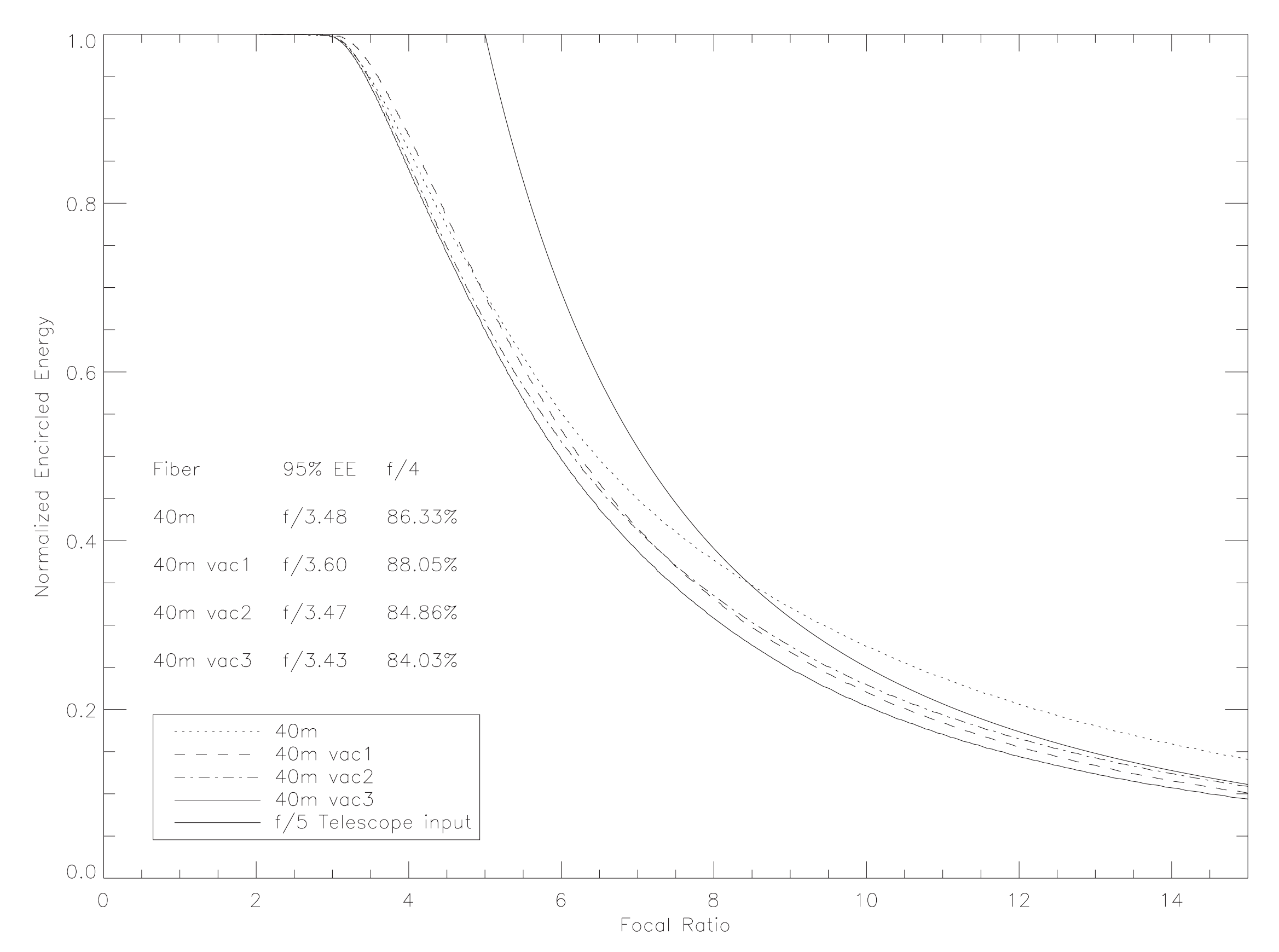}
\caption{\edit1{Normalized encircled energy v. $f/\#$ for the far-field output beam of three $40\,\rm{m}$ test fibers with prototype vacuum feedthroughs and one test fiber without a feedthrough that were illuminated with an $f/5$ beam with uniform illumination.  The results conform with expectations of radial scattering due to FRD.}  The $95\,\%$ encircled energy at $f/3.5$ was adopted as the expected beam to be input into the spectrograph.  This information was used in the final optical design and aperture sizing in the spectrograph.}\label{fig_frd}
\end{figure*}



\begin{figure}
\epsscale{1.05}
\plotone{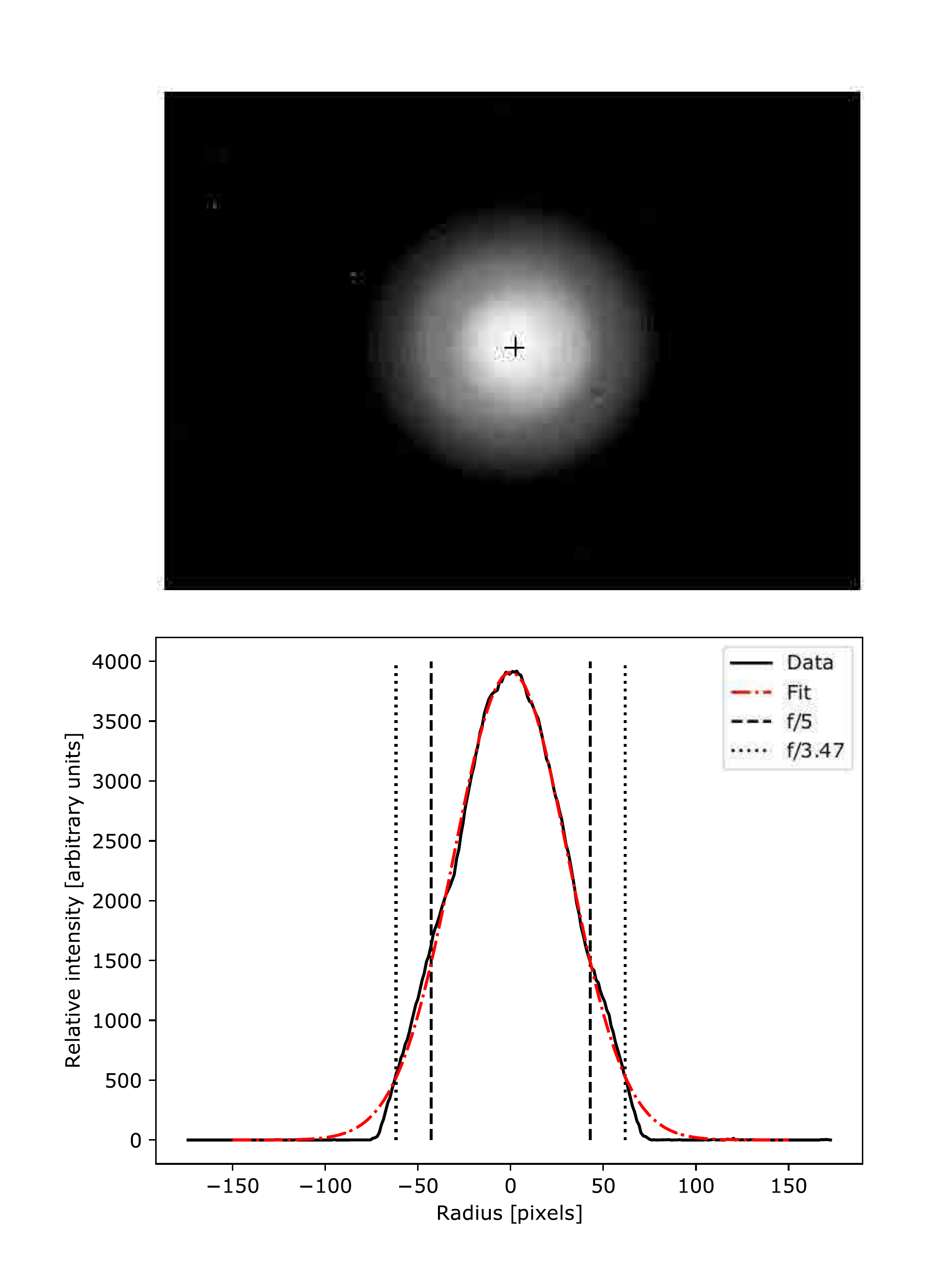}
\caption{\edit1{(Top) The output far field illumination of ``40m vac2'', one of the $40\,\rm{m}$ test fiber assemblies with prototype feedthrough that were tested for FRD.  Figure~\ref{fig_frd} shows the output encircled energy as a function of $f/\#$ for this test fiber when input with uniform $f/5$ illumination.  (Bottom) The average ouput far field illumination profile of multiple radial cuts along with a Gaussian fit, plotted in the same way as Figure 6 of \citet{cla89}.  The central peak and wings of the profile conform to expectations of radial scattering due to FRD.}}\label{fig_farfield}
\end{figure}


Measurements of $40\,\rm{m}$ test fibers with prototype feedthroughs allowed the estimation of the likely $f/\#$ of the light \edit1{exiting the slithead within the spectrograph}.  Figure~\ref{fig_frd} shows normalized encircled energy as a function of $f/\#$ measured for three test fiber assemblies with prototype feedthroughs and a test fiber without a feedthrough, all of which were illuminated with an $f/5$ beam with uniform illumination.  There was $95\,\%$ encircled energy measured within the $f/3.5$ output beam for assemblies with a feedthrough.

\edit1{Figure~\ref{fig_farfield} shows the output far field illumination for ``40m vac2'', one of the three test fiber assemblies with a prototype feedthrough, along with a Gaussian fit to the average of multiple radial cuts.  The profile is well fit by a Gaussian.  The far-field encircled energy measurements and radial profile conform to expectations of radial scattering due to FRD \citep[see, e.g.,][]{eig12} in which light with a uniform input profile is redistributed to produce an output far-field profile with a central peak and wings.  The radial profile in Figure~\ref{fig_farfield} is formatted like Figure 6 of \citet{cla89}, which showed output far field illumination profiles, similarly Gaussian-like to varying extent, as a function of macrobend radius.}

\edit1{Our test results were} used for the final optical design and sizing of apertures within the spectrograph.  \edit1{It should be noted that these tests did not include the MTP\textsuperscript{\textregistered} connector coupling, which will also contribute FRD.  The connectors had not been procured at the time of the tests.  Losses from those connectors are discussed below.}

\subsection{Fiber Assembly Production}

\subsubsection{Fiber Throughput Test Stand}

\begin{figure}
\epsscale{1.2}
\plotone{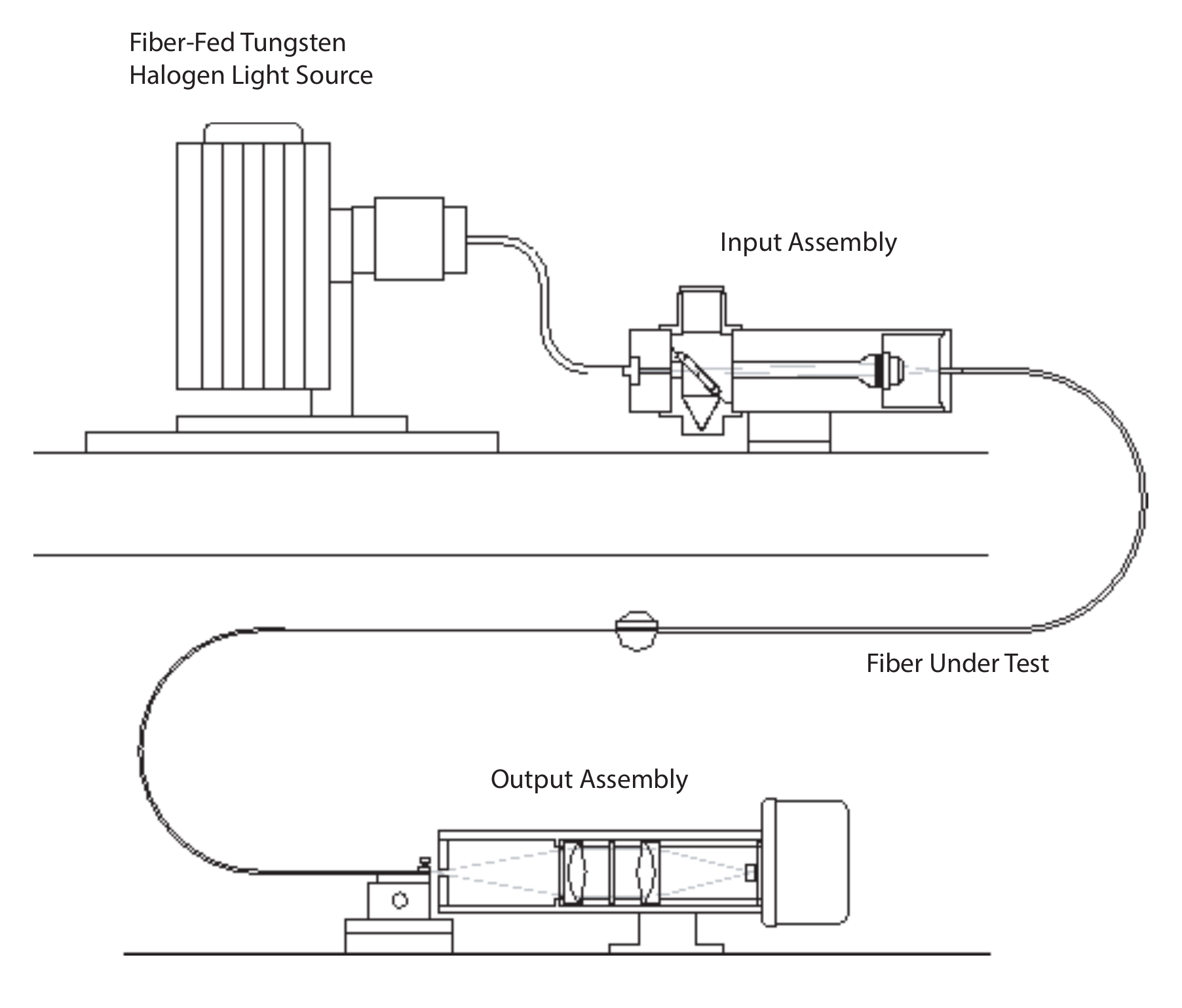}
\caption{A schematic of the SDSS fiber test stand developed for testing fibers for the optical spectrographs.  One test stand was used by the vendor who fabricated the fiber assemblies and one was used by SDSS.  The schematic is from the SDSS Project book \url{http://www.astro.princeton.edu/PBOOK/welcome.htm}.  These testers were also used for testing fibers for APOGEE.}\label{fig_fiber_tester}
\end{figure}

To ensure maximum throughput, all fibers within each harness and fiber link were first tested by \edit1{C-Technologies} to check for compliance with specifications, and then by the APOGEE project, prior to acceptance.  Both the vendor and the APOGEE project used identical re-imaging fiber test stands originally fabricated by the University of Washington for testing the SDSS \edit1{visual} spectrographs \citep{sme13}.  Figure~\ref{fig_fiber_tester} shows the test stand schematic from the original SDSS Project Book.\footnote{\url{http://www.astro.princeton.edu/PBOOK/welcome.htm}}  A Newport model 780 intensity-stabilized tungsten halogen light source provides the illumination.  A front-end unit images the light onto the fiber tip of the fiber under test with an $f/5$ beam, again without central obscuration.  The back-end unit includes an aperture that sets the far-field measurement cone and a Schott BG38 filter.  Reimaging optics refocus the output of the fiber onto a silicon photodiode.  The system response peaks at $\sim 0.6\,\micron$ with $\sim 0.2\,\micron$ FWHM.  The stands were originally designed to test the \edit1{visual} spectrograph fiber harnesses which had sets of single fiber ferrules on the input side (plug plate side) and v-groove blocks on the output end (optical spectrograph slithead side).


The fiber test stands are differential systems.  First, the input and output assemblies are connected and an $f/5$ input cone of visual light illuminates the test stand without a fiber for calibration.  The fiber under test is then measured and the ratio of fiber and calibration illumination is reported.  \edit1{During APOGEE fiber testing it was assumed that the aperture in the back-end unit mentioned above was sized to provide an $f/4$ far-field cone \citep{sme13}.  After-the-fact, the aperture was measured in support of the MANGA survey for SDSS-IV and found to be sized to give an $f/3.2$ far-field cone \citep{dro15}.  So in hindsight, instead of measuring throughput within an $f/4$ cone, APOGEE throughput measurements using the test stand were for the larger $f/3.2$ cone.  Since throughput specifications for the production run of harnesses and fiber links were established based upon the test bench measured performance of the ``first-articles'' produced by the vendor, this measurement discrepancy is not important in the relative sense.  Of course it is important when making judgements about FRD from the data.}

\subsubsection{First Article Testing}

\edit1{\it{Harnesses} ---} Transmission through the fibers in the first set of five harnesses had an average throughput of $92 \pm 0.5\,\%$ using the fiber test stand mentioned above but with special tooling to allow measurements when the output termination was an MTP\textsuperscript{\textregistered} connector instead of a v-groove block.  Ignoring FRD losses, the expected throughput is 93\% after accounting for the Fresnel reflections given the uncoated fiber ends.  The vendor had also fabricated five first articles of ``mapping links'' which are sets of 30 fibers, $3\,\rm{m}$ in length, with an MTP\textsuperscript{\textregistered} termination on one end and a v-groove block on the other.\footnote{\edit1{The mapping links are used during plate plugging in the plug lab to map the correspondence between fiber location in the plug plate and fiber location within the slit \citep{sme13}.  The MTP\textsuperscript{\textregistered}-terminated ends of the mapping links are installed in a gang connector just like the main fiber link for the instrument.  Similarly, the v-groove block ends are arranged in a ``dummy slit'' in the same pattern as used in the instrument.  Laser light sequentially illuminates the fibers at the dummy slit while a camera records the output at the cartridge plug plate.  This ``maps'' the correspondence between fibers at the slit and plate position.  This system eliminates the necessity and tedium of plugging specific fibers into specific plug plate holes.  Instead, the plugging technicians manually mark regions in the plate to which groups of six fibers from each harness can be plugged.  This increases efficiency, because the semi-random plugging can be discovered through the mapping process.  The constraint to general regions results from the fact that, to reduce net fiber length, not all fibers mounted within the cartridge can reach all holes.}}  The average throughput for the first-article mapping links was $91 \pm 1\,\%$.  When the first-article harnesses were coupled in series with the mapping links at the MTP\textsuperscript{\textregistered} connector, the average total throughput was $\sim 84 \pm 2.3\,\%$.  Allowing a small margin, a specification of $82\,\%$ throughput was adopted for all subsequent harnesses produced by the vendor when coupled in series with one of the three best performing first-article mapping links.  The $\sim 84\%$ average throughput for the harness and mapping link coupled together implied FRD losses of $\sim 3\%$ after accounting for losses of about $13\%$ from Fresnel reflections at the four uncoated fiber tip surfaces.

\edit1{\it{Fiber Links} ---}Similarly, first-article fiber links, when tested alone with a modified fiber test stand, had an average throughput of $77.5 \pm 1\,\%$.  Most of the lost light is accounted for by the $13\,\%$ loss due to additional fiber attenuation at the visual bands used by the fiber tester, and $7\,\%$ lost due to the Fresnel reflections at the two fiber tips.  When coupled with the previously tested harnesses, the overall average throughput was $\sim 72.5 \pm 2.5\,\%$.  This result implied there was no additional FRD from the MTP\textsuperscript{\textregistered} coupling.  Thus all subsequent fiber links were specified to meet $70\,\%$ throughput when tested coupled to a first-article harness.

\subsubsection{MTP\textsuperscript{\textregistered} Coupling Performance}\label{MTP_coupling_performance}

The test results discussed above imply the MTP\textsuperscript{\textregistered} connectors did not mitigate the Fresnel losses of the fiber tips being coupled.  They also added small (0 -- $3\,\%$) FRD losses.  This suggests the MTP\textsuperscript{\textregistered} connectors may not be achieving the intended fiber-fiber tip contact.  It should also be kept in mind that the measured results were for a far-field cone of $f/3.2$ --- losses would be larger if measured at $f/3.5$, the expected cone angle input into the instrument.  Given the measurements, we assume a coupling loss of 5\%.  This is supported by the post-facto discovery that the molded fiber holes within the MTP\textsuperscript{\textregistered} \edit1{connectors fabricated for both APOGEE-North and -South} are systematically tilted relative to the normal to the mating face (Gunn et al, in prep.).\footnote{\edit1{The mold for these connectors has since been corrected so the angle will be less than $0.2\,\degr$ on all rows for future APOGEE-like connectors (US Conec, private communication). }}  As discussed in \S~\ref{fiber_tilt_errors}, the mean tilt, after refraction upon exit from each MTP\textsuperscript{\textregistered}, is $0.58\,\degr$.  Given the geometry of coupling MTP\textsuperscript{\textregistered} connectors, this angle is doubled at the connection.  A conservative estimate of the mean loss is $3\,\%$ at the junction, assuming uniform illumination \citep{tsu77}.

In principle, index-matching gels could have been used to reduce connection losses.  But use of gels would have been impractical given the numerous times throughout the night that the connections are changed between cartridges and the high potential for dust contamination.

\subsubsection{Testing Vacuum Integrity}\label{feedthrough_vacuum_integrity}

A long ($3.15\,\rm{m}$) and thin ($38\,\rm{mm}$ outer diameter) custom vacuum chamber was designed and fabricated by U.Va. to verify the vacuum integrity of each hermetic feedthrough prior to installation in the APOGEE cryostat.  This vacuum testing in a separate chamber was important programmatically as it allowed the instrument cryostat to stay open for long periods for optics installation.  Each feedthrough was required to have leakage $< 10^{-8}\,\rm{scc/sec}$ of air at one atmosphere.

In two cases this specification was not met.  The leaks were repaired by applying Masterbond EP29LPSP, a less viscous epoxy, to the fibers on the warm side of the feedthrough.  The epoxy wicked along the fibers and sealed the leaks.

\subsection{Fiber System Robustness}


\begin{figure}
\epsscale{1.2}
\plotone{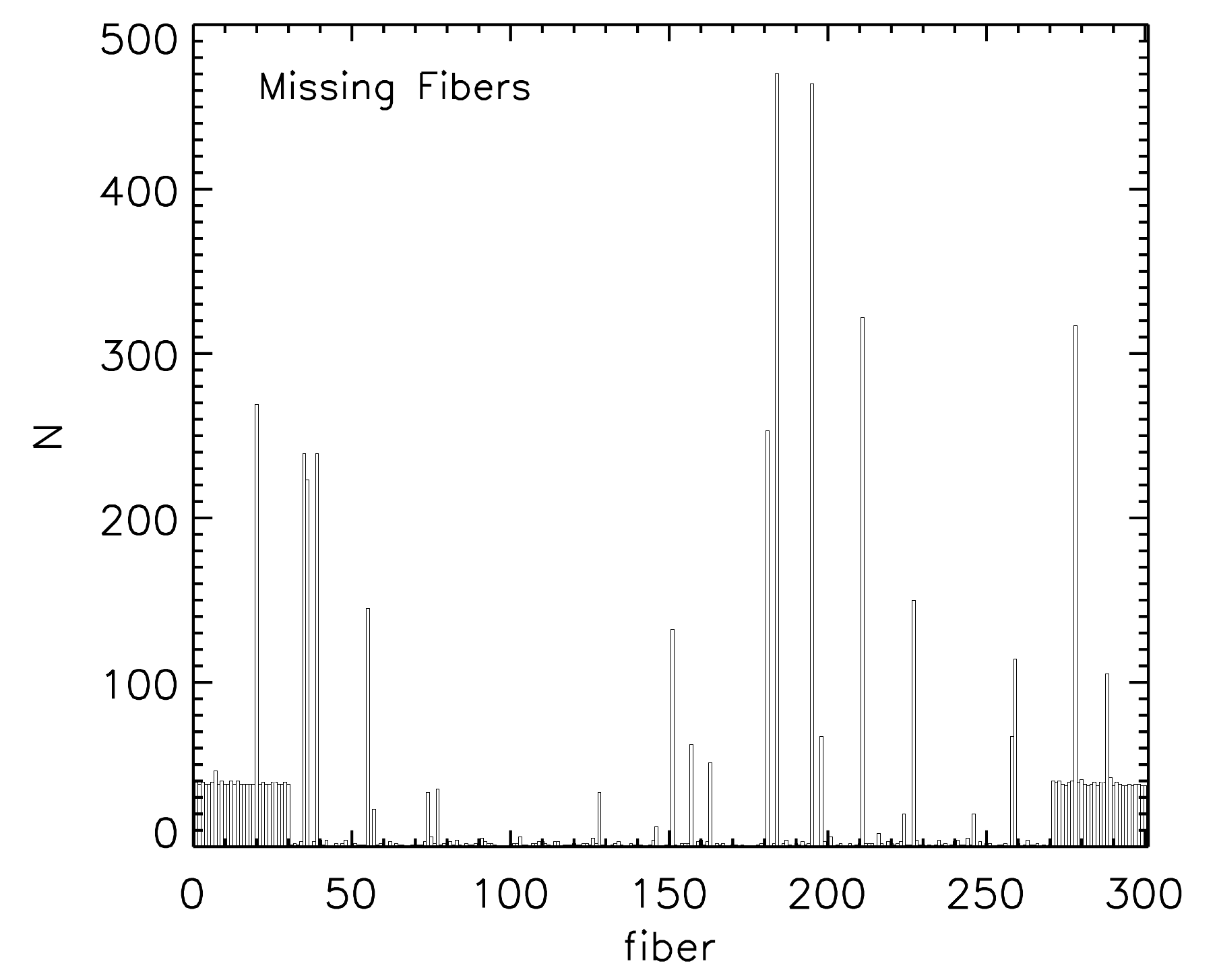}
\caption{The number of times each of the 300 fiber connections to the instrument were missing based on analysis of flat-field observations for each of the $3{,}429$ observed plates between commissioning in 2011 through the start of summer shutdown in 2016.  A fiber is classified as missing if the counts in the spectrum of a telescope flat field taken immediately following a set of observations on the sky is $< 20\%$ of the median flux.  The mean is $\sim 1.25$ missing fibers and the median is 0.  Various events have caused prolonged missing fibers.  E.g., Fiber \# 195 was broken for about six months and MTP\textsuperscript{\textregistered} connectors \# 1 (fibers 1 -- 30) and \# 10 (fibers 271 -- 300) were off-line for several weeks prior to repair during summer shutdown in 2015.}\label{fig_missing_fibers}
\end{figure}

\begin{figure}
\epsscale{1.2}
\plotone{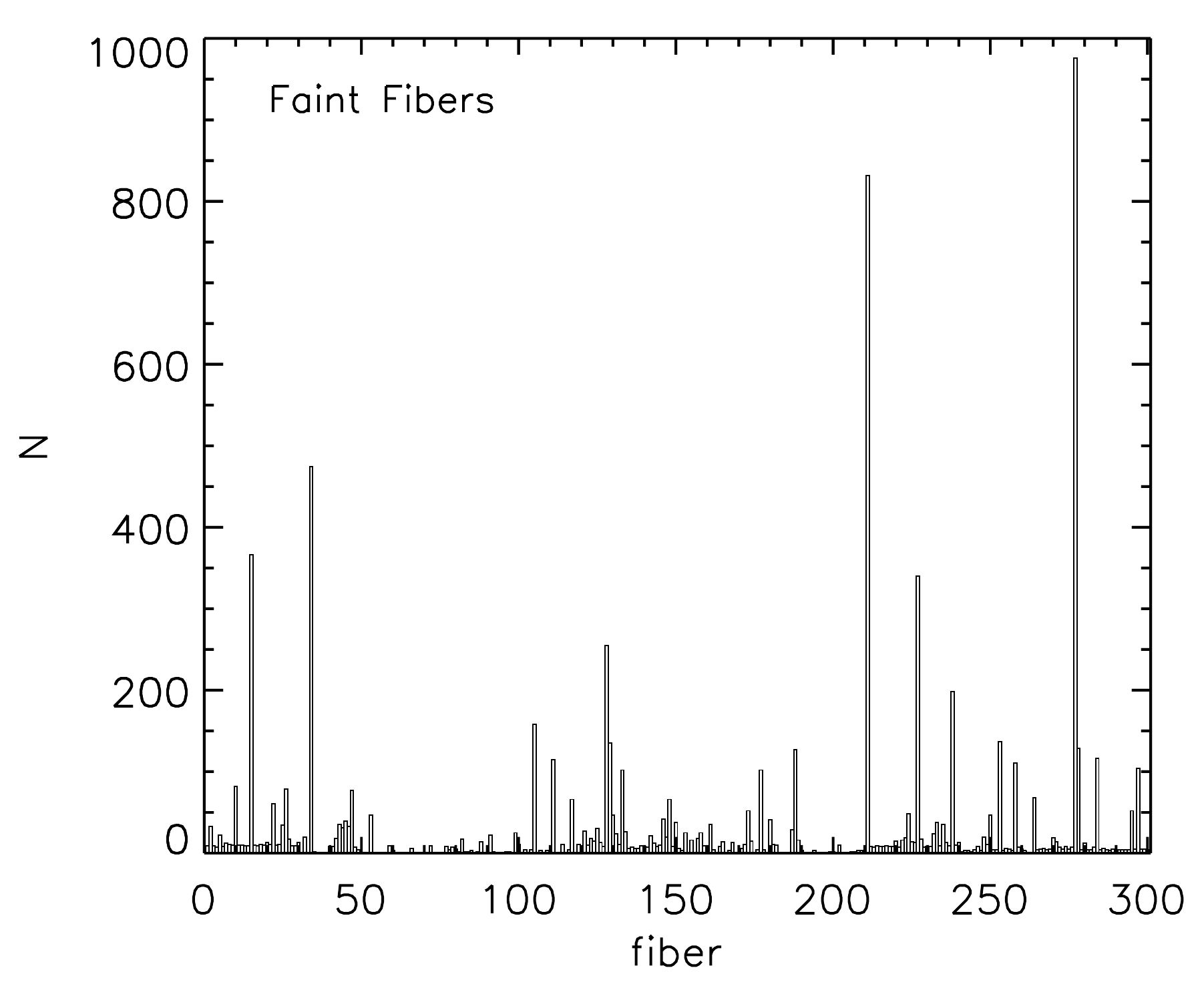}
\caption{The number of times each of the 300 fiber connections to the instrument were faint based on analysis of flat-field observations for each of the $3,429$ observed plates between commissioning in 2011 through the start of summer shutdown in 2016.  A fiber is classified as faint if the counts in the spectrum of a telescope flat field taken immediately following a set of observations on the sky has $\ge 20\%$ and $< 70\%$ of the median flux.  The mean is $\sim 1.52$ faint fibers and the median is 1.}\label{fig_faint_fibers}
\end{figure}

For given observations with one of the nine cartridges\footnote{The APOGEE-1 survey used eight cartridges whereas the APOGEE-2 survey uses nine at APO.}, which together were manually plugged and unplugged $3{,}429$ times from commissioning in 2011 through the start of summer shutdown in 2016, there were a small number of ``missing'' and ``faint'' fibers.  A fiber is classified as missing if the counts in the spectrum of a telescope flat field taken immediately following a set of observations on the sky is $< 20\%$ of the median flux.  Similarly, a faint fiber has relative transmission $\ge 20\%$ and $< 70\%$.  Based on survey statistics for all observations across all cartridges in the period mentioned above, a cartridge with 300 fiber capacity had a mean of $\sim 1.25$ missing fibers and $\sim 1.52$ faint fibers.  The median is 0 missing fibers and 1 faint fiber.  Figure~\ref{fig_missing_fibers} and Figure~\ref{fig_faint_fibers} show the number of times each fiber has been missing and faint, respectively, for an observed plate.


There are at least two reasons for missing fibers.  First, individual fibers within fiber harnesses (plug-plate -- gang connector) can break during plugging if handled improperly or from wear.  Once broken, and until fiber harness replacement, these fibers are consistently identified as missing each time the cartridge is used.  Fibers can be missing for just one or a few days -- sometimes a fiber ferrule will drop out of the plug-plate during cartridge movement to the telescope or during cartridge mounting/un-mounting at the telescope.  As a cartridge can stay plugged for more than one day if it is likely to be re-observed soon, such a dropped fiber can stay flagged as missing until the cartridge is re-plugged.

Similarly, faint fibers can be permanently or frequently faint.  Transient faint fibers, either just one or all fibers of an MTP\textsuperscript{\textregistered} connector, have occurred.  They often occur when dust or debris, either on a connector face or within a connector guide pinhole, prevents sufficiently close connection.  The cleaning of connector guide pinholes often solves cases of recurrent faint fibers in daily calibration exposures across most or all fibers of a given connector.


One fiber connection (\# 195) within the fiber link was inadvertently broken in 2014 December during emergency repairs of the gang connector system after it was discovered that the sheathing protecting the fiber link bundles was splitting close to the $\sim 90\degr$ bend of the fiber link immediately downstream of the gang connector at the podium (see  Figure~\ref{fig_gang_connector_ports}).  The orientation of the gang connector sockets in the podium have since been changed to remove this bend and thus reduce stress on the fiber sheathing.

In the early summer of 2015 a failure of the pin-push assembly of an MTP\textsuperscript{\textregistered} connector in the gang connector pin end caused a systematic failure of two MTP\textsuperscript{\textregistered} connectors shortly before summer shutdown.  One of the stainless steel guide pins pulled out of the MTP\textsuperscript{\textregistered} connector while plugged into a cartridge.  When the gang connector was re-inserted into the cartridge to try and resolve a low throughput connection, the pin wedged into the gap between the MTP\textsuperscript{\textregistered} ferrule and connector housing.  This resulted in a crack in the ferrule of the MTP\textsuperscript{\textregistered} \# 1 of the fiber link, failure of the rear housing of MTP\textsuperscript{\textregistered} \# 10, damage to the face of the ferrule of MTP\textsuperscript{\textregistered} \# 1 in the dense pack socket, and damage to the ferrule face of MTP\textsuperscript{\textregistered} \# 1 on the cartridge 2 socket.

Investigation of the failed pin-push assembly after the incident showed that a component where the pin lands upon mating had broken and that fatigue was a possible cause.  This failure underscored that the MTP\textsuperscript{\textregistered} ferrules in the fiber link had probably undergone $\sim 10,000$ plugging cycles at this point. This is about $10 \times$ more than their design life. In addition, the MTP\textsuperscript{\textregistered} connectors in the dense pack socket had undergone $\sim 5,000$ plugging cycles or about $5 \times$ their design life expectancy.  As a result, all of the connector components that could be replaced in both the fiber link and dense pack socket were replaced with new components during the subsequent summer shutdown in 2015. In addition, MTP\textsuperscript{\textregistered} ferrules were replaced on the fiber link position 1 and 7 (connector 7 had broken fiber \# 195 mentioned above). Position 1 in the dense pack socket was polished to remove any possibility that the collision with the pin had caused any material to protrude above the fiber faces.  Periodic replacement of connector components for preventative maintenance is planned for the future given the high usage of these connectors compared to their design lifetimes.

\section{Spectrograph Optics and Opto-Mechanics}\label{optics_section}

\begin{figure}
\epsscale{1.2}
\plotone{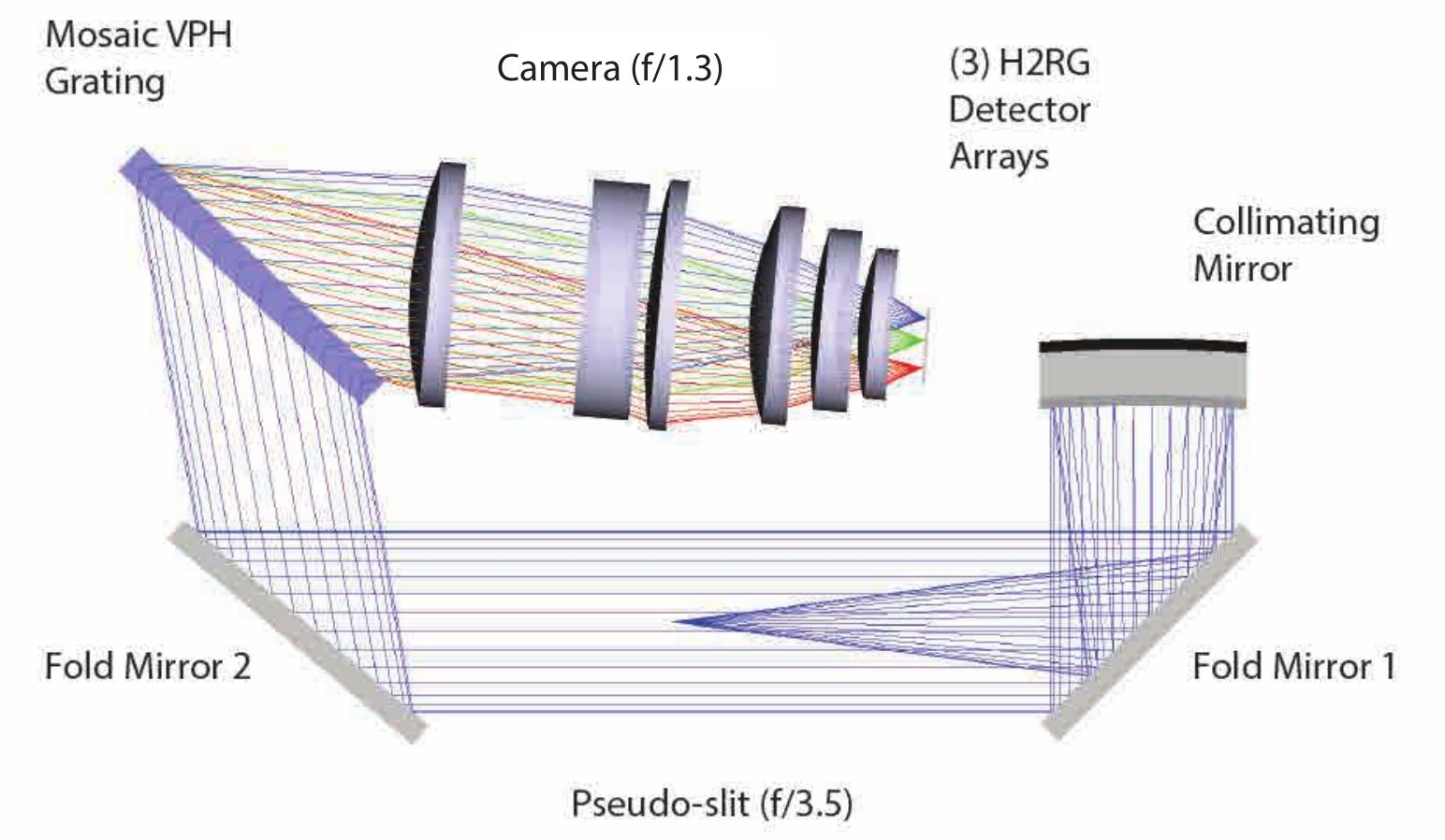}
\caption{The instrument optical schematic.  Diverging light from the fiber tips located at the pseudo-slit is collimated by the spherical collimating mirror.  The collimated light, steered by two fold mirrors, is dispersed by the mosaic VPH grating.  A 6-element refractive camera focuses the resulting spectrum onto three (3) H2RG detector arrays mounted side-by-side. The 300 fibers are stacked together in the direction orthogonal to the plane of the page.}\label{fig_optics_schematic}
\end{figure}

Borrowing heavily from successful spectrograph designs such as the SDSS/BOSS spectrographs \citep{sme13}, APOGEE follows a ``classical'' optical design with an on-axis uncorrected Schmidt camera for a collimator, a dispersive optic located in the collimated beam, and a refractive camera to focus the dispersed light.  The optical prescription is given in Table~\ref{tbl-prescription} and a schematic is shown in Figure~\ref{fig_optics_schematic}.  Fold mirrors are used in the design to efficiently package the optics on a rectangularly-shaped cold plate within the cryostat.

The multiplexing power of the instrument is in its ability to spectrally disperse the light of 300 fibers simultaneously.  While the light from each fiber is simply dispersed in first order across a narrow wavelength range, the relatively high resolution requirement meant that three detector arrays were necessary to record the spectra.  A mosaic volume phase holographic (VPH) grating disperses the light. Successful design and fabrication of a large mosaic VPH was an enabling technology for APOGEE --- use of a large reflection grating would have required an impractically large refractive camera.  Practicality aside, properly designed VPH gratings can give performance that matches the APOGEE science requirements.  With appropriate choices of line frequency and depth of the grating structure (see, e.g., \citealt{bar00}), VPH gratings can provide high efficiencies over narrow wavelength ranges in first order with minimal light dispersed into higher orders.

\edit1{Sized to collect the collimated beams exiting the VPH Grating with a span of angles based on wavelength, the large camera re-images the fiber tips of the pseudo-slit onto the detector arrays.  The camera $f/\#$ can be defined in various ways. For instance, if one pretends all optics preceding the camera are large enough not to vignette any rays, and the camera is illuminated on-axis with collimated light having the maximum diameter that can be accepted by the camera, the $f/\#$ is $\approx f/1.0$.  That is, the camera focal length and the clear aperture diameter of the first lens are comparable.  But the size of the camera elements were not designed for the purpose of collecting a single large diameter collimated beam.  Rather, they are designed to collect smaller collimated beams diffracted in angle by the preceding VPH Grating (as seen in Figure~\ref{fig_optics_schematic}), which is located close to the system pupil.  There is no aperture within the camera that serves as a stop.}

\edit1{Masks at the Collimator and VPH Grating are sized to clip any beams faster than approximately $f/3.7$ and $f/3.85$, respectively, that illuminate the system.  The sizing was purposefully chosen so that most vignetting of fast beams occurs upstream of the camera, with the goal of minimizing vignetting and potential scattering within the camera itself.  When the spectrograph is illuminated with $f/3.85$ the camera has an $f/\#$ of $f/1.3$, which corresponds to camera focal length divided by entrance pupil diameter.  This gives a system demagnification of $3.85/1.30 = 2.96$.}

\edit1{The fiber-fiber spacing of $0.350\,\rm{mm}$ within v-groove blocks at the pseudo-slit is demagnified to $0.118\,\rm{mm}$ (6.6 pixels) at the detector.  With a median FWHM for fiber core images in the spatial direction of 2.1 pixels there is 4.5 pixels of space between fiber ``traces''.  Our aims for fiber-fiber spacing were to maximize the number of fibers that could be imaged onto the detectors, while also following SDSS experience with the visual spectrographs that fibers should be no closer than about two core diameter's separation to keep crosstalk between fibers manageable.  In addition to providing ample spacing between fibers, cross-talk is minimized through the accurate PSF modelling and substraction of flux from adjacent fibers \citep{nid15}.  And programmatically the cartridges are plugged so the relative brightness of targets illuminating adjacent fibers are in the sequence faint, medium, bright, bright, medium, faint, etc.  This ``fiber management'' scheme, along with survey magnitude limits, are discussed in \citet{maj17} and \citet{zas13}.}

There are 8.3 pixels between end fibers of adjacent blocks.  Above the top and below the bottom fiber trace there is an extra $\sim 25$ (10) rows of pixels on the blue (red) end of the detector mosaic, respectively, for alignment contingency.  \edit1{The variation of unused detector area at the top and bottom is due to a change in image scale caused by a variation of focal length and spatial distortion with wavelength.}  The changing focus position with wavelength is accommodated by adjusting the flange focal distances (normal distance from the camera back mounting flange to the detector array center), and individually tilting the detector arrays about the spatial axis, such that the detective area is progressively farther from the camera as wavelength increases.


In the spectral direction the size of the reimaged fiber core varies across the three detectors from $\approx 1.7$ pixels at $1.53\, \micron$ to $\approx 3.0$ pixels at $1.68\, \micron$ given the anamorphic magnification variation with wavelength.  We purposefully accepted undersampled spectra at the blue end of the wavelength coverage to maximize wavelength coverage; undersampling is remedied by spectral dithering in 0.5 pixel steps during observing to halve the sampling period.  The dithering mechanism is discussed in \S~\ref{dither_mechanism}.

As will be mentioned in \S~\ref{array_mount}, there are $2.9\,\rm{mm}$ gaps, the smallest feasible, between adjacent detector arrays.  Therefore, the detector arrays provide the following nominal wavelength coverage:  blue ($1.514$ -- $1.581\,\micron$), green ($1.585$ -- $1.644\,\micron$), and red ($1.647$ -- $1.696\,\micron$).

Zemax (now called OpticStudio\textsuperscript{\textregistered}) optical design software (Zemax LLC, Kirkland, WA) was used to design the instrument optics.


\begin{deluxetable*}{lcccc}
\tabletypesize{\scriptsize}
\tablewidth{0pt}
\tablecaption{APOGEE Spectrograph As-built Prescription\label{tbl-prescription}}
\tablehead{\colhead{Surface} & \colhead{Radius\tablenotemark{a} (mm)} & \colhead{Thickness\tablenotemark{a,b} (mm)} &
\colhead{Material} & \colhead{Nominal Size\tablenotemark{c} (mm)}}

\startdata
Object\tablenotemark{d} & -1051.6 & -1051.6 & \nodata & \nodata \\
Stop\tablenotemark{d} & $\infty$ & 1051.6 & \nodata & \nodata \\
Pseudo-slit & -1051.6\tablenotemark{e} & 709 & 300 Polymicro FIP120170190 fibers & 105.53 (fiber 1 -- fiber 300) \\
Fold Mirror 1\tablenotemark{f} ($1^{\rm{st}}$ reflection) & $\infty$ & -343.281 & Corning 7980 Fused Silica Substrate & $425.45 \times 488.95 \times 38.1 $\\
Collimating Mirror\tablenotemark{g} & -2107.22 & 343.281 & Hextek Gas Fusion Substrate & $311.15 \times 514.35 \times 85 $ \\
Fold Mirror 1\tablenotemark{h} ($2^{\rm{nd}}$ reflection) & $\infty$ & -709 & same mirror as Fold Mirror 1 & \nodata \\
Pseudo-slit & \nodata  & -551.6  &  \nodata & \nodata \\
Fold Mirror 2\tablenotemark{i} & $\infty$ & 500  & Corning 7980 Fused Silica Substrate & $481.08 \times 361.95 \times 38.1 $ \\
VPH Entrance Face\tablenotemark{j} & $\infty$ & 25.4 & Corning 7980 Fused Silica & $305 \times 508 \times 25.4$ \\
VPH Grating\tablenotemark{k} & $\infty$  & 25.4 & Corning 7980 Fused Silica & $305 \times 508 \times 25.4$ \\
VPH Exit Face\tablenotemark{l} & $\infty$ & 210 &  \nodata & \nodata \\
Camera Lens 1 (R1)\tablenotemark{m} & 601.921 & 45.039 & Silicon\tablenotemark{n} & $\phi \,387.0$ \\
Camera Lens 1 (R2)\tablenotemark{o} & 717.838 & 239.190 & \nodata & \nodata \\
Camera Lens 2 (R1) & -625.789 & 40.099 & Corning 7980 Fused Silica\tablenotemark{p} & $\phi \,379.5$ \\
Camera Lens 2 (R2) & 760.746 & 34.073 & \nodata & \nodata \\
Camera Lens 3 (R1) & 1013.919 & 45.099 & Silicon\tablenotemark{n} & $\phi \,393.5$ \\
Camera Lens 3 (R2) & $\infty$ & 109.273 & \nodata & \nodata \\
Camera Lens 4 (R1) & 418.282 & 45.074 & Silicon\tablenotemark{n} & $\phi \,344.5$ \\
Camera Lens 4 (R2) & 490.718 & 49.489 & \nodata & \nodata \\
Camera Lens 5 (R1) & 705.020 & 35.058 & Corning 7980 Fused Silica\tablenotemark{p} & $\phi \,285.0$ \\
Camera Lens 5 (R2) & 261.501 & 35.228 & \nodata & \nodata \\
Camera Lens 6 (R1) & 374.498 & 39.555 & Silicon\tablenotemark{n} & $\phi \,236.5$ \\
Camera Lens 6 (R2) & 532.598 & 61.489\tablenotemark{q} & \nodata & \nodata \\
Detector Arrays\tablenotemark{r} & \nodata & \nodata & (3) HAWAII-2RG & $36.86 \times 36.86$ each \\
\enddata

\tablenotetext{a}{Cold dimensions ($77\,\rm{K}$).}
\tablenotetext{b}{Distance to next optical face or element.  Change of sign indicates a reflection.}
\tablenotetext{c}{Warm dimensions ($293\,\rm{K}$).}
\tablenotetext{d}{Zemax prescription starts with the object and stop.  The stop diameter is controlled by the prescription aperture which is defined as object space numerical aperture = 0.166667 with Gaussian apodization and apodization factor 1.58 to mimic the illumination measured during FRD testing as described in \S~\ref{fno}. Field positions for the middle and ends of the pseudo-slit are defined using the ($\rm{X}$,$\rm{Y}$) coordinates of ($0$,$0$) and ($\pm 52.765$,$- 3.623\,\rm{mm}$), respectively.}
\tablenotetext{e}{The idealized pseudo-slit surface is the intersection of a sphere with radius -1051.6 mm and a cylinder with radius 386 mm, offset laterally by the same amount.  See the text for further details.}
\tablenotetext{f}{Implement in the order Tilt About $\rm{X} = 45\,\deg$, Thickness $= -0.432\,\rm{mm}$, Mirror, Thickness $= 0.432\,\rm{mm}$, Tilt About $\rm{X} = 45\,\deg$.}
\tablenotetext{g}{Implement in the order Tilt About $\rm{X} = 0.0175\,\deg$, Mirror, Tilt About $\rm{X} = -0.0175\,\deg$.  A mask in front of the Collimator has inside dimensions $491.9\,\rm{mm} \times 293.1\,\rm{mm}$ along with rounded internal corners.}
\tablenotetext{h}{Implement in the order Tilt About $\rm{X} = -45\,\deg$, Thickness $= -0.432\,\rm{mm}$, Mirror, Thickness $= 0.432\,\rm{mm}$, Tilt About $\rm{X} = -45\,\deg$.}
\tablenotetext{i}{Implement in the order Tilt About $\rm{X} = -50\,\deg$, Mirror, Tilt About $\rm{X} = -50\,\deg$.}
\tablenotetext{j}{Implement starting with Tilt About $\rm{X} = -54.061\,\deg$.  A mask in front of the VPH Grating has inside dimensions $464.7\,\rm{mm} \times 272.8\,\rm{mm}$ along with complex internal corners.}
\tablenotetext{k}{VPH ``groove'' density: $1009.345\,\rm{lines}\,\rm{mm^{-1}}$}
\tablenotetext{l}{Tilt About $\rm{X} = 54.061\,\deg$ after surface.}
\tablenotetext{m}{Decenter $\rm{Y} = 15\,\rm{mm}$, Tilt About $\rm{X} = -2.963\,\deg$ before first camera surface.}
\tablenotetext{n}{Index of refraction from \citet{fre06} at $-188\,\rm{C}$.}
\tablenotetext{o}{Conic Constant: $k=0$; Aspheric Coefficients: $(0 \times 10^{-5})y^2 + (2.768 \times 10^{-10})y^4 + (1.023 \times 10^{-15})y^6 - (1.756 \times 10^{-22})y^8$.}
\tablenotetext{p}{Index of refraction from \citet{lev08} at $-188\,\rm{C}$.}
\tablenotetext{q}{The center of the middle (green) detector array surface is nominally $49.39\,\rm{mm}$ normal distance from the camera back mounting flange to the detector array center (flange focal distance) and offset laterally $25.69\,\rm{mm}$.  Relative to the green detector array, the flange focal distances for the blue and red detector arrays are $-0.48\,\rm{mm}$ (blue) and $+0.37\,\rm{mm}$ (red).}
\tablenotetext{r}{Detector array tilts: Blue, $-0.81\,\degr$; Green, $-0.59\,\degr$; Red, $-0.50\,\degr$.  Tilts are in a direction such that the long wavelength end of each detector array is farther from Camera Lens 6.}

\end{deluxetable*}


\subsection{Pseudo-slit}\label{pseudo_slit}

\begin{figure}
\epsscale{1.2}
\plotone{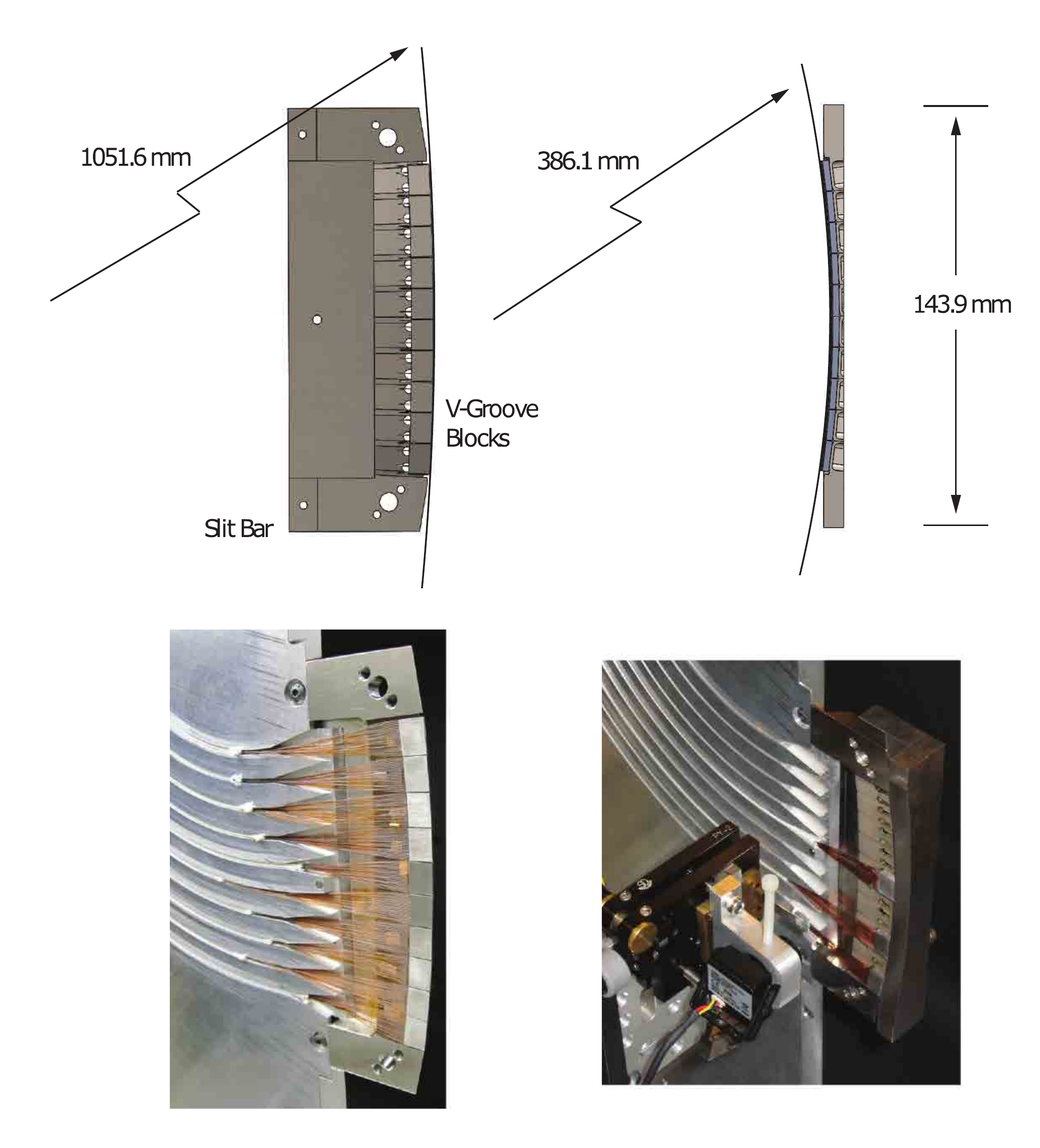}
\caption{The APOGEE pseudo-slit.  (Top Left) Side view showing the radius of curvature of the ideal pseudo-slit curve.  The v-groove blocks are mounted on the slit bar so the tips of Fibers \#8 and \#23  from each MTP\textsuperscript{\textregistered} connector are on this curve.  (Top Right) Front view showing the lateral radius of curvature so the slit image on the detector is linear.  (Bottom Left) As-built pseudo-slit \edit1{with all v-groove blocks installed.}  (Bottom Right) Pseudo-slit after installation of the first \edit1{three} v-groove blocks and fiber assemblies.  The guide for v-groove positioning is temporarily installed in front of the slit bar and the CCD camera used to magnify the v-groove block area is in the foreground.}\label{fig_pseudo_slit}
\end{figure}

Just as photographic plates were bent in Schmidt telescopes to conform to the curved focal surface formed by the spherical primary mirrors, ideally all fiber tips of the pseudo-slit would form a uniform curved surface that has a radius of curvature one-half that of the Schmidt collimating mirror, and each fiber axis would be orthogonal to the pseudo-slit surface, i.e., point back to the center of curvature.  But the complexities of polishing fiber tips on a curved surface, and the practical advantages of fabricating fiber link assemblies in relatively small sets of 30, led to a design (Figure~\ref{fig_pseudo_slit}) in which v-groove blocks with flat polished faces are mounted on a slit bar with ten discrete flat mounting locations to form a polygon approximation of the ideal slit curvature.


When the v-groove blocks are correctly positioned, the fiber tips for fibers \#8 and \#23 of each set of 30 are on the ideal focal curve \edit1{(see Figure~\ref{fig_v-groove_design})} and all other fiber tips are displaced at most $9\,\micron$ (in the focus direction) and are therefore out of focus by $< 2\,\micron$ at the detector arrays given the demagnification between object and image space.  This image-space defocus is very small compared to the pixel-limited depth of focus of \edit1{$\approx 23\,\micron$}, based on $18\,\micron$ pixels and \edit1{$\approx f/1.3$} light at the detector arrays.  Small lateral displacements in fiber positions, which result in spectral traces displaced in the dispersion direction, have no practical impact as each trace is independently wavelength calibrated during data reduction.

Because the v-groove block faces are polished flat, but the fibers lie in grooves that fan out from a distant center of curvature, the fiber tips within each block will be polished with faces that are increasingly tilted relative to the fiber axis.  The tilt angle $\alpha$, in radians, will be approximately $y (\rm{mm})/1051.6\,\rm{mm}$, where $y$ is the distance from the fiber tip to the center of the v-groove block face and the nominal center of curvature is $1051.6\,\rm{mm}$ distant.  Left as is, the emitted light would refract by an angle $1.44 \times \alpha$ away from the fiber axis, where 1.44 is the approximate fiber core index of refraction. These varying angles exiting the v-grooves, largest for fibers at the edge of the v-groove, would take different paths through the instrument optics and create variations in aberration correction and vignetting with fiber position.  To counteract this effect the grooves were machined as if they originated from a center of curvature $\approx 1.44 \times 1051.6\,\rm{mm}$ distant.

The shape of the pseudo-slit also linearizes the image of the slit at the detector.  Normally, the monochromatic image of a straight slit is curved because light that encounters the dispersing optic with non-zero out-of-plane angle, i.e., from slit positions away from the slit center, for a given in-plane angle of incidence, will have a different exit angle in accordance with the general grating equation (\ref{grating_equation}),

\begin{equation}\label{grating_equation}
  \frac{m\lambda}{d} = \cos(\gamma)(\sin(\alpha) + \sin(\beta))
\end{equation}

\noindent
where $m$ is the grating order, $d$ is the groove spacing, and $\alpha$ and $\beta$ are the angles of incidence and exit, respectively, and $\gamma$ is the out-of-plane angle.  In the case of a fiber-fed spectrograph, the slit image will appear curved since $\beta$ must change to compensate for increasing $\gamma$ for fiber numbers away from the center of the pseudo-slit.  Depending on the severity, a curving slit image can waste detector real estate in the dispersion direction.  Were the slit image curved, the maximum wavelength coverage jointly spanned by all fibers for the blue (red) detector array would be reduced by $3\%$ ($4\%$).  Appropriately curving the pseudo-slit laterally could negate this effect.  Noticing that the High Efficiency and Resolution Multi-Element Spectrograph \citep[HERMES;][]{bar08} design included this feature, we followed suit.  Thus the pseudo-slit is designed to have a $386.1\,\rm{mm}$ cold lateral radius of curvature to linearize the slit image at $1.625\,\micron$.  In practice, slit images for all other APOGEE wavelengths are effectively straight as well.

The only tight tolerance for the fiber tip positioning was in the focus direction.  A tolerance of $\pm 0.002\,\rm{in}$ ($\pm 0.050\,\rm{mm}$) was adopted.  Fiber tip displacements of this amount caused spot radii at the detector to increase by $1.3\,\micron$.  In practice this was a tolerance on combined positioning errors of the fibers within v-groove blocks, block position on the slit bar, and slit bar shape.  The accuracy with which the v-groove blocks were actually mounted is discussed in \S~\ref{align&test}.

To match the thermal contraction of the v-groove blocks, the slit bar is also made from A-39.  Since the slit bar attaches to an aluminum slit plate, the bar is pinned to the plate at the optical axis to ensure a correct location when the instrument is cold despite the different materials.  Two other flat indexing surfaces in the slit bar ensure correct spatial orientation.

Overall, the face-on pseudo-slit dimensions are $5.665\,\rm{in} \times 0.269\,\rm{in}$ ($143.89\,\rm{mm} \times 6.83\,\rm{mm}$) attached in front of a $1/4\,\rm{in}$ ($6.35\,\rm{mm}$) thick vertical plate.  Obscuration of the returning collimated beams is at most $\approx 1.7\%$.

\subsection{Collimating Mirror}\label{collimating_mirror}

The spherical collimating mirror is a lightweight Gas-Fusion$\rm{^{TM}}$ mirror from Hextek Corp. (Tucson, AZ).  A lightweight mirror was critical as we wanted to provide tip-tilt-piston functionality using a similar design as the one used for the BOSS Spectrograph collimating mirrors \citep{sme13}.  An important factor in the decision to use Hextek substrates was investigations at NASA Huntsville for JWST ground testing mirrors that showed a 0.25 m diameter Hextek mirror maintained good surface figure (change of only $\approx 25\,\rm{nm}$ RMS) when cooled to cryogenic temperatures \citep{had04}.

The Hextek borosilicate blank is composed of a $6 \times 10$ array of Schott Duran\textsuperscript{\textregistered} glass cylinders sandwiched and fused between Schott Borofloat 33\textsuperscript{\textregistered} face sheets with $9\,\rm{mm}$ ($7.5\,\rm{mm}$) nominal thickness for the front (back).  Material for each face sheet comes from the same parent factory glass plate to ensure a close match in CTE.  For the same reason, all glass cylinders come from the same lot.  The concave $311 \times 514 \times 85\,\rm{mm}$ mirror, with $\approx 2.1\,\rm{m}$ radius of curvature, was slumped during the fusion process to the rough radius of curvature by Hextek and then polished to $< 1/2\,\rm{wave}$ (for $\lambda = 632.8\,\rm{nm}$) PV surface figure across the clear aperture by Nu-Tek Precision Optical Corp. (Aberdeen, MD).  The mirror was coated with a protective gold coating by Infinite Optics (Santa Ana, CA).

The design thickness of the mirror was $89\,\rm{mm}$ so between the fusion and polishing process the mirror overall thickness was $4\,\rm{mm}$ thinner than anticipated.  Nonetheless, no obvious quilting (print-through) due to overly thin face sheets across the cylinders was seen.


\begin{figure}
\epsscale{1.1}
\plotone{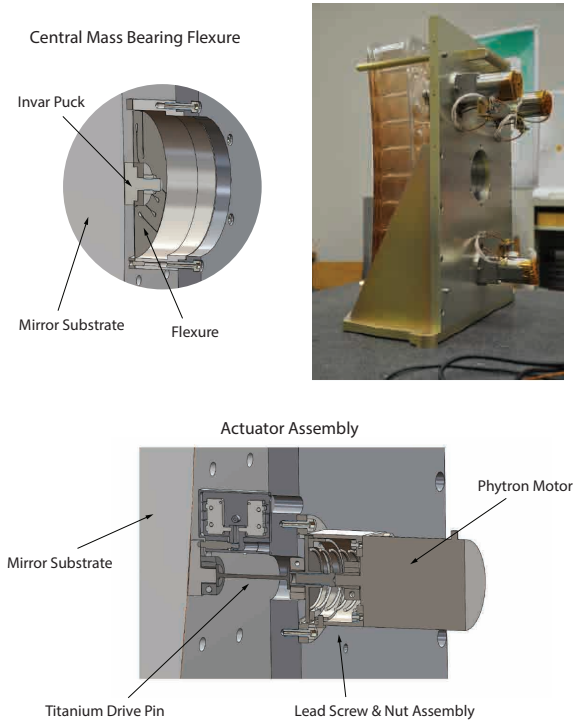}
\caption{The collimator assembly in the lab at Johns Hopkins University (top right).  Also shown is a cutaway of the central mass-bearing flexure (top left) and one of three stepper motor-driven actuators for tip-tilt-piston control (bottom).}\label{fig_collimator_assembly}
\end{figure}


Four locations on the back surface were ground flat and Invar pads were attached with 3M 2216 two-part epoxy adhesive at these locations.  One central pad serves as the mechanical connection to a custom stainless steel membrane flexure ($0.020\,\rm{in}$ thick, $5.0\,\rm{in}$ diameter) that transfers the weight of the mirror to the mount system and allows mirror articulation.  The other three pads, distributed near the mirror perimeter, serve as connection points for actuator-driven titanium pins that together reorient the mirror.  This actuator system is discussed in detail in \S~\ref{collimator_mechanism}.  An image of the collimator assembly is shown in Figure~\ref{fig_collimator_assembly}, along with model cutaways of the central mass-bearing flexure and the actuator assembly.

\subsection{Fold Mirror 1}\label{fold_mirror_1}

Light reflects twice off the surface of Fold Mirror 1 --- both before and after reflection from the collimating mirror.  A simple flat mirror with dimensions 425 x 489 x 38.1 mm, it was fabricated using Corning 7980 Fused Silica and polished to $< \frac{1}{2}\,\rm{wave}$ (for $\lambda = 632.8\,\rm{nm}$) PV surface figure across the clear aperture by Nu-Tek Precision Optical Corp. and coated with a protective gold coating by Infinite Optics.


\begin{figure}
\epsscale{1.1}
\plotone{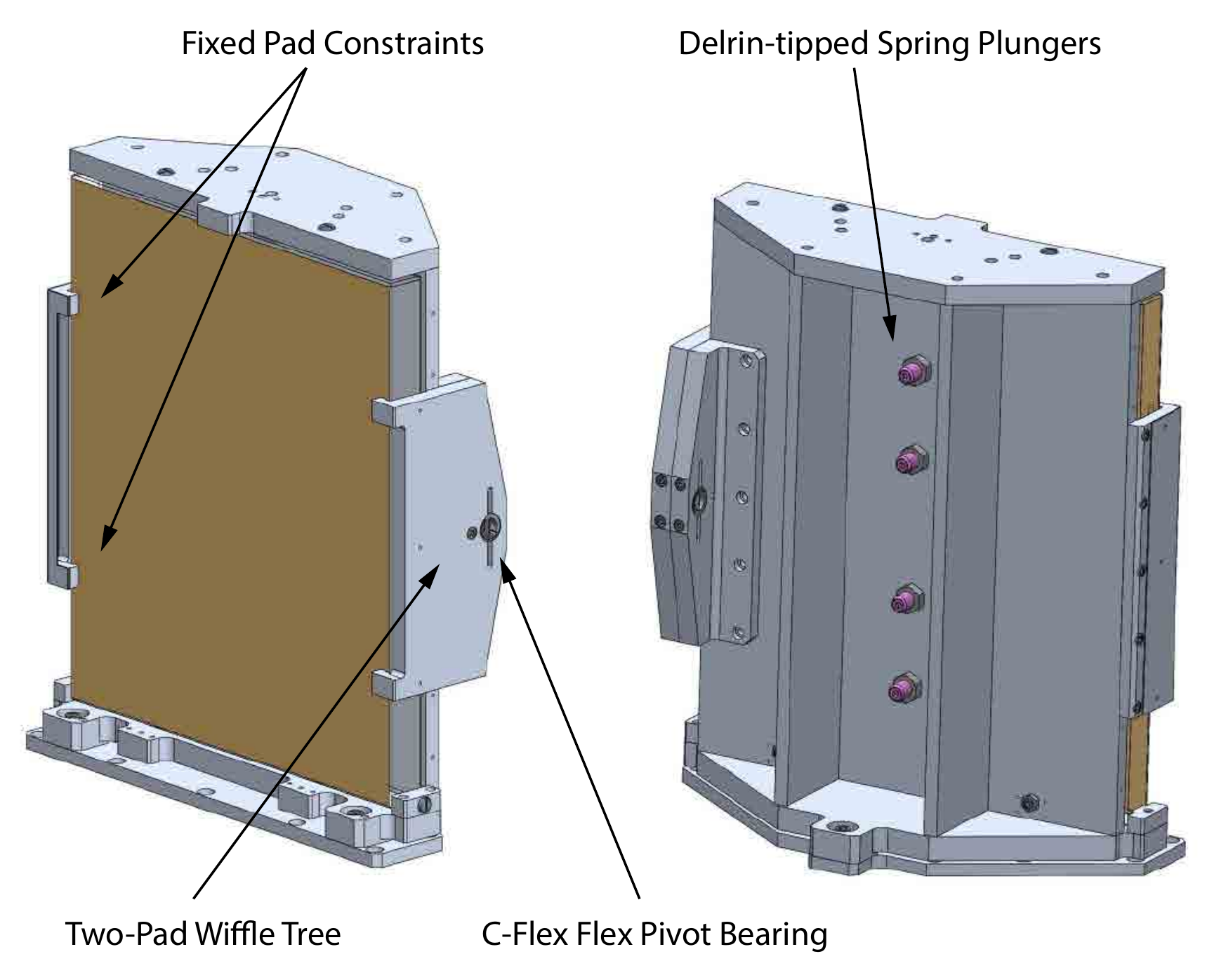}
\caption{The Fold Mirror 1 Mount holds the Corning 7980 fused silica mirror in place vertically.  Four Delrin\textsuperscript{\textregistered}-tipped spring plungers push the back vertical spine of the mirror with a total force of $380\,\rm{N}$ to induce a cylindrical shape with $3\,\micron$ sag to correct for astigmatism from an unknown source.}\label{fig_fold_one}
\end{figure}


As will be discussed in \S~\ref{astigmatism_correction}, astigmatism in the as-built optical system, discovered during end-to-end testing, required a correction that was implemented by bending Fold Mirror 1 into a cylinder where the axis of the cylinder is in the vertical direction, i.e., perpendicular to the cold plate upon which the mirror is mounted.  There is a $3\,\micron$ sag between the sides and middle of the mirror.  This bending was accomplished by applying a total of $\sim 86\,\rm{lbf}$ ($380\,\rm{N}$) vertically with four Delrin\textsuperscript{\textregistered}-tipped spring plungers distributed along the back center spine of the mirror (Figure~\ref{fig_fold_one}).  This correction essentially decreases the spectral power of the fore-optics so the spectral and spatial foci are both in better focus for a given detector array mosaic position.  Serendipitously the beam encounters Fold Mirror 1 twice so its corrective power is roughly doubled.

The front of the mirror is constrained by two fixed pads on one side and a ``two-pad wiffle tree'' on the other side (Figure~\ref{fig_fold_one}) to provide, in effect, three-point support on the mirror face.  Three more supports along the mirror sides, provided by spring plungers opposing fixed pads in the mount, complete the six-position semi-kinematic mount.

\subsection{Fold Mirror 2}\label{fold_mirror_2}

Light that passes the pseudo-slit strikes Fold Mirror 2 with angles of incidence about $50\degr$.  Fold Mirror 2, which reflects the beam towards the VPH grating, has many similarities to Fold Mirror 1.  It was fabricated with the same material and is mounted in a similar manner as Fold Mirror 1 (except it is not purposefully bent).  With a size of $481 \times 362 \times 38.1\,\rm{mm}$, it was polished to $< \frac{1}{2}\,\rm{wave}$ (for $\lambda = 632.8\,\rm{nm}$) PV surface figure across the clear aperture by Nu-Tek Precision Optical Corp.

This optic was also coated by Infinite Optics.  The front surface of Fold Mirror 2 has a dichroic coating to minimize the amount of thermal ($\lambda \ge 2.0 \micron$) light that arrives at the detector arrays.  The dichroic coating is a long-pass filter that transmits $\sim 95\%$ of the light from 2.0 -- $2.6\,\micron$ into the mirror substrate and reflects $> 99\%$ of the light between 1.5 -- $1.7\,\micron$.  The back surface is coated with an AR coating to efficiently transmit the thermal light into a blackened panel, intended to act as a ``light trap,'' integrated into the mirror mount.

\subsection{Mosaic VPH Grating}\label{VPH}


\begin{figure}
\epsscale{1.0}
\plotone{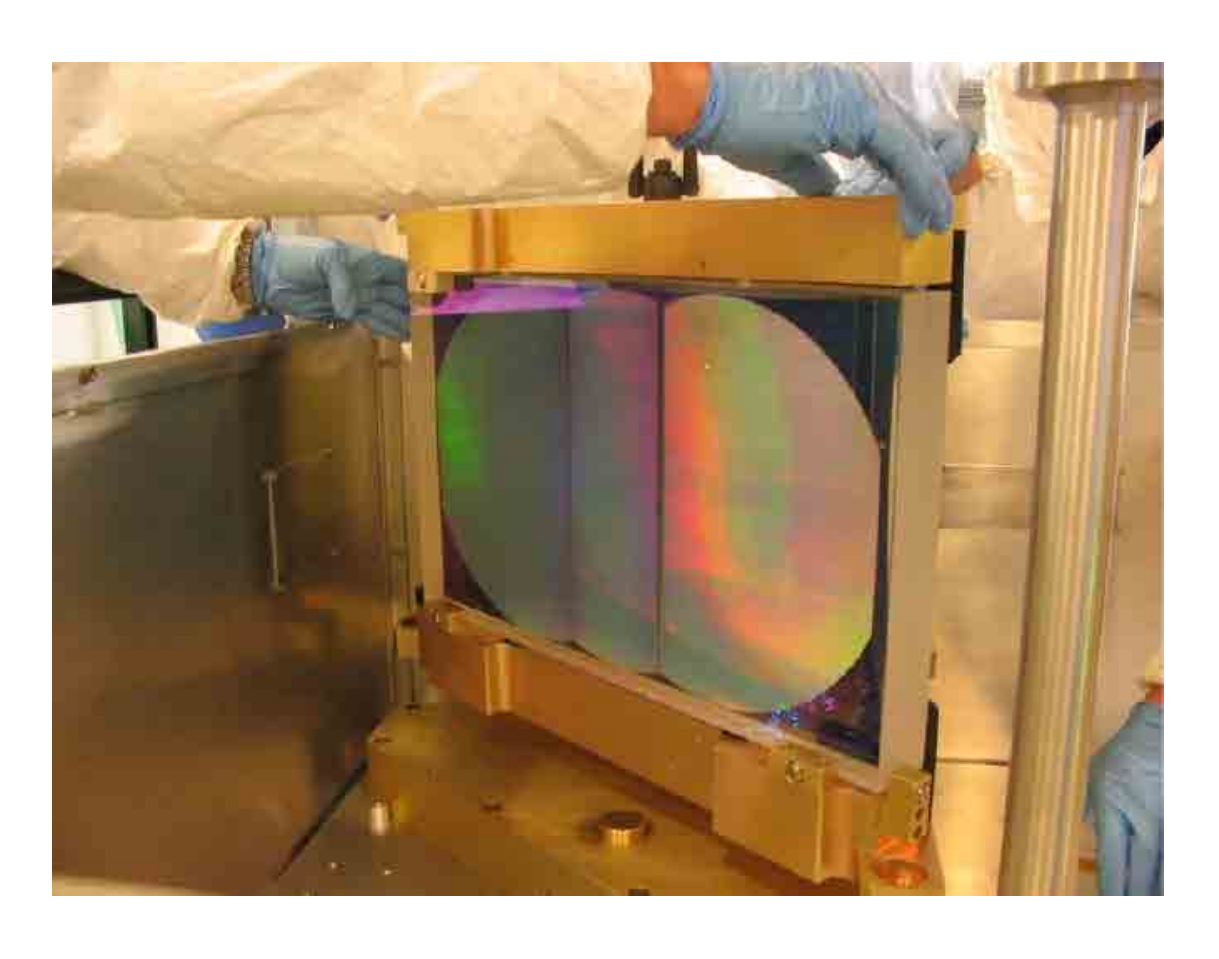}
\caption{The VPH grating as it was lowered into the instrument.  The grating is made up of a mosaic of three segments.}\label{fig_vph}
\end{figure}



\begin{deluxetable*}{lcc}
\tabletypesize{\scriptsize}
\tablewidth{0pt}
\tablecaption{Tightest segment tolerances for mosaic VPH grating \label{tbl-VPH_segments}}
\tablehead{\colhead{Tolerance} & \colhead{Value} & \colhead{Comment}}

\startdata
Differential Groove Density & $\pm 0.0035\,\rm{l mm^{-1}}$ & \tablenotemark{a} \\
Differential Clocking & $\pm 3 \arcsec$ & \tablenotemark{b} \\
Differential Substrate Wedge (relative to x-axis\tablenotemark{c}) & $\pm 3.5 \arcsec$ & \tablenotemark{a} \\
Differential Substrate Wedge (relative to y-axis\tablenotemark{d}) & $\pm 2 \arcsec$ & \tablenotemark{a} \\
\enddata
\tablenotetext{a}{Spot centroid movement of $\pm 0.1$ resolution element.}
\tablenotetext{b}{Spot centroid movement of $\pm 10 \micron$ in spatial direction.}
\tablenotetext{c}{Parallel to the grooves.}
\tablenotetext{d}{Orthogonal to the x-axis and in the plane of the segment.}

\end{deluxetable*}


\citet{arn10} extensively describe the requirements, development process, and performance of the first APOGEE candidate VPH grating manufactured by Kaiser Optical Systems, Inc. (Ann Arbor, MI).  This grating was chosen for deployment in the instrument (Figure~\ref{fig_vph}).  Here we briefly summarize some of the most important aspects of the grating.

Our survey requirements for resolution and wavelength coverage were satisfied with a grating line frequency choice of $1{,}008.6\,{\rm lines\,mm^{-1}}$ and first order operation in Littrow mode where angle of incidence and exit are $54\degr$ for the center wavelength ($1.6042\,\micron$).  We required a clear, elliptically-shaped aperture sized with a minor axis (along the groove length direction) of $280\,\rm{mm}$ and major axis (projected width) of $465\,\rm{mm}$.  As the VPH grating was recorded with an Ar laser ($0.488\, \micron$), the grating equation required a $14.25\degr$ angle of incidence to record an interference pattern with our desired frequency.  However, Kaiser Optical Systems' largest recording beam diameter of $289.6\,\rm{mm}$ could only provide a projected width of $298.7\,\rm{mm}$ for a single recorded VPH grating, well short of our required width.

So development of a mosaic VPH was necessary to meet our grating width requirements.  We greatly benefited from discussion and analysis in \citet{paz08} regarding mosaic VPH grating development in anticipation of the need of large dispersing optics for instruments in the era of extremely large telescopes.  They argued that segment-to-segment differential tolerances must be sufficiently tight such that the dispersed light from each segment is reimaged at the detector sufficiently close in position so the combined point spread function is not overly degraded.  This was found to be the case for APOGEE --- Table \ref{tbl-VPH_segments} lists the three tightest differential tolerances for our VPH grating segments \citep{arn10}.  Informed by these tolerances, we considered the three methods for fabricating mosaic VPH gratings discussed in \citet{paz08}: framed, common-bonded, and step and repeat.

A framed grating uses individual VPH gratings positioned in a mechanical frame.  A common-bonded grating permanently combines individually recorded VPH grating substrates with a common cap.  Lastly, step and repeat gratings use a monolithic VPH grating substrate and the substrate is translated between the recording of adjacent sections of a uniform layer of gelatin applied to the substrate.  We chose the step and repeat process because we thought it offered the best chance of meeting the differential segment tolerances.  The individual segments of a framed (common-bonded) VPH grating would have to be positioned mechanically (during bonding) with accuracies on the order of microns.

Both the substrate and cap layer for the APOGEE VPH Grating are made from Corning 7980 Fused Silica with finished dimensions of $305 \times 508 \times 25.4\,\rm{mm}$.  They were polished on both sides to $< \frac{1}{5}\,\rm{wave}$ (for $\lambda = 632.8\,\rm{nm}$) PV surface figure.  \edit1{This high level of flatness was specified because it is thought to enhance diffracted wavefront performance.  Both Bond Optics (Lebanon, NH) and Zygo Corporation (Middlefield, CT) made two substrates each.  Since the fabrication of VPH Gratings includes a sequence of multiple steps \citep{bar98a}, including holographic recording, wet processing, and testing, that are done iteratively until a grating has achieved the desired parameters, the availability of extra substrates improves process efficiency.}  One side of each substrate and cap was AR-coated by Newport Thin Film Labs (Chino, CA) prior to VPH recording.

No problems with VPH grating integrity have been observed after multiple thermal cycles between room temperature and $80\,\rm{K}$.  Anticipating shrinkage with temperature of the thin ($\sim 10\,\micron$ deep) layer of gelatin to occur at the same rate as that of the fused silica substrate, the room temperature groove frequency target was $1{,}008.42\,{\rm lines\,mm^{-1}}$ to yield $1{,}008.6\,{\rm lines\,mm^{-1}}$ at $77\,\rm{K}$.


\begin{figure*}
\epsscale{1.0}
\plotone{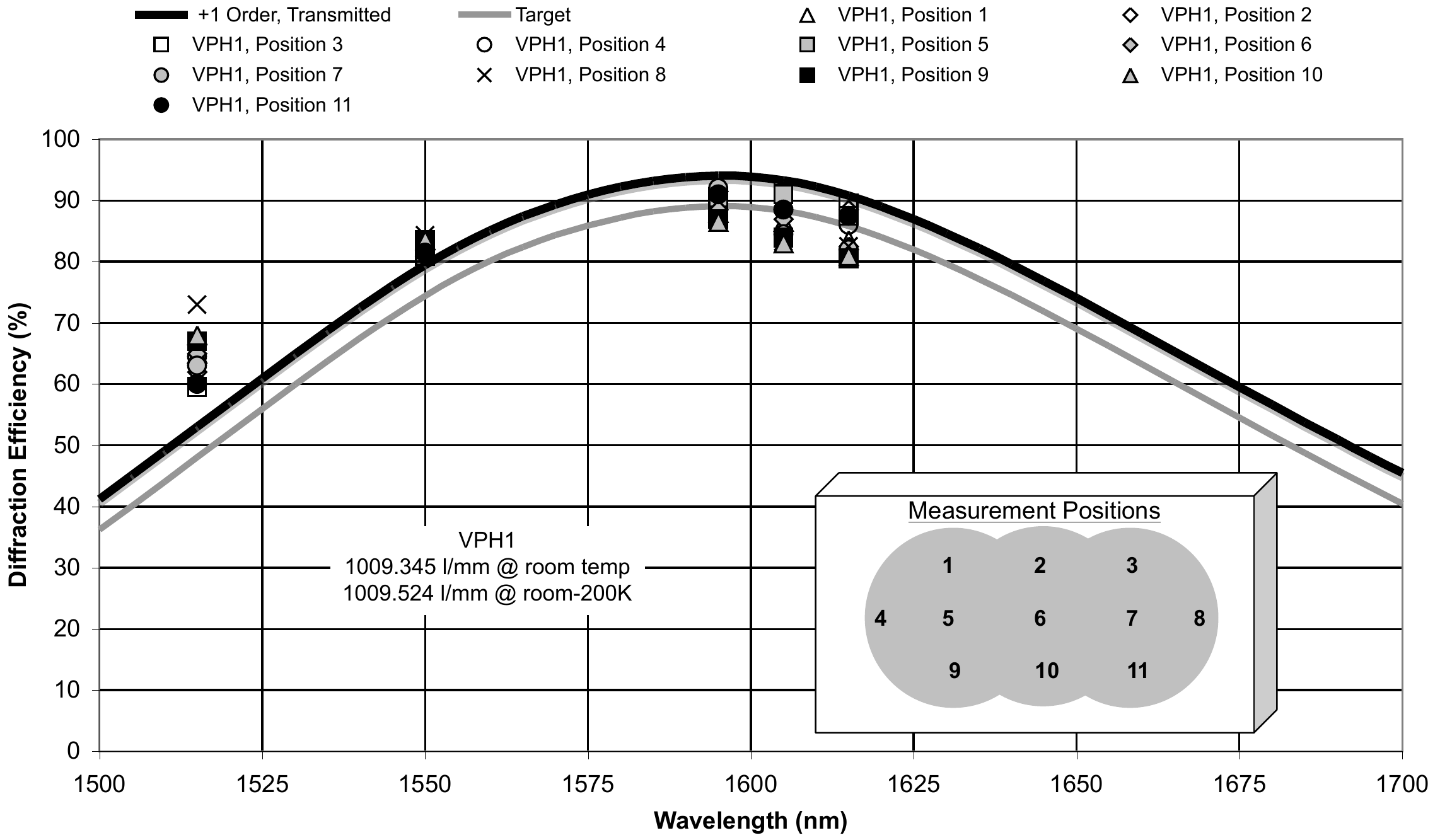}
\caption{Predicted first order transmissive diffraction efficiency (black line) and actual measurements (symbols) for various locations and wavelengths on the mosaic VPH grating.  The prediction is based on Rigourous Coupled Wave Analysis (RCWA).  The target (grey line) is simply 5\% less than the predicted efficiency and is the Kaiser Optical Systems target of minimum performance with production tolerances and real-world losses included.}\label{fig_vph_order1_de}
\end{figure*}


\begin{figure*}
\epsscale{1.0}
\plotone{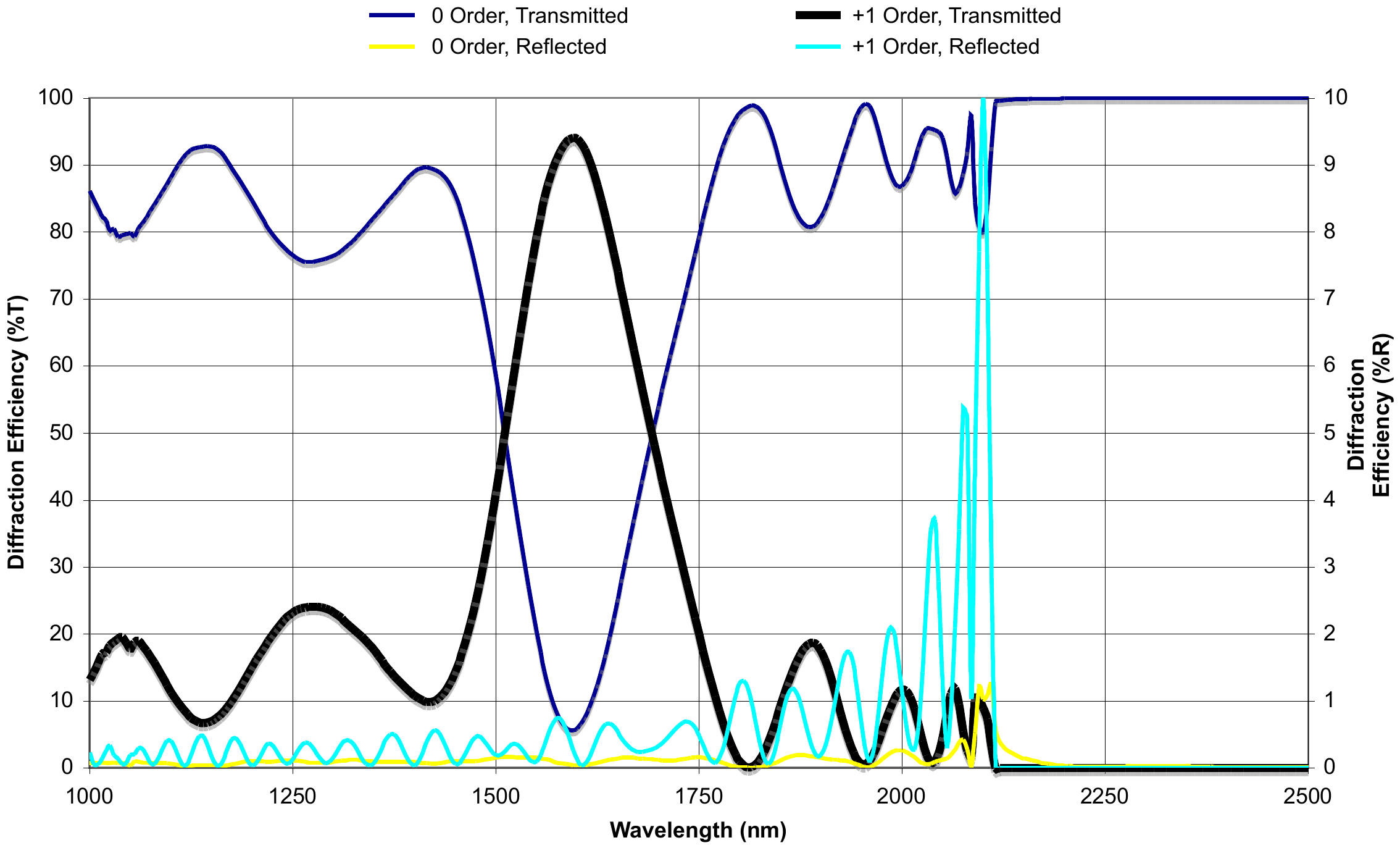}
\caption{The predicted diffraction efficiencies over a broad wavelength range for various orders based on Rigourous Coupled Wave Analysis (RCWA), originally shown in \citet{arn10}.  The left-hand ordinate applies to the transmitted +1 order (black; also plotted in Figure~\ref{fig_vph_order1_de} for the narrower wavelength coverage of the spectrograph) and the transmitted zero order (dark blue).  The right-hand ordinate applies to the reflected first order (light blue) and reflected zero order (yellow).}\label{fig_vph_multi_order_de}
\end{figure*}


Predicted and actual first order grating efficiencies for a variety of positions on the mosaic grating are shown in Figure~\ref{fig_vph_order1_de}.  Figure~\ref{fig_vph_multi_order_de} shows the efficiencies for other orders --- this information guided our strategies for mitigating stray light (\S~\ref{stray_light_mitigation}).

\begin{figure}
\epsscale{1.1}
\plotone{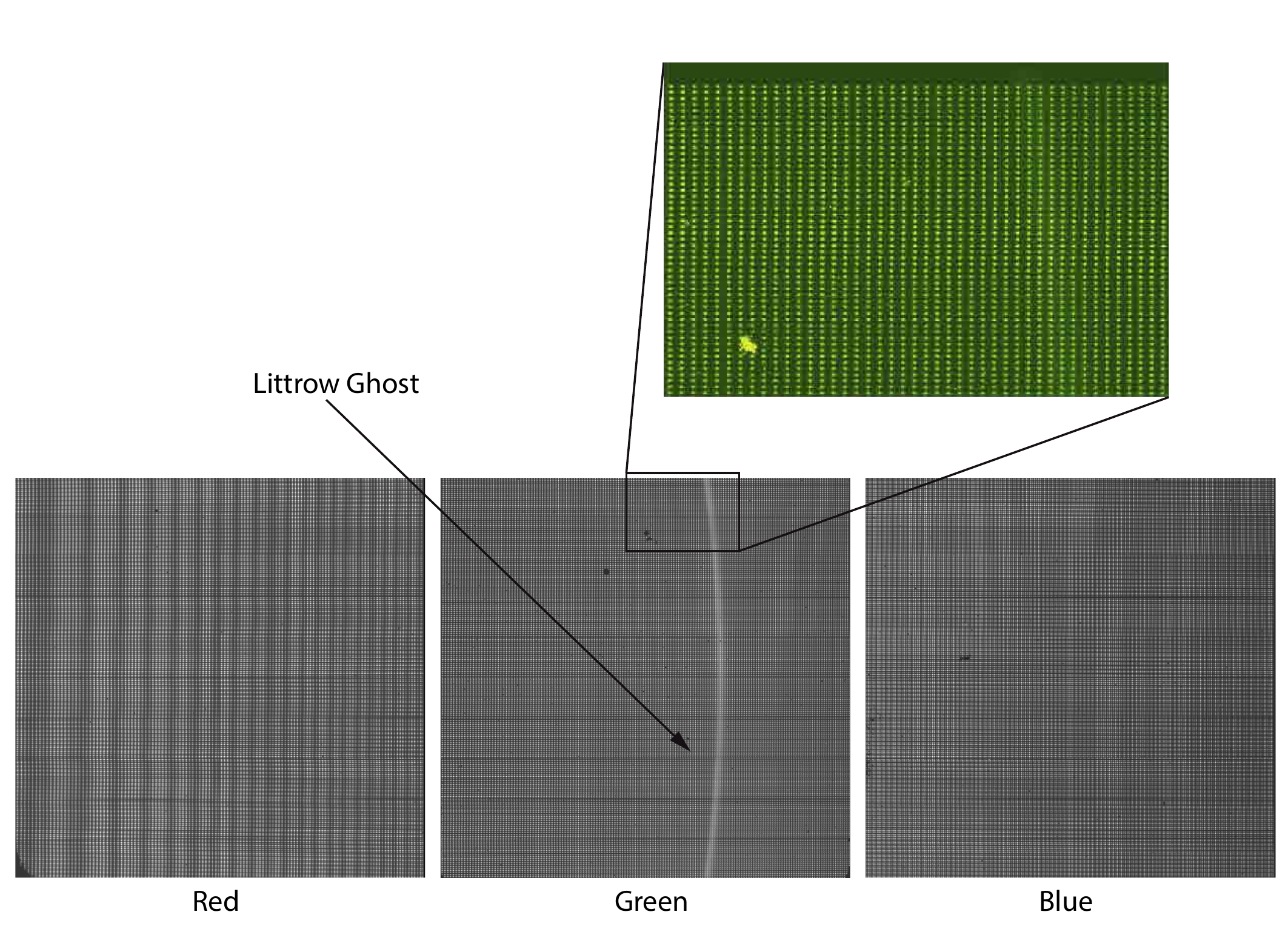}
\caption{The Littrow Ghost, originally shown in \citet{wil12}, created by illuminating the instrument with the Penn State Radial Velocity Group's Fiber Fabry Perot interferometer.  This calibration device provided $\sim 120{,}000$ discrete lines (better seen in the close-up) across the three detector arrays and permitted broadband analysis of the ghost strength.}\label{fig_littrow_ghost}
\end{figure}

An important consideration during instrument design was accommodation of the ``Littrow Ghost,'' a feature comprehensively discussed in \citet{bur07}.  This ghost is formed by dispersed light that is reflected backwards into the camera by the detector arrays and then reflectively recombined by the VPH grating into zeroth order (white light) and finally reimaged by the camera on the detector arrays at the location of the Littrow wavelength.  Its arrival location (in the dispersion direction) can be moved by tilting the VPH grating fringes.  We considered steering this ghost into the gaps between detector arrays.  With a predicted intensity of $1 \times 10^{-3}$ times the combined dispersed spectral intensity, the Littrow Ghost represented a source of scattered light that could affect the APOGEE survey if it fell near important spectral lines.  Fortunately its natural arrival location, $1.6042\,\micron$ (the Littrow wavelength; see Figure~\ref{fig_littrow_ghost}), was an area in the APOGEE spectrum with few important known lines for stellar abundance work.  Thus untilted fringes could be used, easing VPH Grating manufacturability.  The Littrow Ghost as seen in the instrument is discussed further in \S~\ref{observed_littrow_ghost}.


Passive VPH grating rotation about an axis parallel to the grating grooves and centered on the recorded grating is provided by a pin-and-socket arrangement.  The ``VPH Mount Spacer Plate,'' which attaches to the cold plate, has a $\frac{3}{8}\,\rm{in}$ ($10\,\rm{mm}$) tall, $1.5\,\rm{in}$ ($38\,\rm{mm}$) diameter ``pin.''  A complementary, close-fitting, ``socket'' in the ``VPH Base Plate,'' which mounts above the spacer plate, allows rotation.  Oversized bolt clearance holes in the base plate accommodate small angle changes.  This degree of freedom was in fact used during assembly to accommodate the actual recorded groove frequency of $1{,}009.345\,{\rm lines\,mm^{-1}}$ at room temperature: the grating angle (relative to the incident beam) was increased by $0.061\degr$ and the camera was similarly articulated about the grating center by $0.122\degr$.


\begin{figure}
\epsscale{1.2}
\plotone{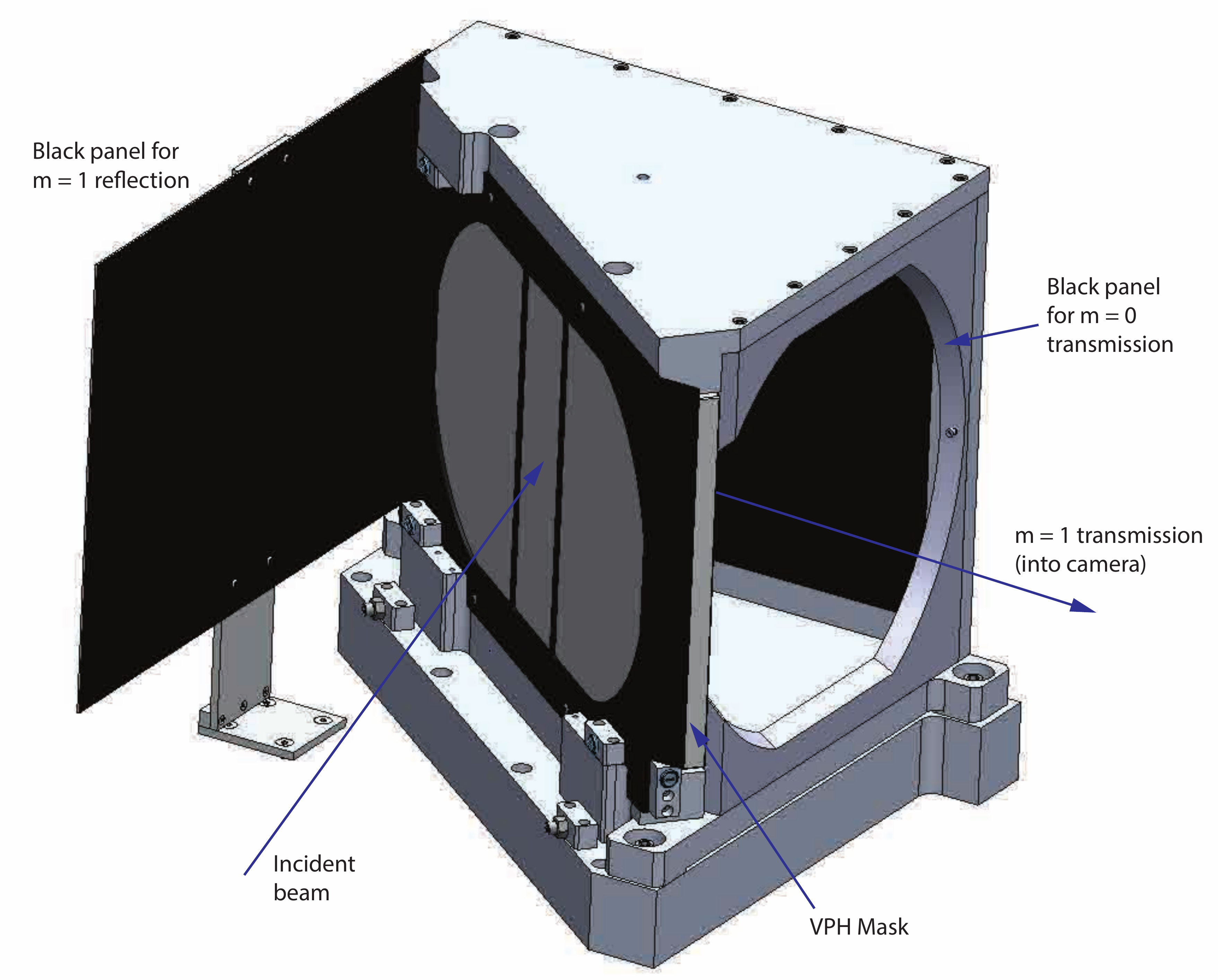}
\caption{The VPH Grating Mount includes a blackened panel on a back wall to intercept the zeroth order light from the VPH grating.  Another panel is positioned to the left of the mount to intercept $m=+1$ reflected light. With the grating operating in Littrow mode, the incident beam enters the VPH grating with a $54\degr$ angle of incidence and exits dispersed in a variety of angles centered on $54\degr$ in $m=+1$ transmission towards the camera.  A blackened panel on the entrance face of the VPH grating serves as the aperture stop.}\label{fig_vph_mount}
\end{figure}


The VPH grating mount (Figure~\ref{fig_vph_mount}), like the mount for Fold Mirror 2, includes a blackened panel that intercepts, at normal incidence, the zeroth order stray light transmitted through the VPH grating.  A blackened mask covers the VPH grating substrate entrance surface and includes $\frac{5}{16}\,\rm{in}$ ($\sim 8\,\rm{mm}$) wide strips to cover the boundaries between recorded segments.  While this mask serves as the aperture stop for all but the reddest wavelengths for the camera and detector, and is located close to the pupil formed by the collimator along the optical axis, it is not strictly positioned at the pupil since the VPH grating and baffle are tilted by $54\degr$ relative to the optical axis whereas the pupil is orthogonal to the optical axis.

\subsection{Camera}

\subsubsection{Optical Design}

\begin{figure}
\epsscale{1.1}
\plotone{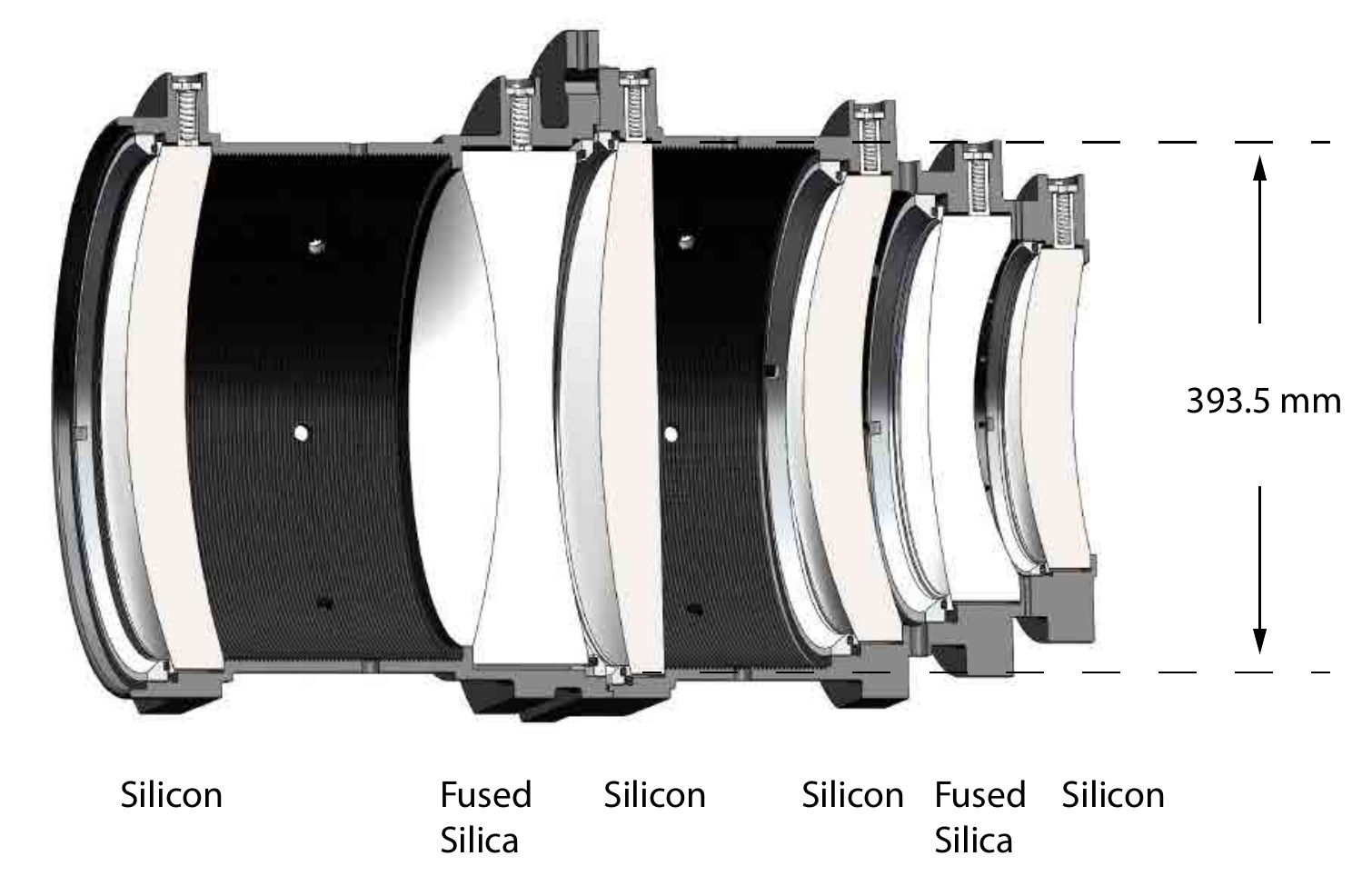}
\caption{A section view of the refractive camera, which features multiple large elements of silicon and fused silica.  Spring plungers push the top of each element against a pair of aluminum pads at radial locations $\pm 45\degr$ from the bottom for radial support.  Axial support for each element is provided by a canted coil spring that pre-loads the element against an annular seat in the barrel.  The inside of the barrel is grooved and painted with Aeroglaze\textsuperscript{\textregistered} Z306 polyurethane coating to mitigate stray light.}\label{fig_camera_cutaway}
\end{figure}

The six-element refractive camera (Figure~\ref{fig_camera_cutaway}), designed and fabricated by New England Optical Systems (NEOS; Marlborough, MA), has a measured focal length of $356\,\rm{mm}$ at room temperature and elements as large as $393.5\,\rm{mm}$ in diameter.  Two materials were utilized for the optical elements: silicon and Corning 7980 Fused Silica.  Chromatic aberration correction was feasible with only two materials because the instrument works in a relatively narrow wavelength range and it was permissible to have longitudinal chromatic aberration (variation of focus with wavelength).  The latter was accommodated by individually pivoting the detector arrays about an axis parallel to the grooves of the dispersive optic.  The number of possible optical materials from which to choose was actually fairly limited given that the largest elements had to be nearly $400\, \rm{mm}$ in diameter.  In addition to size constraints, cost, lead time, and availability of measured cryogenic refractive indices, particularly those obtained with the Cryogenic High Accuracy Refraction Measuring System (CHARMS; \citealt{lev04}) facility at NASA Goddard Space Flight Center, were considered.  Lastly, internal transmittance and resistance to thermal shock \citep[see, e.g.,][]{har98} were evaluated.  Materials considered included $\rm{CaF_2}$, ZnSe, Cleartran, AMTIR-1, silicon, and fused silica.  In the event, the latter two were chosen.  Particularly attractive were the high index of silicon, the low cost of fused silica, and the low CTE and good resistance to thermal shock for both materials.  Moreover, CHARMS measurements of the cryogenic refractive indices for both silicon \citep{fre06} and Corning 7980 Fused Silica \citep{lev08} were available.  While fused silica is available in very large sizes, the growth of silicon boules as large as we required was uncommon and probably at the limit of what is currently practical.


The optical design included four silicon elements with a total center thickness of $175\, \rm{mm}$.  Therefore it was imperative that material specifications were adopted that ensured high silicon internal transmittance and optical quality and low birefringence.  Our baseline requirement for internal transmission (i.e., transmission not including surface reflection losses) through the four silicon elements was $78.5\%$ based on an absorption coefficient of $0.0138\,\rm{{cm}^{-1}}$, typical of optical-grade silicon.  After research and discussions with vendors, we adopted the following material specifications: monocrystalline \hkl<100>; optical grade ``slipfree''; intrinsic or n- or p-type (i.e., the material supplier was free to choose); resistivity $> 65\, \Omega\, \rm{cm}$; fine-annealed; and stress relieved.  Given these specifications, the goal was to realize absorption coefficients of $<0.010\,\rm{{cm}^{-1}}$.  In fact, the silicon internal transmission far exceeded this goal, with measured absorption coefficients that ranged from $0.002\,\rm{{cm}^{-1}}$ -- $0.005\,\rm{{cm}^{-1}}$.  These measurements came from laser calorimetric testing of 10 mm thick witness samples at $1.54\, \micron$ from each boule used to fabricate the silicon elements.  Ultimately the total predicted internal transmission within the four silicon elements was $94.8\%$.

The Corning 7980 Fused Silica was specified to be Grade A and Class 0 to minimize index inhomogeneity and inclusions.  For throughput calculations we assumed an absorption coefficient of $0.001\,\rm{{cm}^{-1}}$ per the Corning data sheet for the two fused silica elements.  When combined with the silicon, overall camera bulk transmission was predicted to be $94.1\%$.

Each element surface has an AR coating with in-band average reflectivity per surface of $< 0.25\%$ and $< 0.5\%$ maximum reflectivity in the APOGEE waveband.  Combining bulk transmission with AR coating performance, the camera throughput is expected to be $93\%$ across the waveband.

As mentioned above, each detector array is individually tilted and positioned for optimal focus (see Table~\ref{tbl-prescription}).  Tilts range from approximately $-0.8\degr$ for the blue detector array to $-0.5\degr$ for the red detector array.  (The tilt is in the sense that the longer wavelength side of each detector is farther from the back of the camera, and a detector array with its surface normal to the camera optical axis would have zero tilt.)

\edit1{A} camera focal ratio of $\approx f/1.4$ implies a pixel-limited depth of focus of $\pm 25\,\micron$.  A detector axial displacement from optimal focus by this amount is predicted to produce a $3\,\micron$ RMS spot radius change. A deviation of detector array tilt from nominal of $1\,\arcmin$ is predicted to produce a $1.2\,\micron$ RMS spot radius change.

Partial (first order) correction of spherical aberration from the uncorrected collimator is provided by a generalized asphere ($0.093\,\rm{mm}$ departure from best fit sphere) on the second surface of the first element.  The prescription of this surface was optimized and figured using computer-controlled polishing techniques after all other camera surfaces had been figured to take into account their as-built parameters.  All other element surfaces are spherical.

\subsubsection{Mechanical Design}

In addition to designing the camera optics, NEOS designed, supervised fabrication, assembled, and tested the camera opto-mechanical assembly.

Particular care was given to the opto-mechanical design at room and operating temperature.  As described in \citet{wil10}, two different cradling schemes provide radial support for the optical elements.  At room temperature, a pair of Delrin\textsuperscript{\textregistered}-- stainless steel ``sandwich'' assemblies, located at radial positions $\pm 55\,\degr$ from the bottom of the barrel, keep each element centered for warm testing.  As the camera cools to cryogenic temperatures, the sandwich assemblies, dominated by the high CTE of Delrin\textsuperscript{\textregistered}, shrink sufficiently that they lose contact with the elements.  In their place a pair of aluminum pads at radial locations $\pm 45\,\degr$ from the bottom of the barrel take up the role of centering the elements.  This system of cold radial centering relies on sufficiently accurate knowledge of the CTEs for 6061-T6 aluminum and the optical element materials.  In both the warm and cold regimes a compression spring at the 12 o'clock (top) position exerts a downward force equal to the element weight.

Axial restraints for each element are provided by canted coil springs, pre-loaded by threaded rings, to exert a force equal to the weight of each element.  A Delrin\textsuperscript{\textregistered} axial ring sits between the canted coil spring and lens surface.  The opposite side of each element is located flat against an aluminum flange machined into each barrel.

Because the instrument has been designed as if the optics were ``bench-mounted'' in a quiescent gravity environment, the assumption is made that the gravity vector will always point downward.  The combination of radial and axial support described above is predicted to keep Lens 2, the most shock-sensitive element due to its size and mass, seated in the barrel for accelerations up to $1.91\,g$ (vertical), $1.95\,g$ (lateral) and $0.94\,g$ (axial).


\begin{figure}
\epsscale{1.0}
\plotone{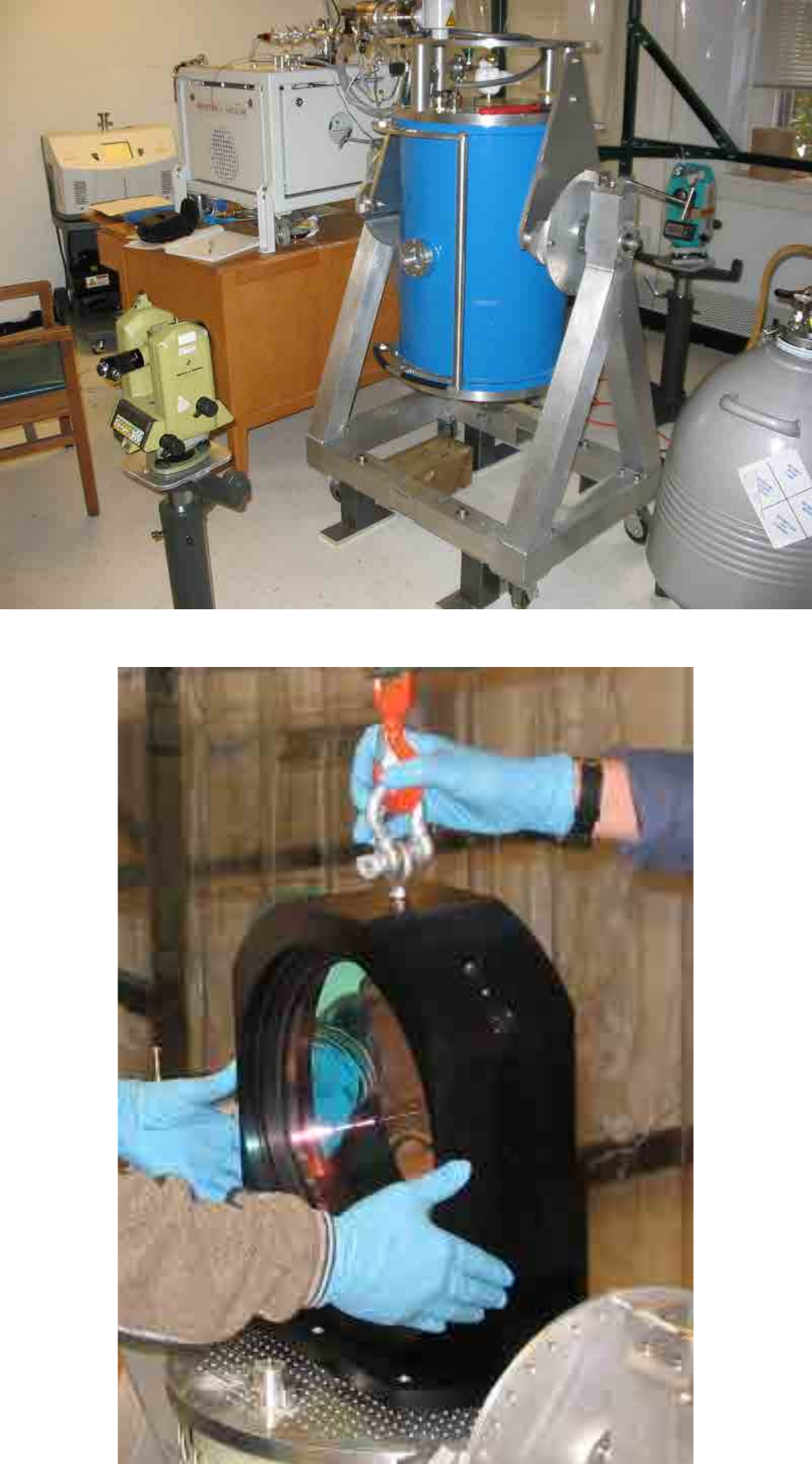}
\caption{The camera radial support scheme was tested using a scale model of the L5 element.  (Top) Reticles on the element and the cell were each monitored through cryostat windows using theodolites during a cryogenic cycle.  (Bottom)  The test cell as it was lowered onto the test cryostat cold plate.}\label{fig_L5_test}
\end{figure}


The dual-radial cradling scheme and axial restraint system were tested in a scale test assembly (Figure~\ref{fig_L5_test}) sized to mimic Lens 5.  The assembly included reticles on both the test lens cell and fused silica dummy element.  Both reticles were observed with theodolites through test dewar windows as the assembly was cryogenically cycled to ensure sufficient centering when cold.

The camera barrel is made up of two sections, each of which is fabricated from single billets of 6061-T6 aluminium.  The sections bolt together near the camera center of gravity.  The total camera weight is $265\,\rm{lb}$ ($120\,\rm{kg}$) so it must be positioned using an engine hoist.  Inside surfaces of the barrel are grooved and painted black to minimize stray light.  The overall mechanical length of the camera is $29.0\,\rm{in}$ ($737\,\rm{mm}$) and the outside diameter of the largest flange is $22.0\,\rm{in}$ ($559\,\rm{mm}$).  Flats are milled into the bottoms of the largest flanges so the bottom of the barrel is only $8.45\,\rm{in}$ ($215\,\rm{mm}$) from the optical axis to reduce the necessary height of the cryostat cold volume.  The camera is supported on the cold plate with a leg assembly that is pinned to the camera in two places and is formed with two triangular-shaped legs and cross-bracing.

\subsubsection{Thermal Considerations}\label{camera_temp}

{\it Camera Design Temperature} --  Because of the long lead time for the camera and the fast-paced instrument development schedule, we had to choose a final camera design temperature
before definitive thermal analysis had been conducted for the cryostat and opto-mechanical system.  We chose the conservative value of $110\,\rm{K}$, reasoning that if the camera did in fact naturally cool below this value the camera could be actively heated to reach this temperature.  In the end, the camera did in fact cool to $79\,\rm{K}$, as thermal analysis completed later in the design process predicted.  Fortunately, camera optical performance at this temperature is minimally degraded compared to $110\,\rm{K}$ performance as long as detector positions and tilts are adjusted; spots are predicted to increase by $\lesssim 0.5\,\micron$ RMS spot radius after the detectors are repositioned to account for the $\approx 0.37\,\rm{mm}$ increased back working distance (BWD) and detector tilts are re-optimized.  In fact RMS spot radii are much more sensitive to small variations of the camera temperature without re-optimizing the BWD and detector array tilts.  For instance, a $\pm 1\,\rm{K}$ change in camera temperature, given fixed detector positions optimized for a nominal camera temperature, can result in an increase of nearly $4\,\micron$ RMS spot radii.  Fortunately, as will be discussed in \S~\ref{thermal_performance}, the instrument maintained good thermal stability.  For example, during the 34 months of the SDSS-III survey, mean camera temperatures were maintained at $78.9\,\rm{K}$ (front half of the barrel) and $77.9\,\rm{K}$ (back half), each with standard deviations of $\pm 0.2\,\rm{K}$.

{\it Camera Temperature Control} --  An active temperature control system is described here for completeness even though it has not been used thus far.  Two resistive heaters ($250\,\Omega$/$25\,\rm{W}$) wired in parallel are mounted on the outside camera barrel (one on each side) adjacent to the third lens element.  The heaters are controlled using a Lake Shore controller based on feedback from Lake Shore Cryotronics, Inc. Cernox\textsuperscript{\texttrademark} temperature sensors mounted near the heaters.  A thermal switch is wired in series with the heaters to prevent overheating.  Aluminum spacers between the camera legs and barrel partially thermally isolate the camera from the cold plate to facilitate heating.

{\it Thermal Shock Susceptibility} --  Transient temperature analysis was conducted to ensure the large fused silica and silicon elements were robust to thermal shock.  Starting with a worst-case simulation, the edge of a $350\,\rm{mm}$ diameter, $40\,\rm{mm}$ thick element was subjected to a linear temperature ramp equal to the maximum predicted rate of change of the lens mounting block ($1.8\,\rm{K/min}$).  The analysis assumed the element edge cooled at this rate indefinitely, establishing a parabolic radial temperature distribution in the element and permitting use of closed-form solutions.  Furthermore, the analysis ignored radiative cooling and convective cooling from the element surface and assumed room temperature values for thermal conductivity and specific heat.  Calculated tensile stress with this simulation implied worst-case safety factors of $\sim 10$ ($\sim 150$) for fused silica (silicon).

A real lens edge cooling scenario was then considered by using an exponential ($2\,\rm{hour}$ time constant) cooling of the element outer rim.  Because the rate of cooling decreases with time, the lens never fully develops the worst-case temperature distribution analyzed previously.  As before, the analysis conservatively ignored radiant and convective cooling, and assumed room temperature values for thermal conductivity and specific heat.  This analysis gave a safety factor of $\sim 30$ for fused silica.  Silicon was not analyzed in this more realistic scenario given its even better resistance to thermal shock.

\subsubsection{Testing}

A laser unequal-path interferometer (LUPI) with a custom null lens on the test arm was used to verify camera performance at room temperature prior to delivery, integration, and first thermal cycle at U.Va.  The camera was tested in-band using a $1.52\,\micron$ HeNe laser and a Goodrich room temperature InGaAs camera to record the interferograms.

Several differences between the test environment and actual cryogenic usage of the camera necessitated the null lens to correct third-order spherical aberration:  (1) the test beam from the LUPI did not have the spherically aberrated beam typical of the beam delivered by the APOGEE fore-optics; (2) the camera was tested in double-pass; and (3), the generalized aspheric surface on the first element gave a different correction at room temperature than at the $110\,\rm{K}$ design temperature.

On and off-axis imaging performance was determined by comparing the resultant double-pass, null-corrected interferogram with expectations from the Zemax optical design for the test conditions.  The as-built effective focal length (EFL) at room temperature was $356.26\,\rm{mm}$, within the specified acceptable deviations of $\pm 0.5\,\%$ about the nominal EFL of  $357.5\,\rm{mm}$.  Lastly, it was verified that the object line-of-sight was co-aligned with the camera mechanical axis to within $0.6\,\rm{mrad}$ in accordance with specifications.  This line-of-sight check indicated that the stack-up errors from, e.g., wedge and decentration, were within tolerance.

\subsection{Vignetting}\label{vignetting}

As mentioned in \S~\ref{fno}, early lab measurements of test fiber assemblies gave $95\%$ encircled energy within an $f/3.5$ output beam using $40\,\rm{m}$ test fibers with prototype feedthroughs.  And the far-field illumination from the fibers was \edit1{well fit with a Gaussian profile}.  We modeled the illumination within Zemax by using $f/3.0$ illumination of the spectrograph optics with Gaussian Apodization and an apodization factor of 1.58.  Thus, according to the Zemax manual, the modeled illumination was described by

\begin{equation}\label{gaussian_illumination}
  A(\rho) = \exp(-1.58\rho^2)
\end{equation}

\noindent
where $A$ is the illumination and $\rho$ is the normalized pupil coordinate.  As it turned out, the camera, \edit1{with a challenging fabrication schedule and elements already near the limit in size of what could be realistically fabricated,} had already been designed assuming the spectrograph optics were illuminated with an $f/4$ beam.  Fortunately, the camera design still gave sufficient image quality with the illumination described above.  The edges of some of the camera elements start to vignette for wavelengths $\ge 1.67\,\micron$ (on-axis fibers) and $\ge 1.66\,\micron$  (top/bottom fibers).  By $1.695\,\micron$, close to the red edge of the red detector array, about $30\%$ of the light on-axis and about $40\%$ of the light at the top/bottom fibers is internally vignetted within the camera.

Within the fore-optics, a center strip installed along the vertical bisector of the Collimating Mirror to mitigate fiber tip ghosts (see \S~\ref{ghosts}) causes the most vignetting --- nearly $12\,\%$ of the beam.  Taking into account the downstream pseudo-slit, also centered on the beam, which alone would also vignette up to $4\,\%$, the net vignetting of the Collimating Mirror center strip is approximately $8\,\%$.  The two masks covering the internal boundaries between recorded VPH Grating segments, discussed above, combine to vignette nearly $6\,\%$ of the beam.  All totaled, approximately $20\,\%$ of the light is vignetted by these strips and masks.

\edit1{Table~\ref{tbl-throughput} lists the total vignetting expected within the spectrograph as part of a comprehensive listing of system throughput.}

\subsection{Spectrograph Alignment and Testing}\label{align&test}

Instrument assembly, alignment, and testing occurred at U.Va. prior to deployment.  The fore-optics and the orientation of the VPH grating were sequentially aligned using a small HeNe $0.6328\,\micron$ laser on a 5-axis mount along with multiple alignment targets consisting of $0.020\,\rm{in}$ ($0.5\,\rm{mm}$) diameter pinholes machined into custom-built precision stands.  The pinholes of each stand were measured to be at a height of $11.250\,\rm{in}$ ($285.8\,\rm{mm}$), our adopted height of the optical axis above the cold plate.  Various sets of tapped holes were machined into the cold plate at strategic locations so the targets could be temporarily positioned along the optical axis.

With the exception of the Fold Mirror 1 mount, which was pinned to the cold plate, each opto-mechanical mount was positioned on the cold plate with three sets of ``bumpers.''  Usually one side of a mount would be placed flush against two different bumpers, and a second side would sit flush against a third bumper, to constrain lateral position.  This system allowed very repeatable positioning of the large optics.  The outside diameters of specific bumpers could be changed as needed to accommodate the ``as-built'' optics.  Oversized clearance holes in the mounts permitted modest position changes on the cold plate.

For warm alignment, a ``dummy pseudo-slit'' was fabricated using 6061-T6 aluminum.  Use of this dummy assembly allowed illumination of the system for fiber locations $\#\,1$, $\#\,150$, and $\#\,300$.  We simply plugged fibers terminated with standard plug-plate ferrules into the dummy pseudo-slit.  The other ends of the fibers were illuminated with arc lamps on a portable breadboard.

A Goodrich room temperature InGaAs camera (hereafter just called a detector, because we removed the C-mounted lenses), with sensitivity from $\sim 0.95\,\micron$ to $1.7\,\micron$, was positioned at the instrument focus on a four-axis stage to record warm images.  The InGaAs detector could be bolted at three different locations on the stage corresponding to the three different locations of the detector arrays.  While the InGaAs detector has large ($40\,\micron$) pixels, and the APOGEE optical design was optimized for cryogenic operation, this testing still enabled verification of gross alignment and first order optics evaluation.  A CCD could not have been used given the visual light absorption by the silicon elements in the camera.

This system was used to check for gross differences in arrival locations of light illuminated from various portions of the instrument pupil --- it was essentially a first order check of how well the three VPH segments acted uniformly.  A Hartmann mask with an array of fourteen $2\,\rm{in}$ ($50.8\,\rm{mm}$) diameter holes was placed in front of the VPH to sample portions of the pupil.  A through-focus sweep produced by moving the InGaAs detector along the optical axis showed that light from the various pupil segments did overlap at minimum blur at the various wavelengths illuminated by the light source across the three detector array positions.  This test was repeated after the first thermal cycle of the instrument optics to verify the integrity of the camera elements.

In practice the collimator mechanism tip-tilt is adjusted warm based on laser alignment of the Fold 1--Collimator--Fold 1 sequence of optics using precision targets.  The collimator focus (and tip-tilt if necessary) is adjusted once the instrument is cooled to near its final temperature based on a through-focus scan to locate the global best focus.  If necessary, tip-tilt is adjusted to fine-tune the position of all 300 fiber traces on the detector arrays.

The v-groove blocks were positioned iteratively in a choreographed evolution that also included installation of the fiber links into the instrument.  A ``guide'' tool, temporarily pinned in place (using pre-drilled holes) in front of the slit bar, and a CCD camera, which provided a magnified view of the v-groove blocks and guide in real-time, aided the installation process (see the bottom left image of Figure~\ref{fig_pseudo_slit}).  The v-groove blocks were positioned by pressing the front of each block flush against the guide.  The guide had ten discrete flats machined at the optimal positions for each v-groove block face.  Slots were also machined into the guide at the edges of the discrete flats and served as visual cues for vertical alignment of the v-groove blocks.  Unfortunately, the depth of field of the CCD system was too small to allow good resolution of both the guide surface and the v-groove block cover simultaneously.


\begin{figure}
\epsscale{1.1}
\plotone{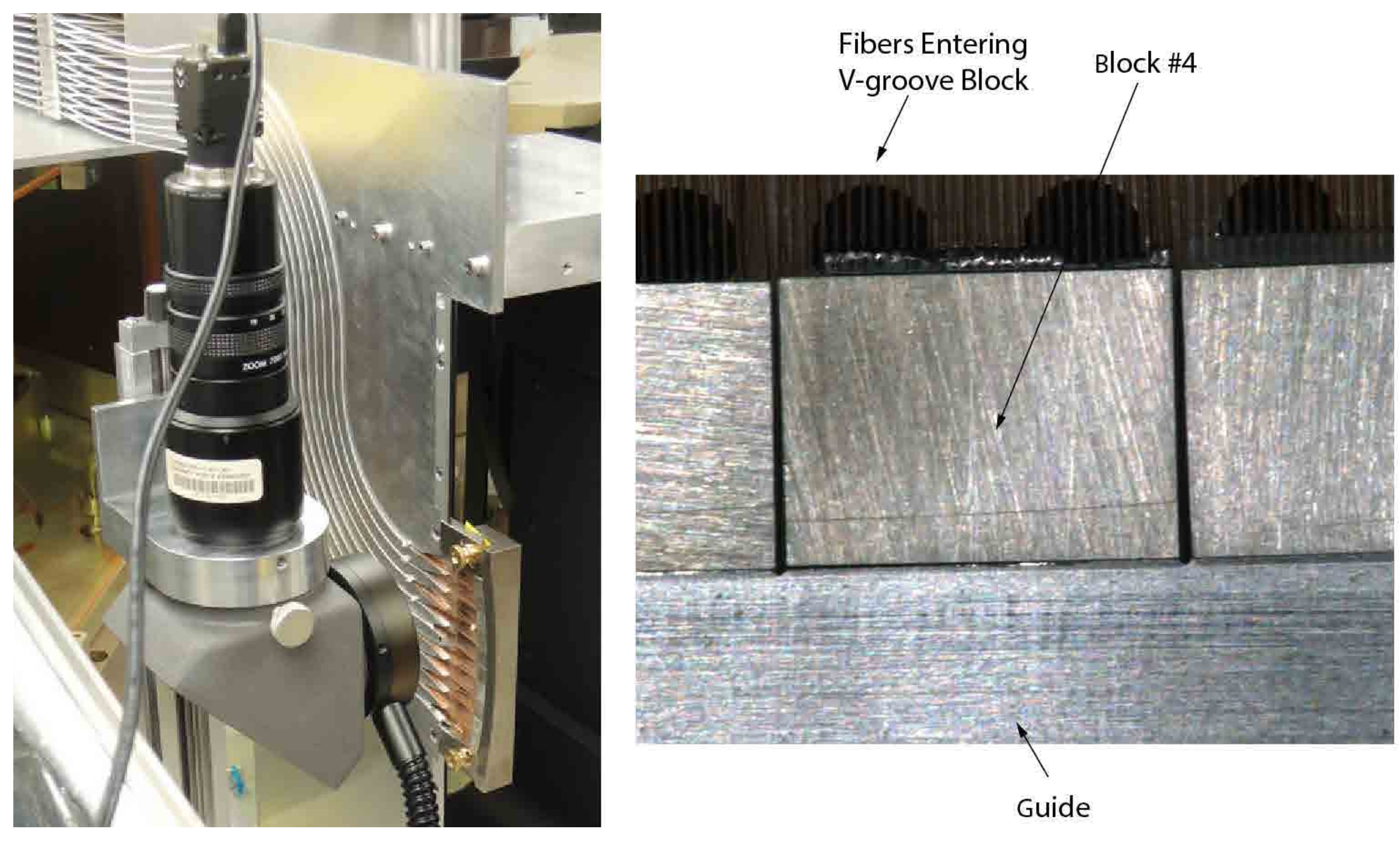}
\caption{(Left) An improved v-groove block position imaging system used in summer 2014.  (Right) \edit1{A magnified image of v-groove block \#4 using this improved system and with the guide block temporarily installed.}  This v-groove block had the largest discrepancy in positioning --- it is tilted by $> 0.3\,\degr$ and the bottom left corner is displaced from the guide by $\sim 0.004\,\rm{in}$.}\label{fig_block4_closeup}
\end{figure}


Based on locations of spectral traces at the detector, the v-groove blocks were vertically positioned with an accuracy of about $\pm 0.037\,\rm{mm}$ ($\pm 0.0015\,\rm{in}$).  For comparison, the design gap between v-groove blocks is $0.311\,\rm{mm}$.  Positioning in the focus direction, the more important characteristic, was difficult because the v-groove block had to be pressed up against the guide with a finger while another person tightened difficult-to-reach screws from the back.  \edit1{Analysis of magnified images of the v-groove blocks taken with a redesigned v-groove imaging system\footnote{This redesigned system was developed for positioning v-groove blocks for the Mapping Nearby Galaxies at Apache Point Observatory \citep[MaNGA;][]{bun15,dro15} Survey for SDSS-IV.} in 2014 July (after completion of SDSS-III) and with the guide mounted showed that the blocks are typically displaced away from the guide by about $0.001\,\rm{in}$ -- $0.002\,\rm{in}$ ($25\,\micron$ -- $50\,\micron$) and have tilts that vary from 0 -- $0.35\degr$.}  Figure~\ref{fig_block4_closeup} shows the placement of v-groove block \#\,4 which has the worst alignment relative to the guide: it is tilted by $> 0.3\degr$ and the bottom left corner is displaced from the guide by $\sim 0.004\,\rm{in}$.  Ideally the v-groove blocks would be flush against the guide.

\subsubsection{10-inch Telescope Observations}

As part of the early lab commissioning, it was helpful to obtain data on astronomical sources, which was possible by making use of the fiber feeds.  The resulting data were useful for assessing the instrument performance, \edit1{testing the data reduction software pipelines}, and understanding the properties of the near-infrared sky background at APOGEE's high resolution.  Initial observations of the daytime sky were taken in 2011 January by pointing the warm end of the fiber train to diffusely projected images of sunlight into the lab.

Subsequent observations of stars were obtained by connecting a bundle of 30 APOGEE fibers to a modified 10-inch Newtonian telescope just outside the lab.  The 10-inch telescope modification involved replacing the Newtonian flat with a dichroic to split visual light, sent to the Newtonian focus for acquiring and guiding objects by eye, from infrared light, sent to the prime focus where fibers in a special ``mini plug-plate'' received the light to send on to the instrument.  The telescope is a scaled-down version of the Sloan 2.5-m telescope, with the same $f$-ratio but $\frac{1}{10}$ the diameter.  With this arrangement, the 10-inch telescope observed the same diffuse background flux as expected at APO and could, in principle, observe stars 5 magnitudes brighter than the 2.5-m telescope to the same signal-to-noise.
With this setup, spectra of Jupiter, the Orion Nebula, and stars from the ``Winter Hexagon'' were observed in late 2011 January. Interestingly, the Winter Hexagon (plus Betelgeuse, in the center of the Hexagon) provided a remarkably diverse range of spectral types (from B8 to M2) useful for sampling a broad range of stellar temperatures.

\subsection{Stray Light and  Ghost Mitigation}\label{stray_light_mitigation}

Careful attention was given to mitigating stray light and ghosts as the survey required $\rm{S}/\rm{N} = 100$ spectra throughout the wavelength coverage.  This was especially important because the chosen wavelength coverage represents a small portion of the broad responsivity (visible to beyond $2.5\,\micron$) of the detector arrays (see \S~\ref{detector_section}).  Observed stray light and ghosts are discussed in \S~\ref{stray_light_performance}.

\subsubsection{Stray Light}

\edit1{Normal attention was paid to minimizing optical surface roughness and macroscopic imperfections, both of which scatter light.  Scattering from surface roughness dominates in the ultra-violet and visual wavelengths since it is inversely proportional to wavelength, whereas macroscopic imperfections such as scratches and digs (pits) become important in the near-infrared and infrared as wavelength becomes comparable in size to scatch width and dig diameter \citep[see, e.g.,][]{ben78}.  All fore optics (mirrors, VPH Grating) surfaces were specified to have $< 5\,\rm{nm}$ RMS surface roughness and the camera optics were specified to have $< 3\,\rm{nm}$ RMS roughness.  (The achieved RMS surface roughness for most optical surfaces was $0.5\,\rm{nm}$ or better.)  A scratch-dig of 60-40 (i.e., a $6\,\micron$ wide scratch and a $400\,\micron$ dig diameter\footnote{Following the commonly used Military Specification MIL-PRF-13830B.}) was specified for all surfaces.  We also took pains to wear clean room garments and keep the instrument, when open, in a clean tent to minimize dust accumulation on the optics.  Cleanliness is important because scattering by particulates also becomes an important source of scattering at near-infrared and infrared wavelengths \citep{ben78}.}

The design of instruments with dispersive optics also requires careful attention to the efficiencies of the various orders and the directions of dispersed (and reflected) light for various wavelengths.  As shown in Figure~\ref{fig_vph_multi_order_de}, $\ga 99\%$ of the light for  $1.0 < \lambda < 1.79\,\micron$ that interacts with the VPH grating is transmitted in either first order transmission ($m = +1$) or zeroth order ($m = 0$), which is diffused by the black-painted panel in the VPH grating mount (Figure~\ref{fig_vph_mount}).

While first order transmission spans all visual wavelengths through $1.79\,\micron$, where the latter is emitted essentially parallel to the grating surface (exit angle at $90\degr$ relative to the grating normal), wavelengths shortward of $1.0\,\micron$ are absorbed by the silicon elements, as are all second order wavelengths.  For $1.0 < \lambda < 1.79\,\micron$, the primary defense against stray light is the use of machined grooves, painted with infrared black, on the inside surfaces of the camera barrel.  There are also multiple blackened baffles between the camera exit and detector surface.

A blackened panel (Figure~\ref{fig_vph_mount}) is deployed to intercept, at normal incidence, the $< 2\%$ of the light from the incident beam reflected by the AR-coated VPH grating substrate.  This baffle also intercepts light reflected by the VPH grating in zeroth order.

Breault Research Organization, Inc. (Tucson, AZ) was contracted to conduct stray light and ghost analysis of our opto-mechanical design.  Their analysis included tracking unwanted VPH grating order and wavelength combinations.  For instance, they showed that light in first order reflection returns to the vicinity of Fold 2 --- thus blackened panels were deployed on both sides of this mirror.  By tracing rays backwards through the system from the detector, they also identified critical objects that could be seen by the detector.  Blackened baffles were also deployed at these locations.

We used both Lord Corporations's Aeroglaze\textsuperscript{\textregistered} Z306 and PPG PRC-DeSoto's Desothane\textsuperscript{\textregistered} HS Military \& Defense Topcoat (CA 8211/F37038 Camouflage Black) polyurethane coatings (with appropriate primers) within the instrument for blackening surfaces.

\subsubsection{Thermal Light Mitigation}

Thermal light (taken to be $\lambda > 2.0\,\micron$) emitted and transmitted by the fibers is mitigated in multiple ways.  First, the fibers partially cool over the $\sim 2\,\rm{m}$ length of fiber between the cryostat wall (room temperature) and pseudo-slit ($\sim 77\,\rm{K}$).  This cooling reduces the black body radiation emitted by the fibers (it is important to recall that fused silica fiber transmission decreases rapidly for $\lambda > 2\,\micron$ --- hence absorption and emissivity increase).  Second, as mentioned above, Fold Mirror 2 is coated with a dichroic coating that transmits light longward of $2\,\micron$ into the mirror substrate.  An AR coating on the opposite face transmits most of this light into a blackened panel within the mirror mount.  Based on coating witness sample measurements, we expect $94\%$ of the light with $2.0 \leq \lambda \leq 2.5 \,\micron$ is transmitted through the two optical surfaces and thus removed from the optical train.  Third, nearly $100\%$ of the light longward of $2.15\,\micron$ is transmitted by the VPH in zeroth order (see Figure~\ref{fig_vph_multi_order_de}) and thus strikes the darkened baffle in the VPH mount.  Lastly, the AR coatings on the four silicon elements in the camera provide increasingly effective thermal light blocking as the coatings were optimized for high throughput for the wavelengths 1.51 -- $1.70 \micron$.  Surface reflectivity of the coatings increases from $3\%$ ($2 \micron$) to $60\%$ ($2.5 \micron$).  Transmission through the four lenses, including AR coating performance and bulk transmission, decreases from $54\%$ ($2.0 \micron$) to $3\%$ ($2.5 \micron$).

\subsubsection{Fiber Tip Ghosts}\label{ghosts}

In addition to the Littrow Ghost (discussed generally in \S~\ref{VPH} and as observed in \S~\ref{observed_littrow_ghost}), ghost images were anticipated near very bright lines due mainly to fiber tip ghosts.  Fiber tip ghosts are produced by focused light that reflects off a detector array, traverses the optical path in reverse, reflects off the fiber tip (or more likely the v-groove block material surrounding the fiber tip), and then traverses the optical path forwards a second time and comes to a second focus at the detector array.  Raytracing in Zemax suggested the ghost images would have spot RMS radii $\sim 2$ -- 5 times larger and fall within one pixel of the primary images.  And intensities were expected to vary between a worst case of 1.5\% (for the center wavelength, middle fiber) to 0.06\% (bluest wavelength, end fiber) relative to primary images at the detector array based on stray light analysis by Breault.  This prediction assumes the ghost light, traversing backwards through the instrument, specularly reflects off the polished nickel alloy steel (A-39) surrounding the fiber tips.  The reflectivity of the A-39 surrounding the fibers, polished at the same time, was not measured.  Thus the approximate reflectivity of mechanically polished aluminum \citep{smi85}, 75\% at $1.6\,\micron$, was adopted as a conservative estimate.  The ghost intensities were predicted to vary with wavelength, primarily driven by the amount of vignetting that occurs during the two extra passes through the instrument after the initial reflection at the detector array.  As discussed in \S~\ref{fiber_tip_ghosts}, fiber tip ghosts have not been definitively identified in spectral images.

There are two means of mitigating this ghost but neither were implemented.  The fiber tips can be AR-coated and the material of the v-groove face surrounding the fibers can be designed to have low reflectivity, perhaps with a black-painted mask or use of a material that absorbs light.

To minimize ghosting from light that reflects off the collimator and then strikes the pseudo-slit, a blackened strip was placed immediately in front of the collimator along the vertical bisector.  The width of the strip, $0.75\,\rm{in}$ ($19.1\,\rm{mm}$), was based on the $\approx 0.64\,\rm{in}$ ($16.3\,\rm{mm}$) wide span of chief ray arrival locations at the collimator illuminated by the fibers at the pseudo-slit.  This strip was added to the instrument design after the stray light analysis was done by Breault.  As mentioned in \S~\ref{vignetting}, it vignettes nearly $12\,\%$ of the light reflected from the collimator.  Since the downstream pseudo-slit, which also obscures the center of the beam, vignettes up to $4\,\%$ of the light, the net vignetting of this collimator center strip is approximately $8\,\%$ of the light.


\section{Detector Arrays and Electronic Control}\label{detector_section}

\edit1{The tips of the fibers positioned at the pseudo-slit are reimaged by the optics described in the previous section onto a mosaic of three Teledyne H2RG detector arrays mounted side-by-side in the detector array mosaic assembly.  Mounted to a flange at the back of the camera, the assembly includes an intricate mechanical design to accurately position each detector array, with their individually tailored pistons and tilts, so they are located at optimal positions to record the blue, green, and red portions of the APOGEE wavelength coverage.  Furthermore, the detector arrays are mounted on a common platform that can be dithered back and forth in the plane of focus between exposures to provide optimal sampling at blue wavelengths.  Materials were chosen and thermal connections were designed to safely permit passive thermal control of the detector arrays during cooldown, operations, and warm-up phases.}

\edit1{As each array is operated in 4-channel mode, a single Leach controller with a 12-channel video card is used to electronically control all detector arrays simultaneously.  Similarly, common bias and clock signals are provided to each detector array for uniformity.  Sample-up-the-ramp readout mode is used to optimize $\rm{S/N}$ for the 47-read (nearly $500\,\rm{sec}$) exposures typically used for the survey while providing low effective readout noise.  These mechanical, thermal, and electronic design features are described below along with a comprehensive discussion of the achieved performance of the detector arrays consisting of measured readout noise, gain, dark current, linearity, bad pixels, and the abnormally bad persistence of the original blue detector array for APOGEE-North which has since been replaced.}

\subsection{Detector Array Mosaic Assembly}\label{array_mount}


\begin{figure}
\epsscale{1.2}
\plotone{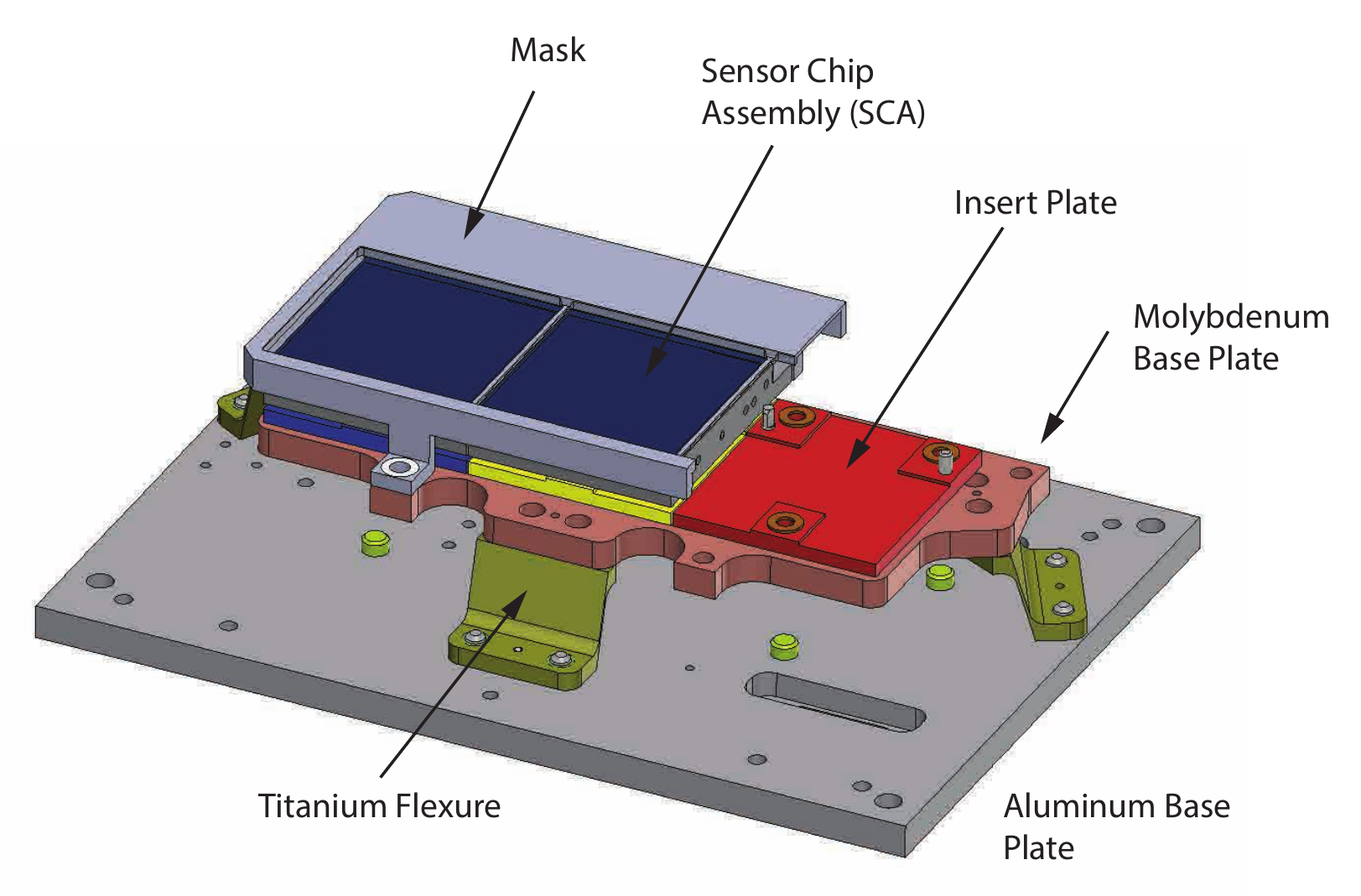}
\caption{Following the JWST NIRCam array mount design, the SCAs are mounted on a molybdenum base plate.  Three titanium flexures accommodate the differential shrinkage between the molybdenum and aluminum base plates.  Individual SCA tip-tilt and piston are set by the use of custom molybdenum insert plates bonded to the molybdenum base plate. Lastly, the thickness of precision washers between the SCA and insert plate are used to fine-tune SCA position.}\label{fig_det_closeup}
\end{figure}


\begin{figure}
\epsscale{1.15}
\plotone{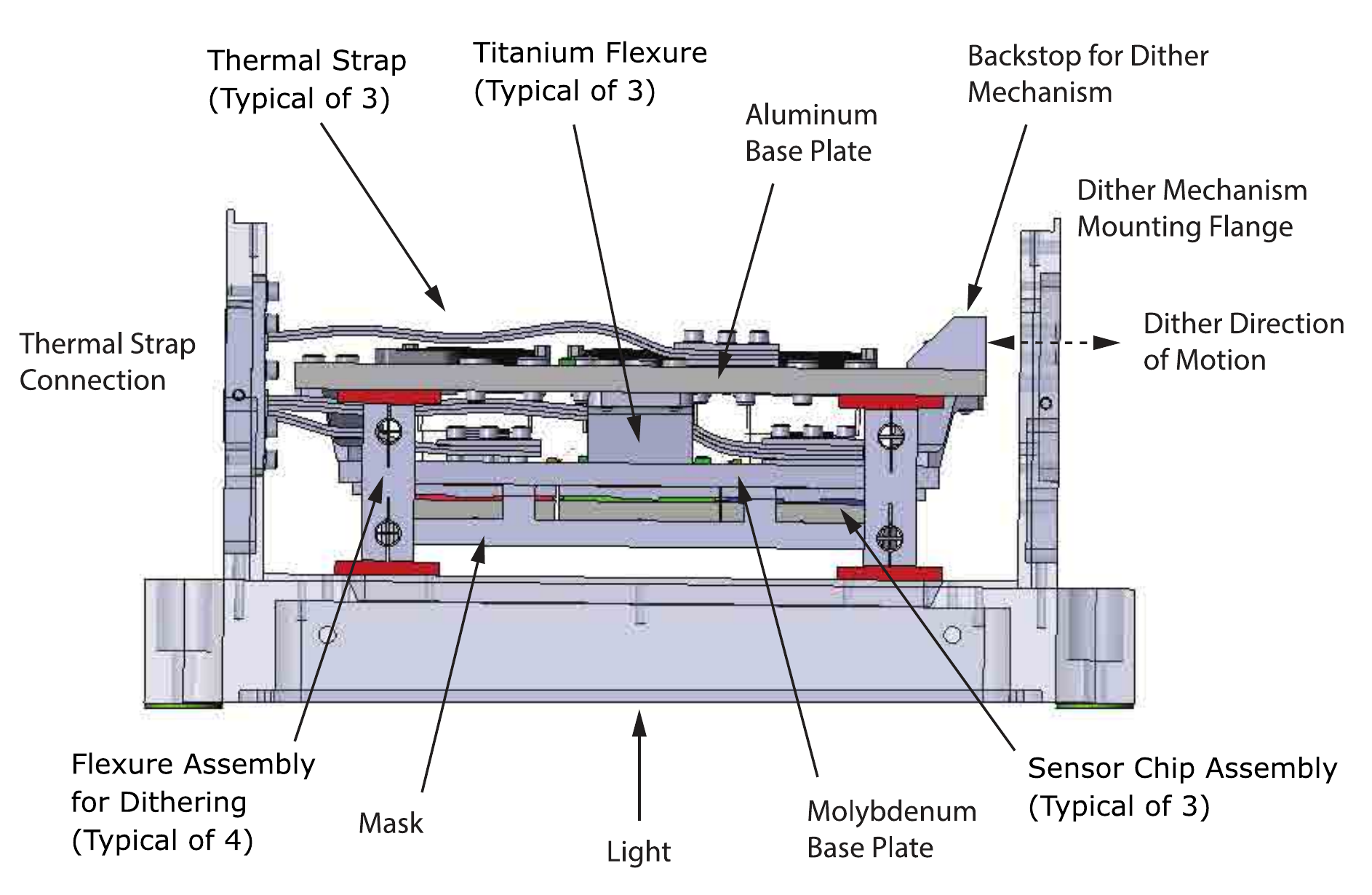}
\caption{Sideview of the detector array mosaic assembly, which shows the flexible thermal straps and the flexure assemblies that permit the mosaic of SCAs to dither in the spectral direction.}\label{fig_det_assembly_sideview}
\end{figure}


Three Teledyne 2048 x 2048 pixel, $2.5\,\micron$ cutoff, HgCdTe H2RG detector arrays with $18\,\micron$ pixel pitch \citep{hod04}, mounted side-by-side in the dispersion direction, are positioned at the spectrograph camera focal plane.  On loan from the University of Arizona, the devices are James Webb Space Telescope Near Infrared Camera (NIRCam) flight reject arrays.  They are substrate-removed devices, with good responsivity at visual wavelengths, and are AR coated.  The detector array mosaic mount system (Figures~\ref{fig_det_closeup} and \ref{fig_det_assembly_sideview}) borrows significant heritage from the NIRCam array mount \citep{gar04,rie07}.

The Sensor Chip Assemblies (SCAs), made up of the hybridized arrays mounted on a molybdenum carrier base with dedicated 37-conductor electronic cabling, are butted together such there is the minimal $\approx 2.9\,\rm{mm}$ gap between adjacent detective areas.  The SCAs mount to a molybdenum mosaic base plate just as in the NIRCam design.  This molybdenum base plate is in turn connected to an aluminum plate via three titanium flexures which accommodate the differential contraction between aluminum and molybdenum.  Titanium was chosen for its low thermal conductivity (see below).

To set the focus and tip-tilt orientation of each SCA there are extra array-specific molybdenum ``insert-plates'' between the SCAs and the mosaic base plates (Figure~\ref{fig_det_closeup}).  The thicknesses and wedge angles of the insert plates set the detector arrays in a ``polygon approximation'' of the focal surface; the spectrograph camera focal length increases with increasing wavelength.  The insert plates are bonded to the mosaic plate with 3M EC 2216 two-part epoxy.  Each SCA sits on three pads on the insert plate and is pinned in two places in keeping with a bolt-and-go approach.  Precision brass washers between the SCAs and insert plates fine-tune the SCA orientations.  The orientations were first set using a precision depth gauge to position dummy chip carriers on the mounts.  After installation of the real detector arrays and focus measurements in the APOGEE-North instrument while cold, the detector array mosaic assembly was removed and the detector array positions were fine-tuned.  Fine-tuning was not required for APOGEE-South.

The mosaic assembly is enclosed within an $8.0 \times 6.0 \times 3.55\,\rm{in}$ ($203.2 \times 152.4 \times 90.2\,\rm{mm}$) aluminum housing which mounts to a flange secured to the back of the camera barrel.  The housing position can be fine-tuned relative to the camera using precision shims to adjust global tip-tilt and piston of the detector array mosaic.  Re-entrant features in the housing wall make the edges and corners light-tight.

To permit movement of the detector array mosaic along the dispersion direction for spectral dithering, the aluminum base plate is attached to the aluminum housing back wall via struts and Riverhawk 5008-600 precision flex pivots (Figure~\ref{fig_det_assembly_sideview}).  Eight flex pivots, one each at the ends of the pivot arms, provide a parallelogram arrangement which allows swinging of the detector array mosaic along the dispersion axis.  A dithering mechanism, driven by a cryogenic stepper motor with fine-pitched leadscrew and lever arm, is described in \S~\ref{dither_mechanism}.  A hemispheric-tipped ``drive pin'' contacts a flat pad secured to the edge of the aluminum base plate.  The pad is kept in contact with the mechanism drive pin through the pre-load of the copper thermal straps (described below).

\subsection{Thermal Control}

For several reasons (explained below) the detector arrays are thermally isolated from the camera barrel and directly cooled by the $\rm{LN_2}$ tank through a carefully designed copper connection.  Thermal isolation is provided by the low thermal conductivity titanium flexures, which help isolate the molybdenum base plate from the aluminum plate, and the flex pivots, which resist thermal conductivity from the camera barrel through the aluminum detector array mosaic housing.  A copper pole, bolted to the top of the $\rm{LN_2}$ tank, protrudes through a hole in the cold plate and rises to a point adjacent to the detector housing.  Flexible copper straps in turn connect the top of the pole to the molybdenum and aluminum plates through a G-10 insert in the wall housing (Figure~\ref{fig_det_assembly_sideview}).

The RC time constant, which consists of the thermal resistance of the copper link assembly to the $\rm{LN_2}$ tank and the thermal capacitance of the molybdenum mosaic base plate, controls the thermal performance and allows completely passive control of the detector array temperature during cool-down, normal operation, and warm-up.  During cool-down the detector arrays cool at a maximum rate of $\sim 0.6 \, \rm{K \, {min}^{-1}}$, well below the recommended maximum of $1.0 \, \rm{K \, {min}^{-1}}$.  During operation the detector arrays typically operate at $\approx 75\,\rm{K}$, about $3\,\rm{K}$ lower than the cold plate.\footnote{Because of a likely inadvertent mismatch between the two Lake Shore Cernox\textsuperscript{\texttrademark} sensor serial numbers and calibration curves, the reported temperatures on the molybdenum base plate below the detectors are $\sim 74\,\rm{K}$.  But this is unphysical --- given an observatory altitude of $2.78\,\rm{km}$ (typical site pressure of 0.69 atmospheres), the predicted ambient $\rm{LN_2}$ temperature should be $74.25\,\rm{K}$ according to equation 1 in \citet{fri50}.  Furthermore, the housekeeping temperature sensing diodes (see \S~\ref{housekeeping}) attached to the bottom and top of the copper pole described above had mean temperatures throughout SDSS-III of approx $75.2\,\rm{K}$ and $74.8\,\rm{K}$, resp.  (The apparent temperature inversion between these two copper pole temperatures --- the bottom of the copper pole, closer to the $\rm{LN_2}$ tank, should be colder --- is probably indicative of sensor calibration inaccuracies.)}  During warm-up, after the $\rm{LN_2}$ tank has been purged, the detector arrays warm quicker than the large camera until close to room temperature.  This ensures the detector arrays will not be at the coldest temperature within the instrument when the getters release adsorbed gases.

While it has not been implemented, the thermal system also includes the ability to actively control detector array temperatures.  Resistive heaters, mounted to the bottom of the molybdenum baseplate, can be controlled by a Lake Shore controller using feedback from precision temperature diodes (also mounted on the bottom of the plate).  We have not used active temperature control to keep dark current constant as the dark current noise penalty during exposures is negligible compared to the source noise for survey target observations.  Operationally, temperature fluctuations are minimized by continuous array readout and automated cryogen fills every morning after observations are complete such that cryogen level and temperature profiles are repeatable every evening.  The maximum detector array temperature excursion over the 34 months of the survey in SDSS-III was $0.28\,\rm{K}$.  Temperature excursions are largely driven by atmospheric pressure variability which directly affects the $\rm{LN_2}$ boil-off temperature.

\subsection{Electronic Control}

\edit1{The detector arrays are operated in four-channel output mode at slow ($100\,\rm{kHz}$ pixel rate) readout speed with the read-out channels oriented perpendicular to the spectral dispersion direction.}  Thus all three detector arrays \edit1{can be controlled by one} Leach Generation III controller as if there were a single detector array with twelve video channels: they are provided the same biases, are clocked synchronously, and are read out simultaneously.  The controller includes two 8-channel IR video boards and one clock board.  A custom fanout board mounted on the cold plate adjacent to the detector housing interconnects the three functional cables (video, clock, bias) from the Leach controller with dedicated device cables going to each detector array.  All cryogenic cabling between the cryostat wall and detector array mosaic housing were fabricated by CryoConnect (Etruria, Stoke-on-Trent, UK) and feature Nomex\textsuperscript{\textregistered} weaved through the cable conductors to ensure cable integrity while retaining bending compliance.  Also, the cable ends are potted into the connectors.  A combination of twisted-pair phosphor bronze (32 AWG), single-conductor phosphor bronze (32 AWG), and single-conductor copper lines is used in the cryogenic cabling.  Biases are supplied by low output impedence voltage sources.

The standard survey exposure \edit1{is taken in sample-up-the-ramp (SUTR) mode composed of a global reset followed by} 47 non-destructive reads, each $\approx 10.6\,\rm{sec}$.  Eight of these $500\, \rm{sec}$ exposures are taken in dithered mode where $\sim 0.5$ pixels separate dither positions.  The dither scheme is typically 2 x ABBA.

The Instrument Control Computer (ICC), described in \S~\ref{icc}, communicates with the Leach controller over fiber optic lines in the typical manner with an ARC-64 (PCI Interface Board) in the computer and an ARC-22 250 MHz Gen III Timing Board in the Leach controller.  The ICC delivers a $2048 \times 8192$ frame for each read of the ramp.  Each frame consists of the three $2048 \times 2048$ detector array images as well as the three different $512 \times 2048$ reference output results for each detector array combined into one extra $2048 \times 2048$ pixel frame \citep[see][]{nid15}.  The ICC has the ability to shorten or extend the number of ramp reads, but the read timing is fixed.

\subsection{Detector Array Performance}

\edit1{In the following section readout noise, gain, dark current, linearity, bad pixels, and the abnormally high persistence of the original APOGEE-North blue detector array, which has since been replaced, are addressed.  (\citet{nid15} also discuss performance details of the detector arrays.)  Readout noise and gain are discussed extensively since an accurate determination of both are important ingredients in an instrument noise model and calculations of achieved S/N.  For infrared detectors that can be read out non-destructively, readout noise depends on the properties of the detector as well as the chosen readout method.  Furthermore, readout noise becomes an important noise contributor for targets at the faint end of the survey (see \S~\ref{noise_model}).  Gain, the relationship between electrons detected and digitized counts delivered by the detector control electronics, is also fundamental to determining overall instrument throughput.}

\subsubsection{Readout Noise}


\begin{figure*}
\epsscale{1.05}
\plotone{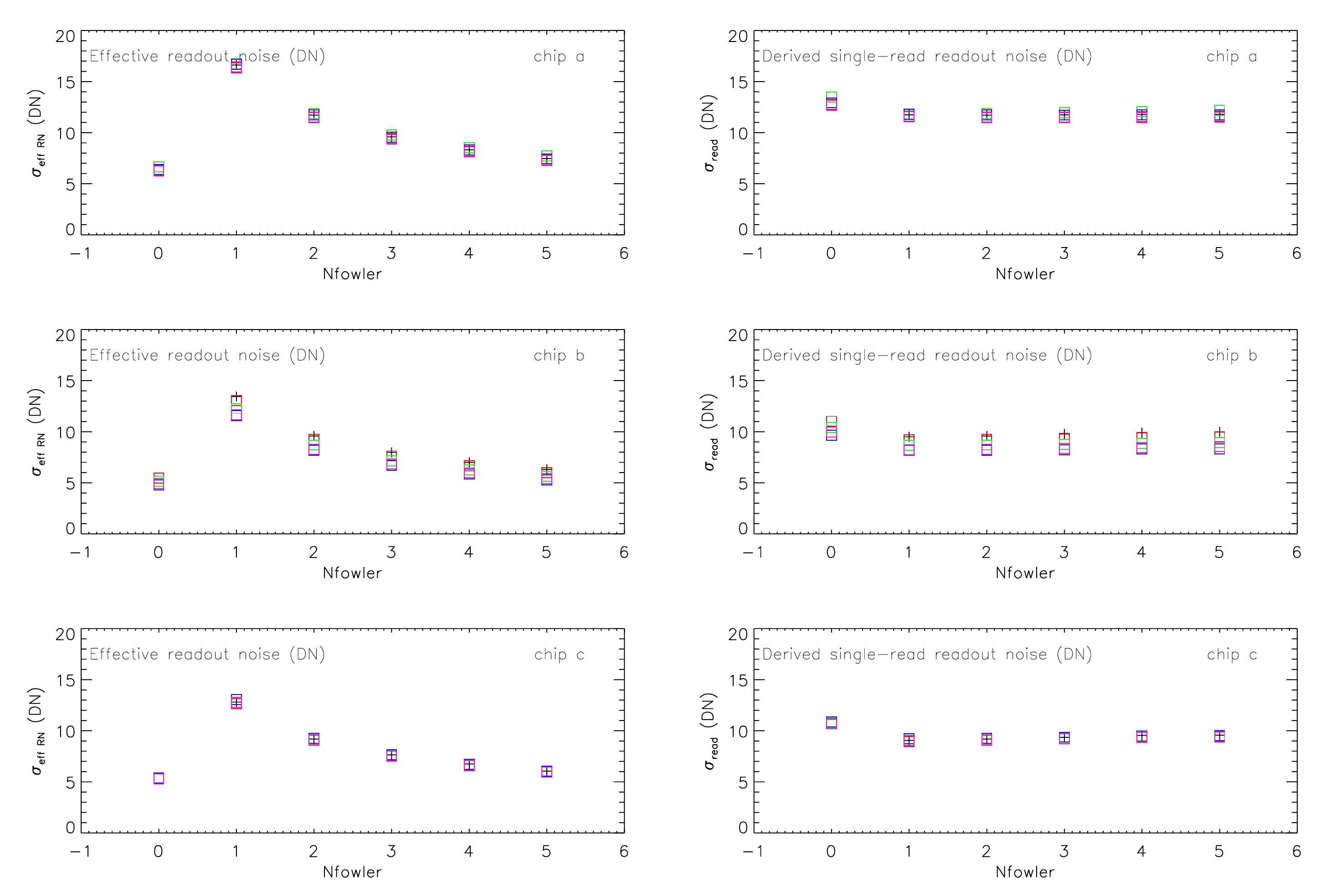}
\caption{\edit1{Readout noise analysis for the three detector arrays of the APOGEE-North instrument \edit1{for SUTR and Fowler Sampling}.  The readout noise was calculated in the four 512-column stripes separately.  Colors correspond to the different detector array output channels.  The abscissa coordinate of $\rm{N_{Fowler}} = 0$ is used to show results for SUTR; $\rm{N_{Fowler}} = 1$ corresponds to CDS. (Left) Effective readout noise.  (Right) Single-read readout noise derived from equation~\ref{readout_noise_equation}, i.e., $\sigma_{\rm{read}}$.  The plots on the left show that effective read noise decreases as $\rm{N_{Fowler}}$ increases, as expected, but is lowest for SUTR.  While the readout noise would be even lower for larger $\rm{N_{Fowler}}$, such a choice would lead to a shorter effective exposure time and lower overall $\rm{S/N}$ in most cases.  The plots on the right show that nearly consistent values for $\sigma_{\rm{read}}$ are recovered; we adopt $\sigma_{read}$ as derived from the SUTR analysis, since this is what we use for normal exposures.}}\label{fig_readout_noise_analysis}
\end{figure*}


\edit1{Readout noise is a critical parameter for a noise model, as mentioned above.  For infrared detector arrays with non-destructive readout, the multiple reads of an exposure can be analyzed in several ways, including Fowler Sampling and SUTR, which give differences in the effective (observed) readout noise.}  This is discussed in detail in \citet{rau07}, who find that the variance of \edit1{the noise of an exposure}, in the absence of photonic noise and dark current, is

\begin{equation}\label{readout_noise_equation}
  \sigma^2_{\rm{eff\:RN}} = \frac{12(n-1)}{mn(n+1)} \sigma^2_{\rm{read}}
\end{equation}

\noindent
where $n$ is the total number of group reads, $m$ is the number of reads per group, and $\sigma_{\rm{read}}$ is the readout noise per read.

\edit1{Since we take exposures with different total numbers of readouts ($n$), we need to determine $\sigma_{\rm{read}}$ to be able to calculate the effective readout noise.  In the analysis that follows, different readout methods (SUTR ($n = \rm{n_{\rm{reads}}}$, $m = 1$) and Fowler Sampling ($n = 2$, $m = \rm{N_{Fowler}}$) with $1 \le \rm{N_{Fowler}} \le 5$) are used with data from the APOGEE detector arrays to demonstrate this variation of $\sigma_{\rm{eff\:RN}}$ and that application of the equation leads to consistent values for $\sigma_{\rm{read}}$, a parameter that should remain unchanged regardless of readout method.}

\edit1{Using the difference image of two reduced dark images, the readout noise of the detector arrays was estimated for the various readout methods by calculating the robust RMS noise of pixel-to-pixel fluctuations. After division by $\sqrt{2}$ to give $\sigma_{\rm{eff\:RN}}$, equation~\ref{readout_noise_equation} was used to correct to $\sigma_{\rm{read}}$.}  Figure~\ref{fig_readout_noise_analysis} shows the results for the three detector arrays of the APOGEE-North instrument based on an image preceding the summer shutdown of 2014.  (It was during this shutdown that the original blue detector array, which suffered from abnormally high persistence (see below), was swapped with a more typically performing detector.)  The readout noise was calculated in each of the four 512-column stripes separately. The left panel shows the effective readout noise, while the right panel shows the readout noise per read derived from equation~\ref{readout_noise_equation}.  SUTR mode is shown at $\rm{N_{Fowler}} = 0$ and correlated double sampling (CDS) at $\rm{N_{Fowler}} = 1$.  Colors correspond to the different detector array output channels.

The results show that the effective readout noise decreases as $\rm{N_{Fowler}}$ increases as expected from equation~\ref{readout_noise_equation}.  And they suggest that the effective readout noise for $\rm{N_{Fowler}} > 8$ may be lower than the effective readout noise for SUTR, as expected from first principles.

Ultimately, though, it is $\rm{S/N}$ that must be maximized.  So SUTR mode is implemented in the APOGEE pipeline \citep{nid15} because it provides the highest $\rm{S/N}$ for fainter signals by virtue of its low effective readout noise and efficient use of exposure time compared to Fowler sampling.  For example, implementation of Fowler-8 sampling would result in the loss of the contribution of 8 reads ($> 80\,\rm{sec}$) to the integrated signal since the integration time for Fowler sampling is defined from the end of the first group of reads (to measure pedestal) to the end of the last group of reads (to measure signal).  SUTR also allows cosmic ray rejection and monitoring for saturation.

Table~\ref{tbl-readout_noise} gives $\sigma_{\rm{read}}$ for SUTR and CDS, in DN and $e^{-}$, adopting the gains described below, for both instruments.  Since we want the best estimate of $\sigma_{\rm{eff\:RN}}$, we adopt the $\sigma_{\rm{read}}$ for SUTR (i.e., the third column of Table~\ref{tbl-readout_noise}), which is then propagated through equation~\ref{readout_noise_equation} in the reduction pipeline.


\begin{deluxetable*}{lccc}
\tabletypesize{\scriptsize}
\tablewidth{0pt}
\tablecaption{Effective Single-Frame Readout Noise \label{tbl-readout_noise}}
\tablehead{\colhead{Instrument} & \colhead{Detector Array} & \colhead{$\sigma_{\rm{read}}$ for SUTR DN (e-)} & \colhead{$\sigma_{\rm{read}}$ for CDS DN (e-)}}

\startdata
APOGEE-N & a (red) & 13 (24.7) &   12 (22.8) \\
 & b (green) & 11 (20.9) & 9 (17.1) \\
 & c (blue) & 10 (19) & 8 (15.2) \\
APOGEE-S & a (red) & 7 (21) & 4 (12) \\
 & b (green) & 8 (24) & 5 (15) \\
 & c (blue) & 4 (12) & 3 (9) \\
\enddata

\end{deluxetable*}


\subsubsection{Gain}\label{gain}

We calculated detector array gain by using the photon transfer technique \citep{jan87}.  After taking the difference between two images, calculating the variance in regions of approximately the same illumination, and subtracting the variance due to read noise, if non-negligible, gain was calculated by finding the ratio of signal to variance.  Note that this can be accomplished across a broad range of illumination levels even with a single image since the internal flat fields span a large range of illumination (assuming there is a single gain value).  Gain can also be determined by looking at the variance of individual pixels in a SUTR image stack and taking the expectation value across many reads.

In either case, the measured variance depends on how the counts are extracted from the image cube.  According to \citet{rau07}, in the limit of no readout noise, the photonic noise is

\begin{eqnarray}\label{gain_equation}
  \sigma^2_{\rm{total\:shot}} & = & \frac{6(n^2+1)}{5n(n+1)}(n-1) t_g f \nonumber \\
  & - & \frac{2(2m-1)(n-1)}{mn(n+1)} (m-1) t_f f
\end{eqnarray}

\noindent
where $n$ and $m$ are defined as above, $t_g$ is the group time, $t_f$ is the frame time, and $f$ is flux in ${e^{-}}\,{s^{-1}}\,{\rm{pixel}^{-1}}$.  For CDS ($n=2$, $m=1$),  equation~\ref{gain_equation} reduces to the familiar equation for shot noise, but this is not the case for SUTR, which gives a slightly higher variance (which would be interpreted as a lower gain if not recognized).


\begin{figure*}
\epsscale{1.0}
\plotone{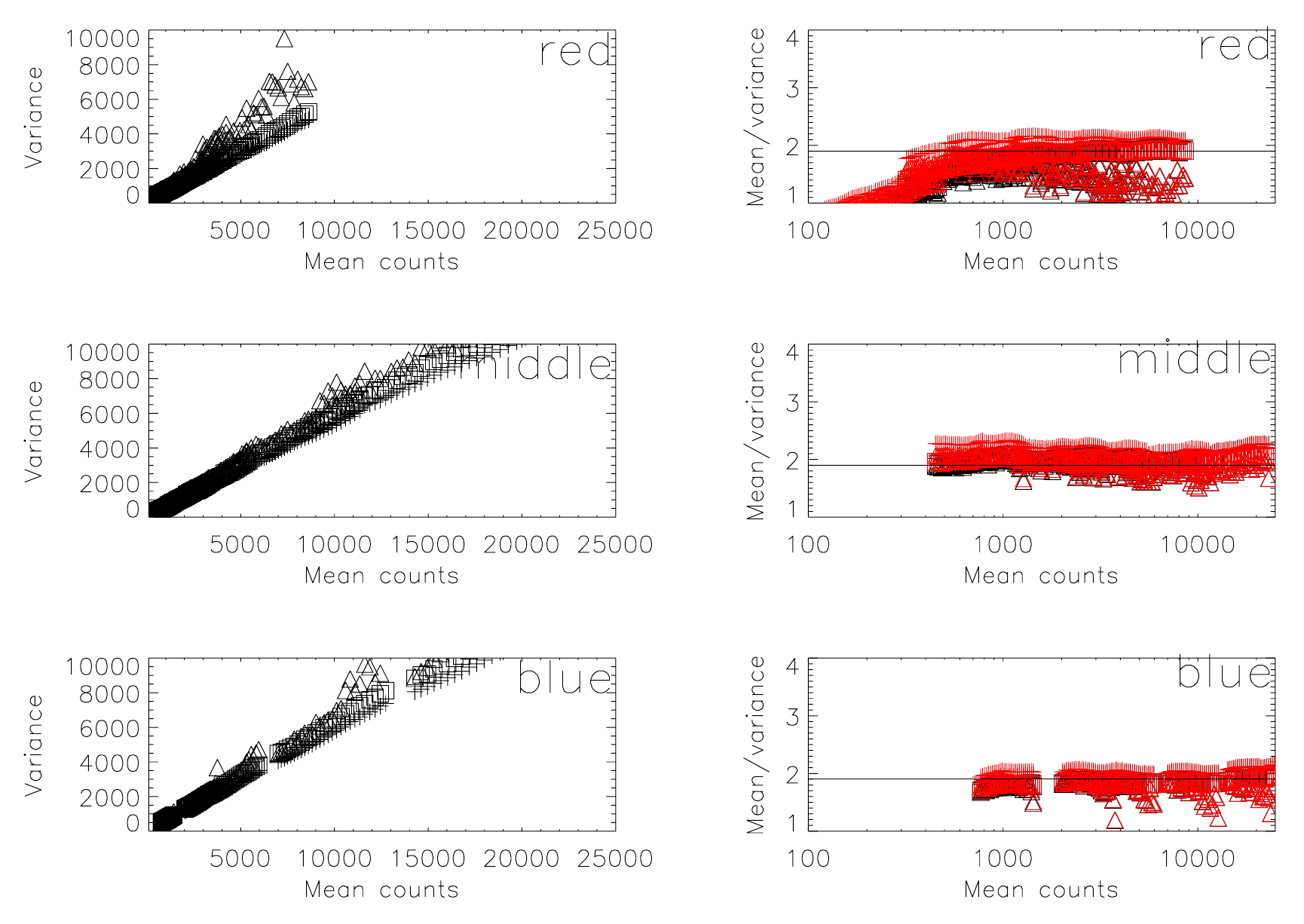}
\caption{Gain analysis for the APOGEE-North instrument based upon a few stacks of internal flat frames. To increase the illumination range, stacks of several different numbers of reads (hence the groups of points, each of which represents a different number of reads in the stack) were considered. (Left) Variance as a function of mean counts.  (Right) Derived gains as a function of mean counts. The different point types come from the difference analysis (squares) and the stack analysis using the mean of the variances (triangles).  Black data are for SUTR uncorrected for readout noise while red have had the the readout noise component of the variance subtracted.}\label{fig_gain_apogee_north}
\end{figure*}


Figure~\ref{fig_gain_apogee_north} shows the results for the APOGEE-North instrument based upon a few stacks of internal flat frames which provide a range of illumination levels as mentioned above. To increase the illumination range, we consider the stacks at several different numbers of reads (hence the groups of points, each of which represents a different number of reads in the stack). Left panels show variance as a function of mean counts, while the right panels show the derived gains as a function of mean counts. The different point types come from the difference analysis (squares) and the stack analysis using the mean of the variances (triangles).  Black data are for SUTR uncorrected for readout noise while red data have had the the readout noise component of the variance subtracted.

The results are challenging to fully interpret, and may reflect incorrect methodology or assumptions.  The gain appears to change with illumination level, and is not totally consistent between detector arrays.  All three detector arrays are read through the same set of electronics, so in principle the gain on each chip should not be independent.  Also, the method using individual pixel stacks (triangles) gives somewhat different results from the difference method (squares).

It is difficult to accurately measure the gain of infrared detector arrays for multiple reasons, including their non-linear response and inter-pixel capacitance (IPC).  For instance, the commonly used photon transfer technique \citep{jan87}, which assumes ideal detector behavior, can give gain values that are incorrect by about 10 -- $15\,\%$ for $2.5\,\micron$ cut-off H2RG detector arrays due to IPC \citep{fin06}.

Based on Figure~\ref{fig_gain_apogee_north}, and similar plots for the APOGEE-South instrument, gains of $1.9\,{e^{-}}/\rm{DN}$ for APOGEE-North and $3.0\,{e^{-}}/\rm{DN}$ for APOGEE-South were adopted.  \edit1{The different gains were surprising since a $0.25\,\rm{V}$ reverse bias is applied to the detector arrays of both instruments.  With the different realized gains, full well in DNs is reached at $\approx 75\,\%$ of 16-bit A/D saturation for APOGEE-South whereas A/D saturation occurs for counts slightly lower than full well for APOGEE-North.  Since the typical exposure levels for APOGEE survey targets are in the hundreds to thousands of counts, well below full well and A/D saturation, the gains were not adjusted.  Lastly,} no adjustment has been made for the effect of IPC to date; including such a correction would lower the calculated S/N by $\sim 7$ -- $8\,\%$.

\subsubsection{Dark Current}\label{dark_current}

Dark current was stable throughout the survey with a rate of $< 0.5\,\rm{DN}$ per read ($< 0.1\,{e^{-}}\,\rm{s^{-1}}$) for most pixels.  Furthermore, $> 90\%$ of the pixels in the blue and red detector arrays, and $\sim 80\%$ of the pixels in the green detector arrays, had rates of $0.1\,\rm{DN}$ per read ($< 0.02\,{e^{-}}\,\rm{s^{-1}}$).

\subsubsection{Linearity}


\begin{figure}
\epsscale{1.15}
\plotone{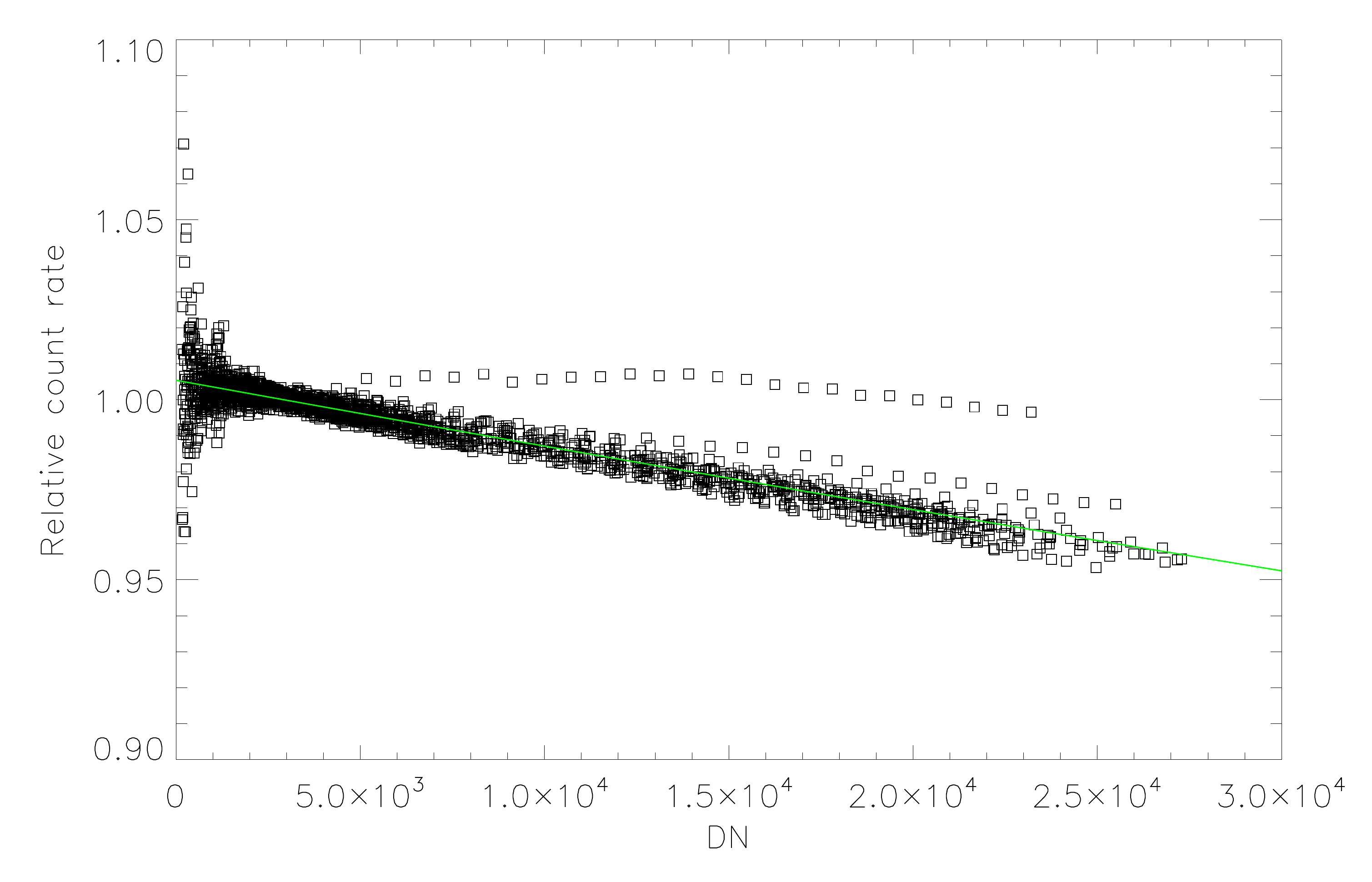}
\caption{Counts accumulated during internal flat field exposures across all three detector arrays before the blue detector array replacement in the APOGEE-North instrument.  Data from regions with significant persistence (i.e, the green detector array and a portion of the blue detector array) have been excluded. The results have been normalized to unity at $3000\,\rm{DN}$. The non-linearity is 6 -- $7\,\%$ up to $20000\,\rm{DN}$.}\label{fig_apogee_north_linearity}
\end{figure}


Linearity has been analyzed based on the accumulation of counts during internal flat field exposures, under the assumption that the illumination level is constant, i.e., comparing the number of accumulated counts in successive reads in SUTR mode. Figure~\ref{fig_apogee_north_linearity} shows data across all three detector arrays before the blue detector array replacement in the APOGEE-North instrument.  Data from regions with significant persistence (i.e, the green detector array and a portion of the blue detector array) have been excluded. The results have been normalized to unity at $3000\,\rm{DN}$. The non-linearity is 6 -- $7\,\%$ up to $20000\,\rm{DN}$. However, the bulk of the survey exposures have counts in the hundreds to thousands of DN, and the range within a given spectrum is generally (significantly) less than a factor of two.  Thus a linearity correction has not been implemented in the data reduction pipeline as of 2018.

\subsubsection{Bad Pixels}


\begin{figure*}
\epsscale{1.15}
\plotone{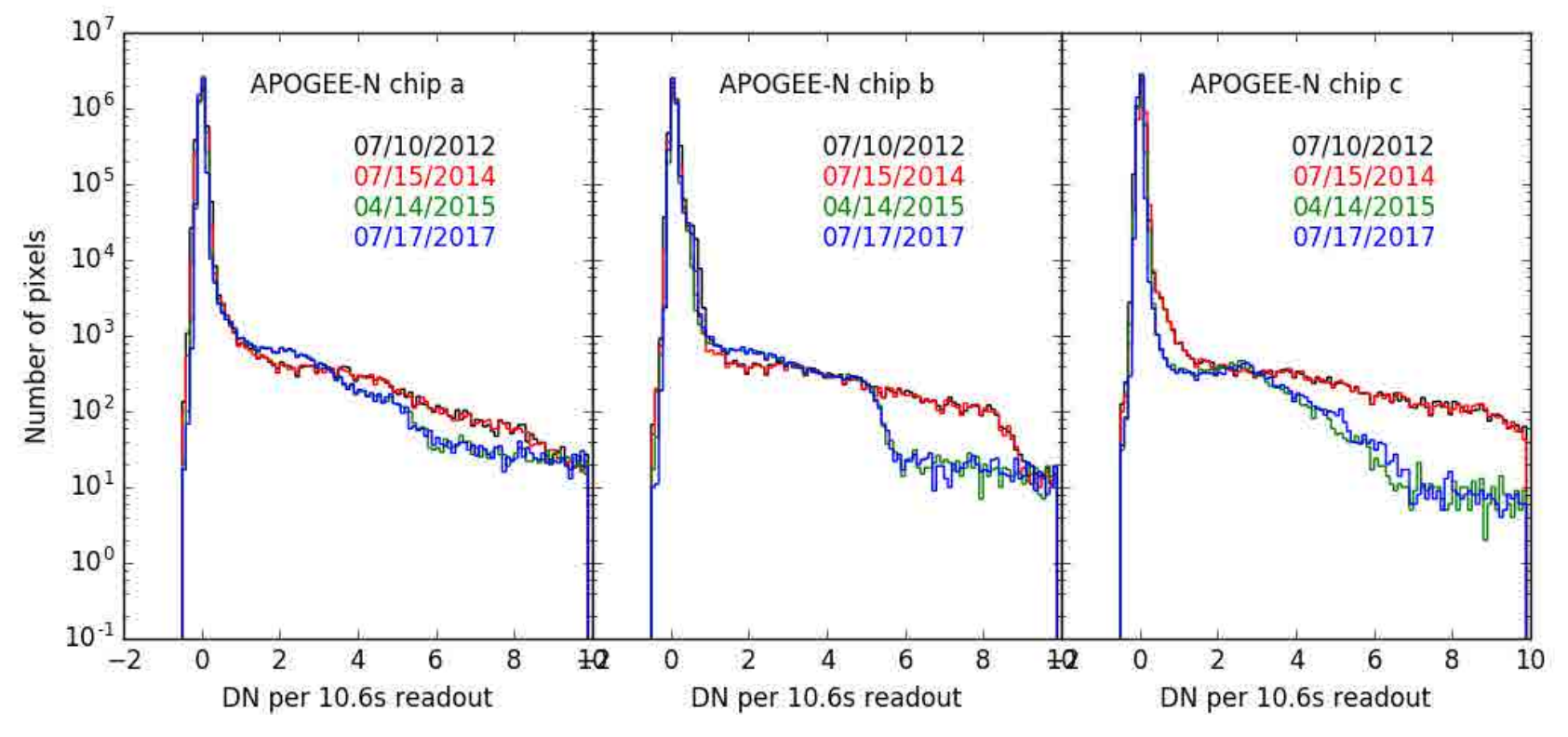}
\caption{Histograms of dark counts for each detector array (red, middle, blue, from left to right) spanning a five year period.  ``chip C'' refers to the blue detector array so the right-hand panel shows data for two different detector arrays since the swap occurred during the summer shutdown of 2014.}\label{fig_darkrate}
\end{figure*}


The detector arrays are quite stable. A few calibration frames are taken daily, and long series of dark and flat frames are taken periodically. Figure~\ref{fig_darkrate} shows histograms of the dark count from four sets of frames taken over a five year span, including after two cryostat warm-ups for servicing. The histograms are fairly similar, and show no indication of growth in the number of hot pixels.  Note that ``chip C'' refers to the blue detector array thus the right-hand panel shows data for two different detector arrays since the swap occurred during the summer shutdown of 2014.

\subsubsection{Abnormally High Persistence}\label{blue_persistence}

The JWST NIRCam detector arrays exhibit $< 1\,\%$ persistence counts compared to full well stimulus \citep[see Figure 3 and Table 1 of][]{lei16}.  Most of the APOGEE detector arrays have regions with similar persistence behavior.\footnote{Note, however, that the JWST NIRCam detector arrays were tested at $\sim 40\,\rm{K}$, whereas APOGEE detector arrays operate at close to $77\,\rm{K}$.}  But approximately one-third of the blue detector array, and portions of the green, suffer from a phenomenon which we called ``super persistence'' in \citet{ahn14}.  With signal levels typically achieved with APOGEE exposures, this phenomenon manifests itself as abnormally high latent images --- $\sim10$--20 times the persistence measured in ``normal'' regions.  See \citet{nid15} for further discussion, including the impact of this persistence on the data.

The cumulative signal level from super-persistence can be modelled with the following double-exponential function (\ref{persistence_model}):

\begin{equation}\label{persistence_model}
p(t) = C_0 + C_1 e ^ {C_2 t} + C_3 e ^ {C_4 t}
\end{equation}

\noindent
where $t$ is time and $C_n$ are model coefficients. In this functional form, the model coefficients represent real-world values with respect to cumulative super-persistence: $C_0$ is the asymptotic ceiling; $C_1$ and $C_2$ are the long-term amplitude and timescale; and $C_3$ and $C_4$ are the short-term amplitude and timescale, respectively.  The coefficients $C_2$ and $C_4$ are properties of the detector and do not change under different conditions. The remaining coefficients depend on stimulus, the depth to which individual pixels are exposed, and exposure time.


\begin{figure}
\epsscale{1.15}
\plotone{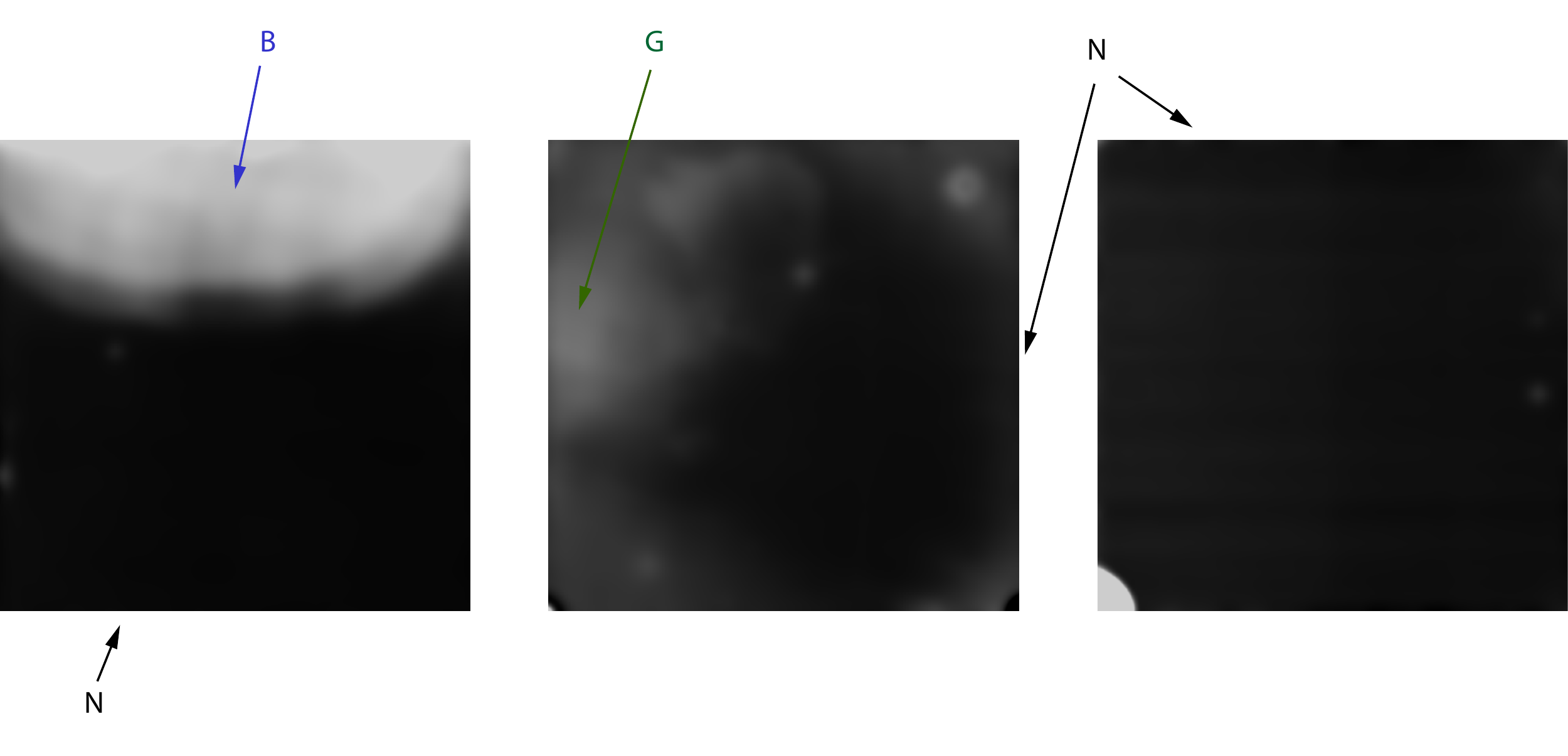}
\caption{Various portions of the blue and green detector arrays shown above, denoted by ``B'' and ``G,'' are affected by an abnormally high persistence.  Areas of typical persistence are denoted by ``N.''}\label{fig_persistence_a}
\end{figure}


\begin{figure}
\epsscale{1.2}
\plotone{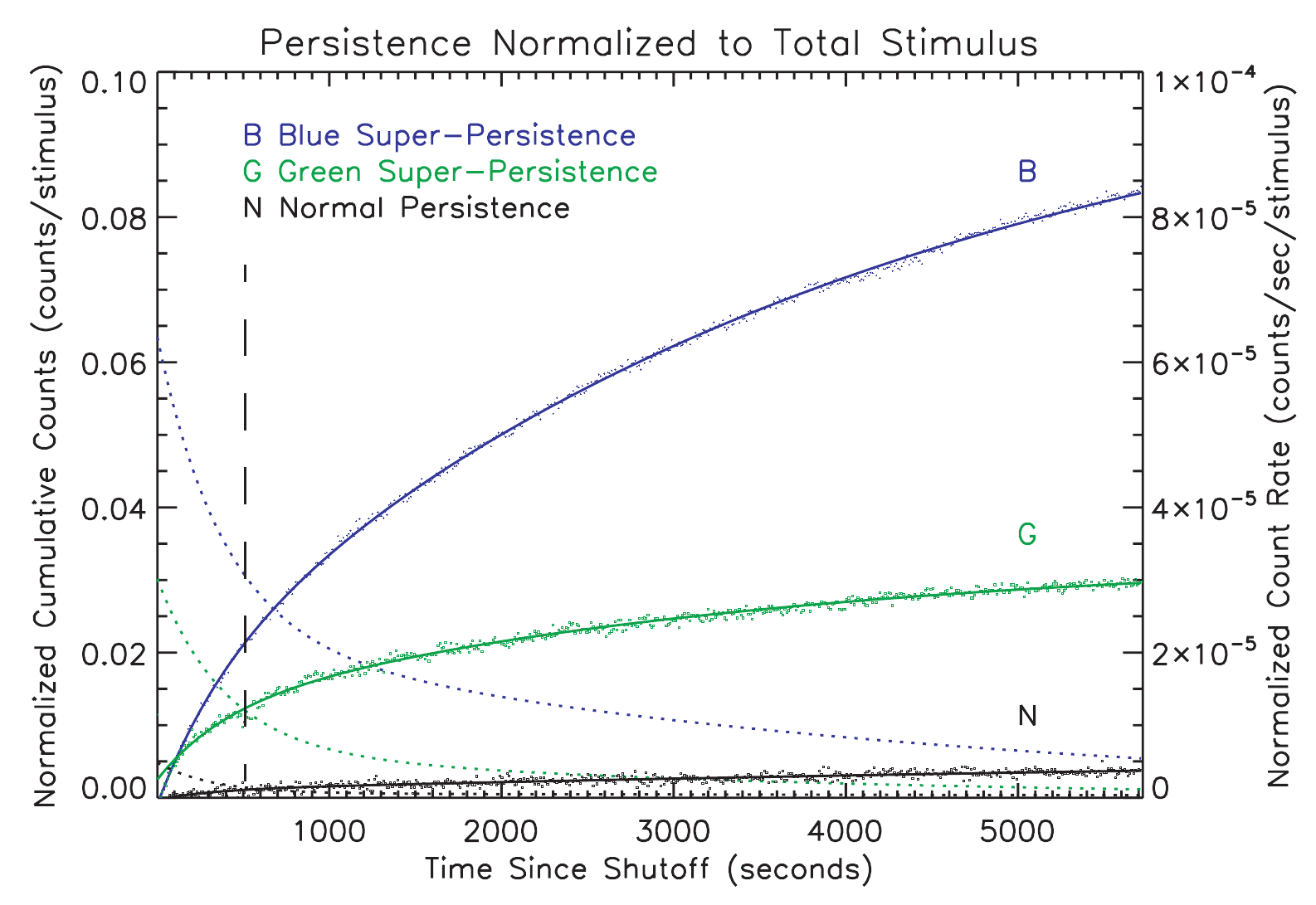}
\caption{The plot shows comparative decays derived from three 180-read (1908 second) exposures which followed a 60-read ($636\,\rm{sec}$) stimulus in which one-half of the internal LEDs illuminated the detector arrays.  Table~\ref{tbl-persist_coeff} gives the coefficients for the fits assuming the decay follows the model given by Equation~\ref{persistence_model}. The typical science frame integration time of $500\,\rm{sec}$ is marked.}\label{fig_persistence_b}
\end{figure}


Specific areas of the blue and green detector arrays affected by this abnormally high persistence (Figure~\ref{fig_persistence_a}), and the resultant decays of pixels representative of those areas are shown in Figure~\ref{fig_persistence_b}, along with the decay of normal persistence for comparison.  The following exposure sequence was used for this analysis: one-half of the internal LEDs illuminated the detector arrays for 60 reads ($636\,\rm{sec}$), followed by three 180-read (1908-second) exposures.  Illumination rates were 289.5, 163.5, and $339.4\,\rm{DN\,s^{-1}}$ for selected pixels in the blue super-persistence, green super-persistence, and normal persistence regions, respectively, which corresponds to 175, 99, and $205\,\%$ of full-well after 60 reads.  The decay curves were created by concatenating counts from the three exposures following the stimulus.  Table~\ref{tbl-persist_coeff} gives the normalized coefficients for the model fits to the curves.  The projected asymptotic output from persistence is $\sim 10.5\%$ from blue, $\sim 3.3\%$ from green, and $0.8\%$ from normal persistence.  Note that these coefficients were determined without first subtracting dark current.


\begin{deluxetable}{lccccc}
\tabletypesize{\scriptsize}
\tablewidth{0pt}
\tablecaption{Detector Array Persistence Fit Coefficients \label{tbl-persist_coeff}}
\tablehead{\colhead{Det Array} & \colhead{$\rm{C_0}$} & \colhead{$\rm{C_1}$} &
\colhead{$\rm{C_2}$} & \colhead{$\rm{C_3}$} & \colhead{$\rm{C_4}$}}

\startdata
Blue & 0.10523 & -0.09054 & -0.00025 & -0.01572 & -0.00262 \\
Green & 0.03342 & -0.02177 & -0.00030 & -0.00919 & -0.00257  \\
Red & 0.00777 & -0.00671 & -0.00009 & -0.00135 &  -0.00316  \\
\enddata

\end{deluxetable}


The modeling above suggests the persistence of a given exposure can be corrected as follows: the flux from a previous exposure or exposures (i.e., stimulus) is used to estimate the amplitude coefficients of the persistence time response function, after which, the resultant persistence time response function is used to estimate the cumulative amount of persistence present at a given time during the exposure, and this amount of flux is subtracted from the exposure.  Some effort at correcting persistence has been made starting with DR13, but in practice it is difficult to fully correct.  During the visit combination stage of data reduction, the pixels from spectra that land in persistence regions are down-weighted \citep{hol18}.

During the 2014 summer shutdown, between the end of SDSS-III and the start of SDSS-IV, the blue detector array was replaced with one with more typical persistence.  But the middle detector array has not been replaced.


\section{Cryostat}\label{mechanics section}

\subsection{Cryostat Design}

\begin{figure}
\epsscale{1.2}
\plotone{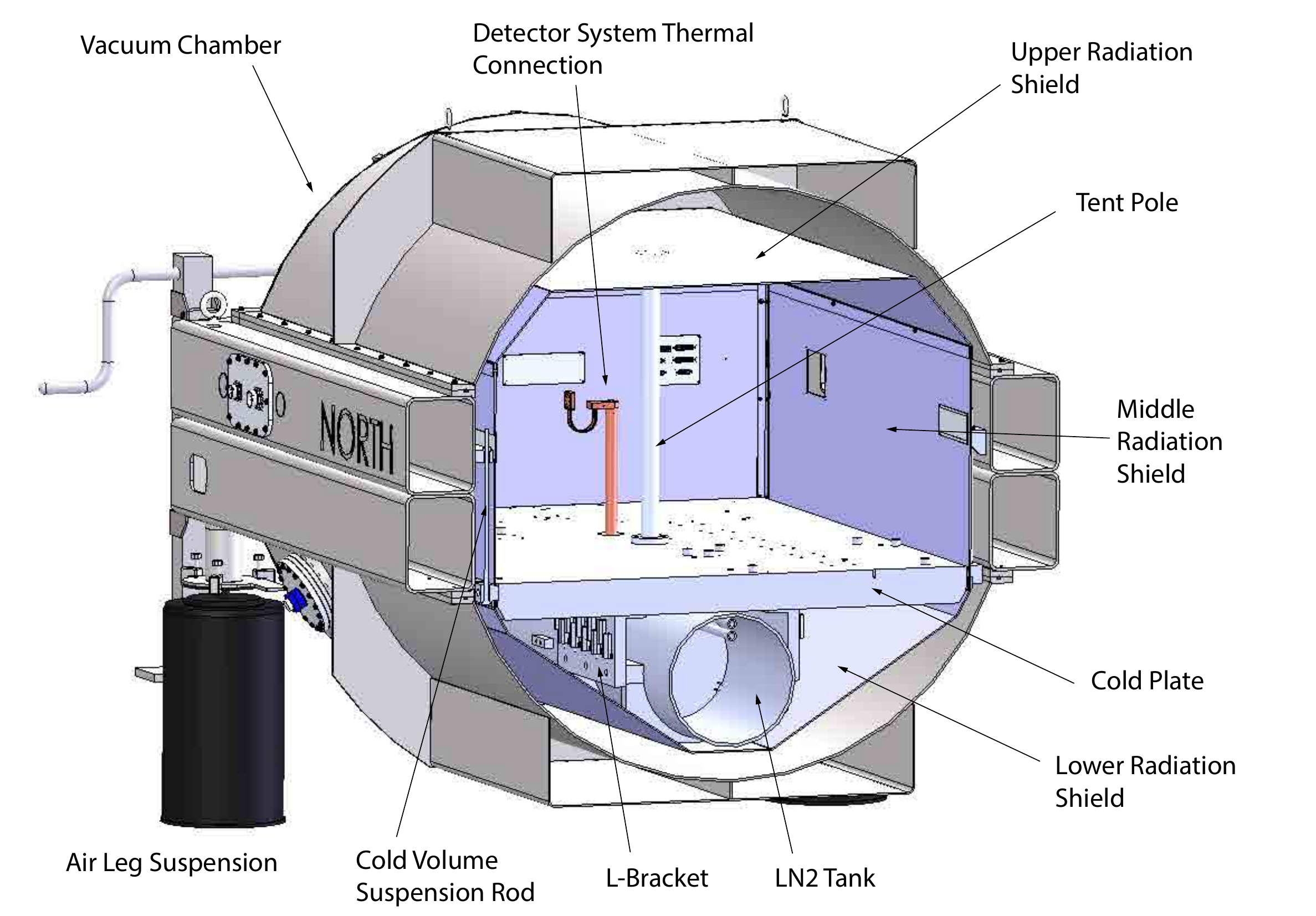}
\caption{A cross-section of the instrument cryostat.  The cryostat chamber is composed of a mid-section constructed from box beams and top and bottom lids.  A system of radiation shields surrounds the cold plate and $\rm{LN_2}$ tank.  The cold volume is suspended by stainless steel rods to accommodate thermal contraction and decouple it from outside vibrations.  Air leg suspension supports the cryostat to mitigate transmitted vibration from the lab floor.}\label{fig_cryostat_cross_section}
\end{figure}

The rectangular-shaped optical train mounts on a cold plate suspended within a large ($1.4\,\rm{m} \times 2.3\,\rm{m} \times 1.3\,\rm{m}$) and massive ($\approx 1{,}800\,\rm{kg}$) cylindrically shaped custom cryostat \citep{bla10} designed and constructed by PulseRay (Beaver Dams, NY).  Fabricated using low-carbon (304L) stainless steel, it is composed of a rectangular mid-section weldment capped with half-cylindrical weldments ('lids') above and below.  Figure~\ref{fig_cryostat_cross_section} shows a cross-section of the cryostat.


The mid-section is formed with $6\,\rm{in} \times 8\,\rm{in}$ ($0.15\,\rm{m} \times 0.2\,\rm{m}$) box beams welded into a frame.  The box beams have a wall thickness of $0.25\,\rm{in}$ ($6.4\,\rm{mm}$).  The lids are $18\,\rm{in}$ ($457\,\rm{mm}$) in height and are constructed with $\frac{3}{16} \,\rm{in}$ ($4.8 \,\rm{mm}$) thick sheet and appropriately positioned strength members.  Finite Element Analysis (FEA) of the cryostat design determined a maximum stress when under vacuum of $\approx 8{,}200\,\rm{kpsi}$ at the corners of the box beam weldment.  This stress implies a safety factor of 3.6 assuming a stainless steel yield strength of $30{,}000\,\rm{kpsi}$.  The maximum deformation of the cryostat under vacuum was expected to be $0.008\,\rm{in}$ ($0.2\,\rm{mm}$).

Stainless steel was selected instead of aluminum because of its superior weldability --- an important consideration given that $\approx 110\,\rm{m}$ of welds were necessary to fabricate the cryostat.  All welds include stitch welds on the outside for strength and continuous vacuum welds on the inside.

Two Viton\textsuperscript{\textregistered} O-rings provide the vacuum seals between the lids and mid-section.  The o-rings are $\frac{1}{4}\,\rm{in}$ ($6.4\,\rm{mm}$) diameter and $\sim 80\,\rm{in}$ ($2\,\rm{m}$) in length.

As the lids weigh $550\,\rm{lb}$ ($250\,\rm{kg}$) each, an engine hoist must be used to install and remove the top lid and a floor jack is used to manipulate the lower lid.  The latter has been attached to the mid-section since the cryostat left the PulseRay shop.

Permanently attached to each end of the cryostat are Atwood $5^{\rm{th}}$ Wheel Landing Gear (P/N 75333) assemblies, devices typically used to jack the front of recreational vehicle trailers.  Each leg of these two-legged jacks supports a corner of the instrument.  Each pair of legs can be manually jacked through a $20\,\rm{in}$ ($508\,\rm{mm}$) range while supporting a $6{,}000\,\rm{lb}$ ($2{,}700\,\rm{kg}$) load.  The jacks permit lowering the cryostat during lower lid and shipping cradle installation as well as final seating of the instrument on vibration isolators (see \S \ref{vibration_isolation}).

\subsection{Cold Plate Design}\label{cold_plate_design}

\begin{figure}
\epsscale{1.1}
\plotone{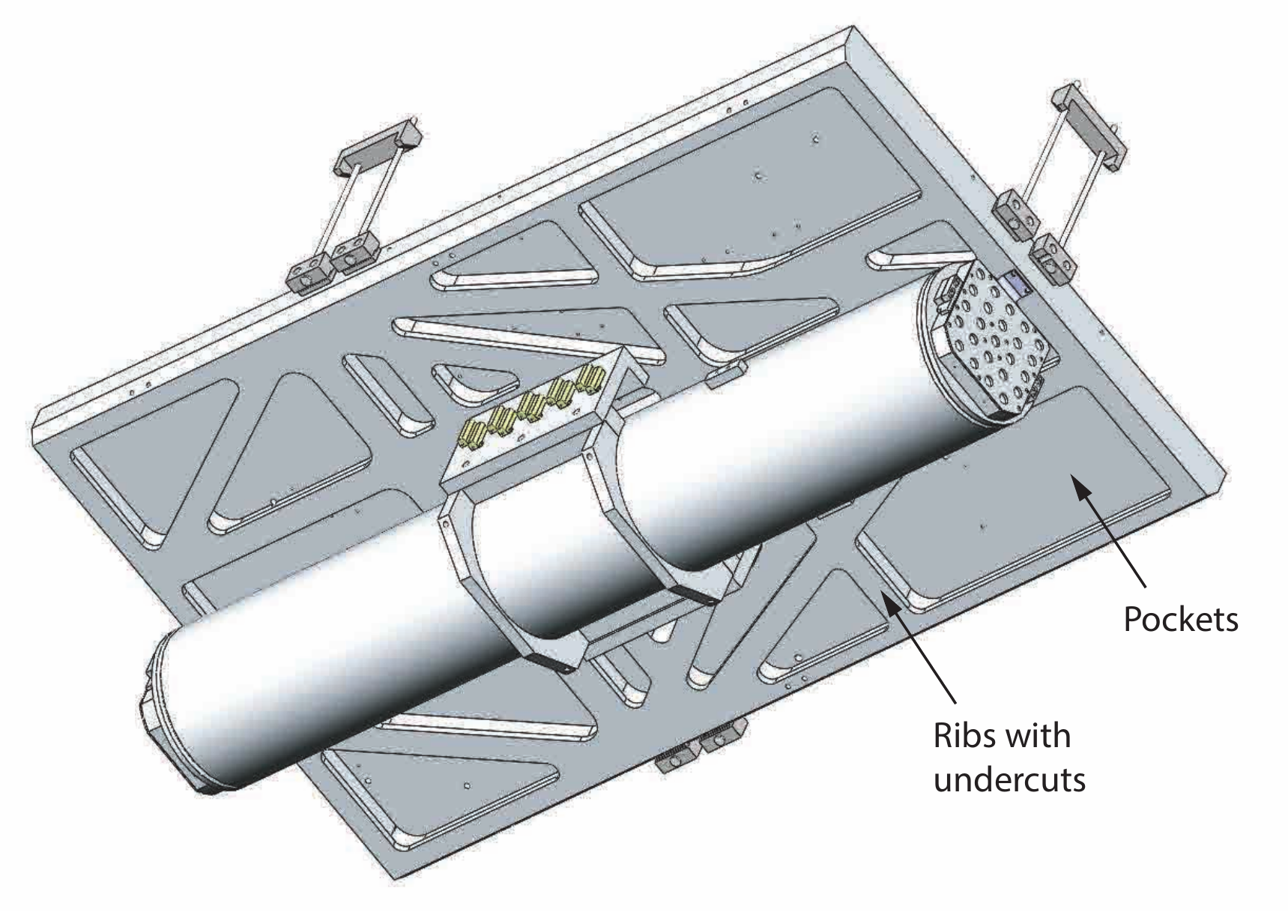}
\caption{The underside of the suspended $3\,\rm{in}$ ($76\,\rm{mm}$) thick cold plate is lightweighted with an isogrid-style scheme that reduced the weight by a factor of 2.3 to $376\,\rm{lb}$ ($171\,\rm{kg}$) while minimizing static plate bending.  The lightweighting scheme consists of aggressive pockets milled to a depth of $2.5\,\rm{in}$ ($63.5\,\rm{mm}$) and ribs with undercuts for strength.}\label{fig_isogrid_coldplate}
\end{figure}

The cold plate, with a size of $73.5\,\rm{in} \times 40.5\,\rm{in} \times 3.0\,\rm{in}$ ($1867\,\rm{mm} \times 1029\,\rm{mm} \times 76\,\rm{mm}$) was fabricated from a single billet of $3\,\rm{in}$ ($76\,\rm{mm}$) thick 6061-T6 precision flat-rolled aluminum plate.  An aggressive isogrid-style lightweighting pattern on the underside of the cold plate (Figure~\ref{fig_isogrid_coldplate}) reduced the weight by a factor of 2.3 to $376\,\rm{lb}$ ($171\,\rm{kg}$) while minimizing static plate bending given its suspension by the three hanger pairs (see below) along the edges.  Despite cold plate self-weight, the weight of the opto-mechanical assemblies, and a full $\rm{LN_2}$ tank suspended below the plate, FEA analysis predicts a maximum deformation (essentially a sag) of the cold plate of $\sim 0.0035\,\rm{in}$ ($0.09\,\rm{mm}$) in the vicinity of the pseudo-slit.  After fabrication, lightweighting, and drilling and tapping of some of the holes, the plate was cooled in a bath of $\rm{LN_2}$ for stress relieving before the top surface was skim cut and remaining holes were drilled and tapped.


Three pairs of $\frac{3}{8} \,\rm{in}$ ($9.5\, \rm{mm}$) diameter 304 stainless steel rods suspend the cold plate within the cryostat.  The lower ends of the rods have conical extrusions which fit into complementary conical cups on receiver blocks.  The upper ends of the rods are threaded and pass through similar blocks and hang from nuts and Belleville washers to allow adjustment of rod heights to evenly distribute the load amongst all rod pairs.  The upper blocks are secured to the cryostat mid-section and the lower blocks are secured to the edge of the cold plate. Rod pair locations were chosen to minimize cold plate deflection under the opto-mechanical assemblies.

\subsection{Cryogenic System}\label{cryogenic_system}

A $97\,\rm{liter}$ $\rm{LN_2}$ tank is fastened to the bottom of the cold plate.  Fabricated from 6061-T6 aluminum, the tank is $70\,\rm{in}$ ($1.78\,\rm{m}$) long and has an $11\,\rm{in}$ ($280\,\rm{mm}$) outer diameter.  Two $1.0\,\rm{in}$ ($25.4\,\rm{mm}$) thick ``L-brackets'' straddle the tank center and provide the structural and primary thermal connection between the tank and cold plate and thus all opto-mechanics mounted to the plate.  Secondary (mechanically compliant) thermal connections between the tank and cold plate are provided by copper straps at the tank end caps.

Inside the tank, the fill tube outlet is at the center bottom while the vent tubing is located at the top at one end of the tank.  Three ``slosh'' plates hinder movement of $\rm{LN_2}$ between tank sections by only allowing flow around an outer $\frac{1}{4}\,\rm{in}$ ($6.4\,\rm{mm}$) annulus surrounding each plate.

$\rm{LN_2}$ fill and venting are accomplished with dedicated stainless steel flexible hoses which run from the cryostat wall (hermetically sealed with Conflat\textsuperscript{\textregistered} style flanges), penetrate the cold-shielding (through a light-tight assembly), pass through the cold volume in the vicinity of Fold Mirror 2, and connect into the top of the tank, through holes in the cold plate, at another pair of Conflat\textsuperscript{\textregistered} style flanges.  Atlas Technologies' explosion-welded, bimetallic construction Conflat\textsuperscript{\textregistered} style flanges are used to transition from the 6061-T6 aluminum used for the tank construction and the stainless steel used for the flanges.

Tank level is continuously monitored and controlled with a liquid level controller from American Magnetics, Inc. (Oak Ridge, TN).  Based on cryogen level reported by a capacitance-based liquid level sensor in the tank, a Model 286 Liquid Level Controller triggers daily tank fills by remotely opening and closing a cryogenic fill valve in-line with an $\rm{LN_2}$ storage dewar adjacent to the instrument.  This storage dewar is, in turn, filled twice a week through a remotely controlled fill system within the building.  Together, these systems minimize the need for entry into the APOGEE room and enhance instrument stability.  Both the $\rm{LN_2}$ tank and the storage dewar vent through the wall to the outside for safety.

Activated charcoal getters are attached to each end of the $\rm{LN_2}$ tank.  Each of the two getters contains $300\,\rm{grams}$ of charcoal distributed within 24 cylindrical cups with $1.5\,\rm{in}$ ($38.1\,\rm{mm}$) diameter formed within a monolithic piece of aluminum to maximize uniform charcoal cooling.  The virtual volume model for getters developed by \citet{atw00} was used to calculate the effect of O-ring diffusion and the influence of charcoal getters in maximizing predicted time between vacuum servicing.  One side effect of using so much charcoal is that there is tangible water adsorption onto the charcoal at room temperature.  Much of this water gets removed during each pumpdown as the cryostat pressure passes $\lesssim 5\,\rm{Torr}$.  A temperature drop at the getter as large as $6\,\rm{K}$, presumably because of charcoal evaporative cooling, has been measured at these pressures.

The charcoal getters, mounted on the end caps of the $\rm{LN_2}$ tank, are in ideal locations from the standpoint of their ability to cool to nearly the $\rm{LN_2}$ bath temperature and their physical separation from the optics to avoid contamination during material release on warm-up.  However, the getters are inaccessible without the removal of the bottom cryostat lid and bottom radiation shield assembly.  In practice this has meant the getters have not been touched since their original installation.  While easy access would have allowed baking and/or periodic getter material changes, the ability of activated charcoal to largely ``regenerate'' at room temperature has mostly mitigated the lack of practical access.

\subsection{Thermal System and Radiation Shielding}\label{radiation_shielding}

A radiation shield system completely encloses the opto-mechanics, cold plate, and $\rm{LN_2}$ tank.  The system was designed to effectively shunt radiative energy from the cryostat walls to the $\rm{LN_2}$ tank while minimizing thermally induced distortion of the cold plate.  Three assemblies make up the system: a top lid weldment, a four-walled assembly that surrounds the cold plate, and a bottom lid weldment that covers the $\rm{LN_2}$ tank from below.  The assemblies mate in a ``tongue-and-groove'' fashion as the top and bottom edges of the mid-section walls feature extruded ``H'' beams.  The shield walls are made from $\frac{1}{8}\,\rm{in}$ ($3.2\,\rm{mm}$) thick 3003 aluminum as this material has higher thermal conductivity than 6061 aluminum.\footnote{See \url{http://cryogenics.nist.gov}}

The system's mechanical connections promote similar shield temperatures above and below the cold plate to minimize differential thermal loads along the plate thickness direction.  The bottom shield bolts directly to two bands welded to the $\rm{LN_2}$ tank.  The shield mid-section is bolted around the edge of the cold plate.  Lastly, two ``tent poles,'' $1.5\,\rm{in}$ ($38.1\,\rm{mm}$) diameter 6061 aluminum poles, directly connect the top shield lid to the cold plate in two places which do not interfere with the light path.  These tent poles are meant to give the top lid a strong connection to the $\rm{LN_2}$ tank, albeit through the cold plate, to balance the direct connection of the bottom lid to the $\rm{LN_2}$ tank.

All sections of the radiation shielding are covered with an intricate blanket system which acts as an effective floating shield.  Both the top and bottom shield weldments have dedicated blankets that overlap the blankets covering the mid-section walls.  The blankets are composed of ten layers of double-sided aluminized Mylar ($\rm{Sheldahl^{TM}}$ P/N G4052) interspersed with ten layers of nylon tulle netting.  The tulle reduces thermal shorts between Mylar layers.  The layers are sewn together for blanket integrity but the stitching is not so tight and complete as to inhibit airflow between layers.  Velcro pads, sewn onto the blankets, allow straightforward attachment to complementary pads on the shield surface or other blankets.

To reduce emissivity of the room-temperature cryostat walls, we followed the lead of \citet{per13} and fastened $\rm{Sheldahl^{TM}}$ double-sided aluminized mylar sheet ($0.003\,\rm{in}$ thick) onto the interior walls with $\rm{3M^{TM}}$ Y966 adhesive transfer tape.

While the interlocking system of radiation shield assemblies described above promotes light-tightness, it inhibits the flow of air across the shield boundaries during cryostat pump-out and return to room pressure.  To aid air flow during these pressure cycles there are four ``air diffusion ports'' distributed around the cold plate at cutouts in the radiation shield system.  Composed of re-entrant fixtures to ensure light-tightness, airflow takes a circuitous path through the ports.

\subsection{Vibration Isolation}\label{vibration_isolation}

Vibration of the optical bench is minimized both mechanically and environmentally.  Mechanically, the cold plate, suspended from the three sets of stainless steel rods, is otherwise unconstrained and free to swing in response to vibrational input.  Given modest vibrational input, the cold plate can swing without disturbing the opto-mechanics.  The spectrograph is supported by four Newport S-2000 vibration isolators regulated to $40\,\rm{psi}$ ($275\,\rm{kPa}$).  Lastly, the instrument sits on the bottom floor of the warm support building on a simple $4\,\rm{in}$ ($100\,\rm{mm}$) thick reinforced concrete slab.

From an environmental standpoint, seismic activity near APO has been relatively benign in the past 100 years.  APO is close to the Alamogordo fault in the Sacramento Mountains within the seismically active Rio Grande Rift.  Nonetheless, the largest recorded New Mexico earthquake was a magnitude 6.2 event in Socorro in 1906 \citep{won09}. Lastly, people infrequently enter the APOGEE room given the automated cryogenic fill systems, described in the previous section, and remote monitoring of instrument telemetry, further reducing optical bench vibration.

\subsection{Cryostat Performance}

\subsubsection{Vacuum}\label{vac_performance}

With the exception of two maintenance periods during summer shutdowns in 2014 and 2017, the instrument has been evacuated and at $\rm{LN_2}$ temperatures since the end of commissioning in 2011 September.  Using a dry scroll rough pump and a $60\,{\rm liter\,s^{-1}}$ capacity turbo molecular pump mounted directly to the cryostat gate valve for the pumpdown, a pressure of $\sim 5 \times 10^{-4}\,\rm{Torr}$ can be achieved in 1.5 -- 2 days, at which point the $\rm{LN_2}$ fill is started.  The pumps are typically run for another day or two until they no longer make a material difference in the pressure.  Once cold, the instrument pressure slowly increases over nearly three years from $\approx 0.6 \times 10^{-6}$ to $\approx 1.3 \times 10^{-6}\, \rm{Torr}$, albeit with superposed seasonal pressure variations driven by instrument room temperature.  On a daily basis, changes in heat load from, e.g., room temperature and whether or not room lights are on, can affect instrument pressure.  Warm-up and cooldown is planned every three years to regenerate the getters.

\subsubsection{Cool-Down Sequence}

Cool-downs are completely passive --- no active temperature control is necessary as the fused silica and silicon optical elements are robust to thermal shock.  And, as discussed above, the thermal connections to the detector arrays were designed to ensure passive cool-down rates of $< 1\,\rm{K\,min^{-1}}$.  The camera, which requires the most time to cool, reaches steady temperatures in 9 days.  Manual $\rm{LN_2}$ filling is needed for the first few days, after which we use the $\rm{LN_2}$ controller to command once-a-day fills.

\subsubsection{Hold Time and Thermal Performance}\label{thermal_performance}

Approximately 16.5 liters of $\rm{LN_2}$ are consumed daily which implies a thermal load of 33 W and a hold time of nearly 6 days.  This is consistent with the measured thermal load during cryostat testing at PulseRay of $28\,\rm{W}$ with a wet-test meter.  Thermal conduction through the cold plate hanger system accounts for about $5\,\rm{W}$ of the thermal load.  The balance, $28\,\rm{W}$, is due to the radiative load.  The shield system is thus $\approx 93\%$ efficient as the cryostat walls ($8.8\,\rm{m^2}$) radiate $\approx 420\,\rm{W}$ (assuming the unpolished interior cryostat walls have an emissivity of 0.1).

There is a $2\,\rm{K}$ difference between the steady-state temperatures of the middle and corner (SDSS-III survey means were $78.2\,\rm{K}$ and $80.0\,\rm{K}$, respectively) of the top of the cold plate.  The inside corner of the mid-shield had an $87.9\,\rm{K}$ mean temperature which further indicates that the blanket system is efficiently minimizing thermal radiation from the cryostat walls.  Over the 34 months of the survey in SDSS-III, the middle cold plate temperature had a maximum temperature excursion of $0.8\,\rm{K}$ from the mean.  The camera mean temperatures of $78.9\,\rm{K}$ (front) and $77.9\,\rm{K}$ (back) had about the same maximum excursions.  But the cryostat is not instrumented with sufficient sensors to robustly assess detailed performance of the shield system such as comparisons between top and bottom shield temperatures and top and bottom cold plate temperatures.

While cryostat hold times are good, the rate of cool-downs could have been improved.  A transient thermal model predicted the cold plate would reach $80.3\,\rm{K}$ within five days after cool-down start whereas the actual temperature was $84.4\,\rm{K}$ at this point in the first cool-down with most of the optics installed.  The model was brought into conformance with the empirical data by including thermal resistance at both the bolted connections between the $\rm{LN_2}$ tank and L-bracket and between the L-bracket and cold plate.  These bolted connections were not athermalized and consequently the clamping force was likely reduced at low temperatures.  (We simply relied upon developing sufficient bolt pre-loads when tightening the bolts ``by hand'' to ensure bolt compression was not lost through differential thermal contraction between the aluminum cryostat parts and stainless steel bolts.)  The suspicion of contact resistance is supported by the measured $\sim 1 - 2\,\rm{K}$ difference in temperature between sensors located on the ends of the L-brackets bolted to the bottom of the cold plate and a sensor in the middle of the top of the cold plate.  A better thermal connection would minimize the contribution of contact resistance to this temperature difference.

The thermal model was also used to predict camera cool-down performance and recommend an athermal mounting scheme between the camera and camera legs to speed up the camera cool-down rate.

\subsubsection{Warm-Up Sequence}

After flushing the $\rm{LN_2}$ tank of excess cryogens, we use two banks of ten $200\,\Omega$, $100\,\rm{W}$, Dale-Vishay resistive heaters wired in parallel to input heat and warm the instrument.  The banks are secured to the sides of the L-brackets.  A total of $\approx 25\,{\rm kW\,h}$ must be input to warm the instrument to room temperature.  While each bank is capable of supplying $0.5\,\rm{kW}$ and thus the instrument could be warmed in one day, we conservatively run the heaters at $\sim 30\%$ power so the warm-up takes 3.5 -- 4 days.

Each resistor bank is protected by several safety mechanisms.  First, thermal fuses set to open at $\sim 60\,^{\circ}\,\rm{C}$ are wired in series with the heater circuitry.  Secondly, thermal switches, tested experimentally to open in the range $34 - 39\,^{\circ}\,\rm{C}$, are bolted to each L-bracket and can trip the heater system in an overheat situation.  Lastly, there is no electronic control of the warm-up --- one person is responsible for monitoring the pace and adjusting voltages across the resistor banks as needed over the warm-up operation.


\section{Moving Parts}\label{mechanism_design}

\subsection{Dither Mechanism}\label{dither_mechanism}

Detector translation in the spectral direction is provided by a custom dithering mechanism with a range of $\approx 18$ pixels and a nominal step size of $0.125\,\micron$ for each step of a 200 step per revolution motor.  The mechanism includes a Phytron VSS 32 cryogenic stepper motor which turns a fine-pitched 1/4-80 ($\frac{1}{4}\,\rm{in}$ diameter, 80 thread per inch) leadscrew.  Both the screw and a complementary traveling nut were fabricated from Electralloy Nitronic 60\textsuperscript{\textregistered} Stainless Steel.  This material has excellent resistance to galling at cryogenic temperatures.\footnote{\url{http://www.electralloy.com}}  Both the leadscrew and nut underwent processing to apply Dicronite\textsuperscript{\textregistered} (a proprietary type of tungsten disulphide dry lubricant) to minimize friction.  A tungsten disulphide lubricant was chosen instead of molybdenum disulphide as the latter is hygroscopic \citep{roh04}.


\begin{figure}
\epsscale{1.0}
\plotone{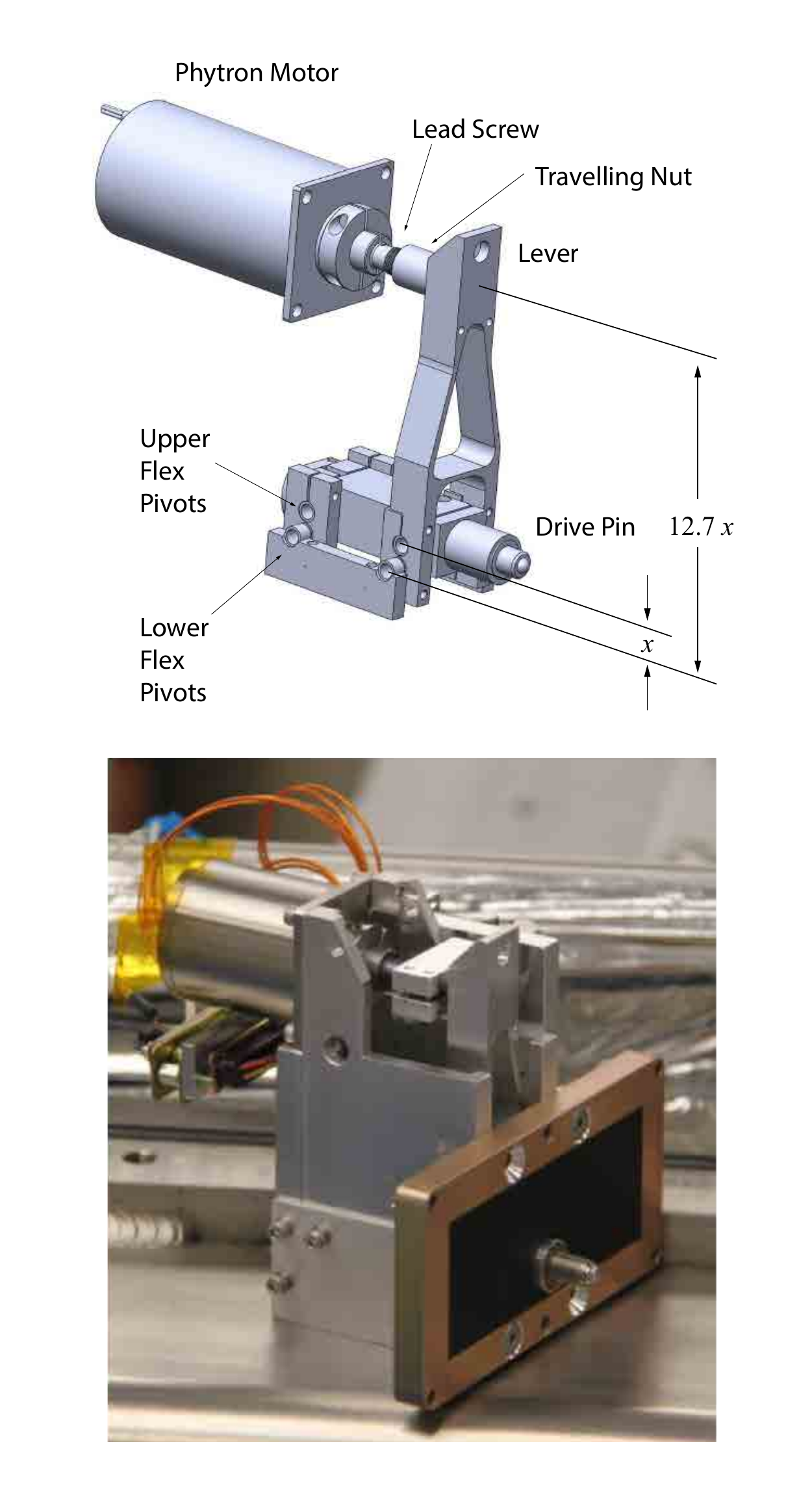}
\caption{The dither mechanism used to translate the detector mosaic in 0.5 pixel moves.  (Top) Model of the essential parts of the mechanism.  (Bottom) Finished assembly prior to installation during commissioning.}\label{fig_dither_mechanism}
\end{figure}


A lever arm maintains contact with the travelling nut through opposed pre-loads of the system (discussed below).  The lever provides 12.7:1 motion reduction and directly translates a ``drive pin'' in an assembly that resembles a battering ram (Figure~\ref{fig_dither_mechanism}).  C-Flex flex pivot bearings constrain mechanism motion along one axis.  The drive pin is a Thor Labs 1/4-80 303 stainless steel leadscrew with a hemispheric tip screwed into a complementary phosphor bronze bushing.  This Thor Labs screw was manually adjusted during assembly to fine-tune the dither mechanism position relative to the detector array mosaic assembly.  When the dither mechanism is bolted to the side of the detector array mosaic housing, the drive pin protrudes through a hole in the wall and the hemispheric tip contacts a flat pad bolted to the aluminum plate of the detector array mosaic assembly.

Two opposing pre-loads ensure that contact is maintained between the drive pin and the flat pad, within the detector array mosaic assembly, and the lever and traveling nut, within the dither mechanism.  First, the stepper motor pre-load direction was reversed by the manufacturer so the motor shaft and thus leadscrew had an outward (relative to the motor case) restoring force of 10 -- $15\,\rm{N}$.  Secondly, the copper thermal straps within the detector array mosaic assembly provide a restoring force of $\approx 15\,\rm{N}$ which tends to resist mosaic movement away from the dither mechanism (this force is reduced to $\sim 1\,\rm{N}$ when transmitted through the dither mechanism lever to the motor shaft).  Two Baumer My Com BS75 precision switches, coupled to the traveling nut, serve as limit switches.

The final design is the culmination of an iterative process of testing (both warm and cold) and design modifications.  Testing showed that the leadscrew/nut interaction had to be as uniform as possible throughout the mechanism range.  This was addressed with precision machining to ensure threads on the leadscrew and nut were coaxial; designs which provided 100\% thread engagement with the nut through the entire range of motion; choice of C-Flex flex pivots to minimize torque requirements; and design features which minimized potential build-up of ``sluffed-off'' solid particulates (debris) from the Dicronite process.

This debris build-up was seen during lifetime testing of both the dither mechanism and collimator actuators at Johns Hopkins University (JHU) prior to instrument deployment.  Debris build-up decreased with time, presumably after the most loosely bound particulates were rubbed off the threads through use of the mechanisms.  The deployed dither mechanism leadscrew and nut set was cooled cryogenically and operated in a test cryostat and then cleaned at JHU prior to the survey start with the expectation of removing this initial debris.  Dither mechanism reliability is discussed in \S~\ref{dith_mech_perf}.

\subsection{Collimator Tip-Tilt-Focus}\label{collimator_mechanism}

As discussed in \S~\ref{collimating_mirror}, a central titanium flexure, bolted to the center Invar pad on the back of the Collimating Mirror (Figure~\ref{fig_collimator_assembly}), transfers the weight of the mirror to the mount.  The compliance of the flexure allows three mirror actuators surrounding the center to push and pull the mirror into the desired orientation.  The actuators provide $\gtrsim 4.7\,\rm{mm}$ focus travel (piston) and $\approx \pm 1.6\,\degr$ ($\pm 2.8\,\degr$) angular movement in the spatial (spectral) directions.

The collimator actuator design borrows essential features of the cryogenic linear actuator (without planetary gear) described in \citet{roh04}, including use of a Phytron VSS 52 cryogenic stepper motor.  Each motor turns a fine-pitched (1/4-80) leadscrew and a complementary nut travels along the screw.  This nut is connected to the motor frame by a custom stainless steel bellows, as done by \citet{roh04}, to prevent nut rotation as the leadscrew turns.  The leadscrew and nut are both fabricated from Electralloy Nitronic 60 stainless steel and lubricated with the Dicronite process.

Lastly, titanium pins are connected between the travelling nuts and Invar pads on the back of the mirror to locally push and pull the mirror.  The pins, with $0.078\,\rm{in}$ ($\sim 2\,\rm{mm}$) diameter cross-section, are designed to provide sufficient bending compliance.  Each actuator includes a pair of SAIA Burgess V4LT7 switches which serve as front and back limit switches.

\subsection{Cold Shutter}\label{cold_shutter}

\begin{figure}
\epsscale{1.1}
\plotone{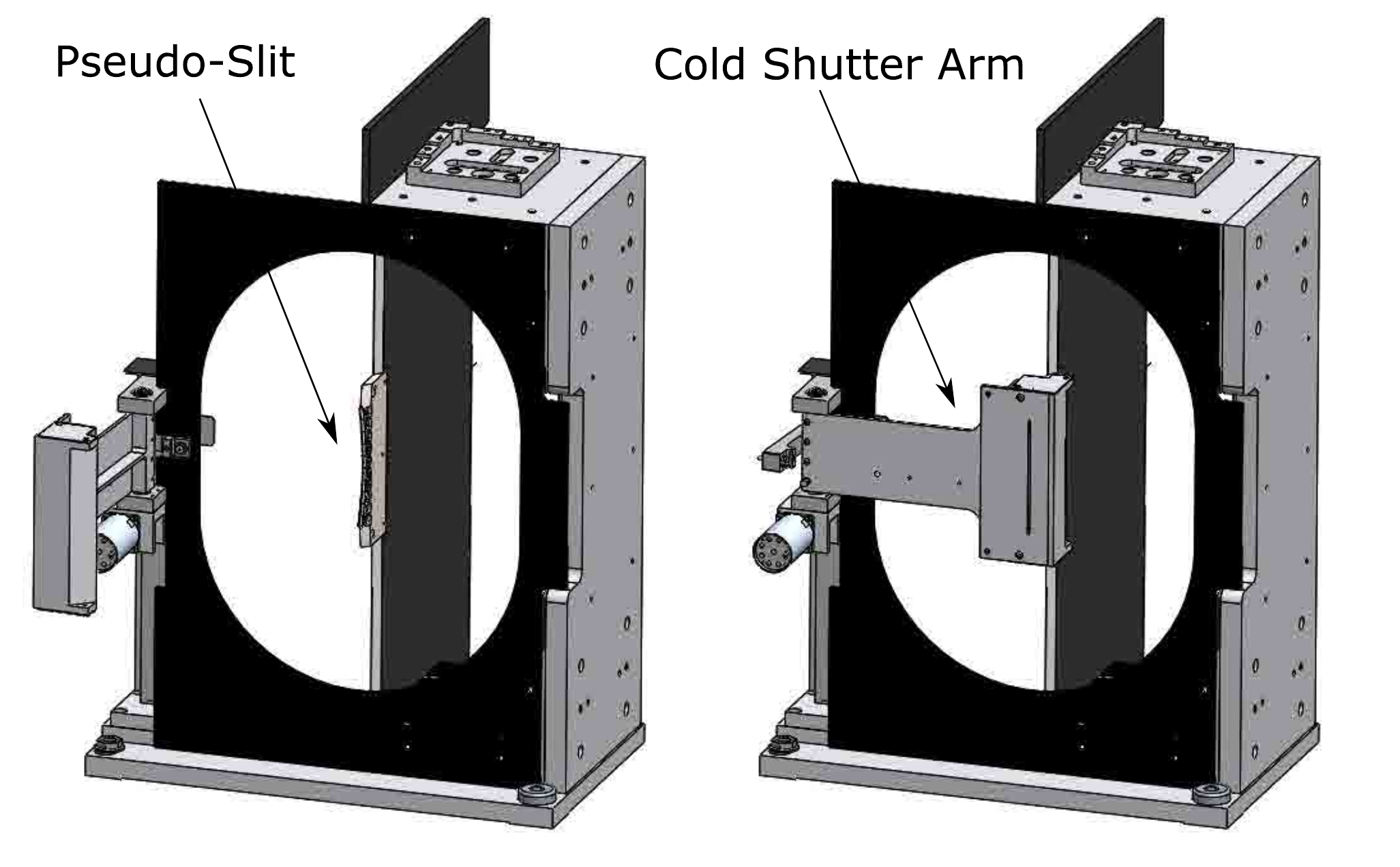}
\caption{The cold shutter in its open (left) and closed (right) positions.}\label{fig_cold_shutter}
\end{figure}

After discovering the abnormally high persistence in a portion of the blue detector array (\S~\ref{blue_persistence}) during lab testing, a cold shutter mechanism was added to cover the pseudo-slit when the spectrograph was not in use for observations as a precaution against inadvertent illumination.  It was installed during instrument commissioning at the telescope.  The shutter (Figure~\ref{fig_cold_shutter}) is simply composed of a rectangular-shaped cap with a light trap at the end of a cantilevered swing arm.  During observational sequences the arm is in the open position, i.e., moved out of the beam so it is adjacent to the radiation shield wall.  When closed, the arm closely covers the pseudo-slit.  It takes $\sim 10\,\rm{sec}$ to change the arm position.


The arm is actuated by a Phytron VSS 33 cryogenic stepper motor which couples to the arm through a 30:1 worm gear.  This gear reduction provides sufficient torque to drive the arm smoothly at low speed.  The worm is fabricated from 303 stainless steel and the worm gear is fabricated from Dupont\textsuperscript{\texttrademark} Vespel\textsuperscript{\textregistered} SP-3.  This material combination was chosen based on its successful use in the Gemini Telescope Near Infrared Imager (NIRI) flexure focus stage \citep{dou98}, the Infrared Multi-Object Spectrograph (IRMOS) Focus Mechanism \citep{sch03}, and subsequent instrument mechanisms designed at JHU.

Custom ``pogo stick'' switches are used as limit switches.  When the arm makes contact with the stick an electrical circuit is closed.  Evidence of high-frequency jitter in the switch closure signal has been seen and is attributed to a rough surface where the switch contact is made.

With the cold shutter closed, any light inadvertently emitted by the fibers is reflected farther into the light trap by a strip of aluminum tilted at $45\degr$ and coated with glossy black paint (PPG Aerospace PRC-DeSoto Desothane\textsuperscript{\textregistered} HS Military and Defense Top Coat CA 8201/F17038 Hi-Gloss Black).  The rest of the light trap walls are painted with diffuse infrared black paint.  The cold shutter is very effective --- nighttime tests with the shutter closed and telescope flat lamps illuminated showed no obvious flux accumulation on the detector arrays.  Additionally, no light is detected with the shutter closed and the gang connector exposed to the bright incandescent lights of the dome.

Housed in the cap behind the light baffle is a compartment that has infrared LEDs used to provide internal illumination for flat fielding.  This is discussed in \S~\ref{internal_leds}.


\section{Instrument Control}\label{instrument_control}

\subsection{Architecture}

\begin{figure*}
\epsscale{1.0}
\plotone{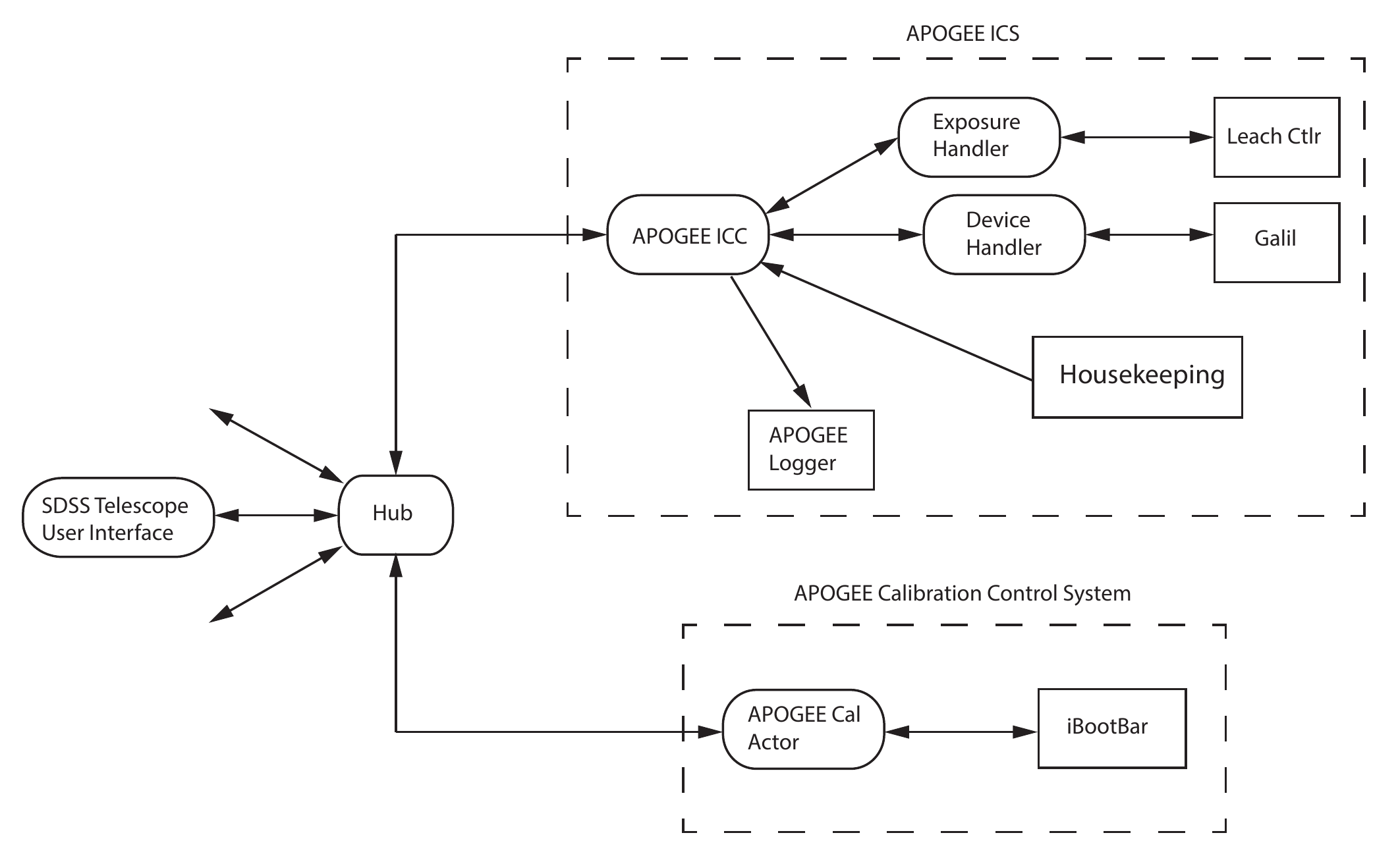}
\caption{A schematic of the distributed processes of the Instrument Control System.  The Sloan Foundation Telescope control system ``Hub'' routes communications to the instrument and forwards replies to the proper process.}\label{fig_ics_diagram}
\end{figure*}

The SDSS control system architecture consists of a central ``hub'' that routes all communications between distributed processes as shown in Figure~\ref{fig_ics_diagram}.  This architecture allows for multiple command and control points to access the APOGEE instrument control system (ICS), along with all other distributed processes in the system, without fear of command collision or confusion about where feedback should go.  The hub process routes all communications to the instrument, forwards replies to the proper process, and keeps a log.  This structure has proven very reliable and was ported from the Apache Point Observatory 3.5-m control system.


Internally, the instrument computer runs a single application which handles all hub communications, and is responsible for routing traffic to a variety of processes which represent the various subsystems that comprise APOGEE.  These processes, known as ``handlers,'' understand the higher level commands and are able to translate them into the needed commanding native to that subsystem's controller.  For instance, there is an exposure handler to coordinate exposures and adjust SUTR frames as required.  There is an electro-mechanical device handler that initiates device movements and provides feedback on motion status.  Feedback from handlers regarding execution of any commanding being processed are fed back to the hub.

Finally, there is an independent process responsible for internal logging of information not necessarily contained in feedback through the standard hub connection.  This allows detailed information from various processes to be archived.

\subsection{Instrument Control Computer}\label{icc}

The ICS resides in a 19-inch computer rack adjacent to the instrument and consists of an instrument control computer (ICC) hosting the main software applications described above.  The rack also contains an uninterruptible power supply (UPS), the $\rm{LN_2}$ level controller, the Lake Shore temperature controller, and the pressure sensor controller.

\subsection{Mechanism Control}

A single Galil DMC-4080 8-axis Stand Alone Controller is used for the five different stepper motors.  The Galil controller is mounted on a rack hanging from the end of the instrument so stepper motor drive lines are as short as possible to minimize broadcast electromagnetic interference.  All stepper motors are driven in $\frac{1}{16}$ micro-step mode and are de-energized between use.

\subsection{Housekeeping}\label{housekeeping}

In addition to the Cernox\textsuperscript{\texttrademark} sensors used to measure the detector array and camera temperatures, inexpensive 1N4148 diodes potted within copper lugs with 3M 2216 epoxy serve as housekeeping temperature sensors in numerous other places where monitoring trends is important yet absolute accuracy is less important.  Diode locations include the cold plate, radiation shield walls, getters, and the L-brackets.  These sensors are simply calibrated at two points:  room temperature and $\rm{LN_2}$ boiling temperature ($77\,\rm{K}$).  All housekeeping temperatures, collected by a custom temperature monitoring circuit board, are logged and available for remote monitoring.  Lastly, for redundancy, both a Pfeiffer Full Range Pirani/Cold Cathode gauge and an MKS 972 $\rm{DualMag^{TM}}$ Cold Cathode/Micro $\rm{Pirani^{TM}}$ Vacuum Transducer are employed to measure cryostat pressure.

\subsection{Lightning Protection}

To prevent electrical damage from lightning, all electronic devices that have grounds connected to the cryostat are powered through Brick Wall Series Mode Surge Filters.  Three of these devices are deployed throughout the APOGEE room to filter AC power to the ICC and the $\rm{LN_2}$ transfer and intermediate tank control electronics.  Also, all cryogenic supply lines which enter the room and vent lines which exit the room have G-10 fiberglass isolators in-line with the piping to minimize conductivity to the cryostat.  Also, vent lines on the building exterior are made of plastic.  Lastly, all communication signals penetrating the room go through optical fiber rather than copper wire.


\section{Calibration System}\label{calibration section}

Another aspect of integrating the spectrograph with the Sloan Foundation Telescope was provision of a dedicated calibration lamp suite separate from the calibration system already integrated into the telescope.  Since the BOSS spectrographs are mounted on the telescope, visual calibration images are taken by closing and illuminating the insides of the enclosure cover segments at the top of the telescope with flat field and arc lamps mounted adjacent to the primary mirror.  While APOGEE does observe flat fields with this telescope system after each exposure (\S~\ref{tele_flat_fields}), its primary calibration lamps are housed in a separate box and fiber-fed into the instrument.  This is necessary for two reasons:  first, the calibration lamps need to be available while the telescope is in use by other SDSS surveys.  Second, hollow-cathode lamps (HCLs) provide the higher order wavelength solution for the instrument (OH airglow lines provide the first-order solution).  Since HCL illumination through the telescope, including reflection from the telescope petals, was apt to be too dim for detection by APOGEE with sufficient counts, a separate calibration box assembly with an integrating sphere was designed to provide APOGEE-specific calibration light.  The gang connector system, mentioned above, allows selection of where APOGEE receives its light -- either from a cartridge in use on the telescope, or a socket linked to the calibration box.

\subsection{Calibration Box}

The calibration box is located in the room underneath the telescope in an electronics rack that hangs beneath the rotating floor surrounding the telescope.  It contains a Labsphere integrating sphere with a $5.3\,\rm{in}$ ($135\,\rm{mm}$) interior diameter with three entrance ports and one exit port.  The interior is coated with Labsphere's Spectralon\textsuperscript{\textregistered} diffuse reflective coating to maximize sphere radiance in the near-infrared.  Each entrance port is baffled so the light source cannot be directly viewed from the exit.  Mounted at the entrance ports are a $51\,\rm{mm}$ ThArNe HCL (Photron P958AN), a $37\,\rm{mm}$ UNe HCL (Photron P863), and a fiber-fed Tungsten Halogen Light Source (Ocean Optics HL-2000-HP-FHSA).  The exit port has an $1.0\,\rm{in}$ ($25.4\,\rm{mm}$) diameter aperture.

A simple imaging assembly is mounted to the exit port.  It consists of two fiber bundles placed at the end of an approximately $5.0\,\rm{in}$ ($127.0\,\rm{mm}$) long tube assembly such that fibers are geometrically illuminated with light with focal ratios of $f/5$ and slower to mimic the light entering science fibers at the telescope.  Field lenses placed immediately before the fibers make the light entering the calibration fibers approximately telecentric.  An electronically-controlled shutter (Uniblitz DSS25) mounted midway along the tube assembly is used with the control scheme to prevent inadvertent illumination of the spectrograph detectors.

The bundles are each mounted $3.5\degr$ off-center.  One bundle (``DensePak'') has a full complement of 300 fibers.  The other bundle (``SparsePak'') only illuminates every sixth fiber for a total of 50 fibers to allow better analysis of the wings of the PSF profile in the spatial direction. There is a dedicated gang connector receiver in the console adjacent to the telescope for each fiber bundle.

The HCLs are operated with $\sim 10\,\rm{mA}$ current using a Photron P209 power supply.  No warm-up time is provided.  The use of HCLs in series with an integrating sphere has posed no problems for the generation of sufficient illumination in a reasonable time --- 12 (40) reads are typically acquired for a total integration time of 128 (426) seconds for ThArNe (UNe) calibration images.

Overall control of the calibration system hardware is accomplished through software commands to turn on/off specific network-controlled AC outlets on a Dataprobe ``iBootBar'' located in the same electronics rack as the calibration box.  AC power cords for the lamp power supplies are plugged into the iBootBar along with a custom designed shutter trigger generator to trigger the calibration shutter driver.  The iBootBar is itself plugged into an UPS.

A set of flat field, HCL, and dark images are taken with the DensePak configuration using automated scripts prior to starting evening observations to allow checks of instrument and lamp health.  More robust sets are taken every morning following completion of observations.  The morning calibrations occur regardless of whether APOGEE was used the preceding evening and serve as our primary source for instrument monitoring.  The automated $\rm{LN_2}$ fills (\S~\ref{cryogenic_system}) are timed to occur after these morning calibrations are completed.

The ThArNe line list developed to support the CRIRES instrument on VLT \citep{ker08} is used for APOGEE.  The UNe HCL was included in the suite of lamps based on a suggestion from the Radial Velocity group at Penn State.  They demonstrated \citep[see, e.g.,][]{red12} the increased number of suitable uranium calibration lines for high resolution spectrographs in the NIR compared to thorium lines.

\subsection{Telescope Flat Fields}\label{tele_flat_fields}

In the same manner as BOSS, spectroscopic flat field observations are interleaved between science observations by observing a flat field screen composed of eight panels which together cover the top of the telescope when closed.  The light of four low-voltage quartz halogen capsule lamps, equally spaced around the primary, is projected onto the underside of the panels.  The panels are painted with Labsphere Duraflect\textsuperscript{\textregistered} white reflectance coating \citep{gun06}.  These flat fields, taken immediately after observations with a plate, provide cartridge-specific flats while the telescope is still pointed at the target field.  They provide the spatial position of each fiber trace on the detector arrays to support the extraction of 1D spectra from 2D images.  They also provide plugging specific fiber-fiber throughput information for use during sky subtraction (see \citealt{nid15}).

\subsection{Spatially Dithered Flat Fields}\label{spatially_dithered_flats}

Spatially-dithered flat fields can be produced by exposing and co-adding a sequence of ``quartz frames'' using the SparsePak calibration fiber bundle.  The collimator is sequentially tipped between $-7$ and $+7.5$ pixel pitch in the spatial direction to ``fill in'' the inter-trace regions for improved flat fielding, particularly at the edges of the trace.  Installation of the internal LED's, discussed below, largely eliminated the need to move the collimator which aids instrument stability.

\subsection{Internal LEDs}\label{internal_leds}

As mentioned in \S~\ref{cold_shutter}, the cold shutter cap includes a compartment which contains infrared LED's to provide an internal source of flat illumination.  It is important to note that these LEDs provide an internal illumination across {\it all} portions of the detector arrays --- external light sources illuminate the detector arrays through the fibers so illumination of pixels between fiber traces is limited to light in the faint wings of the spatial profile and stray light.

There are two sets of four Hamamatsu infrared LEDs:  series L10660 and L10823 which have peak emission at $1.45\,\micron$ and $1.65\,\micron$, respectively.  The LEDs are rated for use down to $-30\,^{\circ}\,\rm{C}$, and they have windows.  The LEDs are oriented such that their light first reflects off a bare, bead-blasted, aluminum surface before leaving the cap through a long, rectangular slit.  We intended to illuminate the LEDs with the cold shutter in the closed position; however, there is significant structure in the illumination pattern on the detector arrays when used this way.  We attribute the structure to insufficient diffusion of the LED illumination.  Fortunately, internal flat fields taken with the cold shutter in the {\it open} position provide much more uniform illumination and are effective for calculating pixel-pixel responsivity for all detector array pixels.  Apparently the unpolished 3003 aluminum radiation shield material sufficiently diffuses the LED light and provides quasi-uniform illumination of the detector arrays.  Pairs of LEDs are powered by a simple LED circuit consisting of $5\,\rm{V}$ applied across a current limiting resistor (to provide $1\,\rm{mA}$ current) and two LEDs, of identical type, in series.

Unfortunately, as described in \S~\ref{led_operability}, all of the LEDs in the APOGEE-North instrument have become inoperable over time, with subsets breaking during pumpdown/cooldown sequences.  We will be replacing the LEDs in the future with the cryogenic-rated LEDs used in the APOGEE-South instrument (\S~\ref{apogee_south_leds}).


\section{Shipping, Installation, and Instrument Room}\label{logistics section}

\subsection{Shipping}\label{apogee-north_shipping}

All decisions regarding the safety and packaging of individual optics and instrument systems were predicated on the use of an air-ride truck with door-to-door shipping service between U.Va. and APO to eliminate extraneous, unmonitored shipping dock movements of the instrument as trucks were changed at intermediate destinations.  A custom cradle for the instrument with air shocks and shock absorbers was also fabricated to minimize vibration transmitted from the truck.

The cradle and door-to-door air ride truck were critical as the cold plate was secured relative to the cryostat walls with four ``shipping restraints'' so it could not freely swing during instrument movement.  The cold plate shipping restraints are composed as follows:  the lower lid of the cryostat includes four vacuum ports consisting of Conflat\textsuperscript{\textregistered} style flanges ($6.75\,\rm{in}$ nominal OD) located near the four corners.  Normally blanked off, for shipment we install stainless steel rods at these locations.  The rods pass through dedicated cutouts in the radiation shield (the same cutouts where the ``air diffusion ports'' are installed; see \S \ref{radiation_shielding}) and bolt to the edge of the cold plate.  The rods are secured to the port with Fenner Drives\textsuperscript{\textregistered} Trantorque\textsuperscript{\textregistered} Keyless bushings.  The assembly includes an O-ring to surround the rod and form a vacuum seal so the instrument can be shipped slightly pressurized for cleanliness purposes.

With these precautions and restraints, only opto-mechanical assemblies deemed to have a risk of breakage during shipment were removed from the instrument to minimize optical re-alignment upon arrival.  Given the dither mechanism system reliance on pre-loads, there was no built-in protection against excessive detector array mosaic movement during the shipping process.  Thus the dither mechanism was removed and the detector array mosaic was bolted from the outside of the aluminum housing in a fixed position.  Vibration of the detector assembly was not a worry as its design heritage for JWST included rocket launch and deployment survivability.

The collimator was removed as it was straightforward to reinstall and reposition with laser alignment (while warm) and with tip-tilt-piston actuation (while cold).  Fold Mirror 1, Fold Mirror 2 and the VPH grating were immobilized by inserting plastic shims between the substrates and their mounts in the vicinity of the various spring plungers.  Lastly, the fiber links emanating from the instrument were coiled on reels and secured to the top of the cryostat.

No breakage was sustained during the shipment.  But the shipping cradle, structurally formed with wooden framing, suffered cracks due to excessive stress from the weight of the instrument. Any future movement of the instrument will require a new cradle with stronger structural members.

\subsection{Instrument Installation}

A professional rigging company removed the instrument, supported by the shipping cradle, from the integration lab at U.Va. and loaded it into the air ride truck with a fork lift.  Hilman Rollers proved effective for moving the instrument along horizontal surfaces.  Similarly, a rigging company removed the instrument from the truck upon arrival at the observatory.  They used a crane to hoist the instrument using dedicated lifting lugs at the corners of the cryostat and lowered it from the bridge between the telescope and warm support building down to the concrete apron outside the APOGEE instrument room.  Hilman rollers were again used to position the instrument within its room.

Upon arrival, optics that had been removed for shipment were reinstalled, the shipping restraints and various shims were removed, and the molecular diffusion ports were installed.

The fiber links were carefully flaked on the ground outside the room and pulled through cable trays and floor/wall conduits to the telescope.  Lastly, the fiber link MTP\textsuperscript{\textregistered} terminations at the telescope end were installed within the gang connector (activity seen in Figure~\ref{fig_gang_connector} (Top)).

\subsection{Instrument Room}

The APOGEE instrument room is a $164\,\rm{in} \times 154.5\,\rm{in}$ ($4.2\,\rm{m} \times 3.9\,\rm{m}$) space accessed by a set of double-doors.  About half the room is set up as a quasi-clean room with curtains and positive pressurization from two HEPA-filtered fan units mounted in a drop ceiling.  Also, the floor and walls are covered with clean room paints.  A clean environment allowed internal work on the instrument without having to move it from the room.  A particle counter was used to verify that the clean area is at least $10{,}000$ times cleaner than outside the building.

An I-beam fixed to the middle of the ceiling allows the upper lid to be raised with a chain hoist and moved to the edge of the room and lowered onto a dedicated stand while the instrument is open.

Temperature within the room is controlled by an in-room thermostat (located outside the clean curtains) which controls a zone of the building's HVAC system dedicated to the APOGEE room.  The zone simply consists of one supply and one exhaust duct located outside the clean curtains.  The mean of the daily maximum temperature deviation of a sensor on a warm electronics circuit board in a housing on the outside of the instrument is $0.42\,\rm{K}$.  No attempt is made to control room humidity.


\section{Instrument Optical Performance and Stability}\label{performance section}

This section describes instrument performance and its stability from an optical and opto-mechanical standpoint.  \citet{nid15} discusses performance from the perspective of reducing the raw data to combined 1-d spectra.  Appendix~\ref{appendix_commissioning_status} describes the instrument configuration during commissioning (2011 May -- 2011 July) prior to the official start of survey operations.

\subsection{Throughput}

Total observed throughput, from the top of the atmosphere to photon detection, is estimated to be $18\%$ at $1.60\,\micron$.  This is based on the best median zero point compared to the predicted flux of Vega observed through the collecting area of the Sloan Foundation Telescope using the photometric system of \citet{cam85}. The best median zero point is $H \sim 15.0$ (the 2MASS magnitude of a star that would give $1\,\rm{DN}\,\rm{pix^{-1}}$ in one $10.6\,\rm{sec}$ read assuming a gain of $1.9\,{e^{-}}/\rm{DN}$ and a quantum efficiency of $100\,\%$) across the green detector array and taking the median across an observed plate.  A zero point of $H \sim 15.0$ corresponds to a best seeing of $\sim 0.95\,\arcsec$ visual FWHM ($0.92\,\arcsec$ RMS diameter at $1.6\,\micron$) based on a scatter plot of past observations.  The uncertainty in observed throughput may be as large as $15\,\%$ based on uncertainties in gain determination discussed in \S~\ref{gain}.


\begin{deluxetable*}{lcccc}
\tabletypesize{\scriptsize}
\tablewidth{0pt}
\tablecaption{APOGEE Spectrograph Throughput \label{tbl-throughput}}
\tablehead{
\colhead{} & \colhead{$1.54\,\micron$} & \colhead{$1.61\,\micron$} & \colhead{$1.66\,\micron$} & \colhead{Comment} \\
\colhead{} & \colhead{Blue} & \colhead{Green} & \colhead{Red} & \colhead{}
}

\startdata
\cutinhead{Atmosphere \& Telescope}
Atmosphere & 0.95 & 0.95 & 0.95 & Assume $5\,\rm{mm}$ Average transmission using ATRAN\tablenotemark{a} \\
Telescope Vignetting & 0.71 & 0.71 & 0.71 & On-axis; see \S~4.3 of \citet{gun06} \\
Telescope Mirror Coatings & 0.96 & 0.96 & 0.96 & Assume $98\%$ reflectance per mirror (fresh coatings) \\
Telescope Correctors & 0.63 & 0.61 & 0.59 & AR Coatings optimized for visual\tablenotemark{b} \\
\cutinhead{Plug Plate \& Fiber System}
Encircled Energy (EE) into Fibers & 0.77 & 0.77 & 0.77 & Area-weighted average EE for 0.28 arcsec offset case, Fig~\ref{fig_ee} \\
Uncoated Fiber Tips & 0.87 & 0.87 & 0.87 & Fresnel losses at four surfaces \\
Coupling Loss at MTP\textsuperscript{\textregistered} Connector & 0.95 & 0.95 & 0.95 & See discussion \S~\ref{MTP_coupling_performance} \\
Fiber Transmission & 0.99 & 0.99 & 0.99 & Through $\sim 46\,\rm{m}$ of fiber \\
FRD & 0.95 & 0.95 & 0.95 & Assume 95\% EE at $f/3.5$ for beam into spectrograph\\
\cutinhead{Instrument}
Fore-optics Coatings & 0.96 & 0.97 & 0.97 & Fold 1 (twice), Collimator and Fold 2 Coatings \\
Fore-optics \& VPH Obscurations & 0.81 & 0.81 & 0.81 & On-axis for $f/3.5$ Gaussian-like profile \\
VPH & 0.76 & 0.84 & 0.59 & Efficiencies, Coatings \\
Camera Transmission & 0.93 & 0.93 & 0.93 & Internal Transmission, Coatings, Vignetting \\
Detector Arrays & 0.85 & 0.85 & 0.85 & Quantum Efficiency \\
\cutinhead{Total}
\nodata & 0.11 & 0.12 & 0.08 & \\
\enddata

\tablenotetext{a}{ATRAN calculation assumes one airmass, $5\,\rm{mm}$ precipitable water vapor, and $R = 22,500$.}
\tablenotetext{b}{Throughput of each corrector was measured individually using a Goodrich room temperature InGaAs camera.}

\end{deluxetable*}


The predicted throughput (from the top of the atmosphere through the detector array) for $1.61\,\micron$ light on the green detector array is 12\%.  Table~\ref{tbl-throughput} lists the various contributions to throughput at three wavelengths within the following three major components:

\begin{itemize}
\item Atmosphere \& Telescope -- Throughput of $\sim 40\%$, mainly reduced by the $\sim 60\%$ throughput at APOGEE wavelengths for the telescope correctors due to visual-optimized coatings (see \S~\ref{corrector_throughput})
\item Plug Plate \& Fiber System -- Throughput of $\sim 60\%$.  The biggest single loss is the encircled energy accepted by the fibers after the various contributors to image size degradation and fiber-image offset are considered (see \S~\ref{ee} and \S~\ref{appendix_errors}).  Throughput lost in the rest of the fiber system is comparable to the amount lost at the telescope focal plane.  Fiber system throughput could be improved by AR coating the fiber tips and minimizing coupling losses at the MTP\textsuperscript{\textregistered} connector.
\item Instrument -- Throughput of $\sim 52\%$ in the middle of the wavelength range but the throughput drops to $\sim 47\%$ and $\sim 37\%$ for the blue and red detector array portions, resp., due to the VPH Grating efficiency variation with wavelength (Figure~\ref{fig_vph_order1_de}).
\end{itemize}

Items with the largest predicted throughput uncertainty are likely the corrector throughput, fiber insertion EE, MTP coupling losses, and FRD.  Fiber-fiber throughput variations calculated from telescope flat fields are $\sim 10\%$ \citep{nid15}.

\subsection{Imaging Performance}

\begin{figure}
\epsscale{1.0}
\plotone{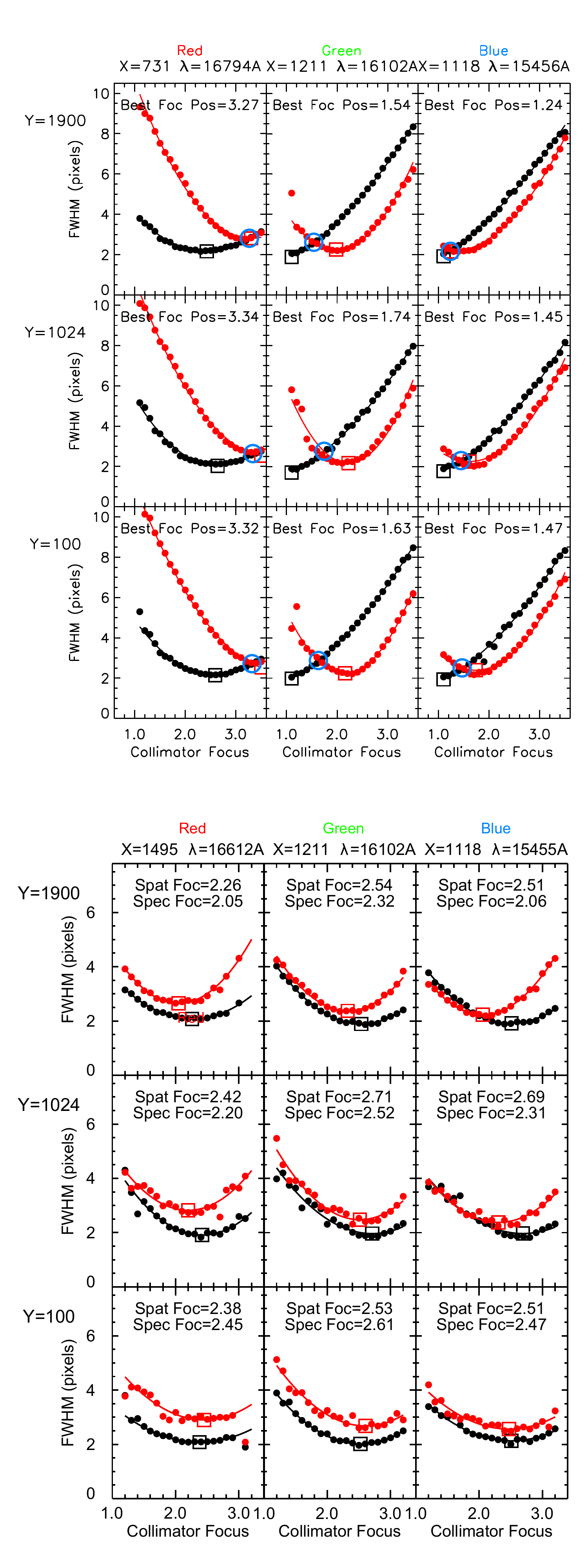}
\caption{Two sets of focus curves, produced by varying the collimator piston, for three wavelengths and three field (fiber) positions.  Red (black) curves correspond to the spectral (spatial) FWHM, \edit1{with the boxes indicating the minimum of the fit to the curves.}  The top set, taken during commissioning, showed astigmatism and also a red detector that was out of focus.  \edit1{Each panel in the top set includes a blue circle that indicates the best compromise focal position, typically located where the spectral and spatial curves cross.}  The bottom set, taken just prior to the start of SDSS-III survey operations, showed some astigmatism correction and improved red detector array positioning.}\label{fig_focus_curve_montage}
\end{figure}

A montage of focus curves is shown in Figure~\ref{fig_focus_curve_montage}.  The FWHM of the LSF (spectral; red) and PSF (spatial; black) directions for spectral lines at various positions on the detector arrays are shown as a function of collimator piston.  The top set of curves, taken during commissioning, showed astigmatism and an improperly positioned red detector array.  The bottom set of curves, taken at the start of the SDSS-III survey, showed both were improved.  The collimator focus (piston) of $2.4\,\rm{mm}$ was selected for the survey.


Image quality varies spectrally and spatially across the detector arrays for a variety of reasons -- ideal detector array position is a compromise across the wavelength span of each detector array; camera performance worsens redward of $1.68\,\micron$, wavelengths beyond the survey science requirements; imperfect detector array and v-groove block positioning; imperfect astigmatism correction (technically the astigmatism can only be perfectly corrected at one wavelength); apparent asymmetry in the astigmatism correction based on varying spectral focus curves with fiber position; and expected variations in FRD on a fiber-fiber basis.

Image quality variation leads directly to complex LSFs which are poorly fit by a Gaussian.  In fact the data reduction pipeline performs sky correction by modelling the LSF as a sum of Gauss-Hermite functions and a wide Gaussian for the wings \citep{nid15}.

Using data from through-focus curves with spectral lamps taken 2011 August 19 (a subset of which are shown on the bottom of Figure~\ref{fig_focus_curve_montage}), detector array position deviations from optimal can be estimated.  Based on the average of collimator positions for spectral\footnote{We use the spectral direction minima for this determination since it directly affects spectral resolution.} minima for a $5 \times 3$ grid of fiber (spatial) and wavelength (spectral) locations across each detector array, and taking into account the magnification differences between the collimator and the average of the three wavelengths on each detector array, the blue (red) detector arrays are $\approx 16\,\micron$ ($\approx 41\,\micron$) out of focus (piston).  Thus the blue detector array is within the pixel-limited depth of focus ($\approx 25\,\micron$) but the red detector array position error exceeds the depth of focus by $\approx 1.5$ times.


\begin{deluxetable}{lcccc}
\tabletypesize{\scriptsize}
\tablewidth{0pt}
\tablecaption{Detector Tilts from Spatial Spot Size Minima \label{tbl-det_tilts}}
\tablecolumns{5}
\tablehead{
\colhead{} & \multicolumn{2}{c}{Tilt x (arcmin)} & \multicolumn{2}{c}{Tilt y (arcmin)} \\
\colhead{Detector} & \colhead{Fit\tablenotemark{a}} & \colhead{Measured\tablenotemark{b}} &\colhead{Fit\tablenotemark{a}} & \colhead{Measured\tablenotemark{b}}
}

\startdata
Blue & -3.0 & 2.9 & 1.6 & 1.7 \\
Green & -4.5 & -2.8 & 1.0 & 0.6 \\
Red & 0.0 & -2.0 & -0.6 & -0.01 \\
\enddata

\tablenotetext{a}{Derived from the minima of curves fit to the measured spatial through-focus spots of various spatial locations on the detector array.}
\tablenotetext{b}{Derived from the smallest measured spatial through-focus spots of various spatial locations on the detector array.}

\end{deluxetable}


For tip and tilt estimation a fit to the grid of spatial minima was used because our correction for astigmatism in the spectral direction appears to vary spatially (as if the forces exerted by the spring plungers on the back of Fold 1 vary).  Table~\ref{tbl-det_tilts} shows the results of using both measured minimum, and minima of fits, to the spatial focus curves.  Tilt x (y) physically means a tilt about an axis aligned with the spatial (spectral) direction imaged onto the detector arrays.  The results suggest the estimates are more accurate for Tilt y.

\subsection{Resolution}\label{apogee_north_resolution}

\begin{figure*}
\epsscale{1.0}
\plotone{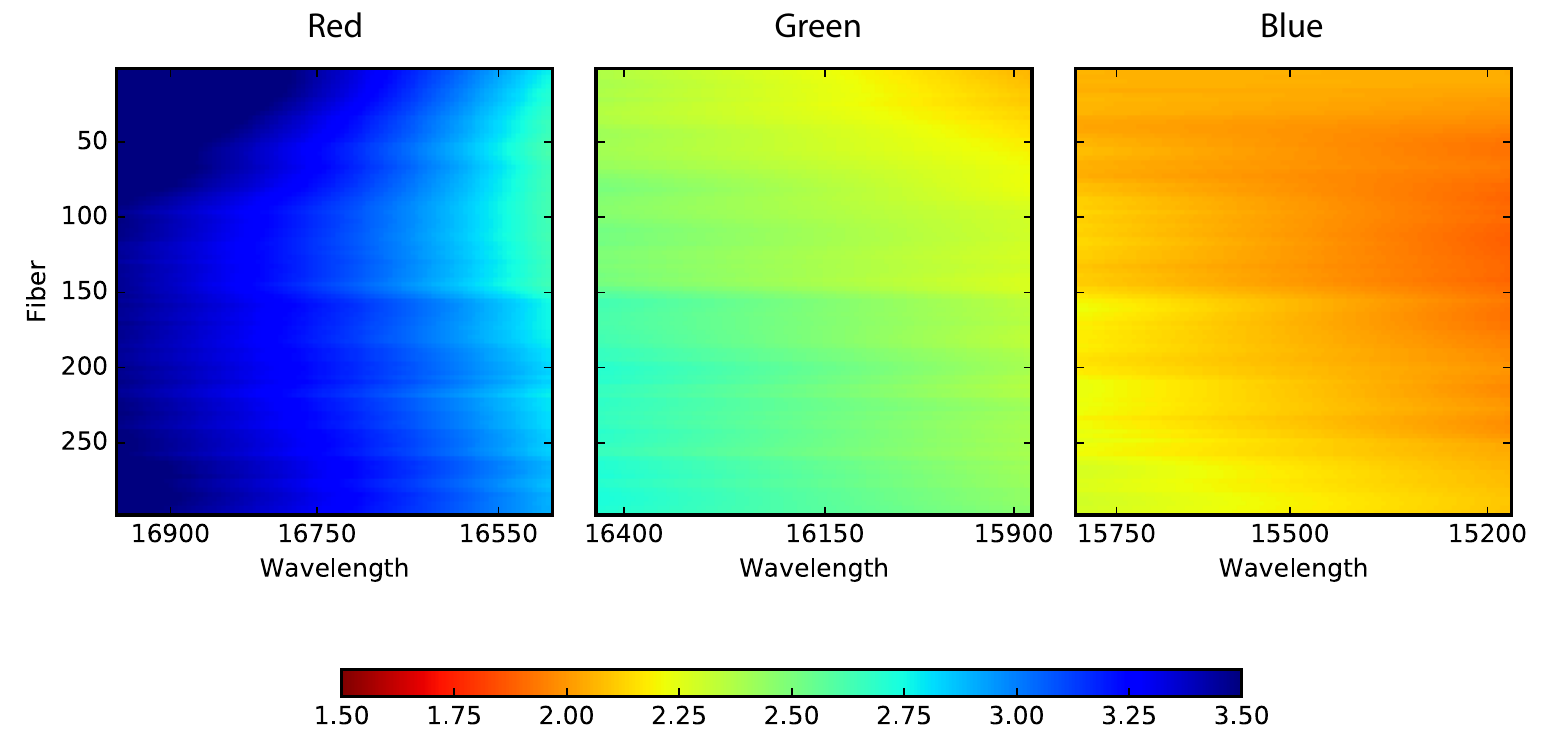}
\caption{FWHM maps for the three APOGEE-North detector arrays.  The FWHMs, color coded by size in pixels, are derived by fitting Gaussians to the LSFs of sky lines characterized using Gauss-Hermite polynomials, smoothed over a range of fibers and wavelengths.}\label{fig_apogee_north_fwhm}
\end{figure*}

\begin{figure*}
\epsscale{1.0}
\plotone{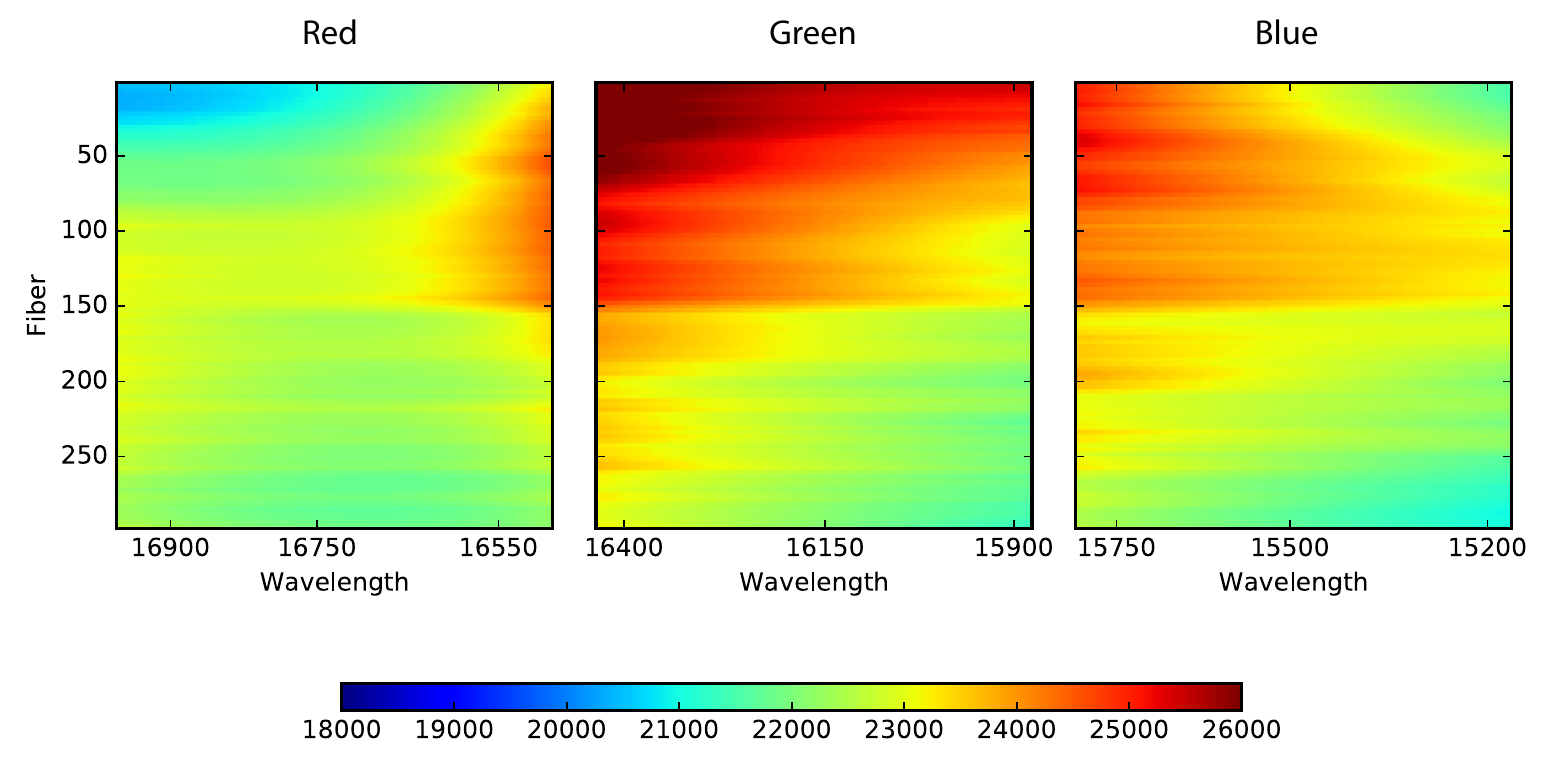}
\caption{{Resolution $R = \lambda/\rm{FWHM}$} mapped across the three APOGEE-North detector arrays using the FWHMs mapped in Figure~\ref{fig_apogee_north_fwhm}.}\label{fig_apogee_north_resolution}
\end{figure*}

Figure~\ref{fig_apogee_north_fwhm} gives a map of the spectral FWHM across the detector arrays.  The plots were derived by first characterizing the LSF using Gauss-Hermite polynomials.  Sky frames were used to derive the coefficients as they provide a good sampling in wavelength, although care needs to be taken for the sky lines that are doublets.  The derived LSFs yielded the FWHM maps, where these were determined by fitting Gaussians to the model profile.  Figure~\ref{fig_apogee_north_resolution} shows a similar map of resolution ($R$), where resolution is defined as $R = \lambda/\rm{FWHM}$.  The variation in resolution is quantified in Table~\ref{tbl-apogee_north_resolution_variation} which gives the median, maximum and minimum resolutions, smoothed over a range of fibers and wavelengths, for each detector array across a small sample of fibers at the top, middle and bottom of the array.  The minima and maxima on each detector array, for a given sub-sample of fibers, generally differ from the sub-sample median by $\la 10\,\%$.


\begin{deluxetable*}{lccccccccc}
\tabletypesize{\scriptsize}
\tablewidth{0pt}
\tablecaption{Resolution Variation \label{tbl-apogee_north_resolution_variation}}
\tablehead{
\colhead{Fiber} & \colhead{Red} & \colhead{} & \colhead{} & \colhead{Green} & \colhead{} & \colhead{} & \colhead{Blue} & \colhead{} & \colhead{} \\
\colhead{Sample} & \colhead{Median} & \colhead{Min\tablenotemark{a}} & \colhead{Max\tablenotemark{a}} & \colhead{Median} & \colhead{Min\tablenotemark{a}} & \colhead{Max\tablenotemark{a}} & \colhead{Median} & \colhead{Min\tablenotemark{a}} & \colhead{Max\tablenotemark{a}}
}
\startdata
Top & 21,200 & 20,400 ($-4\,\%$) & 23,800 ($+12\,\%$) & 25,500 & 24,900 ($-2\,\%$) & 26,500 ($+4\,\%$) & 23,300 & 21,800 ($-6\,\%$) & 25,000 ($+7\,\%$) \\
Middle & 22,900 & 22,800 ($-1\,\%$) & 24,000 ($+5\,\%$) & 23,500 & 23,000 ($-2\,\%$) & 24,400 ($+4\,\%$) & 23,300 & 22,900 ($-2\,\%$) & 23,900 ($+3\,\%$) \\
Bottom & 22,000 & 21,800 ($-1\,\%$) & 22,400 ($+2\,\%$) & 22,200 & 21,500 ($-3\,\%$) & 23,100 ($+4\,\%$) & 21,900 & 21,100 ($-4\,\%$) & 22,800 ($+4\,\%$) \\
\enddata

\tablenotetext{a}{Percentages give the deviation from the median for the detector array and fiber sample.}

\end{deluxetable*}


\subsection{Noise Model Validation \& Performance}\label{noise_model}


\begin{figure*}
\epsscale{1.0}
\plotone{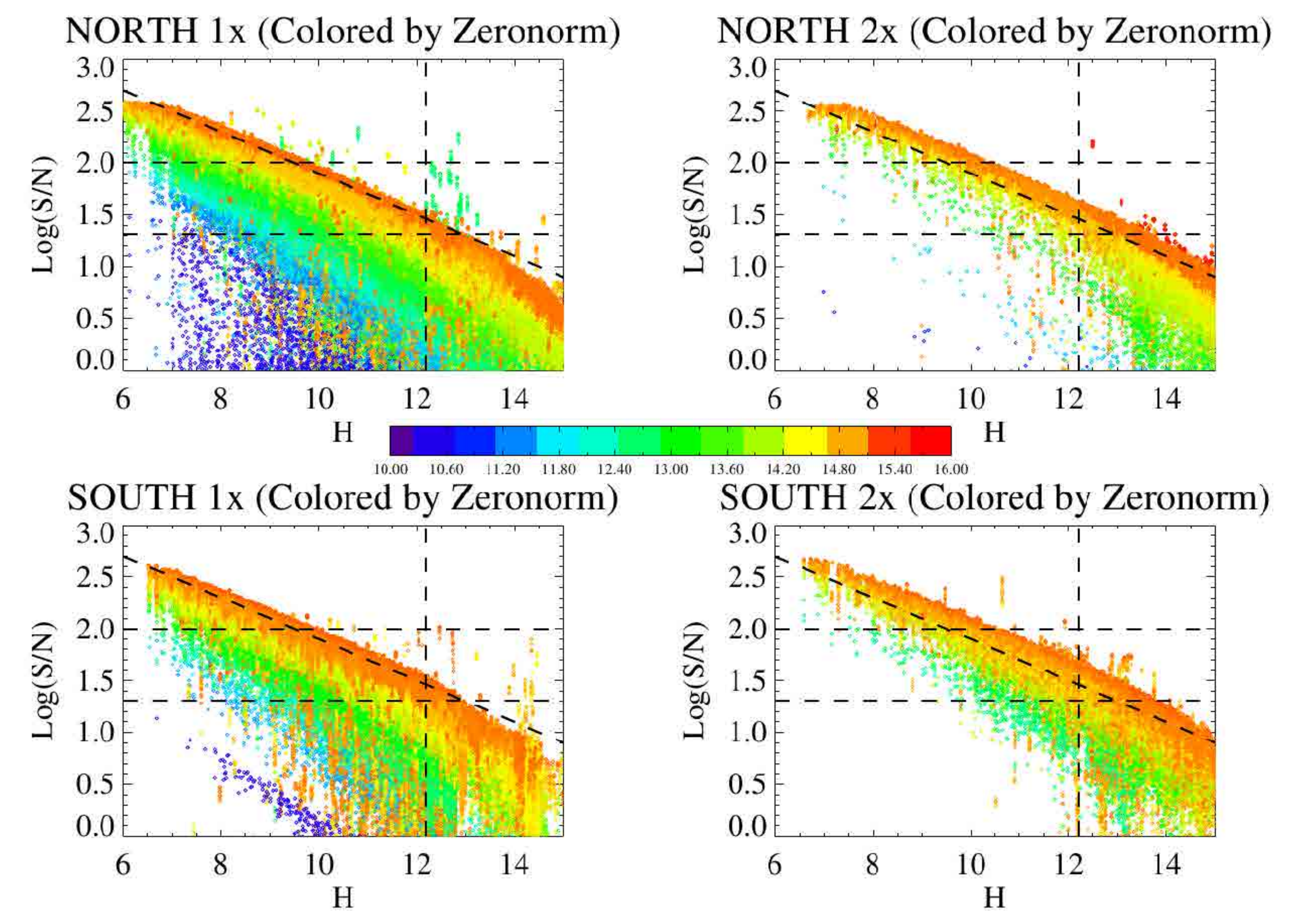}
\caption{Noise model estimates of the S/N for individual exposures taken with the APOGEE-North and APOGEE-South instruments.  The points are color-coded by the exposure throughput expressed as a zeropoint per readout where the zeropoint is the magnitude that gives $1\,\rm{DN}$ with the APOGEE-North instrument in one read ($10.7\,\rm{sec}$). (A 0.496 magnitude offset has been applied to the APOGEE-South zeropoints to account for the different gains).  The upper envelope (orange points) represents the performance under the best conditions.  The left (right) panels are for $500\,\rm{sec}$ ($1000\,\rm{sec}$) exposures.  The two horizontal lines correspond to $\rm{S/N} = 100$, the S/N goal, and $\rm{S/N} = 20.4$, the S/N required per exposure to achieve $\rm{S/N} = 100$ for a 3-visit plate. The diagonal line represents S/N proportional to $\sqrt{\rm{S}}$, as expected in the signal-limited regime.  The vertical line at $H = 12.2$ represents the faint magnitude limit for typical 3-visit survey targets.}\label{fig_snr_plot}
\end{figure*}


The achieved S/N of the instrument is a function of the instrument throughput, detector gain and readout noise, but also varies depending on the atmospheric transmission and seeing. Figure~\ref{fig_snr_plot} shows the noise model estimates of the S/N for individual exposures taken with the APOGEE-North and APOGEE-South instruments.  The points are color-coded by the exposure throughput expressed as a zeropoint per readout, where the zeropoint is the magnitude that gives $1\,\rm{DN}$ with the APOGEE-North instrument in one read ($10.7\,\rm{sec}$). (A 0.496 magnitude offset has been applied to the APOGEE-South zeropoints to account for the different gains.)  Thus the upper envelope (orange points) represents the performance under the best conditions.  The two horizontal lines correspond to $\rm{S/N} = 100$, the S/N goal, and $\rm{S/N} = 20.4$, the S/N required per exposure to achieve $\rm{S/N} = 100$ for a 3-visit plate. The diagonal line represents S/N proportional to $\sqrt{\rm{S}}$, as expected in the signal-limited regime.  The vertical line at $H = 12.2$ represents the faint magnitude limit for typical 3-visit survey targets.  The results suggest the instrument is signal-limited down to $H\sim13$ for $500\,\rm{sec}$ exposures (left panel); for fields with a significant number of fainter stars, $1000\,\rm{sec}$ exposures (right panel) are used to reduce the contribution of readout noise.

\subsection{Fiber--Fiber Spectral Contamination}


\begin{figure}
\epsscale{1.25}
\plotone{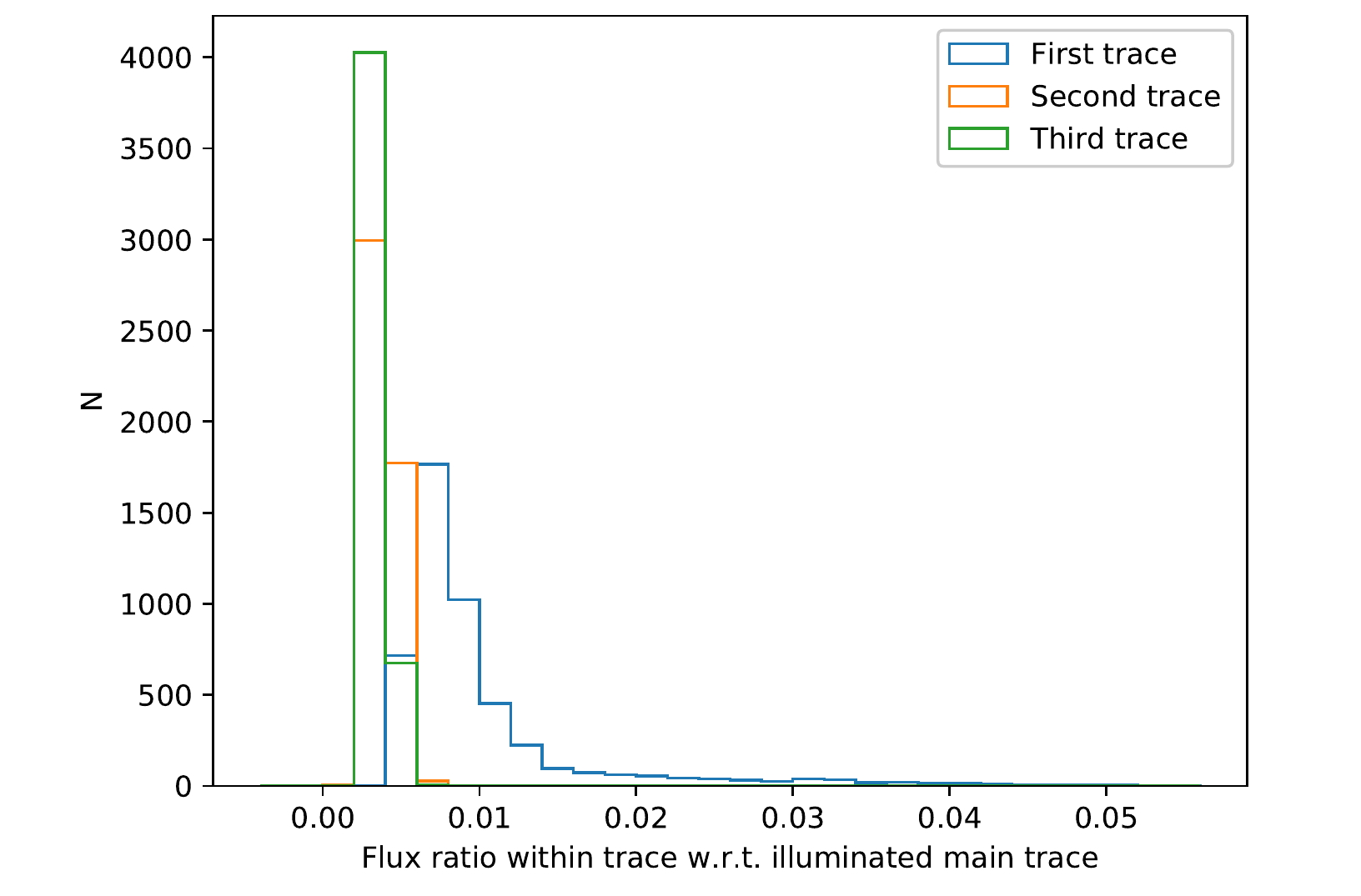}
\caption{A histogram of summed flux (spectral contamination) from the spatial wings of an illuminated main fiber that falls within the boundaries of the first, second, and third adjacent fibers.}\label{fig_fiber_crosstalk}
\end{figure}

\begin{figure*}
\epsscale{0.9}
\plotone{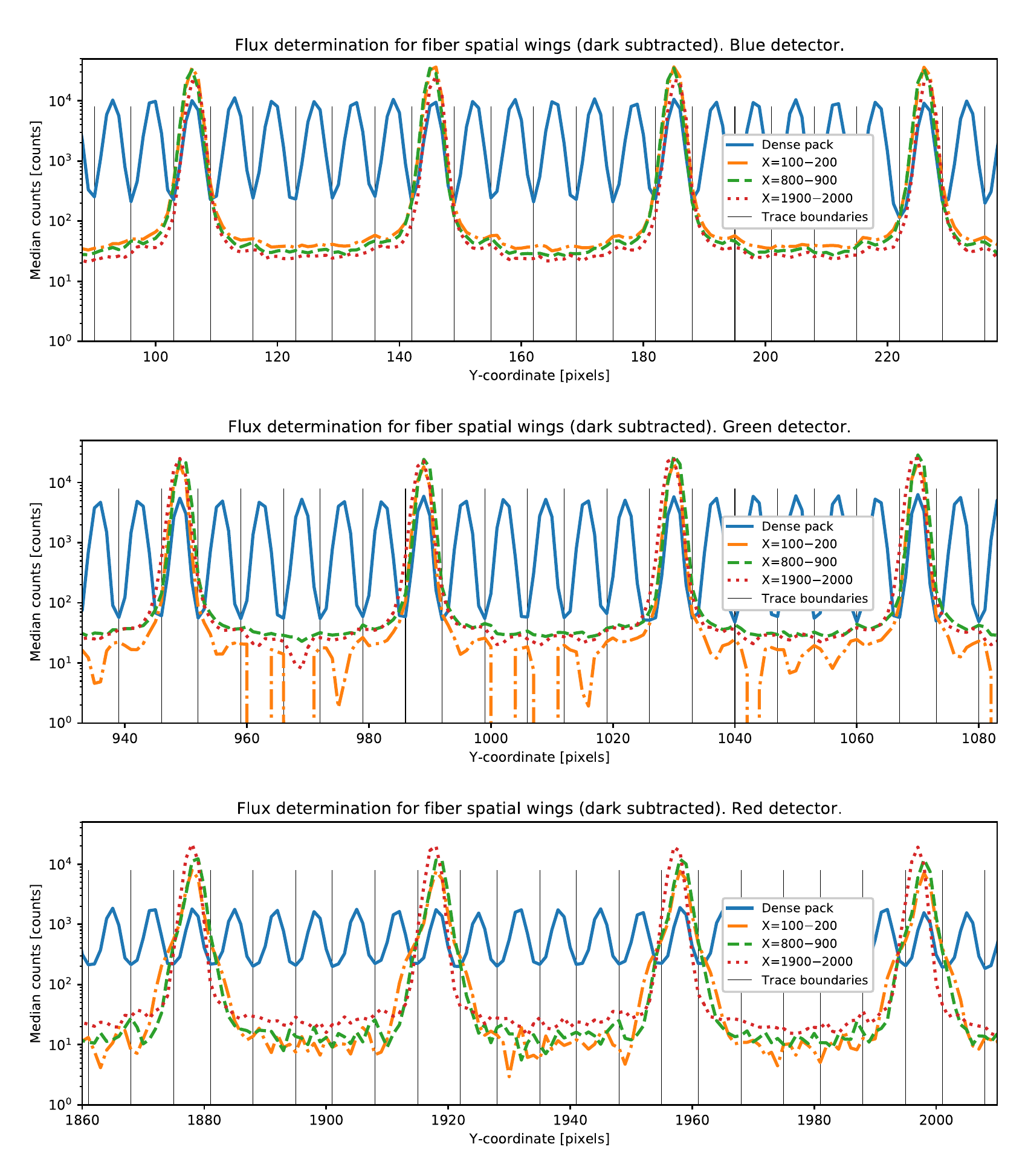}
\caption{Slices along the spatial direction on each of the three detector arrays for SparsePak observations.  Multiple spans of wavelength are shown, with smaller values of x corresponding to redder wavelengths.  Cuts from DensePak observations are superimposed to show fiber boundaries between illuminated SparsePak fibers.}\label{fig_fiber_crosstalk2}
\end{figure*}

The flux contained within the spatial wings of the fiber traces are sufficiently low that the flux from one fiber, the so-called main fiber, measured within the boundaries of the adjacent fiber, is typically $<1\,\%$ of the \edit1{flux within the boundaries of the main fiber}.  Both DensePak and SparsePak observations of the fiber-fed Tungsten Halogen Light Source were used for this analysis.  The DensePak observations were used to determine the locations of the peak and boundaries of all 300 traces at various locations on the three detector arrays.  Then the SparsePak observations were used to measure the summed flux that falls within the boundaries of adjacent fibers due to the spatial wings of a main fiber.  The flux contained within the boundaries of an adjacent fiber is $<1\,\%$ of the flux in the main traces for $73\,\%$ of the measurements ($\rm{N} = 23{,}900$ measurements over the three detector arrays, distributed as (14, 17, 19) 100-pixels-wide stripes running along the spatial axis on the red, middle and blue detector arrays, resp.).  Figure~\ref{fig_fiber_crosstalk} shows a histogram of the measured contamination.  And Figure~\ref{fig_fiber_crosstalk2} shows a slice along the spatial direction on each of the three detector arrays for SparsePak observations.  For each slice, multiple ranges of wavelength are shown, where smaller x values equate to redder wavelengths.  Superimposed on the plot are cuts from DensePak observations to show the location of the fibers between the illuminated SparsePak fibers.


A low cross-talk helps to reduce cross-contamination of adjacent spectra.  Cross-talk is also minimized by managing the plugging of fibers into plates such that bright, medium and faint sources illuminate the slit in an ordered pattern of faint, medium, bright, bright, medium, faint, etc., so the spectra of bright sources do not land adjacent to faint sources.  \citet{maj17} describes this fiber management scheme in further detail.

\subsection{Dithering Accuracy}


\begin{figure}
\epsscale{1.1}
\plotone{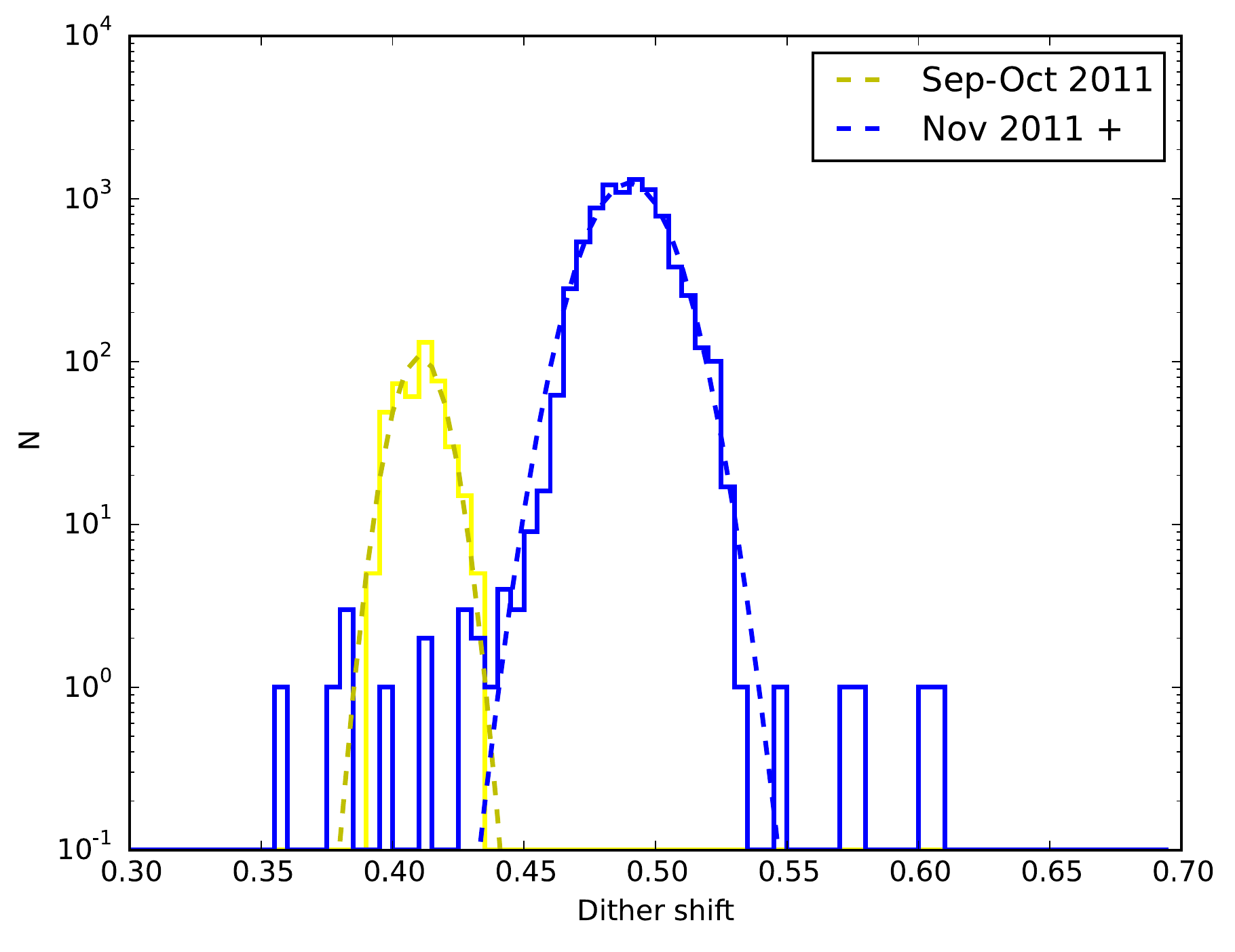}
\caption{A histogram of measured offsets between dither pairings during SDSS-III, along with Gaussian fits to the offsets based on calendar dates.  For commanded 0.5 pixel moves, the mean move is 0.490 pixels for 2011 November through summer shutdown in 2014. After the first two months of the survey, which had a mean move of 0.410 pixels, the positioning algorithm was adjusted to compensate for hysteresis.}\label{fig_dither_histogram}
\end{figure}


A histogram of measured offsets between dither pairings during SDSS-III (\citep[see][]{nid15} regarding the method of dither combination during data reduction) is shown in Figure~\ref{fig_dither_histogram}.  Measurement uncertainty of the offset is $\sim 0.005\,\rm{pixels}$ \citep{nid15}.  Overplotted are Gaussian fits to the offsets based on calendar dates.  For commanded 0.5 pixel moves between dither positions, the mean resultant mechanism move is $0.490$ pixels ($ \pm 0.013$ standard deviation) based on the offsets between $8{,}227$ dither pairings.  The mechanism had a mean offset of $0.410$ pixels ($ \pm 0.008$ standard deviation) for 445 combinations during the first two months of the survey.  After this period the algorithm was adjusted for hysteresis -- the mechanism has a hysteresis of about 0.1 pixel so we always approach new dither positions from the same direction.  More specifically, given dither positions A and B, separated by 0.5 pixels, a move from A to B is simply an advance of 0.5 pixels whereas a move from B to A is accomplished by moving back 1.5 pixels and then advancing 1 pixel.  There were three outlier dither offsets with values $1.3$ -- $1.5$ pixels and four outliers in the range $0.2$ -- $0.3$ which are neglected in the two histograms of Figure~\ref{fig_dither_histogram}.  Periodic glitches in the mechanism during the survey are discussed in \S~\ref{dith_mech_perf}.

\subsection{Observed Stray Light}\label{stray_light_performance}

\edit1{While it is difficult to determine which of the various stray light and ghost mitigation methods described in \S~\ref{stray_light_mitigation} were most effective, in general the APOGEE spectrographs are free of unexpected stray light and ghosts which compromise data quality.  After discussion of the observed Littrow Ghost, the following sub-sections address: the good quality of the VPH Grating as indicated by analysis of the wings of bright laser lines; the lack of evidence of fiber tip ghosts; unexpected ghosts around bright lines which are, however, too faint to affect science quality; and low stray light at the edges of the detector arrays.}

\subsubsection{Littrow Ghost}\label{observed_littrow_ghost}


\begin{figure}
\epsscale{1.1}
\plotone{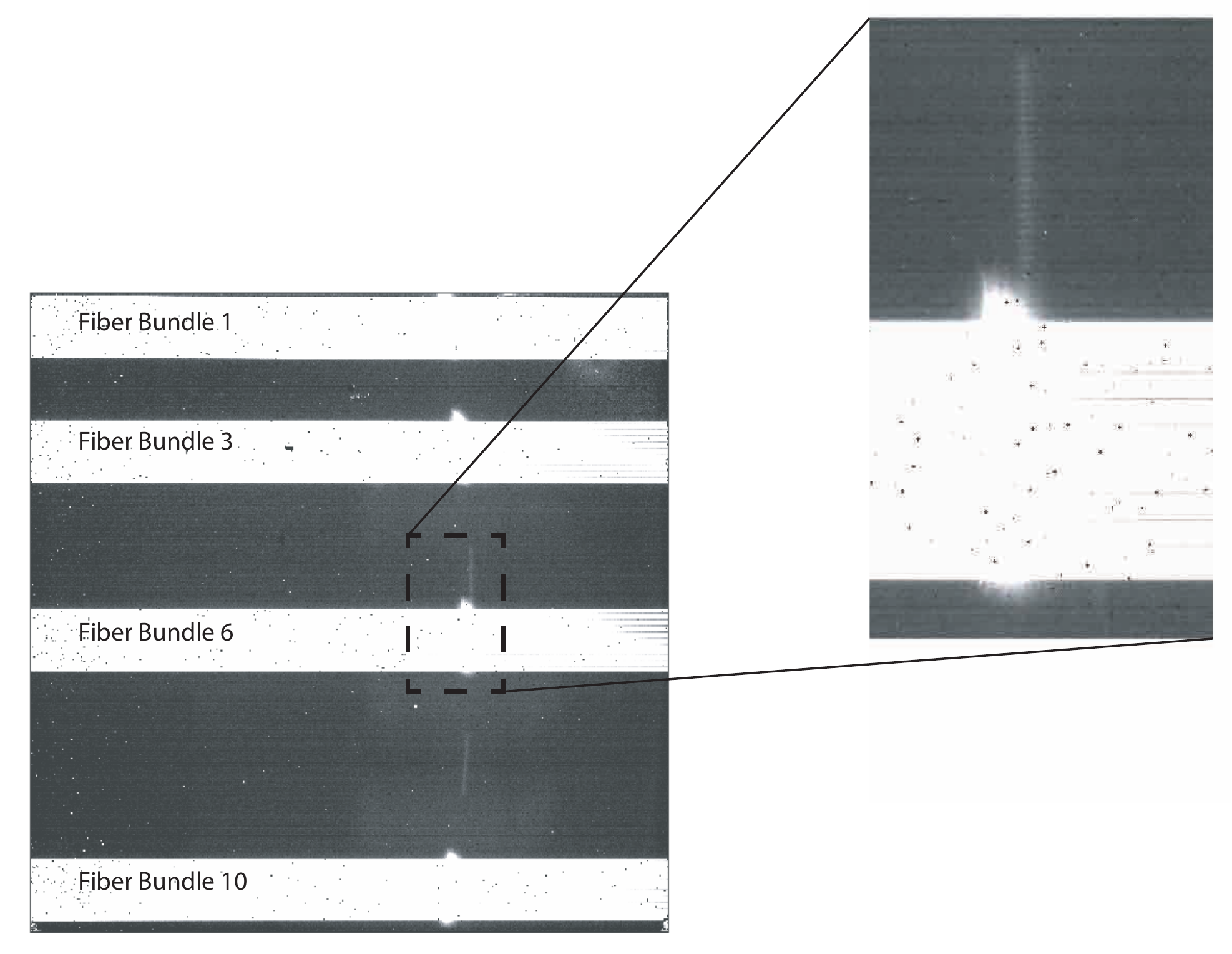}
\caption{A spectral image from the middle detector array during lab testing when the instrument was illuminated with an electronic black-body source.  At the time there were four fiber bundles, each containing 30 fibers, installed in the instrument.  As the close-up image shows, with aggressive stretching, the littrow ghost is made-up of both a spatially inverted and non-spatially inverted ghost.  The source of the latter is not understood.}\label{fig_littrow_ghost_features}
\end{figure}


The Littrow Ghost, discussed in \S~\ref{VPH}, can be seen in Figure~\ref{fig_littrow_ghost}.  It is imaged on the middle detector array in the vicinity of $1.6042$ -- $1.6062\,\micron$, depending upon the fiber.  The ghost is curved since it is the image of straight pseudo-slit lines originally imaged on the detector arrays.  As discussed in \citet{wil12}, images taken during testing of the Penn State Radial Velocity group's Fiber Fabry-Perot (FFP) interferometer \citep{hal14} during 2012 February enabled measurement of the ghost intensity.  A fully photonic doppler reference, it illuminated all fibers across the entire wavelength range with a ``picket fence'' of 400 discrete lines with spacing that varied from $\sim 10\,\rm{pixels}$ at the blue end to $20\,\rm{pixels}$ on the red end of the spectrum, resulting in a total of $\sim 120{,}000$ lines.\footnote{Light from the FFP was transmitted via single-mode fiber to one of the ports of the calibration system integrating sphere.}  A ghost intensity of $0.36\,\%$ relative to the total counts of all FFP lines imaged on the three detector arrays was determined based on overnight FFP testing.

Interestingly, the ghost is composed of two sources, one of which is not well understood.  Figure~\ref{fig_littrow_ghost_features} shows the ghosts created when a black body source was observed with four of the ten fiber bundles installed during APOGEE-North assembly.  The gaps between installed fiber bundles reveals that one set of ghost images is spatially inverted about the middle of the array and one is not.  The spatially inverted ghost, described in \citet{bur07}, is created by the two ``cases'' for generating the Littrow Ghost based on reflection from the detector arrays, subsequent recombination and reflection at the VPH Grating, and reimaging of zero'th order light on the detector arrays at the Littrow wavelength.

The formation scenario for the spatially non-inverted portion, on the other hand, is not understood.  Since it is in close proximity to the spatially inverted ghost and follows its curved shape, it is apt to be an image of the pseudo-slit in zero'th order light.  And its lack of spatial inversion implies a reflection at the detector is not involved.  It is possible it is formed from light that inadvertently passes between the aperture mask in front of the VPH grating assembly and the recorded portion of the grating and subsequently reflects off the interior sides of the substrates which make up the VPH grating assembly.  This could occur if the mask location or shape is incorrect.  But KOSI specified the substrate edges to be polished to 220 grit to mitigate specular reflection from these surfaces, which makes this ghost pathway unlikely to be the cause of fairly well focused ghosts.

\subsubsection{VPH Stray Light}

As described in \citet{wil12}, the instrument was illuminated overnight with light from a $1.55060\,\micron$ laser via the calibration integrating sphere during the FFP testing run in February 2012.  Analysis of the resultant co-added LSF can be used to judge the quality of the grooves in a grating \citep[see, e.g.,][]{woo94,ell08}.  In particular, \citet{wil12} showed the fit of the laser line wings to Equation 6 of \citet{woo94}:

\begin{equation}\label{woods_equation}
  Y_{fit} = \left[\frac{\sinc b}{\sinc b_{\rm{o}}}\right]^2 \left[\frac{0.5}{N^2 \sin^2a} \right] + A_B
\end{equation}

\noindent
where $a = \pi (\lambda/\lambda_{\rm{o}})$, $b = \pi (\lambda - \lambda_{\rm{blaze}})/\lambda_{\rm{o}}) f$, $f$ is the ratio of the groove width to groove spacing, and $N$ is taken to be approximately the smallest consecutive number of grooves without imperfections.


\begin{figure}
\epsscale{1.25}
\plotone{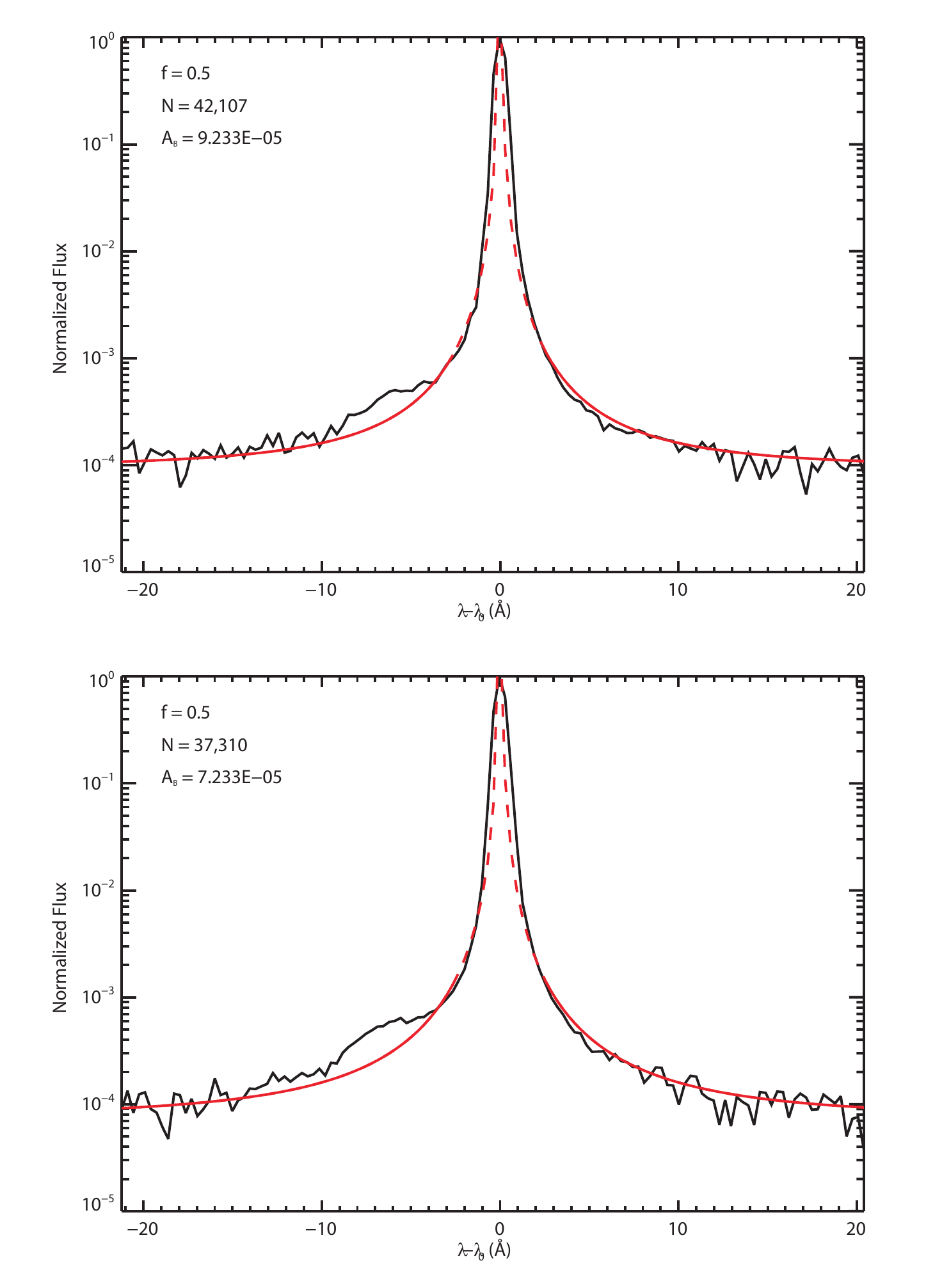}
\caption{Spectra of a $1.55060\,\micron$ laser that illuminated the APOGEE-North instrument via the calibration integration sphere during the FFP testing in February 2012.  (Top) The coadded line and fit to the wings for the top 15 SparsePak fibers which fell within the region of abnormally high persistence on the blue detector array.  (Bottom) Similar data but for fibers at the opposite side of the detector array which are not in the persistence region.  In each case, only data in the wings of the line are used to solve for the best fit marked by the solid fit curve.  The central peak region, arbitrarily defined to be the middle 15 pixels, is excluded from the fit and marked with the dashed fit curve.}\label{fig_vph_laser_testing}
\end{figure}


Since lines at this wavelength are imaged on the blue detector array, and the data was taken before this detector array was replaced due to abnormally high persistence,  we re-analyzed the data in case the persistence affected the results.  Figure~\ref{fig_vph_laser_testing} shows the spectral line and fits for two different regions of the detector array.  The top plot shows the co-added spectra for the top 15 SparsePak fibers which all fall within the persistence region whereas the bottom plot shows the spectra for the opposite side of the detector array unaffected by persistence.  Equation~\ref{woods_equation} is only fit to the wings beyond the central $15\,\rm{pixels}$ because we are only interested in the wings of the line.  Since the ``obliquity factor,'' $f$, is only relevant for addressing asymmetry in the spectral lines, as mentioned in \citet{woo94}, it was set to 0.5.

The central segment of the VPH Grating, seen in Figure~\ref{fig_vph}, is about $90\,\rm{mm}$ wide so it includes about $90,000\,\rm{lines}$.  And given the central obscuration which blocks the middle of this segment (see Figure~\ref{fig_pupil_image}), the minimum contiguous span of lines on the grating is about $30,000$.

The results for $N$, approximately $42,000$ in the persistent region and $37,000$ in the normally behaved area are significantly larger than $N \sim 31,000$ which was reported in \citet{wil12}, but still consistent with the general finding that the groove quality of the VPH Grating is very good.

\subsubsection{Fiber Tip Ghost}\label{fiber_tip_ghosts}

Despite some effort looking for evidence of fiber tip ghosts (see \S~\ref{ghosts}), particularly in the co-added spectral images of the laser line mentioned above, no definitive evidence of this ghost has been found.  It may be that the blackened strip along the vertical bisector of the collimator is effective at mitigating formation of this ghost since it blocks rays that would ordinarily strike the pseudo-slit area at normal incidence.

\subsubsection{Ghosts around Bright Lines}


\begin{figure}
\epsscale{0.90}
\plotone{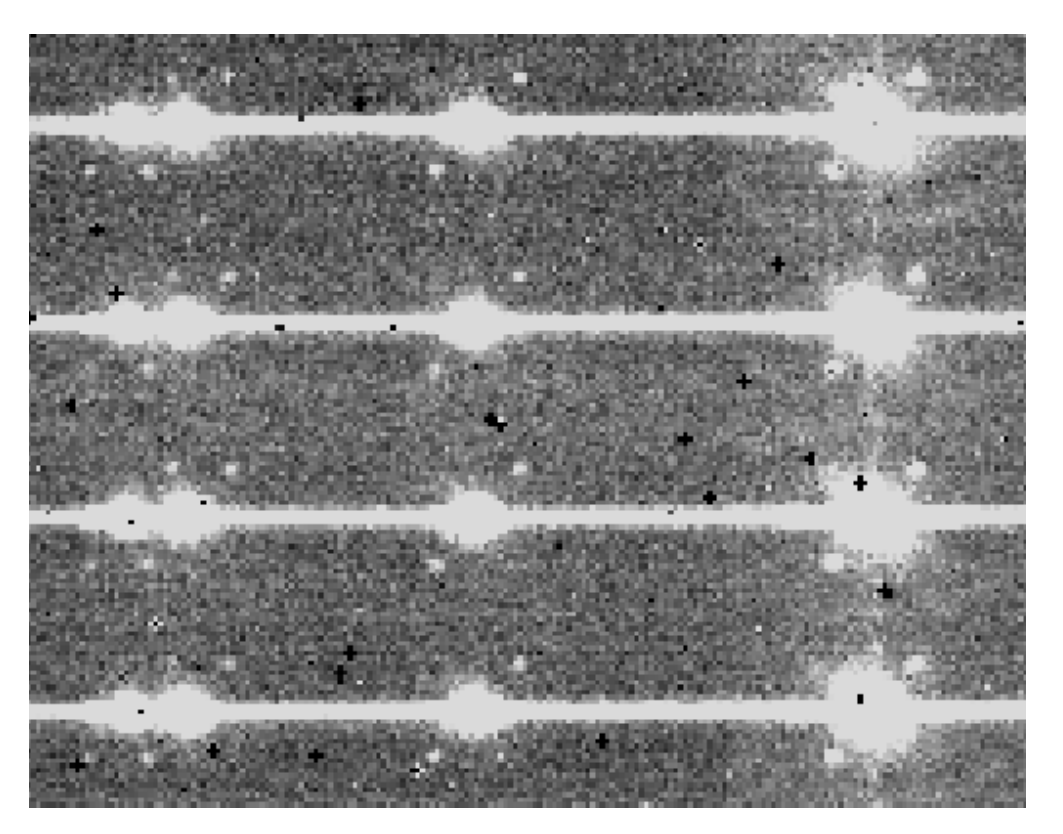}
\caption{Pairs of faint ghosts displaced diagonally by about 15 pixels are produced around bright lines.  The ghosts have peak counts of $\sim 0.035$ to $0.05\,\%$ relative to the peak counts of the original spectral line.}\label{fig_vph_satellite_ghosts}
\end{figure}


Bright spectral lines generate pairs of faint ghosts (Figure~\ref{fig_vph_satellite_ghosts}) that are displaced diagonally by $\sim 15\,\rm{pixels}$ to either side of the original spectral line.  The ghosts have peak counts, consistent across all three detector arrays, of $\sim 0.035$ -- $0.05\,\%$ relative to the peak counts of the original spectral line.  While these diagonally displaced ghosts fall on spectral traces of nearby fibers, they are too faint to affect the science quality of typical observations.

The displacements of the ghosts relative to the original line change slowly and monotonically in the spectral and spatial directions across the mosaic of detector arrays.  Spectrally, the ghost displacements change from $\sim 14\,\rm{pixels}$ at the red end to about $8\,\rm{pixels}$ on the blue end of the wavelength range, an angular change of $\sim 0.017\,\degr$ over the camera focal length.  In the spatial direction, the ghost displacement changes from $\sim 10.5\,\rm{pixels}$ to about $9\,\rm{pixels}$ over the same span.

The good focus of the ghosts suggests a source in collimated light but the origin of these ghosts is not understood.  Two sources have been investigated.  First, internal reflections within the VPH Grating can produce diagonally displaced ghosts. Given the small but finite efficiencies of the $m = 0$ and $m = +1$ reflective diffraction orders (Fig~\ref{fig_vph_multi_order_de}), there will be stray light generated at a small level by multiple pathways involving these orders along with internal reflections (due to imperfect AR coatings) at the substrate outer faces.  Ordinarily these straylight pathways would deliver the light to the same location on the detector arrays as the primary path.  But the ghost arrival location can be displaced from the primary line at the detector array when the relative tilts of the two substrates making up the VPH Grating assembly are taken into account.  The measured wedge is $0.005\,\degr$ ($0.003\,\degr$) in the spectral (spatial) directions based on vendor measurements of the thicknesses at the corners of the VPH Grating assembly after it was capped with the second substrate.  (This wedge is thought to be due to non-uniform thickness of the optical cement that fastens the two substrates, not due to wedge of the individual faces of the substrates.)  But non-sequential modeling of these pathways in Zemax produces asymmetrically displaced ghosts relative to the primary line.  And the individual ghost strengths should be different since they represent pathways made up of steps with different efficiencies.  Lastly, the ghost intensities should have a more marked variation with wavelength since the diffractive efficiencies of the reflective diffractive modes have considerable variation.

Generation of the pair of ghosts is suggestive of a grating-like source.  And it turns out there may be periodic polishing marks present in one or both VPH Grating substrate surfaces.  Two of the four substrates fabricated to support development of the VPH Grating have periodic polishing marks with groove periods of $\sim 30\,\rm{mm}$ along the diagonal of the substrate.  (The vendor apparently used deterministic polishing methods.  We had not specified mid-spatial frequency requirements.  Lastly, the provenance of the substrates was not tracked during grating development.)  Regardless, the groove period is too large to be the source of the pairs of ghosts that are seen around bright lines -- if the outer surfaces of both VPH Grating substrates are modeled as diffraction gratings with $30\,\rm{mm}$ groove periods, first order ghosts are located $\sim 3\,\rm{pixels}$ from zero'th order, not $\sim 15\,\rm{pixels}$.  Such polishing marks are, however, potential causes for broadening the PSF and LSF of the spectral lines.

\subsubsection{Stray Light around Detector Array Borders}


\begin{figure}
\epsscale{1.1}
\plotone{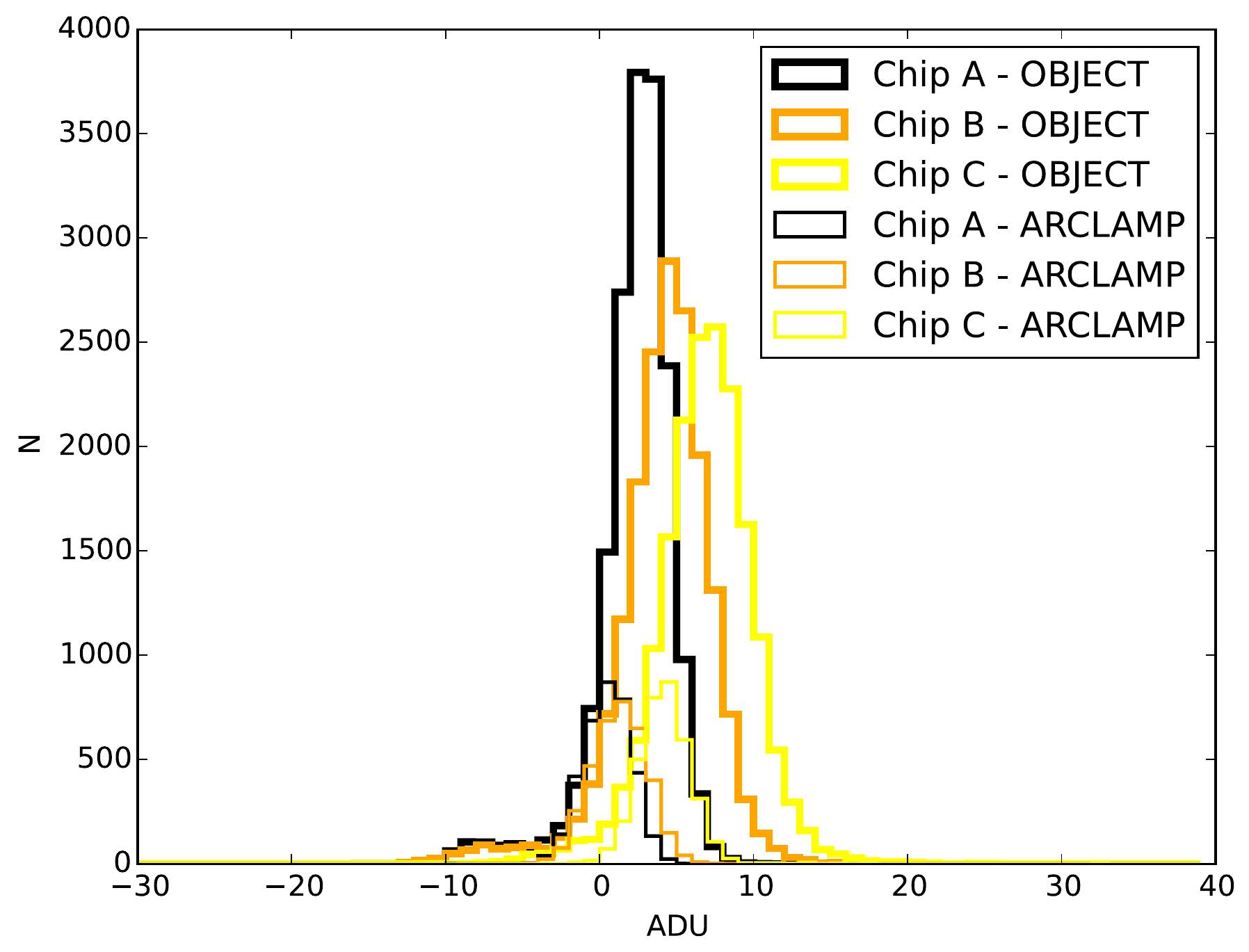}
\caption{A histogram of average pixel illumination during the SDSS-III survey within pixels on the edges of the detector arrays located between the reference pixels on the borders of the detector arrays and the top and bottom of the set of 300 fiber traces.  The counts are the average for columns 100:1947 within rows 4:8 (bottom) and 2039:2043 (top) for each detector array and type of exposure.}\label{fig_border_stray_light}
\end{figure}


Another indicator of stray light is the illumination between the four rows of reference pixels at the top and bottom of the detector arrays and the set of 300 fiber traces.  Figure~\ref{fig_border_stray_light} shows histograms of the average counts for $\sim 498\,\rm{sec}$ object and $424\,\rm{sec}$ arclamp exposures within windows in these areas on the detector arrays during the SDSS-III survey, including commissioning data.  Specifically, the counts are the average for columns 100:1947 within rows 4:8 (bottom) and 2039:2043 (top) for each detector array and type of exposure.  \edit1{The measured counts, mostly between 0 -- $10\,\rm{DN}$, are comparable to the amount expected from dark current (see \S~\ref{dark_current}) and far less than the hundreds to thousands of counts from the targets in most survey exposures.}

\subsection{Stability}\label{stability}

As mentioned in \S~\ref{vac_performance}, the instrument has been under vacuum and at cryogenic temperatures for the vast majority of time since commissioning in 2011 September.  This nearly continuous service promotes instrument stability as exemplified by the following from \citet{nid15}:

\begin{itemize}
  \item Flat fields during SDSS-III remained very stable --- a ratio of two super-flats produced two years apart are fit with Gaussians of $\sigma \sim 0.002$ -- $0.003$, compatible to the fit one would expect from photon statistics alone
  \item Temporal variations in wavelength calibration over year timescales were $\approx 0.015$ \AA ($\sim 0.02 \,\rm{pixels}$)
  \item RMS Gaussian FWHM of Th, Ar, and Ne calibration lines observed over three years varied by $\sim 1$ -- $2\%$
\end{itemize}

\subsubsection{Radial Velocity Stability}\label{rv_stability}

\citet{nid15} describe determination of radial velocity in the data reduction pipeline.  A good representation of internal RV stability is the $\sim 70\,\rm{m\,s^{-1}}$ RV scatter for giant stars with three or more visits of $S/N \ge 20$ for DR12.  The relative RV precision over long-term (1 -- 2 year) baselines is $\sim 100$ -- $200\,\rm{m\,s^{-1}}$ \citep{fle15}.

The strongest known source of instrumental instability is a repeatable movement of the spectra that correlates with $\rm{LN_2}$ tank level.  Given $\rm{LN_2}$ level changes of about $13\%$ over an 18 hour period, spectral traces move by about $0.15\,\rm{pixels}$ in the spatial (fiber-fiber) direction and 0.02 -- $0.03\,\rm{pixels}$ in the dispersion direction.  This $\rm{LN_2}$ level change causes a $22.5\,\rm{lb}$ ($10.2\,\rm{kg}$) reduction in tank weight.  Since the $\rm{LN_2}$ tank is suspended from the bottom center of the cold plate, we suspect the cold plate shape flexes with this cyclic weight change and causes one or more of the optics to move slightly.  The differential movement of the spectra are corrected during data reduction since the wavelength calibration of each individual dither exposure is zero-point corrected.

The FFP testing (described in \S~\ref{observed_littrow_ghost}) included measurements throughout two consecutive nights to monitor radial velocity (RV) stability of both the FFP and APOGEE \citep{hal14}.  Drift varies with detector array and fiber.  The total RV drift of a single fiber was $\sim 60$, $\sim 40$ and $\sim 0\,\rm{m\,s^{-1}}$ for the blue, green, and red detector arrays, respectively, over 12 hours.  After fitting with a low-order polynomial, \citet{hal14} found residual RMS scatter for each fiber of 2 -- $3\,\rm{m\,s^{-1}}$.  FFP testing also identified a small RV drift that correlates with periodic cryostat pressure changes.  Internal pressure changes of $\approx 15.8\,\rm{n Torr}$ ($2.1\,\mu \rm{Pa}$) during the second night of the FFP run correlated with drifts of $\sim 6\,\rm{m\,s^{-1}}$.  This pressure variation during the night is typical although its source is unknown.  As described in \citet{hal14}, after correction of this second instrumental drift, residual fiber-averaged RV scatter was $\sim 80\,\rm{cm\,s^{-1}}$ per night.


\section{Lessons Learned}\label{lessons learned}

\subsection{Teaming}

With a fast-paced development schedule --- two years between critical design review and final commissioning --- the project relied upon the successful design, fabrication, and testing of numerous modules by vendors and collaborators spread across a large geographic area.  In particular, the project strived to foster and maintain close teaming relationships with vendors and university research groups alike to improve chances of successful delivery and performance of modules and to minimize communication errors.  Key to successful teaming was establishing good rapport and communications patterns which were enhanced and enabled by numerous face-to-face visits.

\subsection{Astigmatism}\label{astigmatism_correction}

The source of the astigmatism, which was corrected by bending Fold Mirror 1 (\S~\ref{fold_mirror_1}), is likely the VPH Grating.  Two clues suggest the source is an optic near the pupil.  First, the astigmatism is uniform for all field positions (fibers) at a given wavelength, including on-axis.  Secondly, the system was well aligned compared to the design in both the dispersion direction and spatial direction (all 300 fibers were imaged on the detectors on the first cool-down), thus eliminating a decentered camera optic from consideration.  Also, the astigmatism is manifested by a mismatch in power between the spectral (parallel to the cold bench plane) and spatial (perpendicular to that plane), a directionality which suggests an optic with rectangular shape, i.e., a fore-optic or the VPH Grating.

The astigmatism was measured by pistoning the collimator and producing through-focus curves for both the spectral and spatial directions.  The longitudinal astigmatism (difference in focal position of the sagittal and tangential astigmatic images) was $\sim 100\,\micron$ based on collimator position differences of $\sim 1\,\rm{mm}$ (Figure~\ref{fig_focus_curve_montage}).  The curves were roughly matched in Zemax by modeling the collimator as an irregular surface with an added $12.5\,\micron$ sag at the top and bottom edges such that the mirror had less power in the spatial direction.  After doubling the sag since the collimator is a reflective surface, this sag corresponds to $15.6\,\rm{waves}$ of astigmatism at $1.6\,\micron$.

But this does not mean the collimator is the likely source of the astigmatism:  The Fold Mirror 1--Collimating Mirror--Fold Mirror 1 chain was checked at room temperature with a LUPI using a $0.6328\, \micron$ HeNe laser light and no astigmatism was seen.  Additionally, post-facto FEA analysis by Hofstadter Analytical Services (Tucson, AZ) of the modeled mirror and mount system at cryogenic temperatures did not reveal any reasons to suspect the cold collimator is the source.  They predicted on the order of a $100\,\rm{nm}$ PV departure from a fit for power due to slight differential CTE between the face sheet and glass cylinders that make up the fused mirror, and $\approx 450\,\rm{nm}$ PV distortions due to forces from the tip-tilt actuators with real-world tilts applied.

The semi-kinematic optical mounting schemes were reviewed but no likely opto-mechanical design causes for the astigmatism were identified.  Nonetheless, none of our optics were tested cold in a test cryostat, either individually or in their mounts.  Such tests are difficult and costly since a large cryostat, with large windows, is necessary to test optics with beam sizes of $ \gtrsim 300\,\rm{mm}$ diameter.

Fold Mirror 2 was considered since it has a dichroic coating with 22 dielectric layers on the front surface.  Stresses from differential CTEs between dielectric layers and substrates are known to cause bending if the substrate strength and thickness are insufficient to resist the stresses.  FEA modeling using the CTEs of the coating materials (both the dichroic and the AR coating on the opposite side) and substrate suggested the APOGEE dichroic coating would be in tensile stress such that the substrate would bend into a concave shape on the dichroic coating side.  The sag from middle to corner at cryogenic temperatures was predicted to be $2.5\,\micron$.  When used as a fold mirror, a spherically bent surface produces astigmatism because of differences in projected aperture size between the spatial and spectral direction.  The bending of Fold 2 was modeled in Zemax with a $3.6 \times 10^7\,\rm{mm}$ radius of curvature on the dichroic surface.  This modeling suggested the bent Fold 1 actually implemented would have been an inappropriate correction for the relatively small Fold 2 curvature stated above -- a Fold 1 bent by at least a factor of four less would have been a better correction.  Thus Fold 2 is unlikely to be a primary source of the astigmatism.  Furthermore, the Fold 2 Mirror in the APOGEE-South instrument, as discussed in \S~\ref{apogee_south_fold2}, did not have a dichroic coating and yet it had similar astigmatism to the APOGEE-North instrument.


\begin{figure}
\epsscale{1.2}
\plotone{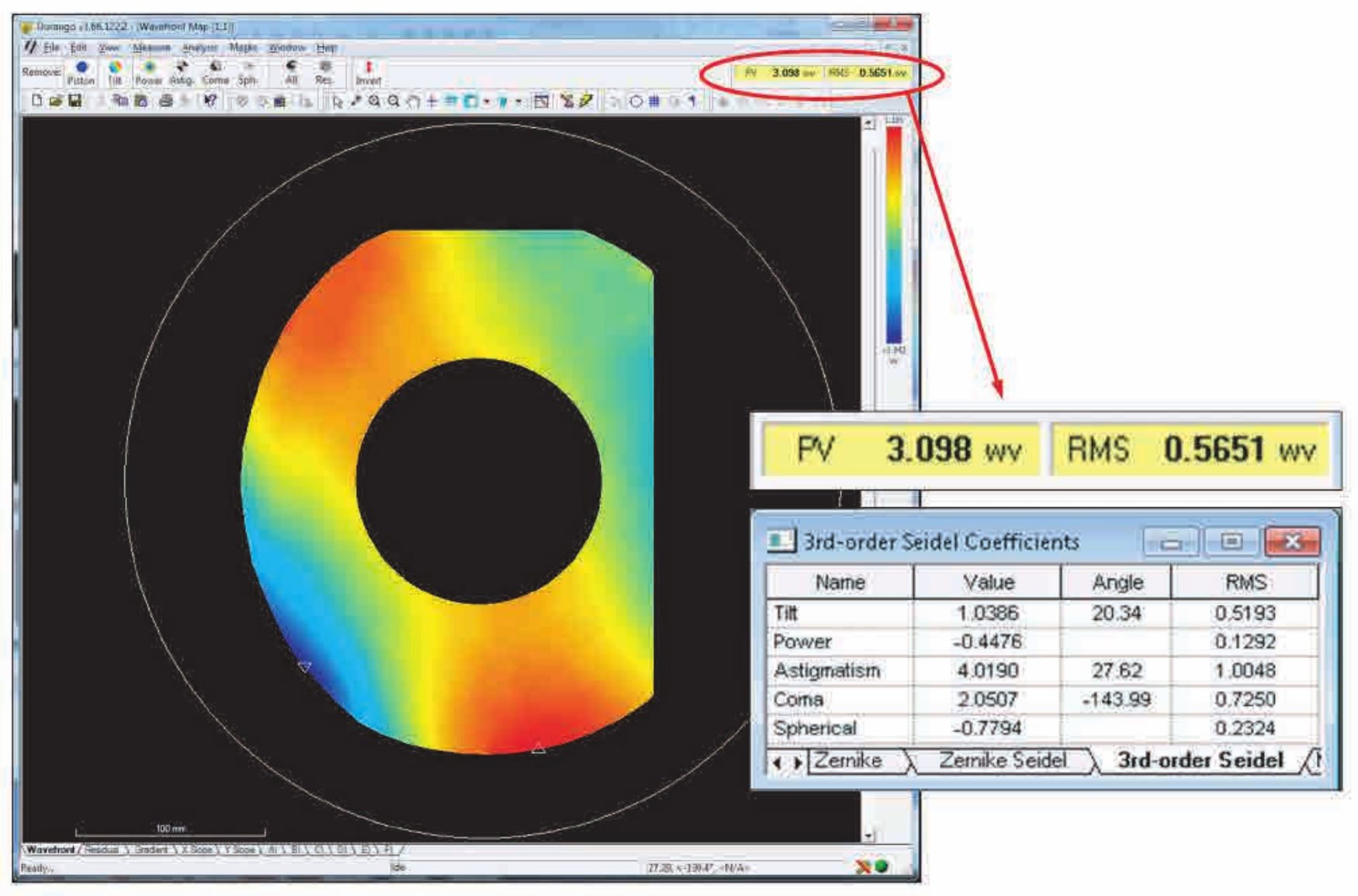}
\caption{The diffracted wavefront error of left-panel of the APOGEE-South VPH Grating when illuminated with $0.6328\,\micron$ light with an $18.61\,\degr$ angle of incidence using the interferometric test set-up at JHU.  The image is the average of six individual interferograms.  The dominant third order Seidel aberration is astigmatism.}\label{fig_vph_interferogram}
\end{figure}


The VPH Grating, located at the pupil, is probably the source.  As will be discussed in \S~\ref{correction_wedge}, the APOGEE-South VPH Grating, featuring a two-panel design instead of three-panels, was interferometrically tested at JHU using the LUPI test set-up described in \citet{bar14}.  Both panels of the grating were tested at $0.6328\,\micron$ wavelength and $18.61\,\degr$ angle of incidence, the Bragg angle for this wavelength.  Given the test set-up constraints, the VPH Grating could be translated laterally such that the left-hand panel could be more fully illuminated with the $15\,\rm{in}$ ($380\,\rm{mm}$) diameter test beam at JHU.  Figure~\ref{fig_vph_interferogram} shows the average first order diffracted wavefront error of six individual interferograms for the left-hand panel along with the resultant third-order Seidel aberrations which include 4 waves of astigmatism.

Zemax modeling was performed to determine how a VPH Grating with 4 waves of astigmatism at $0.6328\,\micron$ used at $18.61\,\degr$ angle of incidence would affect imaging performance when used at $1.6\,\micron$ with $54\,\degr$ angle of incidence.  Specifically, an optically fabricated hologram surface was created in Zemax by interfering two collimated construction beams at $0.488\,\micron$ and the correct geometry to produce the nominal fringe spacing.  One of the construction beams was aberrated by using a fringe Zernike sag surface with a single astigmatism coefficient.  The coefficient was iteratively adjusted until 4 waves astigmatism was measured at $0.6328\,\micron$ when the optically fabricated hologram file was used at $18.61\,\degr$ angle of incidence.  When this optically fabricated hologram surface was illuminated with a collimated beam and the diffracted beam was focused with a paraxial camera with $356\,\rm{mm}$ focal length (the APOGEE camera focal length) and the appropriate wavelength dependent angle of incidence, the longitudinal astigmatism was about $32\,\micron$ at $0.6328\,\micron$ and $145\,\micron$ at $1.6\,\micron$.  The latter result is the same order of magnitude as measured in the APOGEE instruments.  Aberrations were also measured \citep{bar14} in the test VPH Gratings fabricated by KOSI for the Prime Focus Spectrograph in development for the Subaru Telescope.  In that case both trefoil and astigmatism were observed.

\subsection{Unanticipated Cold Plate Flexure}

The cold plate flexure with changing $\rm{LN_2}$ tank level, described in \S \ref{stability}, was simply a design oversight.  FEA analysis predicts that the maximum cold plate deformation, which occurs in the vicinity of the pseudo-slit, changes by $\sim 0.0001\,\rm{in}$ ($3\,\micron$) over the 24 hours during which the $\rm{LN_2}$ has depleted to $\sim 83\%$ full.

\subsection{VPH Segment Design}\label{segment_design}


\begin{figure}
\epsscale{1.0}
\plotone{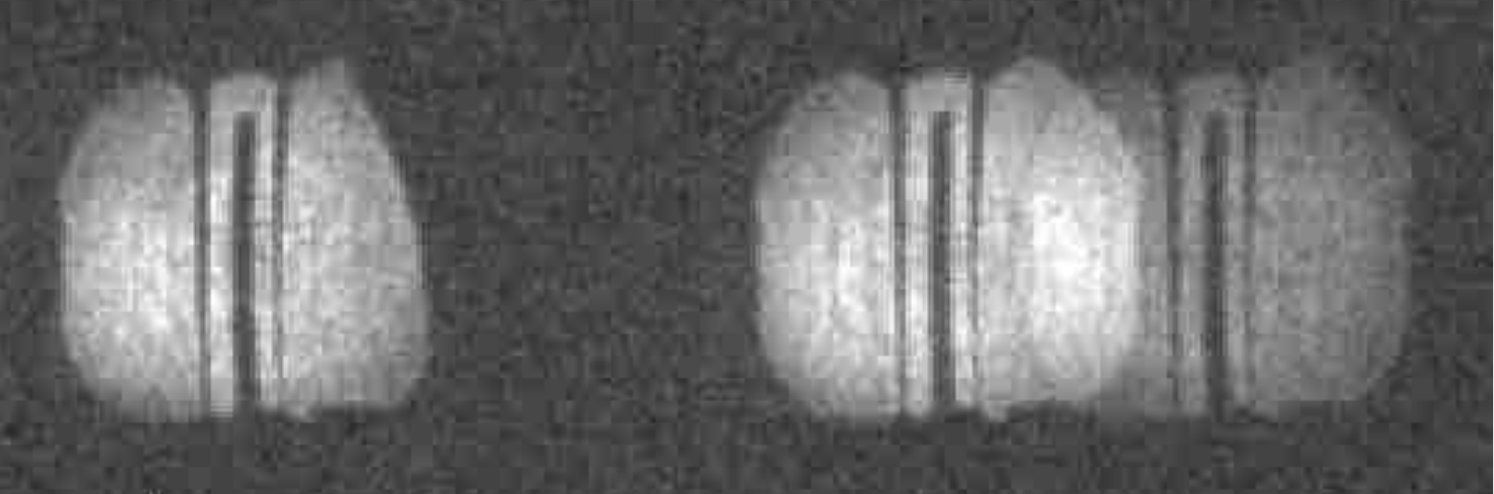}
\caption{Out of focus images taken with a room temperature InGaAs camera which show an image of the pupil, a portion of which was shown in \citet{wil12}.  The three pupil images correspond to three discrete lines from a Krypton arc lamp at the blue edge of the instrument wavelength coverage.}\label{fig_pupil_image}
\end{figure}


Obvious in hindsight, we should have designed the VPH grating with two recorded segments instead of three.  During the design we neglected to consider the position of the pseudo-slit obscuration relative to the segment boundary positions.  As seen in an out-of-focus image taken with a room temperature InGaAs camera (Figure~\ref{fig_pupil_image}), the pseudo-slit obstructs a substantial fraction of the light which would otherwise go through the central rectangular segment.  In a two segment design the throughput should improve since the segment boundary and pseudo-slit obstruction would overlap.

\subsection{Camera Design Temperature}

Despite schedule pressures, we should have completed definitive thermal analysis and derived a robust prediction of final camera temperature prior to adopting a temperature for camera fabrication purposes.  As mentioned in \S~\ref{camera_temp}, we chose a conservatively warmer design temperature to be safe.  Fortunately, the design is fairly accommodating of temperature deviations.

\subsection{Dither Mechanism Glitches}\label{dith_mech_perf}

During the first two years of survey operations there were various glitches, or unexpected moves, of the dither mechanism.  And the mechanism became stuck in 2013 December after an increasing number of glitches.  Increasing the maximum available stepper motor drive current from $1.0\,\rm{A}$ to $1.4\,\rm{A}$ resolved the issue and eliminated the glitches.

Given the build-up of solid lubricant debris seen during lifetime testing (\S~\ref{dither_mechanism}), it was assumed that increasing debris was the cause of these glitches.  Yet after SDSS-III was completed the mechanism was inspected and found to be free of debris.  Aside from simply underestimating the frictional forces between the leadscrew and nut at cryogenic temperatures, there is not an obvious explanation for the glitches at the lower motor current.

\subsection{LED Cryogenic Operability}\label{led_operability}

The Hamamatsu infrared LEDs described in \S~\ref{internal_leds} have all become inoperable over time.  They are only rated for use down to $-30\,^{\circ}\,\rm{C}$, and they have windows. After the last two instrument pump outs and cool downs, subsets of the LEDs have been found inoperable when turned on.  Cryogenic-rated LEDs, without windows, are now in use in the APOGEE-South instrument (\S~\ref{apogee_south_leds}) and this type of LED will be retrofitted into the APOGEE-North instrument the next time it is opened for maintenance.

\subsection{Miscellaneous Design and Fabrication Errors}\label{misc_lessons_learned}

{\it $LN_2$ Tank Bolt Connections} -- As discussed in \S~\ref{thermal_performance}, the bolt connections between the L-brackets and cold plate were simply torqued by hand.  Since differential shrinkage between aluminum and stainless steel bolts will loosen bolt connections and reduce pre-load at cryogenic temperatures, we suspect there is thermal resistance at this location which lengthens the instrument cool-down time.  A better approach would have been to athermalize the bolt connection by including either Belleville washers or Invar sleeves under the bolt heads to maintain bolt pre-load when cold.

{\it Camera Legs} -- The first version of mounts for the camera barrel consisted of separate triangular-shaped legs which connected each side of a camera flange to the cold plate.  But the individual legs were not directly connected to each other such that, as a system, the camera mount was not sufficiently rigid to constrain the camera position in space.  Addition of cross-braces under the camera barrel to connect the legs rectified the situation.

{\it Positioning a Large Camera} -- While the fore-optics and VPH Grating were positioned using visual laser light, the camera could not be positioned in the same manner because of the absorption of visual wavelengths by its silicon elements.  Instead, its position was checked using traditional shop measurement tools such as large calipers and scales.  The meter-sized distances involved in checking the camera position relative to cold plate mechanical features such as the edges with such tools made for imprecise measurements.  For future large instrument assemblies we plan to mount a FARO Prime $8\,\rm{ft}$ ($2.4\,\rm{m}$) portable coordinate measuring machine (CMM) from FARO Technologies, Inc., at a fiducial location on the cold plate to fine-tune the positioning and alignment of opto-mechanical modules to within $\sim 50\,\micron$.  A similar FARO arm was used during alignment \citep[see, e.g.,][]{lee12} of the Visual Integral-Field Replicable Unit Spectrograph (VIRUS) instruments for the Hobby-Eberly Telescope Dark Energy Experiment (HETDEX).

{\it Incorrect Black Paint Primer} -- Many of the aluminum parts which make up the opto-mechanical mounts and baffles underwent the Chem Film (chromate conversion coating) process to prevent oxidation.  Subsequently, some of these parts were painted black for stray light reduction.  At first, Lord Corp. Aeroglaze\textsuperscript{\textregistered} 9924 wash primer was applied to the parts, just as we had done in the past for bare aluminum parts, prior to application of Aeroglaze\textsuperscript{\textregistered} Z306 black paint.  Inadvertently, use of the wash primer led to loss of adhesion and peeling paint on several parts after a cycle between room temperature and $77\,\rm{K}$.  According to Lord Corporation, phosphoric acid in the wash primer etches, via acid pitting, bare metal surfaces to give more adhesive area.  But the Chem Film process blocks this acid etching and thus compromises primer adhesion.  Aeroglaze\textsuperscript{\textregistered} 9929 epoxy primer should be used with parts that have undergone Chem Film processing.

{\it Insufficient LED Diffusion} -- The LED internal illumination system on the cap of the cold shutter was described in \S~\ref{internal_leds}.  To minimize structure seen in the internal flat field exposure with the arm closed, diffusing optics need to be added adjacent to the LEDs.

{\it Difficulty in Positioning V-grooves} -- The difficulty of accurately positioning v-groove blocks was described in \S~\ref{align&test}.  The redesigned camera system that was used in 2014 July to post-facto check the v-groove block positions (which, e.g., was used to image Figure~\ref{fig_block4_closeup}) would aid future adjustments of v-groove block positions.  Also, a new guide block has been designed and fabricated which uses two kinematic pads to define the target v-groove position rather than trying to keep the entire front v-groove face flush with the guide.


\section{Second Spectrograph}\label{second_spectrograph}

The APOGEE-South spectrograph, a near-copy of the first instrument, was commissioned at the $2.5\,\rm{m}$ du Pont Telescope at the Carnegie Observatory's Las Campanas Observatory in early 2017.  Notable design changes for the second instrument include a rigid internal cold plate suspension system and external base isolation system to protect the instrument during seismic events, a revised $\rm{LN_2}$ tank suspension system, a different method of lightweighting the cold plate, use of a two panel VPH Grating, and the necessity of an optical wedge in front of half of the grating to correct for a mismatch in recorded groove density between the two segments of the VPH Grating.  These changes are discussed below, along with a description of the infrastructure developed to support survey operations at the du Pont Telescope.

\subsection{Use of the du Pont Telescope}

The du Pont Telescope was a natural choice as a Southern Hemisphere counterpart to the Sloan Foundation Telescope --- it has the same aperture, a similarly wide field of view ($2.1\,\degr$ v. $3.0\,\degr$), and a similar focal ratio ($f/7.5$ v. $f/5$).  Also, it had already been used for the Las Campanas Redshift Survey \citep{she96, she93}, a wide-field fiber-fed spectrographic survey conducted from 1988 -- 1994.  Yet since the completion of that survey, the telescope had mainly been used with instruments with relatively narrow fields of view.


\begin{figure}
\epsscale{1.1}
\plotone{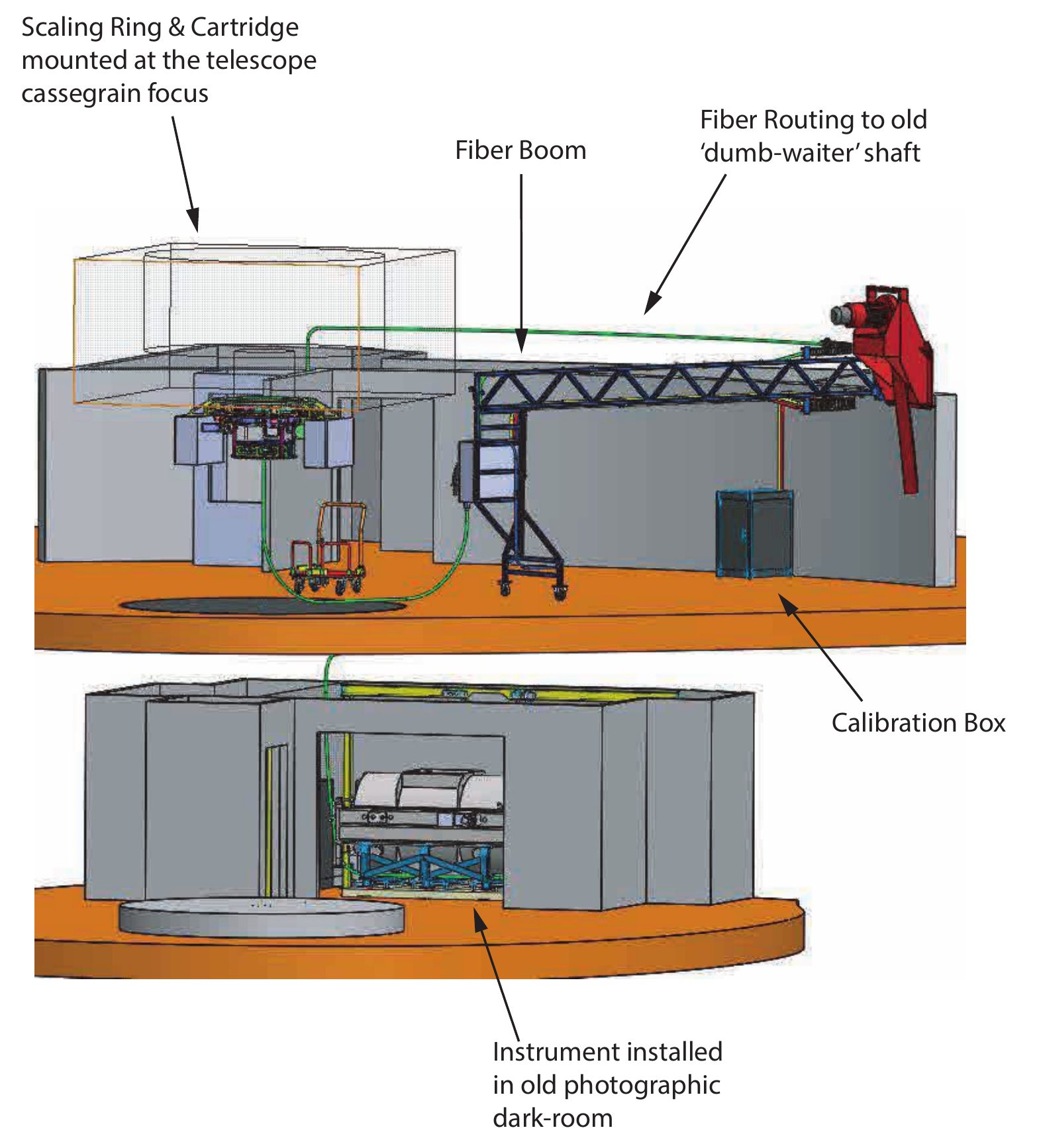}
\caption{The infrastructure developed for the du Pont Telescope at LCO.  The observing floor of the telescope and the old photographic dark room on the floor below are shown in this model. After going along the fiber boom to the dome wall, the fiber routes to the old dumb waiter shaft and down to the instrument in the dark room.}\label{fig_lco_layout1}
\end{figure}

\begin{figure}
\epsscale{1.0}
\plotone{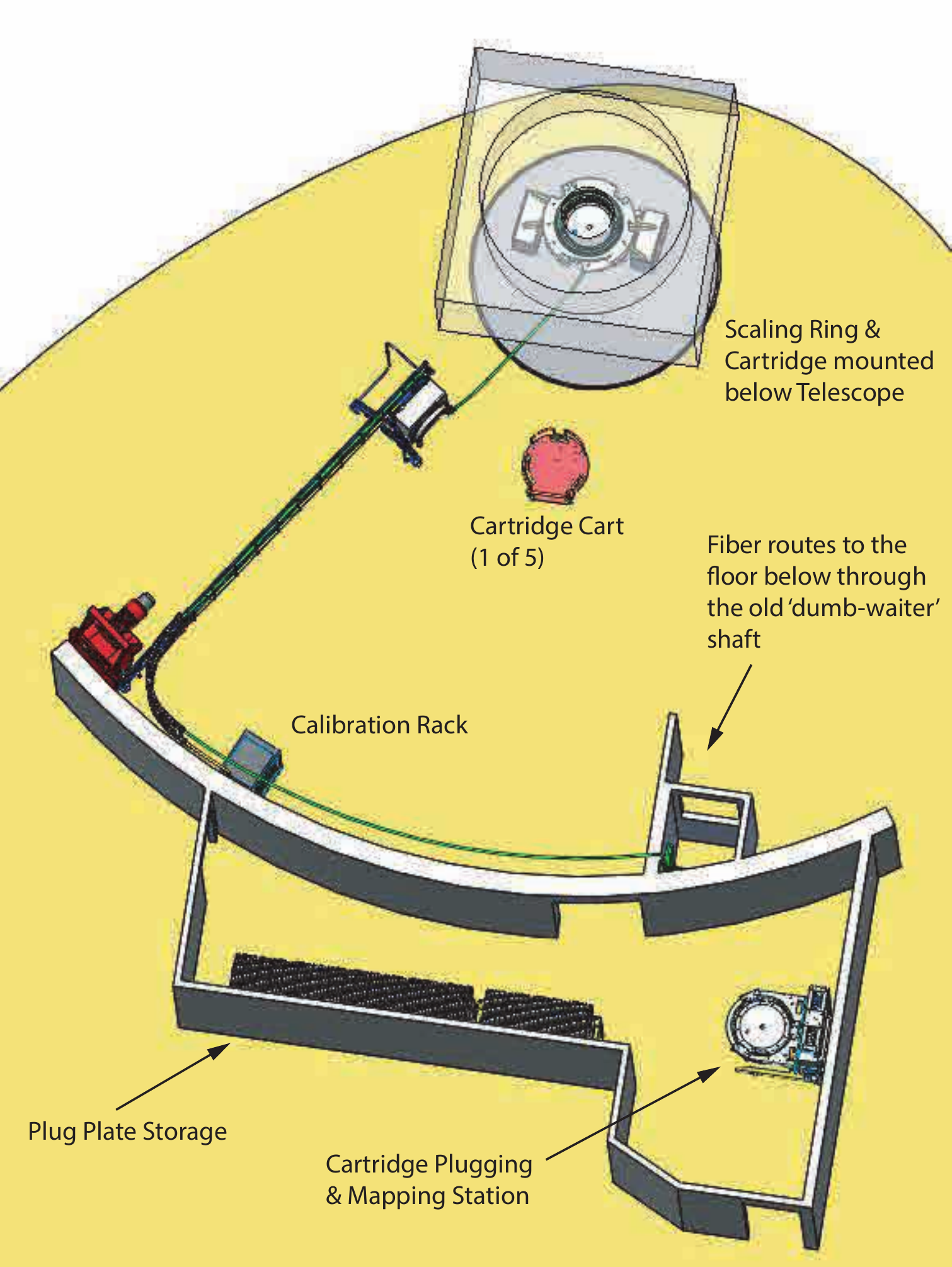}
\caption{A view of the infrastructure from the top of the dome.  The cartridge mapping and plugging station and plug plate storage is in a room adjacent to the telescope on the dome floor.}\label{fig_lco_layout2}
\end{figure}

Significant infrastructure \citep{bla17} was developed to enable survey operations (Figure~\ref{fig_lco_layout1} and Figure~\ref{fig_lco_layout2}).  A ``Scaling Ring'' assembly was fabricated that bolts to the back end of the telescope and positions cartridges at the telescope focal plane.  Cartridges are changed throughout the night through the use of a hoist system integrated into the Scaling Ring.  The hoist raises and lowers the $200\,\rm{lb}$ ($91\,\rm{kg}$) cartridges with three BOB\textsuperscript{\textregistered} rope\footnote{BOB\textsuperscript{\textregistered} rope is a synthetic rope construction available from Cortland, Ltd.} segments driven by a single motorized drum.  The cartridges are kinematically positioned at the bottom of the Scaling Ring with a set of three v-grooves and a latch ring system secures the cartridge in place.  Five cartridges were fabricated for the survey.  As more than five cartridges can be observed in a night, a crew of pluggers does real-time re-plugging after the first cartridges have been observed early in the night.  Dollies are used to move the cartridges between the mapping and plugging station in a room adjacent to the observing floor and the telescope.


Unlike the Sloan Foundation Telescope, which features a primary mirror that translates along the optical axis to adjust image scale to correct for the expansion and contraction of the aluminum plug plates with temperature variations, and partially correct for airmass differences relative to the plate design \citep[see][\S~3.4.1]{gun06}, the du Pont Telescope has no means of changing scale in real time.  Thus the Scaling Ring provides real-time translation of the cartridge along the optical axis with a thread ring driven by a closed loop servo-controller.  The thread ring drive system consists of a $1\,\rm{m}$ diameter nut and screw with 10-pitch Acme threads. The screw is rotated by a servomotor through a friction drive system.  Three die pins with $120\,\degr$ separation restrict rotation of the nut relative to the telescope to constrain focal plane translation along the optical axis.

The $120\,\micron$ core fibers used with the APOGEE instrument have a FOV of $1.31\,\arcsec$ on the sky when used on the du Pont Telescope compared to $2.0\,\arcsec$ at the Sloan Foundation Telescope given the different focal ratios.  This reduction in FOV is largely compensated by the excellent seeing at Las Campanas Observatory: The H-band seeing at LCO was expected to be $ \sim 0.6\,\arcsec$ FWHM.\footnote{Derived from Y, J and H-band seeing measurements on the du Pont Telescope from M. Phillips, private communication, supported by the $0.67\,\arcsec$ site seeing measurements near the du Pont Telescope reported in \citet{tho10}.}  Nonetheless, we considered the use of fiber lenslets to convert the $f/7.5$ beam to $f/5$ upstream of the fiber tips at the telescope focus to allow the input of identical beam sizes into the instrument and permit a larger FOV to accommodate other pointing errors such as from guiding and fiber positioning errors.  But given fiber FRD, which was expected to speed the beam to focal ratios close to what is delivered to the APOGEE-North instrument, focal conversion with lenslets was determined to be unnecessary from an instrument standpoint.  And the native FOV was deemed sufficient for accommodating the aggregate of fiber positioning errors.

\edit1{Using bending rings and central rods that pull down the plate center (in contrast to the central rods in cartridges for the Sloan Foundation Telescope which push up), the plug plates are bent to an $8800\,\rm{mm}$ radius of curvature within the cartrdiges.  This} is the approximate exit pupil distance for the du Pont Telescope so the system is essentially telecentric --- rays from the center of the telescope pupil, for all field positions, arrive orthogonal to the plate. This simplifies fabrication of plug plates used at the du Pont Telescope since the ferrule holes can be drilled perpendicular to the plate surface.

Based on image quality testing at the du Pont Telescope before the survey started (\citet{bla17}, \S~2.2), we chose a nominal focal plane position of $10\,\rm{in}$ ($228.6\,\rm{mm}$) below the instrument mounting plate at the Cassegrain focus.  The nominal corrector position is $39.85\,\rm{in}$ ($1012.3\,\rm{mm}$) above the focal plane.

\subsection{Fiber System}

The fiber connection between the cartridges and instrument are mostly the same as described in \S~\ref{fibers_and_fiber_routing} for APOGEE-North.  A gang connector system at the telescope-end of the fiber train is again used to change fiber connections between the instrument and the various cartridges.  And fiber links with fiber-feedthroughs at the cryostat wall and v-groove block terminations at the slithead are used again to bring the fiber-fed light into the instrument.  The two main differences are the fiber routing and the addition of an $8\,\rm{m}$ fiber segment at the telescope-end that is meant to be a sacrificial link in case of damage from rolling carts.  There were ten of these ``Telescope Links'' added in series and each link contained thirty fibers and were terminated on both ends with appropriate MTP\textsuperscript{\textregistered} connectors.

For fiber routing (Figure~\ref{fig_lco_layout1} and Figure~\ref{fig_lco_layout2}), a boom assembly was installed within the telescope dome.  Anchored and hinged at the dome interior wall, the $6.4\,\rm{m}$ boom secures the fiber run from the dome to a position adjacent to the telescope.  During observing runs the boom is rotated so its end is near the telescope.  When the instrument is not in use, the boom is stowed against the dome wall.  An electronics cabinet (``boom box'') with a user interface panel to allow the operator to control the scaling ring and hoist mechanism during cartridge changes is mounted at the end of the fiber boom.  Also, Stow, DensePak, and SparsePak positions for the gang connector are provided under the bottom of the control panel.  Lastly, MTP\textsuperscript{\textregistered} connections between the telescope links and fiber links are located within the panel.  Starting from the end of the boom, the fiber run follows the boom, then follows the inside wall of the telescope dome, and goes down the photographic plate dumb-waiter shaft that connects a room adjacent to the telescope with the old dark room on the ground floor below the telescope where the instrument is located.

\subsection{Cold Plate Rigid Support Structure and Seismic Isolation}

The cold plate support structure and instrument seismic isolation schemes were changed for the APOGEE-South instrument to minimize potential damage given the markedly increased seismic activity in Chile.  Instead of the free-hanging support of the cold plate used in the first instrument (see \S~\ref{cold_plate_design}), which has a natural frequency of $\sim 1.3\,\rm{Hz}$ ($\sim 0.7\,\rm{Hz}$) along the long (short) directions, a modified hexapod structure was employed to rigidly support the cold plate.  Using an iterative sequence of FEA of the cryostat system, knowledge of the maximum sustainable forces on the optics, and comparison to earthquake predictions for the observatory, an external isolation system was designed to minimize transmission of shocks to the instrument. This isolation system features a custom THK\footnote{THK America, Inc., Schaumburg, IL, (847) 310-1111} seismic isolation module model TGS system which dampens earthquake energy in both horizontal axes, and a custom load cradle which evenly distributes the instrument weight across the TGS system. Lastly, RUBLOC Trisolator pad systems are used to isolate the instrument from small amplitude, high frequency vibrations.

\subsubsection{Seismic Predictions and Isolation Scheme}


\begin{figure}
\epsscale{1.2}
\plotone{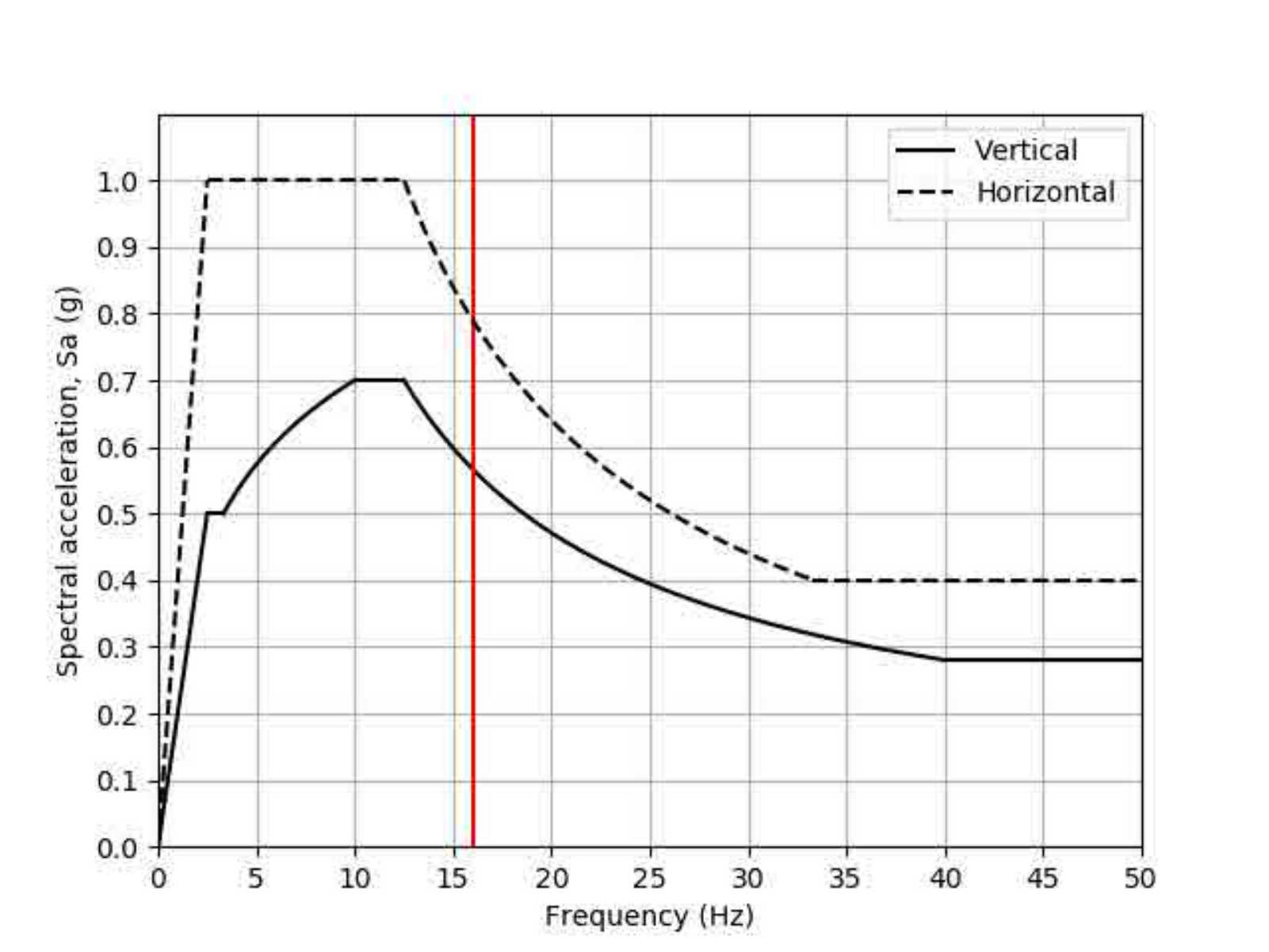}
\caption{The vertical and horizontal Survival Level Earthquake design spectra from \citet{urs11} modified to show acceleration v. frequency instead of acceleration v. period.  The vertical line at 16 Hz identifies the approximately $1^{\rm{st}}$ natural frequency of the instrument plus load cradle.}\label{fig_sle}
\end{figure}


In preparation for the Giant Magellan Telescope (GMT), a Site-Specific Seismic Hazard Assessment \citep{urs11} was prepared for Las Campanas Peak at LCO.  The assessment is described in the GMT System Level Preliminary Design Review documentation\footnote{\url{https://www.gmto.org/resources/slpdr}}.  Since the du Pont Telescope, located on the Manquis Ridge at LCO, is only $\sim 4\,\rm{km}$ away from Las Campanas Peak, we adopted the predicted Survival Level Earthquake (SLE) design spectra (Figure~\ref{fig_sle}) for Las Campanas Peak as the worst case seismic hazard that the APOGEE-South instrument would encounter.  The SLE is estimated to have a rate of occurrence of 100 -- 200 years.\footnote{Frank Kan, Simpson Gumpertz \& Heger, Inc., private communication}

By adopting the SLE design spectra, which applies at the ground level of the telescope, we neglected potential amplification of vertical accelerations during an earthquake by the structure supporting the first-floor room in which the APOGEE-South instrument is located.  Specifically, one side of the room floor is supported by the foundational walls of the dome whereas the other side is supported by columns.


\begin{deluxetable}{lccc}
\tabletypesize{\scriptsize}
\tablewidth{0pt}
\tablecaption{Optical Element Maximum Accelerations before Unseating \label{tbl-optics_accel}}
\tablehead{\colhead{Optic} & \colhead{Axial} & \colhead{Lateral} &
\colhead{Vertical}}

\startdata
Fold 1 & 3.0 & 1.3 & 1.9 \\
Fold 2 & 1.7 & 1.2 & 1.7 \\
VPH Correction Wedge & 5.8 & 3.6 & 4.4 \\
VPH & 1.4 & 1.1 & 1.6 \\
Camera L1 & 1.2 & 2.1 & 1.1 \\
Camera L2 & 0.7 & 2.0 & 0.9 \\
Camera L3 & 1.4 & 2.3 & 1.2 \\
Camera L4 & 1.4 & 2.4 & 1.3 \\
Camera L5 & 1.6 & 2.8 & 1.8 \\
Camera L6 & 2.4 & 4.1 & 3.0 \\
\enddata

\tablecomments{Calculations include the frictional forces resisting optic movement.  A coefficient of friction for Delrin\textsuperscript{\textregistered} on polished (fine ground) fused silica is assumed to be 0.175 (0.3).  Calculations include gravity.}

\end{deluxetable}


The predicted SLE design spectra were compared with the maximum accelerations (Table~\ref{tbl-optics_accel}) that could be accommodated by the restoring forces (e.g., spring-plunger pre-loads, frictional forces and element self-weight) inherent in each opto-mechanical system such that the optic stays seated.  The Collimating Mirror is excluded from Table~\ref{tbl-optics_accel} because the mirror and mount are connected through a central membrane flexure (see \S~\ref{collimating_mirror}) bolted to an Invar post that is epoxied to the back of the mirror.  The Collimating Mirror and mount system was analyzed with FEA using the SLE design spectra and judged to be in no danger of breakage from seismic accelerations in any direction.

Upon realization that all of the optics could accommodate the vertical accelerations expected in an SLE earthquake, but the Camera Lens 2 maximum axial acceleration would be exceeded, we concentrated on providing external isolation only in the horizontal direction.  Products designed for horizontal base isolation of electronics racks were well suited for this application, and we selected the TGS system (described below) given its modularity, customizability, and its ability to support the weight of the instrument.  Also, the use of linear motion guides in the TGS system would constrain movement to the horizontal direction so there would be minimal compliance in the vertical direction.  I.e., the top portion of the TGS system would not separate from the floor mounted bottom portion in case of vertical g-loads.  Lastly, the linear motion guides also support moment loads that can prevent tipping and handle high acceleration movement.

\subsubsection{Modified Hexapod Design}


\begin{figure}
\epsscale{1.2}
\plotone{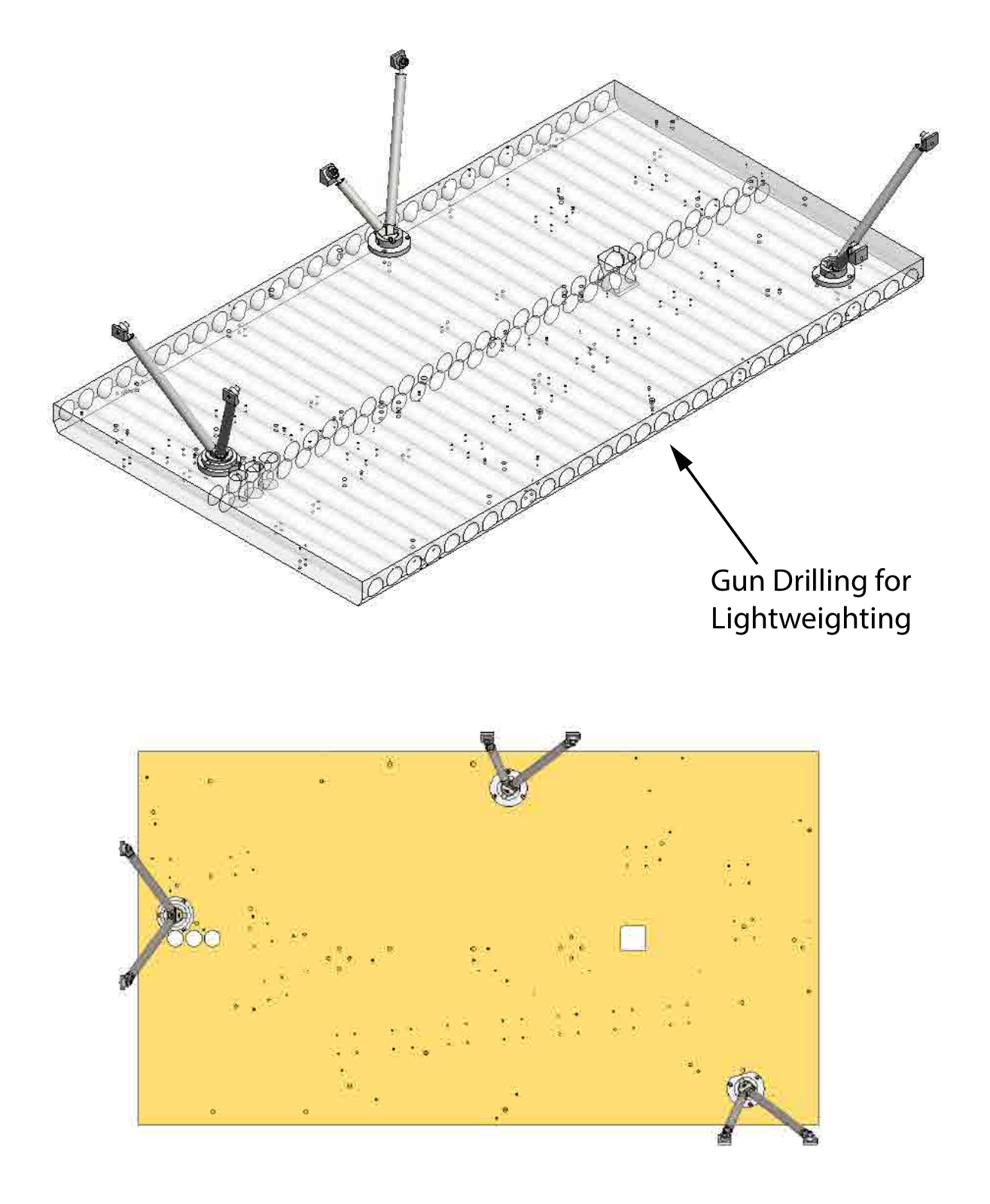}
\caption{The modified hexapod strut system used to support the APOGEE-South cold plate.  The tops of each strut assembly are welded to the inside cryostat wall. The cold plate model in the top image is transparent to highlight the lightweighting scheme of gun drilled blind holes along both plate edges with depths nearly to the plate center.}\label{fig_apogee-south_hexapod}
\end{figure}


The modified hexapod support structure (Figure~\ref{fig_apogee-south_hexapod}) was designed out of necessity to increase the natural frequency of the cold plate system to prevent resonance during an earthquake. This was achieved by stiffening the cold plate hangers and fixing their upper and lower ends to the cryostat inner wall and cold plate, respectively. The mount location of the strut pairs on the cold plate were chosen to distribute the suspended weight between the pairs as much as possible. Additionally, to increase the natural frequency of the system, the mount points were moved inwards from the edges of the cold plate. This strut orientation mitigates a ``wagging mode'' which is the first vibrational mode of the system.  This mode is manifested by a rotation of the cold plate, in the plane of the plate, about the strut mount point near the short end of the cold plate. FEA performed on the final design predicted a natural frequency of $17\,\rm{Hz}$.


\begin{figure}
\epsscale{0.8}
\plotone{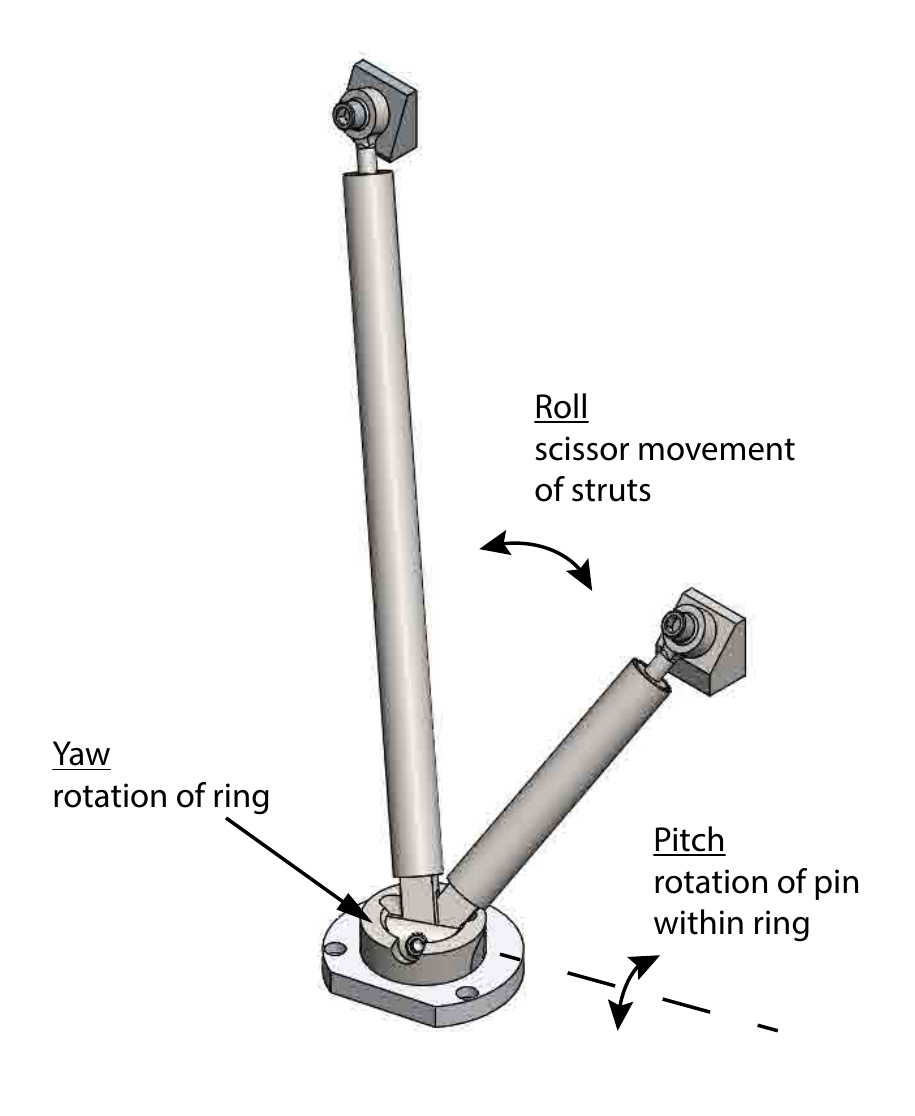}
\caption{A close-up of a one of the three strut pairs which make up the hexapod strut system.  The design permits movement in roll, pitch and yaw, as shown, to compensate for the differential shrinkage between the warm cryostat wall and the cold plate.  The blocks at the end of each strut, to which the tie-rod ends are bolted, are welded to the inside wall of the cryostat.  The bottom mount assembly is bolted to the top of the cold plate.}\label{fig_strut_pair}
\end{figure}


The struts are fabricated from 304 stainless steel tube with $1\,\rm{in}$ ($25.4\,\rm{mm}$) OD and $0.035\,\rm{in}$ ($0.9\,\rm{mm}$) wall thickness (Figure \ref{fig_strut_pair}). Two of the three strut pairs have different length struts. The four long strut tubes are $14.0\,\rm{in}$ ($355\,\rm{mm}$) long while the two short strut tubes are $5.81\,\rm{in}$ ($147.5\,\rm{mm}$) long. Welded inside the top end of all strut tubes are $\frac{3}{8}\,\rm{in}$ ($9.5\,\rm{mm}$) discs with holes tapped to receive Aurora Bearing SM-6E commercial tie rod ends.  Fabricated from 17-4 PH Stainless Steel and with 440C Stainless Steel ball bearings, the tie rod ends are bolted to weldments on the inside of the cryostat.  Welded to the bottom end of each strut are either single or double clevises.  The two clevises of the struts are pinned to a custom universal joint bolted to the top of the cold plate. The universal joint allows pitch and yaw, while the clevis ends of the struts allow roll. This freedom of motion provides the cold plate and strut system the necessary mechanical compliance to accommodate cooling between room and cryogenic temperatures.

\subsubsection{TGS System}


\begin{figure}
\epsscale{1.2}
\plotone{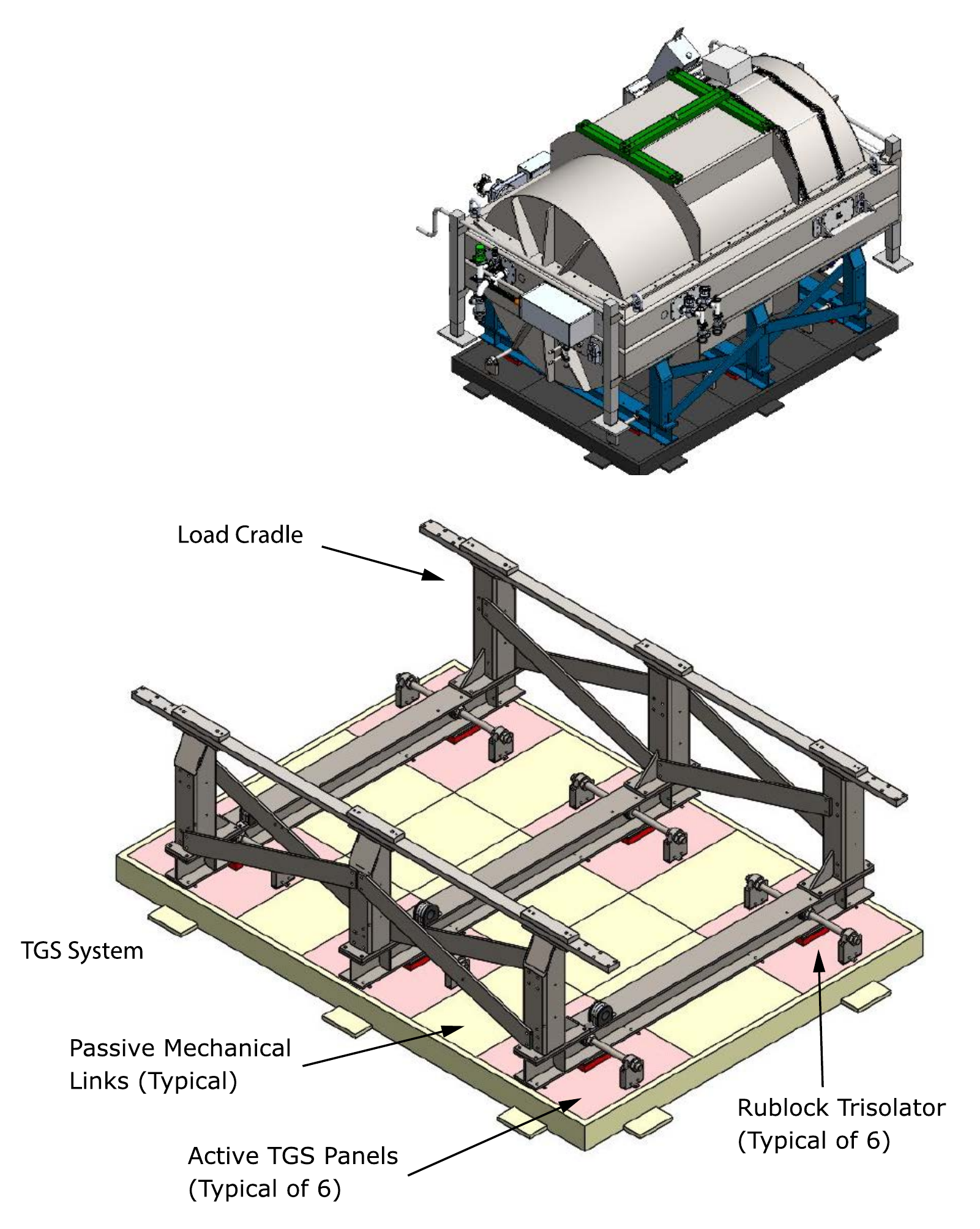}
\caption{The horizontal seismic isolation system for the APOGEE-South instrument.  A load cradle transfers the weight of the instrument to the six active panels of the TGS system.  Environmental vibration is damped by RUBLOC Trisolator vibration isolation pads that are placed between each load cradle foot and the corresponding active TGS system panel. }\label{fig_apogee-south_external_support}
\end{figure}


The THK seismic isolation module model TGS system consists of six interconnected active panels which carry the load of the cryostat (Figure~\ref{fig_apogee-south_external_support}).  Each active panel includes an upper and lower segment connected with THK linear motion guides, a pulley damping system, and tension springs which provide restoring forces.  The lower segment of each active panel is bolted to the instrument room floor while the upper segment carries a portion of the instrument weight and moves with damped motion in the horizontal plane to absorb earthquake energy.  The individual active panels are interconnected with passive mechanical links which bolt to the upper portions of the active panels to provide uniform motion of the entire system.  THK tuned the active panels based on cryostat loads and the presence of the RUBLOC vibration pads to respond to the seismic acceleration time histories\footnote{\citet{urs11}, kindly provided by the Giant Magellan Telescope Organization.} representative of an SLE earthquake.  THK analysis predicted the mean maximum horizontal displacement and acceleration at the base of the TGS system would be approximately $100\,\rm{mm}$ and $0.5\,\it{g}$, resp., based on the modeled system response to the time histories.  Similarly, the isolated APOGEE instrument was predicted to have a mean maximum horizontal displacement and acceleration of approximately $1.5\,\rm{mm}$ and $0.07\,\it{g}$.

\subsubsection{Load Cradle Design}\label{load_cradle}

A custom load cradle distributes the instrument weight equally among the active panels of the TGS system (Figure~\ref{fig_apogee-south_external_support}).  The base of the load cradle is fabricated from three W4 x 13 A36 steel I-beams running the width of the cryostat and resting directly on top of the RUBLOC Trisolators (described below).  At each end of these I-beams is a $1.25\,\rm{in}$ ($31.8\,\rm{mm}$) diameter hole drilled through the vertical web.  An ultra-high-molecular-weight (UHMW) polyethylene bolt and nut assembly is clamped at this hole.  These parts contain a through-hole in the middle to serve as clearance for a hard stop, comprised of a $\frac{7}{8}\,\rm{in}$ ($22.2\,\rm{mm}$) diameter threaded rod with nuts straddling the UHMW assembly. The threaded rods are then bolted to steel blocks which in turn bolt to the individual TGS system active panels. The spacing between the threaded rod and nuts and UHMW parts is $\frac{1}{8} \,\rm{in}$ ($3.1\,\rm{mm}$) to prevent transmission of vibrations to the cryostat through means that bypass the RUBLOC Trisolators. In the event of an earthquake, these hard stops prevent the cryostat from moving laterally off the TGS system.

Bolted on top of the I-beams are six uprights fabricated with $4\,\rm{in}$ x $4\,\rm{in}$ x $1/4\,\rm{in}$ ($100\,\rm{mm}$ x $100\,\rm{mm}$ x $6.4\,\rm{mm}$) square A36 steel tubes. Along the long edge of the cryostat, eight $3\,\rm{in}$ x $\frac{3}{8}\,\rm{in}$ ($76\,\rm{mm}$ x $9.5\,\rm{mm}$) A36 steel plates are bolted between the I-beams in an ``X'' arrangement for cross-bracing. Bolted atop the square tubes are two long 304 stainless steel beams that run the full length of the cryostat. These beams are bolted directly to the cryostat at the four corners. The center pads of these stainless steel beams are held in contact with the cryostat only through gravity.

Additional FEA was performed on the cryostat-load cradle system. The addition of the $620\,\rm{lb}$ ($281\,\rm{kg}$) load cradle reduced the natural frequency of the system from $17\,\rm{Hz}$ to $16\,\rm{Hz}$. This was deemed close enough to the cryostat's natural frequency of $17\,\rm{Hz}$ to be acceptable.

\subsubsection{Vibration Pads}

Installed between the load cradle and the TGS system are a set of six RUBLOC Trisolator\footnote{This product, used for vibration isolation for the HARPS instrument, was recommended by Francesco Pepe, private communication.  The pads were purchased from FMT Acoustics: \url{www.fmtacoustics.com}.} pad systems, model 60.40.60. These pad systems are comprised of three layers: the outer layers are hard rubber and the inner layer is a softer rubber for damping. The density of the inner layer can be matched to the natural frequency and weight of the load to be isolated. For APOGEE-South, the density of the middle layer was calculated to dampen $86\,\%$ of vibrations at $19\,\rm{Hz}$ with close to $100\,\%$ damping at $200\,\rm{Hz}$.

\subsection{\textbf{$\rm{LN_2}$} Tank Suspension}

Rather than mimic the APOGEE-North design and bolt the tank under the center of the cold plate, which inadvertently caused movement of the spectra due to the changing gravitational load of the tank as $\rm{LN_2}$ was depleted, the APOGEE-South tank is supported from the bottom of the cold plate edges opposite where the strut pairs connect on the top of the cold plate. This transfers the mass load to the location of mechanical support to minimize changes in the cold plate shape with $\rm{LN_2}$ level.  It also reduces the thermal paths between the room temperature upper strut mounting points and the locations where the $\rm{LN_2}$ tank connects below the cold plate.  Thus the cold plate system should behave like a capacitive thermal sink and reduce internal temperature changes as the room temperature fluctuates.  Some of the tank mechanical attachments include flexures to minimize changing stress on the cold plate with load. Since these attachments have less cross-sectional area compared to the tank support scheme of the APOGEE-North instrument, thermal coupling between the tank and cold plate is augmented with numerous copper cold straps.


\begin{figure}
\epsscale{1.1}
\plotone{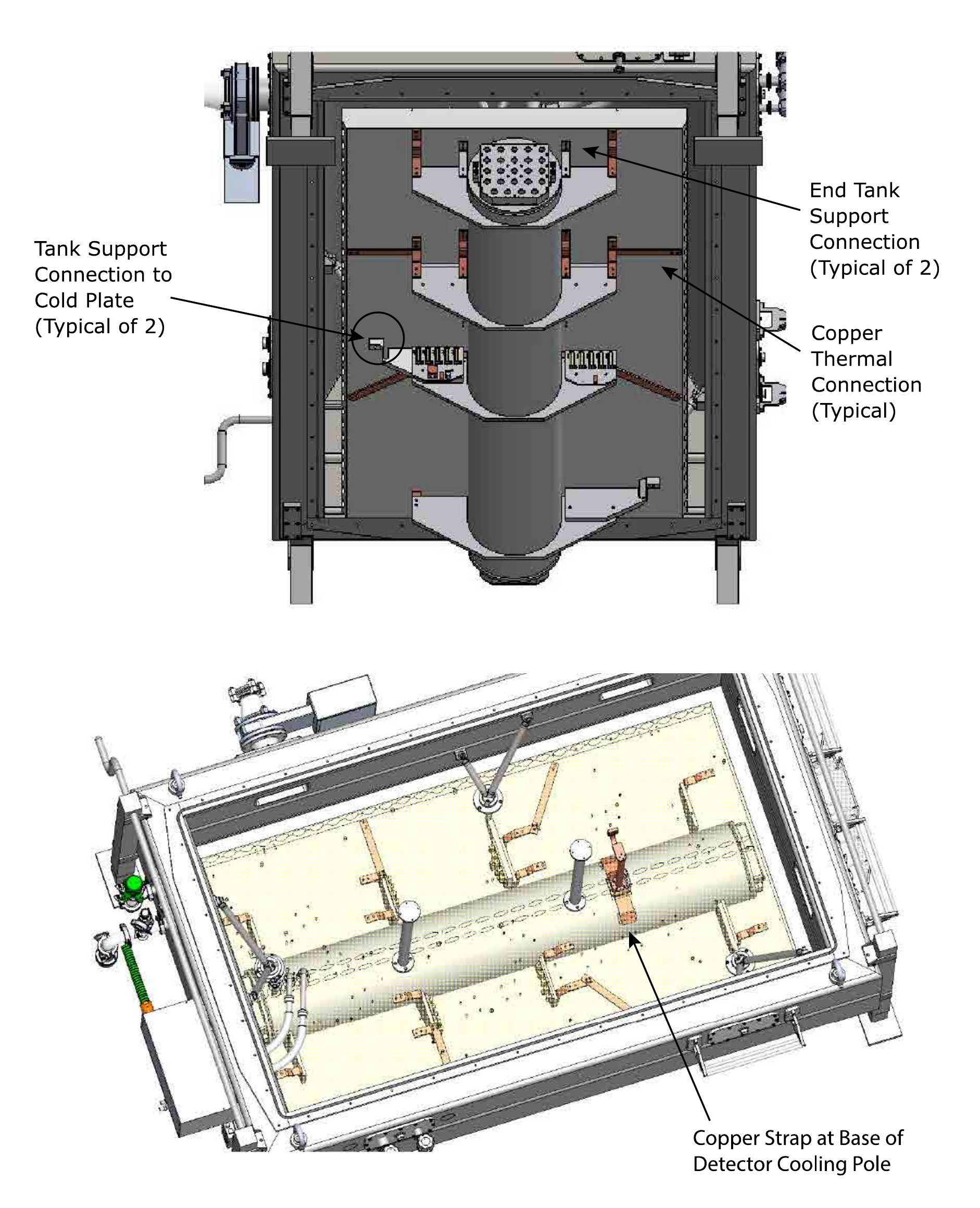}
\caption{(Top) A model rendering of the $\rm{LN_2}$ tank suspended below the cold plate.  Brackets on one end, along with two other flexure mounts, one of which is identified, support the tank.  Numerous copper straps provide additional thermal connections between the gussets and various positions below the cold plate.  (Bottom) A model rendering of the top of the cold plate.  The plate is rendered transparent to show how the strut positions correlate with the tank support scheme.  Also identified are the copper straps below the cold plate that connect to the bottom of the detector cooling pole and wrap a portion of the way around the tank to maintain cooling as the $\rm{LN_2}$ level drops.}\label{fig_apogee-south_tank_support}
\end{figure}


The $\rm{LN_2}$ tank has a series of four gusseted hangers welded to the bottom of the tank (Figure~\ref{fig_apogee-south_tank_support}). These hangers provide structural mounting points for the tank to the cold plate, provide attachment points for thermal straps, and hold the resistive heaters used for cryostat warm-up. Distributed over three of the hangers are four mounting fixtures that attach the tank to the cold plate. The combination of these four fixtures prevent the tank from imparting additional forces on the cold plate as it will be colder than the cold plate.

Two mounting plates are located off the tank axis, underneath the strut pairs on the long edges of the cold plate. These plates are fabricated from $\frac{3}{4}\,\rm{in}$ ($19\,\rm{mm}$) thick 6061-T6 aluminum and are $16.25\,\rm{in}$ ($412.8\,\rm{mm}$) off-axis from the tank. There are flexures cut into these plates which allow motion toward the center of the tank (axially and radially). The other two mounting fixtures are off-axis to the tank underneath the strut pairs on the short edge of the cold plate. These fixtures are fabricated from $\frac{1}{8}\,\rm{in}$ ($3.2\,\rm{mm}$) thick 6061-T6 aluminum and are $6\,\rm{in}$ ($152.4\,\rm{mm}$) off-axis from the tank. These mounts have a $90\,\degr$ bend allowing motion axial to the tank.

In addition to the four mounting points between the tank and the cold plate, there are twelve $\frac{1}{4}\,\rm{in}$ ($6.4\,\rm{mm}$) thick copper straps that terminate directly underneath the optic mounts. Additionally, the boss where the detector cold finger attaches to the top of the $\rm{LN_2}$ tank has a much larger copper strap that extends down both sides of the $\rm{LN_2}$ tank and connects below the $\rm{LN_2}$ level reached when 20 liters have been consumed (Figure~\ref{fig_apogee-south_tank_support}). This allows the cold finger to be in constant contact with the coldest part of the $\rm{LN_2}$ tank.

This redesign provided a marked improvement in the stability of spectral location on the detector arrays compared to the movement seen for the APOGEE-North instrument (\S\ref{rv_stability}).  In $14.7\,\rm{hrs}$, during which the tank level dropped $12\,\rm{\%}$, the median movement was 0.023 -- $0.025\,\rm{pixels}$ in the spatial (fiber-fiber) direction and 0.005 -- $0.009\,\rm{pixels}$ in the dispersion direction.  And at the expense of having cold plate and optics warmer by a few degrees, which has no operational impact, temperature uniformity on the cold plate and camera has improved dramatically.  The difference in mean temperatures between the middle and corner of the cold plate has reduced from $2\,\rm{K}$ to about $0.2\,\rm{K}$, a factor of ten reduction, based on the first eight months that the instrument was cold.  And the difference in mean temperatures between the front and back of the camera has reduced from $1\,\rm{K}$ to $0.03\,\rm{K}$.

From a thermal load standpoint, the APOGEE-South instrument consumes about $20\,\rm{liters}$ of $\rm{LN_2}$ per day compared to approximately $16.5\,\rm{liters}$ for the APOGEE-North instrument.   It may be that the shield and blanket system is not as efficient as expected.

\subsection{Cold Plate Lightweighting}

The lightweighting scheme for the APOGEE-South cold plate differs from the isogrid-style of the APOGEE-North instrument. It is more cost effective to remove material by gun-drilling through the sides of the cold plate along the neutral axis.  So a series of twenty-nine $2\,\rm{in}$ ($50.8\,\rm{mm}$) diameter holes were drilled from both long edges of the cold plate to a depth of $19.5\,\rm{in}$ ($495\,\rm{mm}$), as seen in Figure~\ref{fig_apogee-south_hexapod}. The final weight of the lightweighted cold plate was $510\,\rm{lb}$ ($231\,\rm{kg}$). Compared to the APOGEE-North cold plate weight of $376\,\rm{lb}$ ($170.6\,\rm{kg}$), less material was removed but the plate was significantly faster to fabricate, less design time was spent on the lightweighting scheme, and the cold plate had similar stiffness.

\subsection{VPH Design \& Correction}

\subsubsection{Modified VPH design}


\begin{figure}
\epsscale{1.0}
\plotone{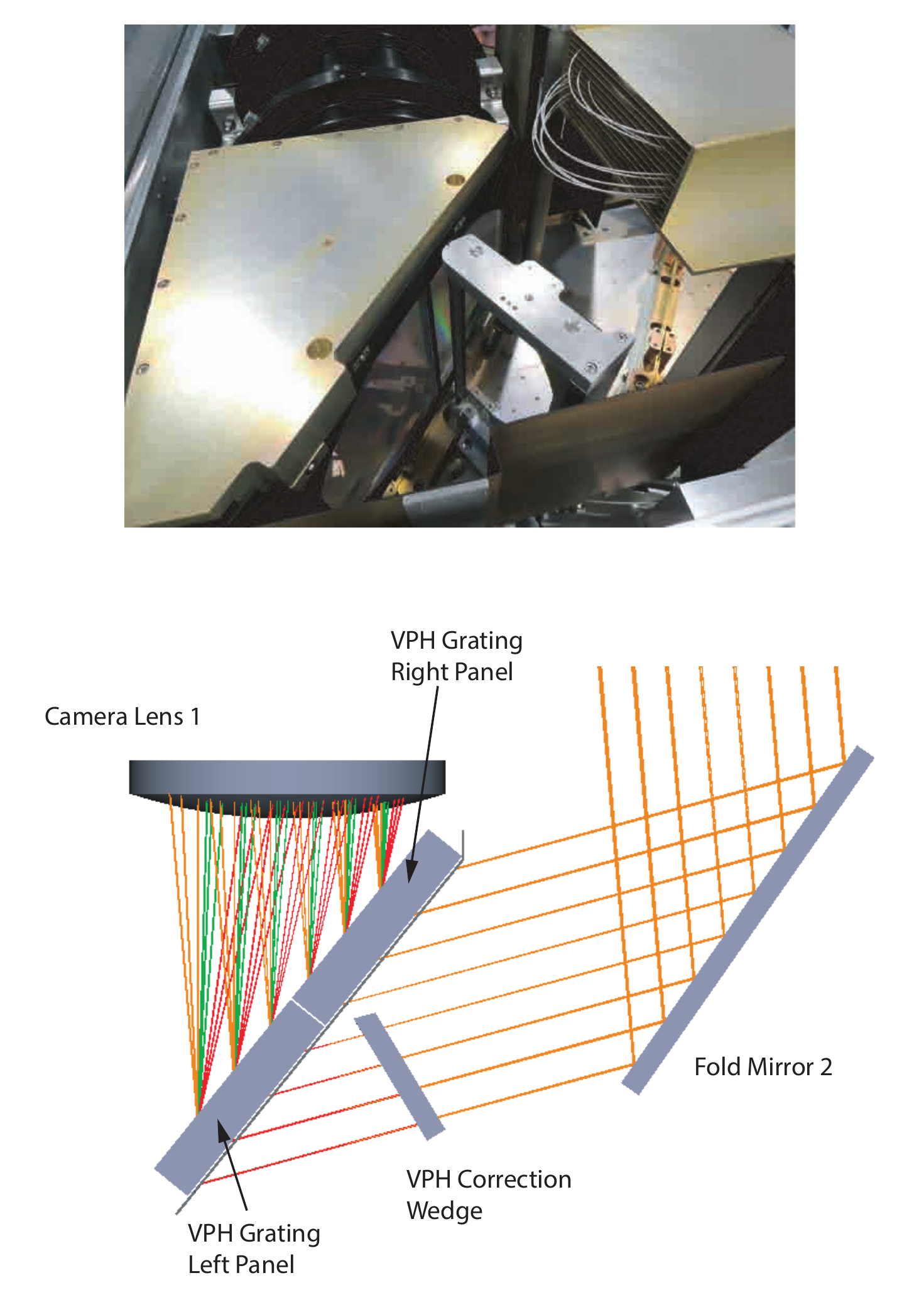}
\caption{(Top) The two-panel VPH Correction Wedge installed in front of the VPH Grating in the APOGEE-South instrument.  (Bottom) A ray trace showing the position of the wedge in front of the left-hand panel.  The wedge slightly reorients the light incident on the left panel to correct differential fringe density and differential clocking between the two panels.}\label{fig_vph_correction_wedge}
\end{figure}


As discussed in \S~\ref{segment_design}, the shadow of the pseudo-slit in the light path preceding the grating falls in the center of the small middle panel of the three-panel APOGEE-North VPH Grating.  A more efficient grating design would have a panel boundary in the center as well.  So a two-panel VPH Grating (Figure~\ref{fig_vph_correction_wedge}) was fabricated for the APOGEE-South instrument instead.  Theoretically, there is $< 0.5\,\%$ additional loss of clear aperture for the two-panel design compared to the three-panel design, neglecting the additional loss from the pseudo-slit obscuration which would affect the three-panel design more than the two-panel design.  Also, use of two-panels was expected to make grating fabrication easier.

\subsubsection{VPH Correction Wedge}\label{correction_wedge}

Unfortunately, the two recorded panels of the APOGEE-South VPH Grating had differential groove frequency and/or differential groove rotation which exceeded specifications.  In practice the two effects cannot be disentangled so hereafter we refer to both as differential groove frequency.  In accordance with the grating equation (equation~\ref{grating_equation}), the two panels with slightly different groove frequencies send light with the same wavelength, incidence angle, and out-of-plane angle in directions with slightly different exit angles.  Since the camera converts angular differences in collimated space into spatial location differences at the focal plane, differential groove frequency causes blurring of the PSF and subsequent loss of spectral resolution.  The specification for differential groove frequency was $0.0035\,{\rm lines\,mm^{-1}}$ \citep{arn10} which gave a spatial location difference at the detector of $\frac{1}{10}$ resolution element or $0.23\,\rm{pixels}$ for the nominal $2.3\,\rm{pixel}$ spot size (resolution element) at $1.6\,\micron$.

Independent measurements were made at JHU using the test setup described in \citet{bar14}.  Both interferometric and geometric (theodolite) measurement techniques gave exit angle differences of 2 -- $3.5\,\arcsec$ at $0.6328\,\micron$ (He-Ne laser) for the two recorded panels given a $\sim 18.6\,\degr$ angle of incidence, the Bragg angle for this wavelength.  These exit angles implied differential groove frequencies of 0.015 -- $0.025\,{\rm lines\,mm^{-1}}$, over $4 \times$ the specification.  A grating with these differential groove frequencies used at the APOGEE nominal wavelength of $1.6\,\micron$ and incidence angle of $\sim 54\,\degr$ would have exit angle differences of 8.4 -- $14.0\,\arcsec$.  With the nominal APOGEE camera focal length of $356\,\rm{mm}$, the spatial arrival location differences would be 14.8 -- $25.4\,\micron$ (0.8 -- $1.4\,\rm{pixels}$).

Fortunately, the panels, with their differing groove frequencies, happened to be placed within the mosaic such that the exit beams converge, and thus cross, as opposed to diverge.  This offered the opportunity to minimize blur by positioning the detector array mosaic assembly so the detective surfaces were near beam crossing locations across the wavelength span.

Testing at JHU also identified a small spatial location difference between arrival spots for individual grating panels which implied there was differential clocking between grooves of the two panels.

We addressed both differential groove frequency and differential clocking by inserting a wedge in the collimated beam preceding the left-hand panel of the VPH Grating (Figure~\ref{fig_vph_correction_wedge}).  The wedge slightly reorients the incidence angle of the light going into the left-hand panel, in-plane and out-of-plane, such that, after dispersion, the exit angle more closely matches that of the right-hand panel.  The wedge magnitude is $29.2\,\arcsec$ and the wedge direction is oriented $\sim 26\,\degr$ from the horizontal direction of the optic so $\sim 90\,\%$ of the bending action of the wedge contributes to changing the in-plane angle of incidence to correct differential groove frequency and the balance corrects the differential clocking.  The wedge was manufactured by Nu-Tek Precision Optical Corp. using a $1\,\rm{in}$ ($25.4\,\rm{mm}$) thick piece of Corning 7980 Fused Silica and it was AR coated by Newport Thin Film Labs.

For stray light purposes, the wedge is oriented at $16\,\degr$ relative to the light coming from the Fold 2 Mirror.  This angle was chosen so light that reflects off the detector arrays and backtracks through the Camera and VPH Grating would reflect off the back face of the wedge in a direction towards the side of the instrument and not retroreflect back into the VPH Grating. Given this wedge orientation, the inboard side of the wedge is chamfered at the same angle to minimize the obscuration of the beam.  Lastly, the clear aperture for polishing and coating purposes was defined to be as close to the inboard edge of the optic as practicable to minimize vignetting.

The final wedge magnitude and direction were determined based on empirical warm testing with a Raptor Photonics Ninox-640 InGaAs camera with $15\,\micron$ pixels positioned at the instrument focal plane.  For illumination, one end of a short fiber with standard SDSS plug plate ferrule terminations was positioned directly in front of an educational Argon arc lamp.  The other end was positioned at the dummy pseudo-slit (described in \S~\ref{align&test}).  During the warm testing we experimented with a $4\,\rm{in}$ ($101.6\,\rm{mm}$) diameter test wedge placed in front of the left-hand panel.  Also manufactured by Nu-Tek Precision Optical Corp., the test wedge magnitude was $40.85\,\arcsec$.   By clocking the wedge orientation, we demonstrated the ability to steer the exit beam of an individual panel and thus our general correction scheme.

This setup also enabled separate illumination of each panel, without use of the test corrective wedge.  With each panel stopped down with cardboard masks to an effective $4\,\rm{in}$ ($101.6\,\rm{mm}$) aperture, and blocking the unused panel, each panel was iteratively illuminated with combinations of three different fiber locations (i.e. top, middle and bottom on the dummy slit) and various bright Argon lines across the wavelength span.  For each combination the collimator was pistoned to produce a focus sweep.

The results were inconsistent between the various combinations so we conservatively chose to specify the final wedge amplitude and clocking based on results for the middle fiber and the $1.60\,\micron$ Argon line.   We reasoned that the results of this warm test would be most reliable for on-axis illumination of the instrument.  Using the Zemax predicted VPH Grating panel differential groove frequency which gave comparable on-axis results, this led to assuming a smaller differential groove frequency and thus smaller wedge angle for correction than was implied by the JHU test results.  In hindsight, the actual differential groove frequency is probably in the range of what was inferred from testing at JHU so we are likely under-correcting with the current wedge.

Differential groove frequency was also mitigated by iterating the collimator piston and the global position of the detector array mosaic assembly over several lab cooldowns so the crossing positions of the two panel exit beams coincided as close as possible with the detector array surfaces (thus minimizing FWHM) across the wavelength range.  To guide these iterative adjustments, a two-panel Hartmann door mechanism was developed for lab cooldowns.  Positioned between the Fold 2 Mirror and the VPH Grating, and driven by a Phytron stepper motor, the door could swing through a range of $\sim 180\,\degr$ thus allowing illumination of either panel (or both panels) during cold testing.  This capability enabled image centroid measurements as a function of wavelength for the two VPH Grating panels independently as a function of Collimating Mirror piston (focus).

In sum, both the corrective wedge, which reduced the difference between exit angles and caused the reimaged spots to be closer together, and iteration of the detector array mosaic assembly position and collimator piston to place the detective surfaces at the beam crossing locations across the wavelength span, mitigated the problem of differential groove frequency.

\subsection{Astigmatism Correction \& Imaging Performance}

\begin{figure*}
\epsscale{1.2}
\plotone{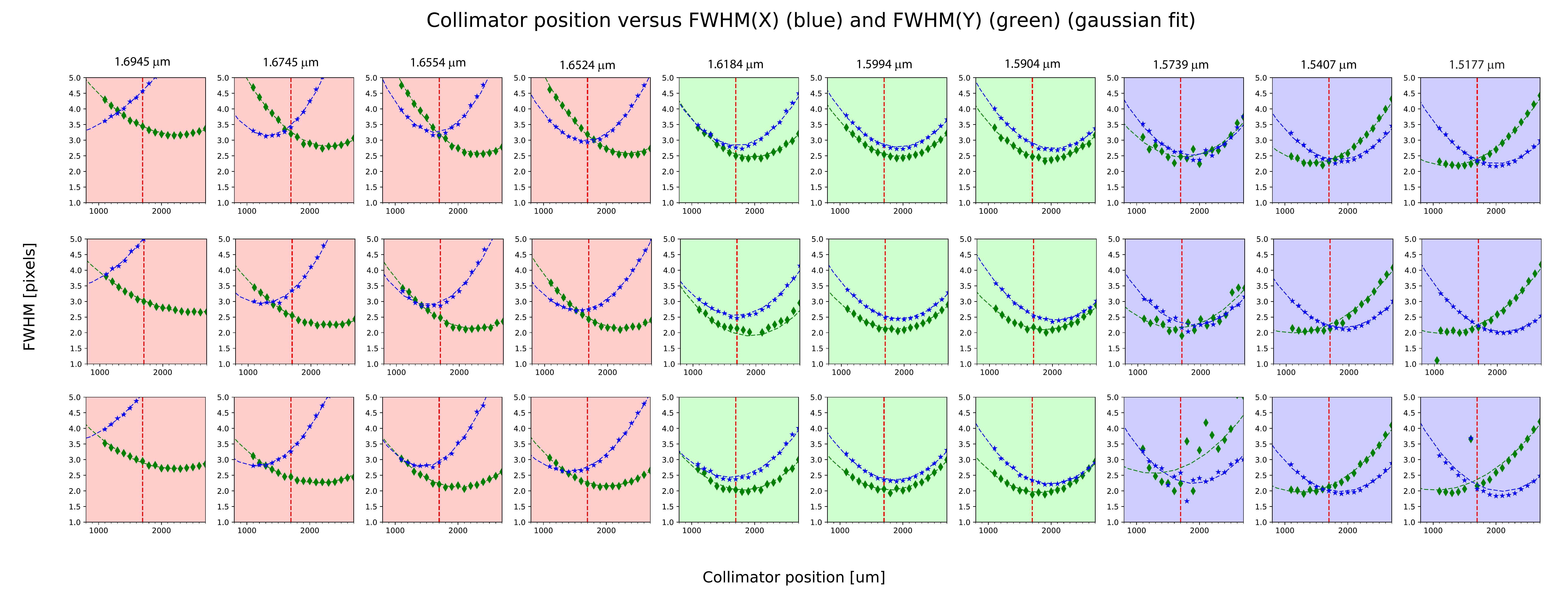}
\caption{Focus curves for the LSF (stars) and PSF (diamonds) for various UNe spectral lines spanning the wavelength coverage of the APOGEE-South instrument.  Taken during commissioning, the top, middle, and bottom rows are for fiber near the top, middle, and bottom of the detector arrays.  The instrument suffers from residual astigmatism at both the blue and red ends.  The vertical line in each panel gives the adopted collimator position of $1700\,\micron$.}\label{fig_apogee_south_focus_curves}
\end{figure*}

As discussed in \S~\ref{astigmatism_correction}, astigmatism was measured interferometrically in the left-hand panel of the APOGEE-South VPH Grating.  It was also observed in end-to-end test cooldowns of the second instrument, just like the first instrument.  And it was similarly mitigated by bending Fold Mirror 1 (\S~\ref{fold_mirror_1}).  Figure~\ref{fig_apogee_south_focus_curves} shows UNe focus curves for the APOGEE-South instrument taken during commissioning at LCO.  Each panel shows the FWHM of the LSF (stars) and PSF (diamonds) for a specific UNe spectral line as a function of collimator piston, along with Gaussian fits.  The spectral lines span the wavelength coverage.  The vertical line at $1700\,\micron$ gives the adopted collimator piston (focus position).  The top, middle and bottom rows correspond to fibers near the top, middle, and bottom of the detector arrays.  There is obvious residual astigmatism at both the blue and red ends of the wavelength coverage.  The choice of focus position ($1700\,\micron$) does not degrade the blue-end image quality too much.  In contrast, at the red-end the LSF is quite large which leads directly to diminished resolution.  In choosing a compromise focus position we were mindful that wavelengths redward of $1.68\,\micron$ were beyond the survey science requirements.


\begin{figure*}
\epsscale{1.0}
\plotone{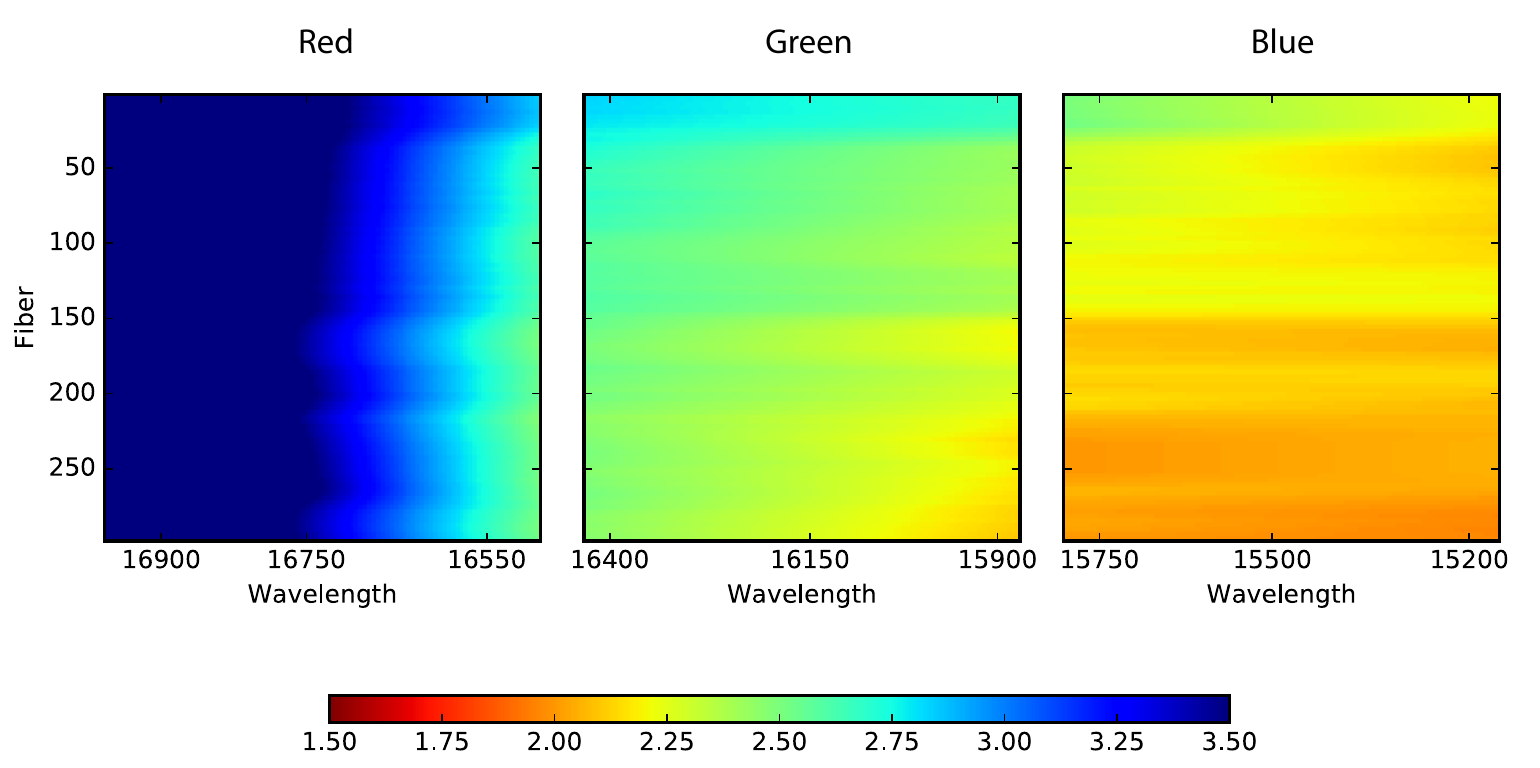}
\caption{FWHM maps for the three APOGEE-South detector arrays.  The FWHMs, color coded by size in pixels, are derived by fitting Gaussians to the LSFs of sky lines characterized using Gauss-Hermite polynomials, smoothed over a range of fibers and wavelengths.}\label{fig_apogee_south_fwhm}
\end{figure*}

\begin{figure*}
\epsscale{1.0}
\plotone{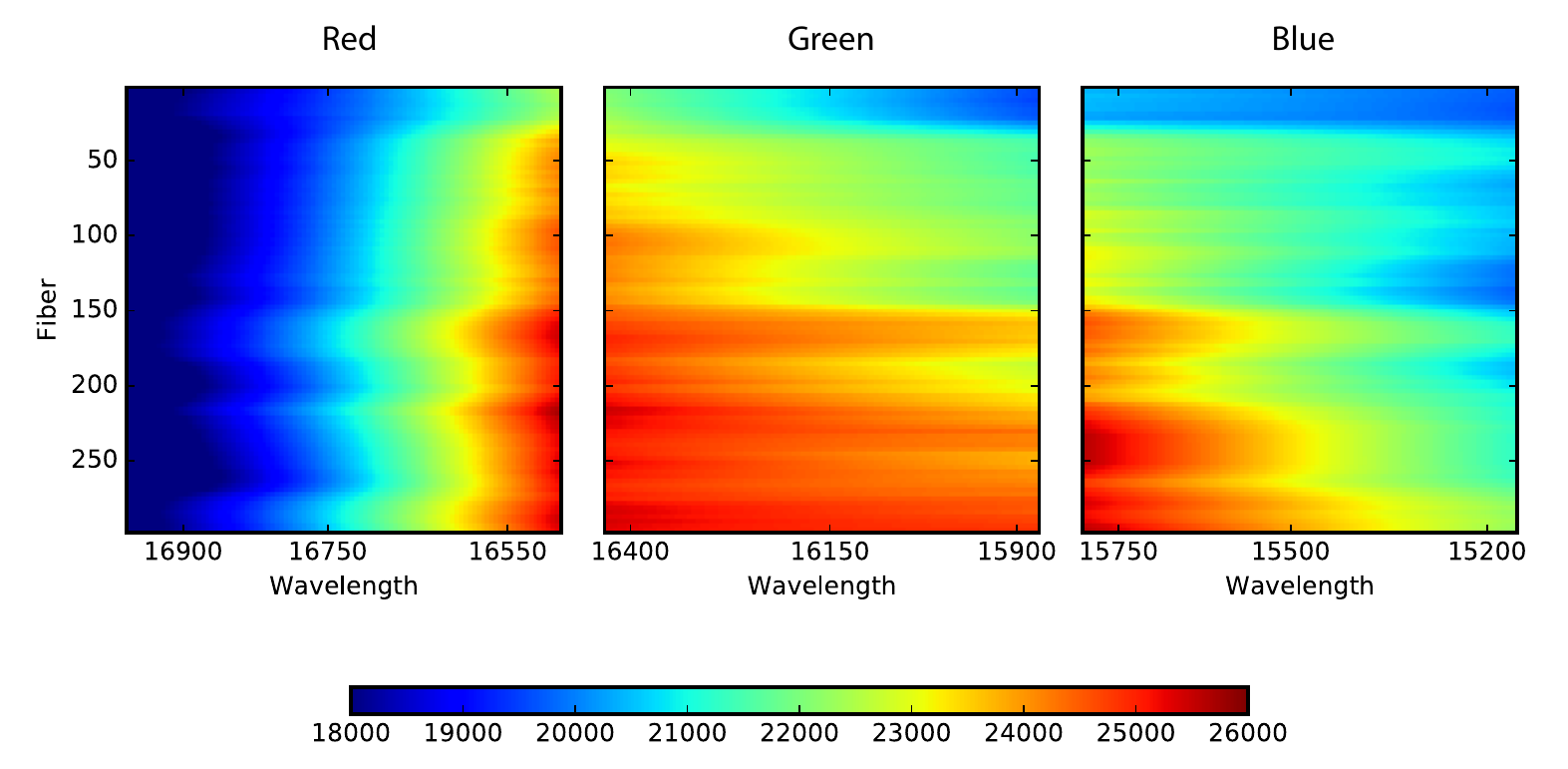}
\caption{{Resolution $R = \lambda/\rm{FWHM}$} mapped across the three APOGEE-South detector arrays using the FWHMs mapped in Figure~\ref{fig_apogee_south_fwhm}.}\label{fig_apogee_south_resolution}
\end{figure*}

Despite the likely undercorrected differential groove frequency, imperfectly compensated with global detector array mosaic assembly placement to minimize FWHM across the wavelength span, and residual astigmatism at both the blue and red ends of the wavelength coverage, the instrument PSF and LSF are sufficiently corrected for scientific purposes.  Figure~\ref{fig_apogee_south_fwhm} shows a map of spectral FWHM across the detector arrays for APOGEE-South, analogous to Figure~\ref{fig_apogee_north_fwhm} for APOGEE-North.  And  Figure~\ref{fig_apogee_south_resolution} shows a similar map of resolution for APOGEE-South, analogous to Figure~\ref{fig_apogee_north_resolution} for APOGEE-North.  Generally the FWHM are slightly larger and thus the delivered resolution is slightly lower, although APOGEE-South image quality is much more degraded at the red end compared to APOGEE-North.  Table~\ref{tbl-apogee_south_resolution_variation} gives the resolution variation for APOGEE-South.  The minima and maxima for each detector array for a given fiber sub-sample differ by $< 10\,\%$ (Blue \& Green) and $\la 20\,\%$ (Red) compared to the median.


\begin{deluxetable*}{lccccccccc}
\tabletypesize{\scriptsize}
\tablewidth{0pt}
\tablecaption{Resolution Variation \label{tbl-apogee_south_resolution_variation}}
\tablehead{
\colhead{Fiber} & \colhead{Red} & \colhead{} & \colhead{} & \colhead{Green} & \colhead{} & \colhead{} & \colhead{Blue} & \colhead{} & \colhead{} \\
\colhead{Sample} & \colhead{Median} & \colhead{Min\tablenotemark{a}} & \colhead{Max\tablenotemark{a}} & \colhead{Median} & \colhead{Min\tablenotemark{a}} & \colhead{Max\tablenotemark{a}} & \colhead{Median} & \colhead{Min\tablenotemark{a}} & \colhead{Max\tablenotemark{a}}
}
\startdata
Top & 19,800 & 17,600 ($-11\,\%$) & 22,400 ($+14\,\%$) & 20,900 & 19,700 ($-6\,\%$) & 22,300 ($+7\,\%$) & 20,100 & 19,700 ($-2\,\%$) & 20,300 ($+1\,\%$) \\
Middle & 20,700 & 17,500 ($-16\,\%$) & 25,000 ($+21\,\%$) & 23,600 & 23,000 ($-3\,\%$) & 24,400 ($+3\,\%$) & 22,200 & 20,500 ($-7\,\%$) & 24,000 ($+8\,\%$) \\
Bottom & 21,000 & 17,600 ($-16\,\%$) & 25,300 ($+21\,\%$) & 24,800 & 24,600 ($-1\,\%$) & 25,300 ($+2\,\%$) & 23,700 & 22,300 ($-6\,\%$) & 25,300 ($+7\,\%$) \\
\enddata

\tablenotetext{a}{Percentages give the deviation from the median for the detector array and fiber sample.}

\end{deluxetable*}


\subsection{Miscellaneous}

\subsubsection{Fold 2 Coating}\label{apogee_south_fold2}

In contrast to the Fold 2 Mirror in the APOGEE-North instrument, which has a dichroic coating on the front side and an AR coating on the back side to remove thermal light from the optical train (\S~\ref{fold_mirror_2}), the Fold 2 Mirror in the APOGEE-South instrument simply has a protected gold coating on the front surface.  This optic was coated by Infinite Optics.  This change occurred because of difficulty in applying the original coating pair without inducing delamination.

\subsubsection{Damaged Camera Lens 2}


\begin{figure}
\epsscale{0.90}
\plotone{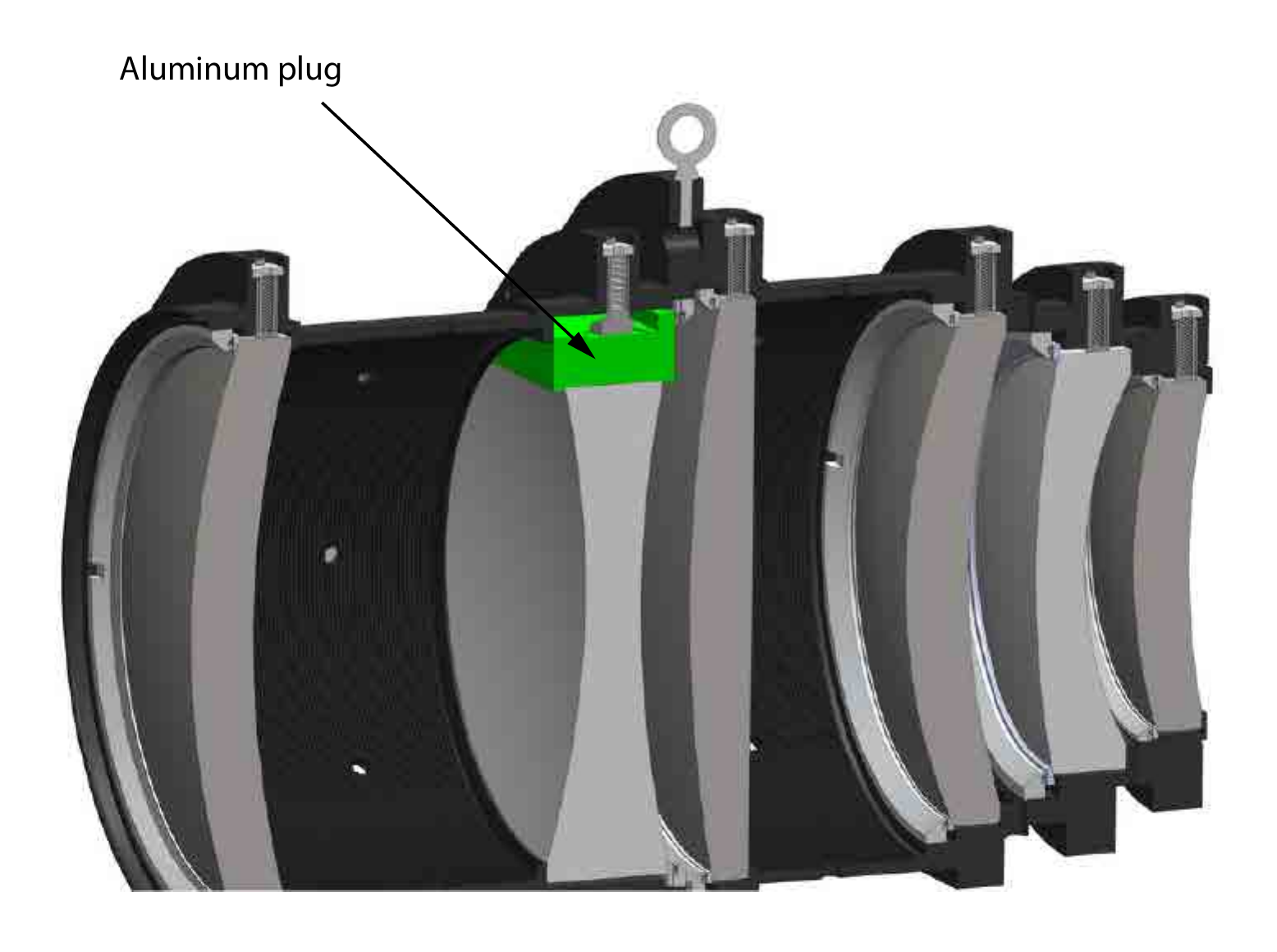}
\caption{Lens 2 of the APOGEE-South Camera was inadvertently chipped during assembly.  Fortunately, the chip was towards the edge of the lens so it could be rotated out of the field of view.  And then the material containing the chip was cut from the lens.  An aluminum plug fills the area where the material was removed.  No deleterious effects have been caused by this modification.}\label{fig_damaged_L2}
\end{figure}


Lens 2 was inadvertently chipped during camera assembly.  The chip ran to a depth of $\sim \frac{1}{4}\,\rm{in}$ ($6\,\rm{mm}$) which made stoning risky.  Fortunately, the illumination pattern on the lens is more rectangular than circular so the lens could be rotated so the chip was out of the field of view.  This, in turn, permitted the truncation of the lens (made from Corning 7980 Fused Silica) with a straight cut to remove the material containing the chip.  Figure~\ref{fig_damaged_L2} shows a model of the as-built Lens 2 within the camera.  An aluminum plug fills the area that was removed from the element.  No deleterious effects have been seen due to this modification.

\subsubsection{Internal LEDs}\label{apogee_south_leds}

As discussed in \S~\ref{led_operability}, over time the Hamamatsu LEDs used as internal flat-field light sources (\S~\ref{internal_leds}) in the APOGEE-North instrument became inoperable at cryogenic temperatures.  It is likely that the LEDs minimum stated operating temperature of $-30\,^{\circ}\,\rm{C}$, and the windows on the LEDs, contribute to the problems.  Given these issues, and the fact that the Hamamatsu L10823 LEDs were discontinued, the APOGEE-South instrument uses cryogenic-rated LEDs, without windows, available from Roithner LaserTechnik GmbH (Vienna, Austria).\footnote{These LEDs are used in the NIRCam detector array test cryostat at the University of Arizona (M. Rieke, private communication).}  Specifically, sets of LED17 (nominal peak emission of $1.73\,\micron$ at $300\,\rm{K}$) and LED18 (nominal peak emission of $1.85\,\micron$ at $300\,\rm{K}$) are installed.  The wavelengths of peak emission shift blueward upon cooling to cryogenic temperatures.

Given the experiences with the LEDs in the northern instrument, lab testing was conducted to ensure operability of these LEDs and continuity of the traces on the printed circuit board to which they are attached given cycling between room and cryogenic temperatures.  Also, care was taken to leave a $\sim 0.1\,\rm{in}$ ($\sim 2.5\,\rm{mm}$) gap between the bottom of the LED casing and the top of the circuit board to minimize the effects of thermal stress during cooling.

\subsubsection{Cold Shutter Inoperable}

Despite satisfactory cold testing at JHU, the cold shutter became stuck in a position that partially obscured the beam when it was first operated at cryogenic temperature within the APOGEE-South instrument in Chile.  This was the first time the cold shutter assembly had been operated cold within the instrument and with the Galil controller.  Upon increasing the maximum current available to the cold shutter Phytron stepper motor from 1.0 to $1.4\,\rm{A}$, the shutter was moveable again and it was fully opened.  We have since left the cold shutter open and unused.  When the instrument is opened again the mechanism will be inspected.  The motor current used during cryogenic testing at JHU was $1.2\,\rm{A}$ so it may simply be that the motor is underpowered when current is limited to $1.0\,\rm{A}$.

\subsection{Shipping}\label{apogee-south_shipping}

The instrument, with all optics removed, was transported by truck and ship between UVa and LCO since it was too large to transport via air freight.  The cryostat, along with most of the rest of the hardware for the instrument and lab, were first transported in a dedicated air-ride truck to the Carnegie Observatories in Pasadena, CA, where the contents were transferred to a dedicated $40\,\rm{ft}$ ocean-going container.  The container was then trucked to Long Beach, CA, and loaded onto an ocean-going freighter.  The voyage to Chile took $\sim 3\,\rm{weeks}$ and included the transfer of the container to another ship in L\'{a}zaro C\'{a}rdenas, Mexico.   Upon arrival in San Antonio, Chile, the container was trucked to LCO.

The discrete optics, with the exception of the camera, were removed from their mounts and shipped via typical air-freight to LCO.  These optics, namely the Fold 1 Mirror, Collimator, Fold 2 Mirror, VPH Grating Correction Wedge, and VPH Grating, were placed in dedicated Pelican Cases.  The Pelican Cases were placed in larger cardboard boxes with ample paper cushioning.  Lastly, the boxes were strapped to pallets.

\subsubsection{Instrument Shipping Pallet}


\begin{figure}
\epsscale{1.1}
\plotone{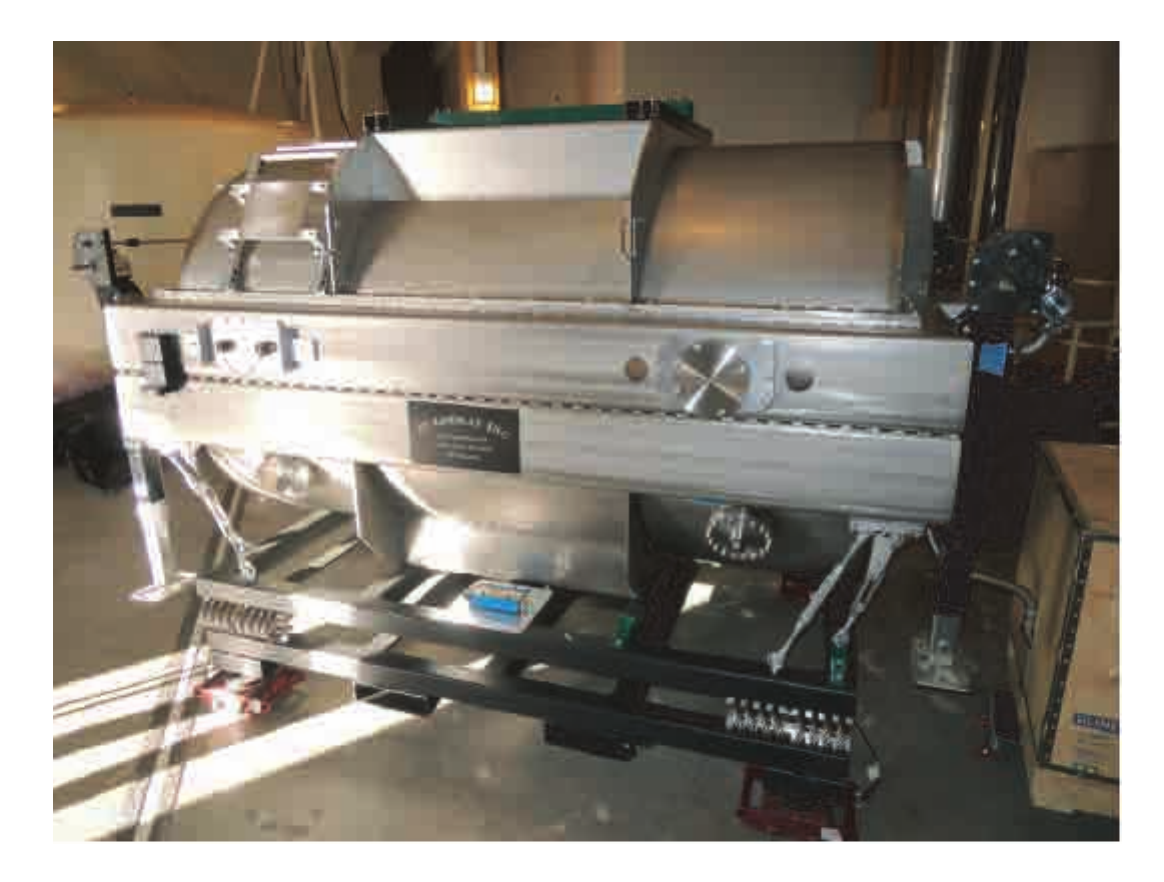}
\caption{The custom steel shipping pallet weldment fabricated for the APOGEE-South Instrument.  Wire spring isolators provide shock and vibration isolation between the lower and upper pallets.  In addition to bolts which attach the lower lid of the instrument to the upper pallet, eight turnbuckles help secure the instrument and resist shear forces.  The fiber spool crate is on the right.}\label{fig_shipping_pallet}
\end{figure}


Since we learned from the APOGEE-North instrument shipment that wood lacked sufficient strength to be used to make an instrument shipping pallet (\S~\ref{apogee-north_shipping}), a custom steel weldment was designed and fabricated for the APOGEE-South instrument (Figure~\ref{fig_shipping_pallet}).  The shipping pallet is $78\,\rm{in} \times  54\,\rm{in}$ and is fabricated from $3\,\rm{in} \times 3\,\rm{in} \times \frac{3}{16}\,\rm{in}$ A36 steel square tubing.  After fabrication, the weldment was powder coated for corrosion resistance.
Four wire rope isolators (Isolation Dynamics Corp. (IDC), M24-480-08), used in compression, isolate the upper pallet from the lower pallet to mitigate shocks and vibrations during transportation.  After consulting with IDC to purchase the isolators, they suggested the more stable and vibration resistant configuration was to use the isolators in a 45 degree compression/roll orientation.  Unfortunately, the shipping pallet was already being fabricated so it was not possible to re-design.  The isolators were chosen given the cryostat weight and anticipated maximum accelerations of $20\,\it{g}$ at frequencies up to $100\,\rm{Hz}$ based on shipping reports provided by Carnegie Observatory for previous instruments shipped to LCO using similar transportation methods.

The lower lid of the cryostat was bolted to the upper pallet with four $\frac{3}{8}$-16 bolts.  Additionally, eight turnbuckles connected the bottom side of each corner of the cryostat box beams to eye bolts installed on the corners of the upper pallet to help resist shear forces since the tapped holes in the lower cryostat lid were not designed to secure the cryostat.

After the cold portion of the fiber links were installed in the instrument, the warm portion, protected with flexible conduit, was wrapped onto a $2\,\rm{ft}$ diameter wooden spool that was placed within a crate.  Due to shipping container ceiling height constraints, the fiber spool crate could not be bolted to the top of the cryostat for the instrument transport.  So a cantilevered platform was designed with Unistrut that extended from the end of the upper pallet to ensure the fiber spool crate was never separated from the cryostat to prevent fiber damage.  However, placing the $\sim 150\,\rm{lb}$ fiber spool crate on a cantilevered platform which extended $47\,\rm{in}$ beyond the edge of the upper pallet would asymmetrically load the wire rope isolators, affecting their performance.  As a result, this cantilevered platform was only used while the instrument was being moved by forklift.  Once the cryostat was loaded into the shipping container, the platform was lowered to the floor and bolted to the unsprung pallet, thus removing the cantilevered load from the isolators and immobilizing the fiber spool crate.

Shock Log accelerometers were attached to both the upper (sprung) and lower (unsprung) portions of the shipping pallet.  The maximum acceleration recorded during shipment was $22.2\,g$ (vertical) with a pulse time of about $0.01\,\rm{sec}$ on the unsprung portion.  Based on the timing, it probably occurred during container movement in Long Beach, CA.  The corresponding acceleration on the sprung portion was $\sim 4\,g$, a shock mitigation of over $5 \times$.  The maximum temperature during shipment was $33.6\,\rm{deg}\,\rm{C}$ recorded by the Shock Log on the unsprung portion while at sea.

\subsubsection{Camera Shipment}

The camera was shipped with special care given its low threshold for mechanical shock.  Masterpiece International was contracted to provide door-door ``white-glove'' transportation service from Charlottesville, VA, to LCO.  A dedicated air-ride truck transported the camera, in a custom crate (described below), to JFK Airport in New York.  The camera crate was then loaded into a dedicated cargo container and transported on a non-stop flight to Santiago, Chile.  Lastly, a delivery van transported the camera from Santiago to LCO.


\begin{figure}
\epsscale{1.1}
\plotone{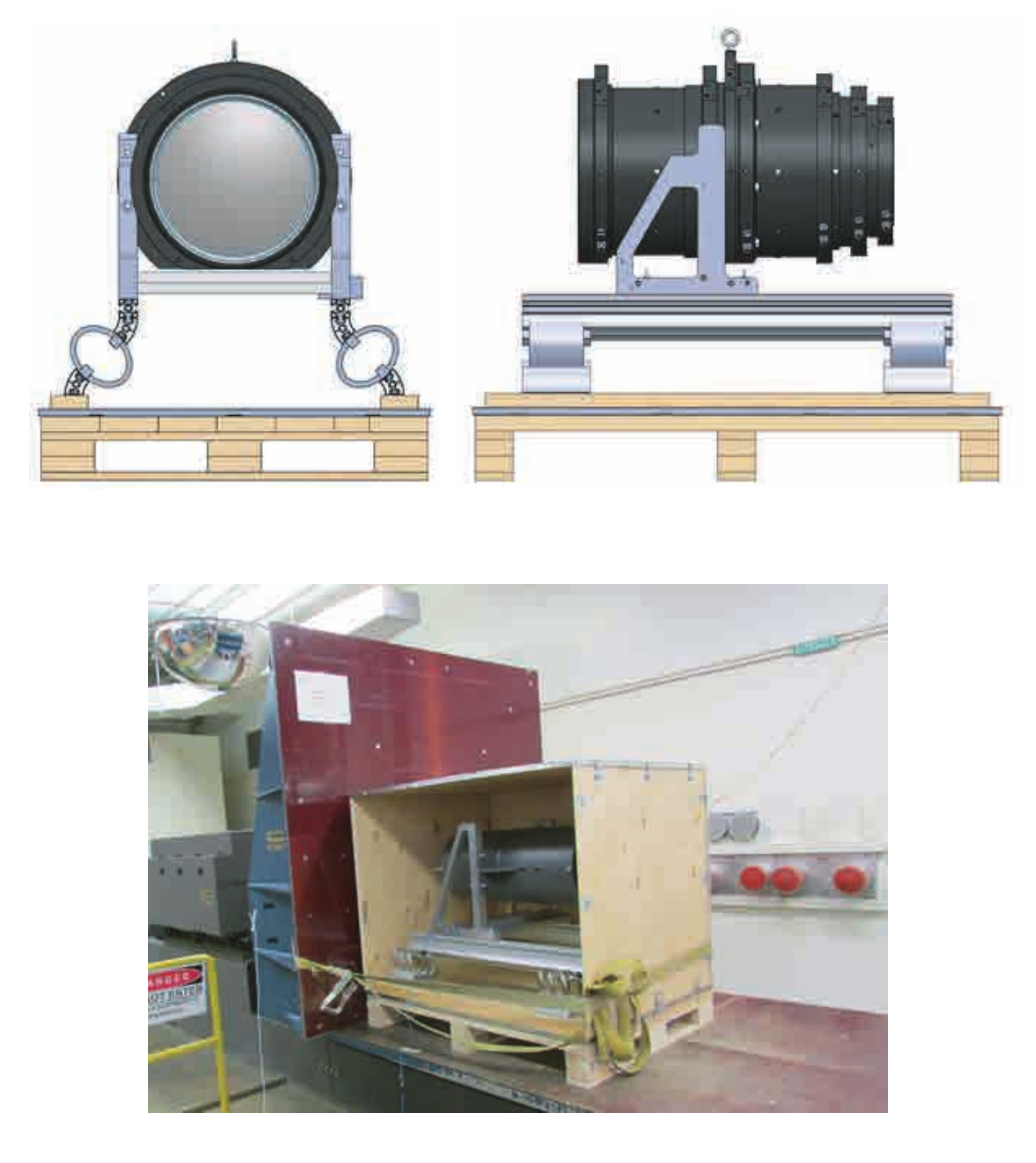}
\caption{(Top) A model of the camera secured to the vibration isolation system of its dedicated shipping crate.  (Bottom) A dummy mass model of the camera inside the shipping crate during shock and vibration testing.}\label{fig_camera_crate}
\end{figure}


Nefab Packaging Inc. designed the camera shipping crate (Figure~\ref{fig_camera_crate}) which included wire rope isolators to provide vibration and shock isolation for the camera.  Each leg of the camera was bolted to a $1\,\rm{m}$ long Bosch Rexroth square aluminum structural frame, oriented parallel to the camera axis. Bolted underneath and parallel was a Bosch Rexroth 45 degree angled aluminum structural frame oriented outward from the camera to provide a wider base.  This assembly was bolted to a set of four IDC SB20-500-03-B wire rope isolators.  Lastly, another set of Bosch Rexroth 45 degree angled structural frame oriented downward was bolted to the bottom of the isolators.  The 45 degree compression/roll orientation was recommended for maximum performance.  Nefab consulted with IDC, the manufacturer of the isolators, to select the correct isolators given the expected load and the planned transportation method via air ride truck. The system was designed to damp accelerations of $10\,\it{g}$ to $2\,\it{g}$ at the camera.

Given the low axial acceleration thresholds for the camera lenses before unseating (Table~\ref{tbl-optics_accel}), we had planned to replace the vertical spring plunger of each lens (Figure~\ref{fig_camera_cutaway}), which provide radial compliance, with dedicated solid pins.  These pins would constrain the elements vertically and significantly increase the axial acceleration threshold by increasing the static friction between the lenses and lower Delrin\textsuperscript{\textregistered} pins.  With solid pins we could comfortably accommodate a $2\,g$ axial force during shipment.  And $2\,g$ was about the lowest acceleration which could be practically provided through normal isolation methods.  So the crate vibration and isolation was designed assuming pins would be used.  But later we realized that the camera lenses would be vulnerable to breakage due to differential thermal expansion in case of excessive temperature increases during shipment.

We decided to go ahead with the shipment despite the likely unseating of the lens elements with the vertical spring plungers in place mainly because the elements, if unseated, could only move $\le 0.014\,\rm{mm}$ axially within the cell.  This small movement would reduce the chance of damage.

The Nefab ExPak crate, fabricated with $12\,\rm{mm}$ 9-ply hardwood plywood, had internal dimension of  $48 \frac{5}{8}\,\rm{in} \times 36 \frac{7}{8}\,\rm{in} \times 42 \frac{3}{8}\,\rm{in}$ ($1.16 \times 0.94 \times 0.42\,\rm{m}$) and provided ample clearance around the camera to prevent impact with the crate walls in case of compression/extension of the wire rope isolators during a shock.

A dummy mass with nearly identical mass, center of gravity and size as the camera was designed and fabricated so the crate and isolation system could undergo the International Safe Transit Association (ISTA) 3H testing sequence to check its isolation performance (Figure~\ref{fig_camera_crate}).  The ISTA 3H tests were performed at Arena Products (Rochester, NY).\footnote{For the ISTA 3H procedure, the required tests for certification were performed but optional tests were skipped.}  One panel of the crate was remvoed for testing to allow the full range of motion of the dummy mass to be observed to ensure the camera would not impact the walls of the crate.

The camera crate with dummy mass were preconditioned in the testing facility at $71$ deg F and $53.6\%$ relative humidity for a duration of 19 hours.  Before the horizontal impact test (sequence \#2), the sled was calibrated to produce a $10\,\it{g}$, $15\,\rm{msec}$ impact. Each side of the crate was positioned against the hard wall on the sled, and the impact was repeated twice per side for a total of eight impacts. Of the four impacts aligned with the camera axis, the average sled impact was $11.25\,\it{g}$ and the average camera response was $1.79\,\it{g}$ which includes one outlier at $2.52\,\it{g}$ (neglecting the outlier the average was $1.54\,\it{g}$).  IDC predicted the maximum axial acceleration to be approximately $1.6\,\it{g}$. Of the four impacts aligned lateral to the camera axis, the average sled impact was $11.09\,\it{g}$ and the average camera response was $2.11\,\it{g}$.  IDC predicted the maximum lateral acceleration to be approximately $2.0\,\it{g}$.

During the $240\,\rm{minute}$ random vibration test (sequence \#8) using an input power spectral density that simulated an air-ride truck, two resonances were detected.  One resonance was at $17.58\,\rm{Hz}$ along the vertical axis (normal to the base of the crate).  The other resonance was at $14.66\,\rm{Hz}$ along the camera axis.

At no point during the testing did the wire rope isolators bottom out, which according to IDC could lead to degradation in the performance of the isolators.  And since no damage occurred to the crate or isolators during testing, all components were reused for the actual shipment.  All bolts between the isolators and the framing were checked and re-tightened before shipping the camera.  A pair of Multifunction Extended Life (MEL-x16) 3-axis $16\,\it{g}$ accelerometer Data Loggers were acquired from Gulf Coast Data Concepts, LLC, for use during shipping.  One Data Logger was bolted to the wood floor of the crate and the other was bolted to the top of the Bosch Rexroth structural framing.  Unfortunately, the accelerometers were not acquired until after the ISTA 3H testing was performed so shipment performance could not be compared with the simulated testing.

After the camera was bolted to the Bosch Rexroth framing for shipping, the camera was wrapped in 4 mil plastic sheeting with six silica gel desiccant packs (sufficient to protect 30-42 cubic feet from moisture) and the plastic sheeting was sealed with packing tape to reduce moisture from reaching the camera.  The crate was sealed and a Uline Cool Shield foil laminated bubble pallet cover was placed over the top and sides of the crate.  The intent was to minimize internal temperature increases due to radiant heating while the crate was on airport tarmacs. The crate was then wrapped with plastic wrap to hold the thermal shield in place and provide some protection from rain.


\begin{deluxetable}{lcc}
\tabletypesize{\scriptsize}
\tablewidth{0pt}
\tablecaption{Camera Accelerations during Shipment \label{tbl_camera_shocks}}
\tablehead{\colhead{Shock Range ($g$)} & \colhead{Lens Affected} & \colhead{Number}}

\startdata
\cutinhead{Axial}
0.7 - 1.1 & L2 & 170 \\
1.1 - 1.3 & L1-L4 & 17 \\
1.5 - 2.2 & L1-L5 & 4 \\
$> 2.2$ & L1-L6 & 0 \\
\cutinhead{Lateral}
$> 2.0$ & any & 0 \\
\cutinhead{Vertical}
1.9 - 2.1 & L2 & 1 \\
2.1 - 2.2 & L1-L2 & 1 \\
2.2 - 2.3 & L1-L3 & 0 \\
2.3 - 2.7 & L1-L4 & 1 \\
2.7 - 4.0 & L1-L5 & 4 \\
$> 4.0$ & L1-L6 & 4 \\
\enddata

\end{deluxetable}


The camera arrived safely at LCO.  Shocks recorded by the accelerometer on the sprung portion are listed in Table~\ref{tbl_camera_shocks}.  The camera certainly encountered multiple accelerations that exceeded the limits (Table~\ref{tbl-optics_accel}) beyond which the individual lenses become unseated.  The maximum sprung accelerations recorded during shipment were $1.8\,g$ (axial), $1.0\,g$ (lateral), and $7.0\,g$ (vertical, downward).  The $7.0\,g$ event occurred at JFK airport, presumably during handling.  Unfortunately, timing between the sprung and unsprung accelerometers is not completely in sync.  Nonetheless, analysis of the data suggests the $7.0\,g$ event on the sprung portion was in response to a $4.0\,g$ event on the un-sprung portion of the crate, implying there was amplification of the shock.  The maximum temperature experienced inside the crate, $28\,\rm{deg}\,\rm{C}$, occurred $\sim10\,\rm{hours}$ after the crate arrived in Santiago, Chile.

\section{Conclusion}

The APOGEE spectrograph is a near-infrared, fiber-fed, multi-object, high resolution spectrograph built for the Apache Point Observatory Galactic Evolution Experiment. There are two instruments in operation, one at the Sloan Foundation Telescope at APO in New Mexico, and the other at the du Pont Telescope at Las Campanas Observatory in Chile.  Table~\ref{tbl-parameters} summarizes the instrument performance.  \citet{maj17}, \S~7.4, gives an extensive discussion of the breadth of potential science applications of the APOGEE data and representative research that has been published.




\acknowledgments

Funding for SDSS-III has been provided by the Alfred P. Sloan Foundation, the Participating Institutions, the National Science Foundation, and the U.S. Department of Energy Office of Science. The SDSS-III web site is \url{http://www.sdss3.org/}.

SDSS-III is managed by the Astrophysical Research Consortium for the Participating Institutions of the SDSS-III Collaboration including the University of Arizona, the Brazilian Participation Group, Brookhaven National Laboratory, University of Cambridge, Carnegie Mellon University, University of Florida, the French Participation Group, the German Participation Group, Harvard University, the Instituto de Astrofisica de Canarias, the Michigan State/Notre Dame/JINA Participation Group, Johns Hopkins University, Lawrence Berkeley National Laboratory, Max Planck Institute for Astrophysics, Max Planck Institute for Extraterrestrial Physics, New Mexico State University, New York University, Ohio State University, Pennsylvania State University, University of Portsmouth, Princeton University, the Spanish Participation Group, University of Tokyo, University of Utah, Vanderbilt University, University of Virginia, University of Washington, and Yale University.

Funding for the Sloan Digital Sky Survey IV has been provided by the Alfred P. Sloan Foundation, the U.S. Department of Energy Office of Science, and the Participating Institutions. SDSS-IV acknowledges support and resources from the Center for High-Performance Computing at
the University of Utah. The SDSS web site is \url{www.sdss.org}.

SDSS-IV is managed by the Astrophysical Research Consortium for the
Participating Institutions of the SDSS Collaboration including the
Brazilian Participation Group, the Carnegie Institution for Science,
Carnegie Mellon University, the Chilean Participation Group, the French Participation Group, Harvard-Smithsonian Center for Astrophysics,
Instituto de Astrof\'isica de Canarias, The Johns Hopkins University,
Kavli Institute for the Physics and Mathematics of the Universe (IPMU) /
University of Tokyo, Lawrence Berkeley National Laboratory,
Leibniz Institut f\"ur Astrophysik Potsdam (AIP),
Max-Planck-Institut f\"ur Astronomie (MPIA Heidelberg),
Max-Planck-Institut f\"ur Astrophysik (MPA Garching),
Max-Planck-Institut f\"ur Extraterrestrische Physik (MPE),
National Astronomical Observatories of China, New Mexico State University,
New York University, University of Notre Dame,
Observat\'ario Nacional / MCTI, The Ohio State University,
Pennsylvania State University, Shanghai Astronomical Observatory,
United Kingdom Participation Group,
Universidad Nacional Aut\'onoma de M\'exico, University of Arizona,
University of Colorado Boulder, University of Oxford, University of Portsmouth,
University of Utah, University of Virginia, University of Washington, University of Wisconsin,
Vanderbilt University, and Yale University.

The University of Virginia Astronomy Department gratefully acknowledges support from the Levinson/Peninsula Foundation and Ed Owens.  We are especially grateful to the JWST NIRCam Team for the loan of their detector arrays and the Stacey Group at Cornell University for the loan of a test cryostat.  We appreciate use of the ATRAN atmospheric model \citep{lor92} made available to the SOFIA program at this website (\url{https://atran.arc.nasa.gov/cgi-bin/atran/atran.cgi}).  We thank Frank Kan, of Simpson Gumpertz \& Heger, for helping us understand the seismic hazard assessments for Las Campanas Observatory.  \edit1{We thank the referee for very helpful comments and suggestions which improved the manuscript.}

We gratefully acknowledge the helpful support and wonderful spirit of Wendell Jordan, Electronics Technician at Apache Point Observatory.



\facilities{Sloan (APOGEE-North), Du Pont (APOGEE-South)}



\appendix


\section{Telescope and Fiber Alignment Errors}\label{appendix_errors}

Numerous errors can affect the amount of light accepted by individual fibers in the telescope focal plane during observations \citep[see, e.g.,][]{new02}.  The errors can be separated into two groups: defocus errors that increase the PSF and positioning errors that offset a fiber relative to the delivered image location.  Also, light loss can occur at fiber terminations and connections due to alignment and position errors and FRD can be induced by fiber tilt errors \citep[see, e.g.,][]{wyn93}.  We discuss contributors to each type for APOGEE-North below.

\subsection{Errors That Increase PSF}

The following defocus errors expand the telescope PSF imaged onto the fiber during an integration and were added in quadrature to the results shown in Figure~\ref{fig_image_quality} prior to calculating predicted encircled energy:

{\it Defocus from Imperfect Plate Curvature} -- The plug plate system is comprehensively discussed in \citet{sme13}.  Profilometry conducted on mounted plug plates in cartridges gives a mean defocus of $30\,\micron$ based on local differences between actual and ideal plate surface locations due to plate curvature imperfections.  The standard deviation of measurements for specific locations on the plate are $\la 100\,\micron$.  We adopt a $130\,\micron$ defocus error for this analysis.

{\it Defocus from Ferrule Shoulder} --  While the holes drilled into the plug plate are aligned with the telecentric angle, the telecentric angle deviates from the local plate normal from 0 -- $2\,\degr$.  This variation causes defocus since the ferrules have a bottom flange which sets the insertion distance into the plug plate hole.  Given a $4\,\rm{mm}$ flange diameter, a $2\,\degr$ tilt induces a $65\,\micron$ piston error (for simplifying encircled energy analysis, we assume the maximal tilt applies regardless of field location).


{\it Defocus from Tip Length Variability} --  The tip length of each ferrule can vary by up to $30\,\micron$ due to the polishing process.

\subsection{Errors That Offset Fibers}\label{appendix_fiber_offset_errors}

Fibers can be offset in the focal plane relative to the telescope image location for a particular object.  A good measure of aggregate offset errors is the measured radial errors calculated by the guider system. (See \citet{sme13} for a discussion of the updated guider system of the Sloan Foundation Telescope).  System performance during APOGEE-North observations suggests a typical 2D RMS radial error is $0.28\,\arcsec$ which is $17\,\micron$ at the focal plane.  The major contributions to the guiding error are the following:

{\it Plug Plate Hole Location Error} --  On average the hole radial position error from the drilling process is $9\,\micron$.

{\it Plug Plate Hole-Fiber Radial Position Errors} --  This $7\,\micron$ error is the convolution of the ferrule OD variability, plug plate hole ID variability, and hole-to-ferrule radial clearance.  The last component, a $6.5\,\micron$ error, dominates.  To prevent fibers from dropping out of the plug plate holes, the ferrules get cocked to one side of the holes by the spring force of the nylon jacketing surrounding the fiber below the plate.  As ferrules are always cocked in this manner, this error is added to the convolution of all other offset error contributors.

{\it Plug Plate Ferrule-Fiber Radial Position Errors} --  This $10.5\,\micron$ error is the convolution of fiber core-to-cladding concentricity, fiber buffer OD variability, ferrule ID variability, ferrule-to-fiber radial clearance, and ferrule OD/ID concentricity.  The last component, a $9\,\micron$ radial error, dominates.

{\it Object Astrometric Precision} -- A position error of $0.080\,\arcsec$ for both RA and Dec is adopted since most target coordinates came from 2MASS \citep{skr06} during SDSS-III.

{\it Object Proper Motion Errors} -- Most proper motions for APOGEE targets during SDSS-III came from the Naval Observatory Merged Astrometric Dataset (NOMAD) and Tycho2 catalogues for visual counterparts identified in 2MASS.  The average NOMAD proper motion error for DR12 cross-matched targets is 3.5 mas/year error for each of RA and Dec.  Assuming the typical DR12 target is observed $12.5\,\rm{years}$ after the epoch 2000.0, the proper motion error results in a positional error of $0.060\,\arcsec$.

There is also a differential refraction correction (quadrupole) error (discussed in \citet{new02} \S~3.1.8).  It varies from zero to $\sim 0.50\,\arcsec$ 2D RMS radial error, roughly increasing linearly with field position on the plate, and as the observation time deviates from the design HA for a given plate design.  Individual patterns of residual error across the plate will vary based on plate design details such as observing hour angle and declination.  But this error is neglected in the estimation of encircled energy since the purpose of the analysis is to compare estimated throughput with zero-points measured with the instrument.  As the latter represent the best case throughput which are apt to occur when observing close to the plate design HA, differential refraction correction error should be minimal.

\subsection{Errors That Tilt Fibers}\label{fiber_tilt_errors}

Various orientation and positioning errors inevitably lead to slightly tilted entrance and exit faces at fiber terminations and connections relative to the nominal light path direction.  These tilts lead to lost light and beam spread (FRD) with subsequent reductions in throughput.  Causes of these tilts within the APOGEE fiber train include:

{\it Ferrule Tilt at Plug Plate} -- The slightly oversized holes drilled in the plug plate that permit daily plugging and re-plugging inevitably lead to a ferrule tilt of $0.34\,\rm{deg}$ relative to the hole axis (ideally aligned with the incoming ray direction).  The tilt amount, controlled by the hole oversize and plate thickness, will be constant since the semi-rigid fiber assembly length, gently bent between the plug plate and anchor points within the cartridge, produces the frictional force that keeps the ferrule in place.  The azimuthal orientation of the tilted face can vary.  The tilt will lead to FRD \citep{wyn93}.

{\it Ferrule Tilt within Ferrule/MTP\textsuperscript{\textregistered}} -- To enable raw fiber insertion into the ferrule/MTP\textsuperscript{\textregistered} holes during manufacturing, the hole within the ferrule/MTP\textsuperscript{\textregistered} must be oversized relative to the fiber diameter.  Thus the fiber can be tilted relative to the ferrule/MTP\textsuperscript{\textregistered} hole with any azimuthal orientation and a magnitude of 0 -- $0.593\,\degr$ (ferrule) and 0 -- $0.32\,\degr$ (MTP\textsuperscript{\textregistered}), where the maxima are based on geometry.  Once the epoxy that secures the fiber within the ferrule/MTP\textsuperscript{\textregistered} has cured, the fiber tilt will not change.  This tilt was specified to be $\le 0.3\,\degr$ for manufacturing.

{\it Ferrule Tilt from Polishing} -- Ideally the direction normal to the polished face of the assembled ferrule/MTP\textsuperscript{\textregistered}/v-groove will be parallel to the fiber holes.  This angular difference was also specified to be $\le 0.3\,\degr$ for manufacturing.

{\it Hole Tilt Errors in Ferrule/MTP\textsuperscript{\textregistered}} -- A tilt error of the fiber hole axis, relative to the mechanical axis, in either the plug plate ferrule or MTP\textsuperscript{\textregistered}, can also be present.  In fact, systematic errors have been found in the molded fiber holes in the MTP\textsuperscript{\textregistered} connectors (Gunn et al., in prep.) by the team developing fiber assemblies for the Subaru Telescope Prime Focus Spectrograph.  They found that the holes in each of the four rows that make up the $4 \times 8$ array are systematically tilted relative to the mating face normal.  The tilts are correlated with row, or ``tier,'' such that all holes of the lowest tier are tilted up by $0.2\,\degr$.  The tilts uniformly increase for subsequent tiers.  The other tilts are 0.33, 0.47, and $0.6\,\degr$.  Thus the mean tilt is $0.4\,\degr$, and upon exit from the fiber the angle becomes $0.58\,\degr$ due to refraction.  Lastly, given the format of the connectors, when two are mated the tilt is doubled.

{\it V-groove Block Tilt Error} -- A tilt error of the v-groove block can directly lead to vignetting if the output beam is not properly aligned with the center of the instrument pupil.  As mentioned in \S~\ref{pseudo_slit}, the tilt of v-grooves were modified to compensate for light refraction at the face of the v-groove.


\section{APOGEE-North Commissioning Configuration}\label{appendix_commissioning_status}

As described in \citet{maj17}, data taken during instrument commissioning (2011 May -- 2011 July) are not of survey quality but have nonetheless been included in public data releases for completeness.  Commissioning data quality was degraded due to:

{\it Uncorrected Astigmatism} --  It was not until summer shutdown (2011 August), which followed commissioning, that the correction for astigmatism (see \S~\ref{fold_mirror_1} and \S~\ref{astigmatism_correction}) was implemented.

{\it Non-optimal Focus} -- The detector arrays were not in optimal focus.  In particular the red detector array was improperly positioned so resolution across red wavelengths were degraded to $\sim 15{,}000$.  During summer shutdown the red detector array tilt and position was refined at the University of Arizona.

{\it No Spectral Dithering} -- Because of imperfect focus the blue wavelengths were sufficiently sampled so commissioning observations were not spectrally dithered.

{\it No Cold Shutter} -- The cold shutter (\S~\ref{cold_shutter}), which mitigates inadvertent illumination of the super-persistent regions, was not installed until summer shutdown.

\clearpage




\end{document}